\documentclass[12pt,a4paper]{book}
\pdfoutput=1
		
\usepackage{framed}
\usepackage[pdftex]{graphicx}
\usepackage{capt-of}
\usepackage[usenames, dvipsnames]{xcolor} 

\usepackage[english]{babel}         
\usepackage[utf8]{inputenc}       
\usepackage[T1]{fontenc}            

\usepackage{mathrsfs}
\usepackage{empheq}
\usepackage{multicol}
\usepackage{commath}
\usepackage{amsmath,amssymb} 
\usepackage{mathtools} 
\usepackage{IEEEtrantools}

\usepackage[final]{pdfpages}
\usepackage{breakcites}
\usepackage{etoolbox}
\usepackage{relsize} 
\usepackage{accents} 
\usepackage{datetime} 
\usepackage{tcolorbox} 
\usepackage{eurosym}  
\usepackage{tikz} 
\usetikzlibrary{decorations}

\usepackage{comment} 
\usepackage{version}

\usepackage{import}

\usepackage{amsthm} 
\newtheorem{thmcnter}{thmcnter}[chapter]

\newtheorem*{Thrm*}{Theorem}
\newtheorem*{Prop*}{Proposition}

\newtheorem{Theorem}[thmcnter]{Theorem}
\newtheorem{Proposition}[thmcnter]{Proposition}
\newtheorem{Lemma}[thmcnter]{Lemma}

\theoremstyle{definition}
\newtheorem*{Def*}{Definition}
\newtheorem{Definition}[thmcnter]{Definition}

\newcommand{\T}{\mathbb{T}}
\newcommand{\PT}{\mathbb{PT}}	
\newcommand{\F}{\mathbb{F}}
\newcommand{\PF}{\mathbb{PF}}
\newcommand{\PTc}{\mathcal{PT}}
\newcommand{\pab}{\bar{\partial}}
\newcommand{\pah}{\hat{\partial}}
\newcommand{\p}{\pi}						
\newcommand{\ph}{\hat{\pi}}	
\newcommand{\pp}{\pi.\hat{\pi}}
\newcommand{\Zh}{\hat{Z}}
\newcommand{\J}{{\textit{J}}}

\newcommand{\iq}{{\mathrm{i}}}
\newcommand{\jq}{{\mathrm{j}}}
\newcommand{\kq}{{\mathrm{k}}}

\newcommand{\jqb}{{\bar{\mathrm{j}}}}

\newcommand{\IObb}{{\rm{Im}\mathbb{O}}}

\newcommand{\M}{\rm{M}}
\newcommand{\Mc}{\mathcal{M}}
\renewcommand{\P}{\rm{P}}

\usepackage{colortbl}

\newcommand{\RC}{\color{Red}}

\usepackage{accents} 
\usepackage{bm}		 

\newcommand{\utilde}[1]{\underaccent{\tilde}{#1}}

\newcommand{\ga}{\alpha} 						
\newcommand{\gab}{\overline{\alpha}}
\newcommand{\gah}{\hat{\alpha}}
\newcommand{\gat}{\widetilde{\alpha}}
\newcommand{\gb}{\beta}							
\newcommand{\gbt}{\widetilde{\beta}}

\newcommand{\gbh}{\hat{\beta}}
\newcommand{\gc}{\gamma}						

\newcommand{\gct}{\tilde{\gamma}}
\newcommand{\gG}{\Gamma}

\newcommand{\gd}{\delta}						

\newcommand{\bdgd}{\bm{\delta}}

\newcommand{\eps}{\epsilon}						

\newcommand{\epst}{\widetilde{\epsilon}}
\newcommand{\epsut}{\utilde{\epsilon}}
\newcommand{\gz}{\zeta}							
\newcommand{\gzb}{\overline{\zeta}}
\newcommand{\tht}{\theta}						

\newcommand{\gi}{\iota}							
\newcommand{\gl}{\lambda}						
\newcommand{\glb}{\overline{\lambda}}

\newcommand{\gL}{\Lambda}
\newcommand{\gLt}{\widetilde{\Lambda}}
\newcommand{\gx}{\xi}							
\newcommand{\oh}{\hat{o}}						

\newcommand{\gpb}{\bar{\pi}}
							
\newcommand{\gr}{\rho}							
\newcommand{\grh}{\hat{\rho}}
\newcommand{\grt}{\widetilde{\rho}}
\newcommand{\gs}{\sigma}						
\newcommand{\gsb}{\bar{\sigma}}
\newcommand{\gst}{\widetilde{\sigma}}
\newcommand{\gS}{\Sigma}

\newcommand{\gSt}{\widetilde{\Sigma}}
\newcommand{\bdgS}{\bm{\Sigma}}
\newcommand{\bdgSt}{\bm{\widetilde{\Sigma}}}
\newcommand{\gt}{\tau}							

\newcommand{\gtb}{\bar{\tau}}

\newcommand{\Psit}{\widetilde{\Psi}}
\newcommand{\go}{\omega}						

\newcommand{\goh}{\hat{\omega}}
\newcommand{\gob}{\bar{\omega}}
\newcommand{\bdgo}{{\bm{\omega}}}
\newcommand{\bdgot}{{\widetilde{\rm{\omega}}}}
\newcommand{\gO}{\Omega}

\newcommand{\gOh}{\hat{\Omega}}
\newcommand{\gOb}{\bar{\Omega}}
\newcommand{\bdgO}{{\bm{\Omega}}}

\newcommand{\ab}{\overline{a}}					

\newcommand{\At}{\widetilde{A}}

\newcommand{\Ac}{\mathcal{A}}

\newcommand{\bda}{{\bm{a}}}
\newcommand{\bdat}{{\bm{\widetilde{a}}}}
\newcommand{\bdA}{{\bm{A}}}
\newcommand{\bdAt}{{\bm{\widetilde{A}}}}

\newcommand{\bdb}{{\bm{b}}}
\newcommand{\bdB}{{\bm{B}}}

\newcommand{\Ch}{\hat{C}}
\newcommand{\Cc}{\mathcal{C}}
\newcommand{\Db}{\overline{D}}					

\newcommand{\Dt}{\widetilde{D}}

\newcommand{\Eb}{\overline{E}}
\newcommand{\Ebb}{\mathbb{E}}
\newcommand{\bde}{{\bm{e}}}
\newcommand{\bdet}{{\bm{\widetilde{e}}}}
\newcommand{\bdE}{{\bm{E}}}

\newcommand{\ft}{\widetilde{f}}					
\newcommand{\bdf}{{\bm{f}}}	
\newcommand{\bdft}{{\bm{\widetilde{f}}}}	
\newcommand{\bdF}{{\bm{F}}}	
\newcommand{\Ft}{\widetilde{F}}	
\newcommand{\bdFt}{{\bm{\widetilde{F}}}}

\newcommand{\frg}{\mathfrak{g}}					
\newcommand{\gti}{\tilde{g}}

\newcommand{\Gt}{\widetilde{G}}			
\newcommand{\hb}{\overline{h}}					

\newcommand{\Hbb}{\mathbb{H}}
\newcommand{\bdH}{{\bm{H}}}
\newcommand{\rmi}{\textrm{i}}

\newcommand{\kt}{\widetilde{k}}
\newcommand{\Kt}{\widetilde{K}}

\newcommand{\bdm}{{\bm{m}}}						
\newcommand{\Mbb}{\mathbb{M}}
		
\renewcommand{\Mc}{\mathcal{M}}
\newcommand{\Mt}{\widetilde{M}}
\newcommand{\Nbb}{\mathbb{N}}
\newcommand{\Oc}{\mathcal{O}}					

\newcommand{\Obb}{\mathbb{O}}
\newcommand{\pb}{\bar{p}}						
\newcommand{\qb}{\bar{q}}						
	
\newcommand{\Tc}{\mathcal{T}}					
\newcommand{\Vc}{\mathcal{V}}					

\newcommand{\bdw}{{\bm{w}}}						
\newcommand{\bdW}{{\bm{W}}}	

\newcommand{\Wt}{\widetilde{W}}
\newcommand{\xb}{\bar{x}}						
\newcommand{\xh}{\hat{x}}					
\newcommand{\Xt}{\widetilde{X}}
\newcommand{\Xb}{\overline{X}}					

\newcommand{\Yb}{\overline{Y}}

\newcommand{\zh}{\hat{z}}

\newcommand{\R}{\mathbb{R}}
\newcommand{\C}{\mathbb{C}}
\newcommand{\Z}{\mathbb{Z}}
\renewcommand{\S}{\bm{S}}

\newcommand{\su}{\mathfrak{su}}
\newcommand{\so}{\mathfrak{so}}
\renewcommand{\sl}{\mathfrak{sl}}
\newcommand{\GL}{{\rm{GL}}}
\newcommand{\SL}{{\rm{SL}}}
\newcommand{\PSL}{{\rm{PSL}}}
\newcommand{\SLt}{{\widetilde{\rm{SL}}}}
\newcommand{\SO}{{\rm{SO}}}
\newcommand{\SU}{{\rm{SU}}}

\newcommand{\U}{{\rm{U}}}
\newcommand{\Sp}{{\rm{Sp}}}
\newcommand{\Spin}{{\rm{Spin}}}

\newcommand{\CP}{\mathbb{C}\bm{P}}
\newcommand{\HP}{\mathbb{H}\bm{P}}
\newcommand{\OP}{\mathbb{O}\bm{P}}

\newcommand{\Tr}{\textrm{Tr}}

\newcommand{\from}{\colon}
\newcommand{\xto}[2][]{\xrightarrow[#1]{#2}}
\newcommand{\inj}{\hookrightarrow}

\newcommand{\circonf }{\text{\textasciicircum}}
\newcommand{\N}{\nabla}
\newcommand{\pa}{\partial}
\newcommand{\W}{\wedge}
\newcommand{\id}{\intprod} 
\newcommand{\Ld}{\mathcal{L}} 
\newcommand{\Id}{\mathbb{Id}}

\newcommand{\bra}{\left\langle}
\newcommand{\ket}{\right\rangle}

\newcommand{\Mtx}[1]{\begin{pmatrix} #1 \end{pmatrix}}
\newcommand{\sMtx}[1]{\begin{bmatrix} #1 \end{bmatrix}}

\newcommand*{\TakeFourierOrnament}[1]{{
		\fontencoding{U}\fontfamily{futs}\selectfont\char#1}}
\newcommand*{\danger}{\TakeFourierOrnament{66}}

\DeclareFontFamily{U}{MnSymbolC}{}				
\DeclareSymbolFont{MnSyC}{U}{MnSymbolC}{m}{n}
\DeclareFontShape{U}{MnSymbolC}{m}{n}{
	<-6>  MnSymbolC5
	<6-7>  MnSymbolC6
	<7-8>  MnSymbolC7
	<8-9>  MnSymbolC8
	<9-10> MnSymbolC9
	<10-12> MnSymbolC10
	<12->   MnSymbolC12}{}
\DeclareMathSymbol{\intprod}{\mathbin}{MnSyC}{'270}

\usepackage{fancyhdr} 
\usepackage{indentfirst} 
\usepackage{layout}
\usepackage{changepage}

\usepackage{setspace}
		
\usepackage[Bjornstrup]{fncychap}
	
\usepackage{chngcntr}
\numberwithin{equation}{section}

\usepackage{tocloft}
\usepackage{minitoc}

\mtcsetfeature{parttoc}{open}{\vspace{-0.5cm} \hrule }
\mtcsetfeature{parttoc}{close}{\vspace{0.5cm} \hrule }

\mtcsettitle{parttoc}{Contents of Part \thepart}
\mtcsetrules{parttoc}{off}

\renewcommand\cftpartpresnum{Part~}
\renewcommand\thepart{\arabic{part}}

\definecolor{chaptercolor}{rgb}{0.4,0,0}
\newcounter{chapcntr}
\newcommand{\chaptercolor}{%
	\ifcase\value{chapcntr}%
	\color{black}			
\or \color{black}			
\or	\color{chaptercolor}	
\or	\color{chaptercolor}	
\or \color{black}			
\or	\color{chaptercolor}	
\or	\color{chaptercolor}	
\or \color{chaptercolor}	
\or \color{black}			
\or \color{chaptercolor}	
\or \color{chaptercolor}	
\or	\color{black} 			
\or	\color{black} 			
\or	\color{black} 			
\or	\color{black} 			
\or \color{black}			
\else \color{chaptercolor}	
	\fi}

\definecolor{linkcolor}{rgb}{0,0,0.4} 
\usepackage[ pdftex,colorlinks=true,
pdfstartview=FitV,
linkcolor= linkcolor,
citecolor= linkcolor,
urlcolor= linkcolor,
hyperindex=true,
hyperfigures=false,
bookmarks=true]
{hyperref} 
\hypersetup{linktocpage}

\usepackage{bookmark}

\usepackage[disable]{todonotes} 
\usepackage[final]{showlabels}	

\excludeversion{ExtraComputation}

\includecomment{Title}
\includecomment{ToC}
\includecomment{Intro}
\includecomment{PartI}
\includecomment{PartII}
\includecomment{PartIII}
\includecomment{Conclusion}
\includecomment{Appendix}
\includecomment{Bibliography}

\begin{document}
\begin{Title}	
\frontmatter


\includepdf[pages=-,offset=0cm -1cm]{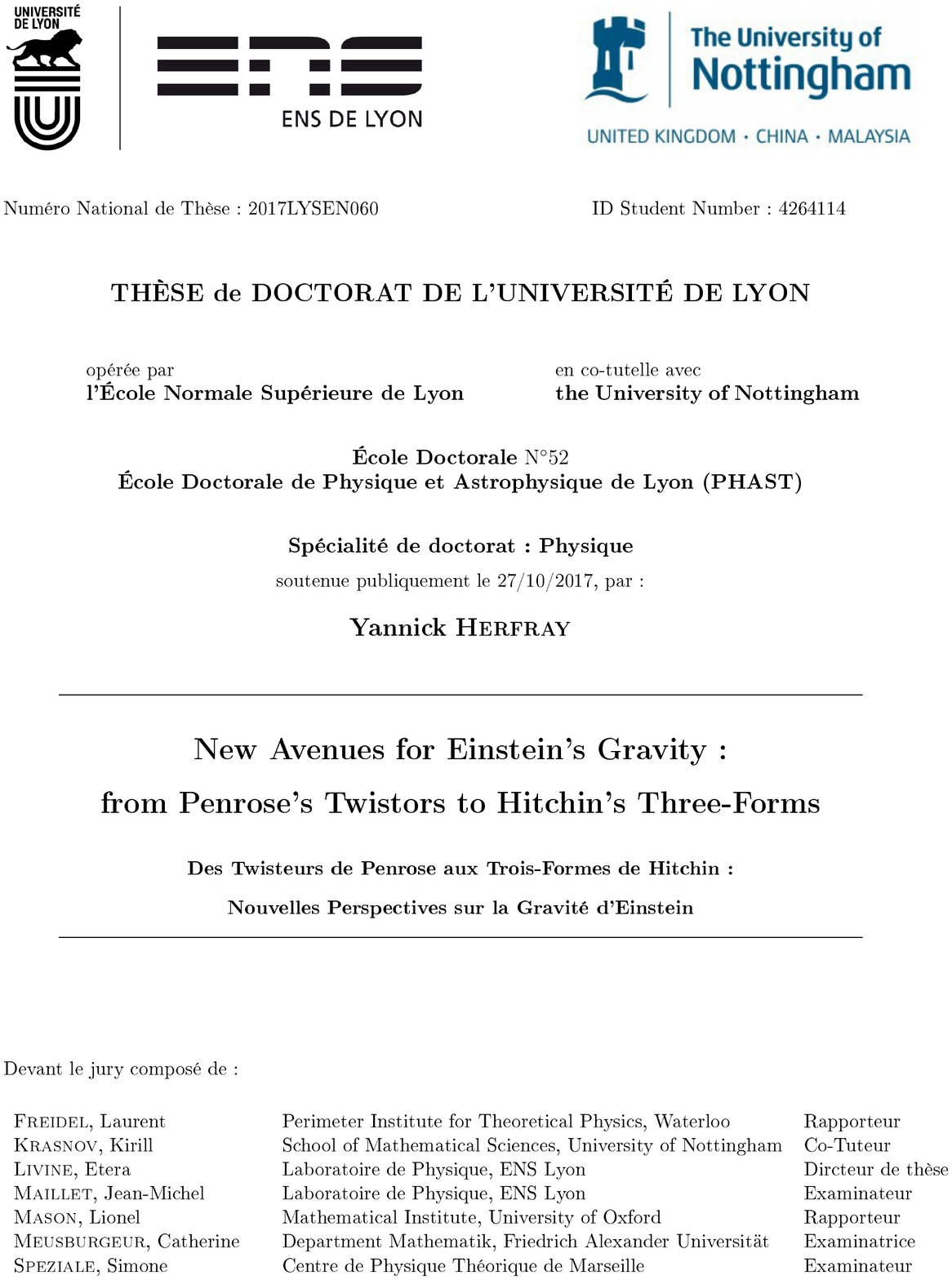} 
\thispagestyle{empty}
\cleardoublepage

\setcounter{page}{1}

\setlength{\parindent}{16pt}

\vspace*{4cm}

 \emph{Pour mon père, pour m'avoir transmis le désir de comprendre et la rigueur qui exclut les chemins de traverse.} \cleardoublepage

\section*{Acknowledgement}

First and foremost I wish to express my gratitude to my supervisors. Etera first welcomed me in the world of theoretical physics and opened gravity's realm for me: I probably would not have dared going any further in that direction if it wasn't for his support and encouragement. Research is a world full of trap, most of them being set by ourselves. Kirill took me as a student and guided me on that tortuous landscape until I could start walking for myself as a young researcher. He appreciated my enthusiasm for geometry and pushed me hard to get the best out of our collaboration but nevertheless always took the time to guess, discuss and understand my concerns. This was an invaluable support. Kirill's eclectic knowledge on the different approaches to quantum gravity also was most inspiring to me: he taught me that one can follow a direction of one's own while keeping an open mind.

From those two years in Nottingham I will also keep vivid and exciting memories of our discussions and work with Yuri Shtanov, Carlos Scarinci and Joel Fine. Research work can be a lot of fun but also really tough. Research work with other people that one admire and appreciate can be even tougher but also incredibly rewarding.

I also wish to thank Laurent Freidel and Lionel Mason for exciting discussions (at least they were to me!) and accepting to be my examiners and consider reading this manuscript. I am also grateful to Jean-Michel Maillet, Catherine Meusburger and Simone Speziale for being part of my jury.

This is my duty to acknowledge all those who sweetened those PhD years, mainly all the post-docs and PhD students that I had the chance to interact with here (in Lyon) and there (in Nottingham). My Nottingham's years would not have been as rewarding and as exciting without the physics reading groups during day-time and discussions in pubs at night in company of Benito, James, Paul, Marco, Carlos, Chi-Hao and Caroline. I learned a lot from our friendship and your diverse experiences. I still miss our late debate at JA's table! I also thank my office-mate in Lyon, Arnaud, Jerome, Valentin and Clement but also Charles-Edouard and Vincent for integrating me so fast and all the others for making life easier, funnier and more interesting: Geraldine, Pauline, Thibault, Samuel, Denis, Brice, Celeste etc 

It is important for me to say here how much I am grateful to my parents. During this whole PhD I couldn't stop but remembering the time where my dad used to help me with my math homework in junior high-school. I can distinctly remember his will to transmit his pleasure for solving geometrical problems. Those were happy moments, at least I remembered them as happy moments and it pushed me forward at times where research work was tough. Even thought this thesis is not written in his mother-tongue nor in a mathematical language that he can understand I would like him to know that the essential pleasure I had while playing with unusual geometrical structures to describe Einstein's gravity is essentially the same I learned with him in those high-school years. This thesis is dedicated to him and those happy memories. Somehow, my mother traded her poor eyes for an acute vision of the social world. In a very literal sense she has the ability to `give sense', with her words, to the world around. I got just a glimpse of that science from her but it `saved' me many times: Bourdieu once said that `sociology is a martial art', in the sense that it can (or should) only be used to protect oneself. In these times of institutionalised individualism I believe basics of this martial art should be part of everyone's survival-kit. 

I also wish to thank those of my friends that did not directly participate in this thesis but in all the rest: Helmy, Nicolas, Denis, Florian, Nils, Guillaume, Louisianne, Masaya, Marin, Karin, Etienne, Audilenz. I cannot forget to cite Fabien here for one never goes as far and deep in life as one can go with a friend that also is a brother.

I finally come to Florence. I do not wish to thank you for anything precise, that would completely miss the point: with you, I am a better man.

\cleardoublepage
\vspace*{-2cm}
\section*{Résumé de la Thèse (English version below)}

Dans cette thèse nous explorons les aspects de la gravité d'Einstein qui sont propres à  la dimension quatre.
L'une des propriétés surprenantes liées à cette dimension est la possibilité de formuler la gravité de manière 'Chirale'.  Dans ce type de reformulations, typiquement, la métrique perd son rôle centrale. La correspondance entre espace-temps et espace des twisteurs est un autre aspect propre à la dimension quatre. Ces formulations, Chirale et Twistorielle,  semblent très différentes. Dans la première partie de cette thèse nous montrons qu'elles sont en fait intimement liées: en particulier nous proposons une nouvelle preuve du `théorème du graviton non-linéaire', due à Penrose, dont le cœur est la géométrie des SU(2)-connections (plutôt qu'une métrique). 
Dans la seconde partie de cette thèse nous montrons que la gravité en trois et quatre dimensions est liée à des théories d'une nature complètement différentes en dimension six et sept. Ces théories, due à Hitchin, sont des théories de trois-formes différentielles invariantes sous difféomorphismes. 
En dimensions sept, nous rencontrons seulement un succès partiel puisque la théorie 4D qui en résulte est une version modifiée de la gravité. Cependant nous prouvons au passage que les solutions d'une déformation particulière de la gravité ont, en 7D, l’interprétation de variétés avec holonomies G2.
Par contre, en réduisant la théorie de six à trois dimensions nous obtenons précisément la gravité 3D. Nous présentons aussi de nouvelles fonctionnelles pour les formes différentielles en six dimensions. Toutes sont invariantes sous difféomorphismes et deux d’entre elles sont topologiques

\section*{Summary of the Thesis}

In this thesis we take Einstein theory in dimension four seriously, and explore the special aspects of gravity in this number of dimension.
Among the many surprising features in dimension four, one of them is the possibility of `Chiral formulations of gravity' - they are surprising as they typically do not  rely on a metric. Another is the existence of the Twistor correspondence. The Chiral and Twistor formulations might seems different in nature. In the first part of this thesis we demonstrate that they are in fact closely related. In particular we give a new proof for Penrose's `non-linear graviton theorem' that relies on the geometry of SU(2)-connections only (rather than on metric).
In the second part of this thesis we describe partial results towards encoding the full GR in the total space of some fibre bundle over space-time. We indeed show that gravity theory in three and  four dimensions can be related to theories of a completely different nature in six and seven dimension respectively. This theories, first advertised by Hitchin, are diffeomorphism invariant theories of differential three-forms.
Starting with seven dimensions, we are only partially succesfull: the resulting theory is some deformed version of gravity.  We however found that solutions to a particular gravity theory in four dimension have a seven dimensional interpretation as G2 holonomy manifold. 
On the other hand by going from six to three dimension we do recover three dimensional gravity. As a bonus, we describe new diffeomorphism invariant functionnals for differential forms in six dimension and prove that two of them are topological.

\end{Title}

\begin{ToC}
\cleardoublepage

\changetext{1cm}{}{}{}{} 
\tableofcontents

\doparttoc

\end{ToC}

\mainmatter
\begin{Intro}
\bookmarksetup{startatroot}
\chapter*{Introduction}
\addstarredchapter{Introduction} \markboth{}{Introduction}
\counterwithout{equation}{section}		\setcounter{equation}{0}
\counterwithout{thmcnter}{chapter}		\setcounter{thmcnter}{0}

\section*{Name the problem: Quantum Gravity}

The essential field equations of the general theory of relativity \cite{Einstein:1915ca} are now more than one hundred years old. Less than a year after the celebration of this centenary, the LIGO cooperation offered the theory its most triumphal confirmation with the first detection of gravitational waves \cite{LIGGO2016}. The other pillar of contemporary physics, quantum field theory, emerged through a longer and more chaotic development but finally attained an impressive maturity. The most salient evidence for this surely was the detection of the Higgs boson in 2012 that crowned the standard model \cite{Chatrchyan:2012xdj, Aad:2012tfa}. This is however only the tree that hides the forest of succeeding high precision tests and large range of application from particles to condensed matter physics. Established on such firm basis, general relativity (GR) and quantum mechanics (QM) altogether form the bedrock of contemporary physics. 

There is a sense in which the overall paradigm of quantum mechanics is, for now, our fundamental theory of dynamics. It tells us what sort of evolution our theories should predict: with the quantum revolution that take place at the beginning of the 20th century - and propagated since then- the very meaning of `determinism' changed. Physicists progressively evolved from an all-mighty dream where they could hope to predict the past and future -were they given precise enough initial data- to a more humble position where evolution is probabilistic. In its trail the quantum revolution however triggered a wave of questions about the respective status of observers and observables which are still unsettled. What is a state? What do we mean by `probing' it ? 

On the other hand, general relativity might just appear as our fundamental kinematical theory.  The essential message of the theory is indeed that physics is intrinsically \emph{relational}. In the framework of general relativity, `observables' only make sense with respect to each others. Put abruptly it just gives a precise sense to the obvious fact that \emph{ one never measures space or time }: One measures correlations between events, e.g between the ticking of a clock, a ruler and the positions of a falling stone. In technical jargon, this is implemented by the \emph{diffeomorphism invariance} of the theory.

It is already fair to say that the above discussion, as for what parts of
QM or GR are fundamental aspects of nature, surely is a polemical one and
that many would have argued differently. It is also easy to come to the conclusion that these types of discussions are `just words', i.e words to which one cannot give a precise content, and there would definitely be some truth in this. In our opinion however, just because the discussion cannot be settled by words - but rather will have to wait for the experimental confirmation of an hypothetical unifying framework for both QM and GR - it does not make it idle. The different answers to this question - what aspects of QM and GR are fundamental principles and what are artefacts of a particular theory? - indeed organise the work of the community towards a theory of \emph{quantum gravity}.

The mere discrepancy in the description of the world that both theories give might indeed be a sufficient motivation for trying to develop an unifying framework: From the QM perspective, matter fields are quantised and there is a subtle distinction between observers and states. From the GR one, the gravity field is a field just like the others and observables are of relational types. There is certainly a tension between these different statements and it leads to problematic questions - What is quantum space-time? What does it mean to observe it? It is hard not to think that solving this tension will have drastic implications as for our understanding of nature. From a more pragmatic perspective, the different infinities (or singularities) that plague both theories are another motivation: singularities seems to be ubiquitous to GR \cite{Penrose:1964wq, Hawking:1969sw} and even though in QFT most infinities were tamed during the tortuous development of the standard model, when it is applied to gravity they proliferate to the point of making the theory non-predictive at high energy (or non-renormalisable, see \cite{Goroff:1985sz, vandeVen:1991gw} for the traditional result, \cite{Bern:2015xsa} for a modern perspective). What is more, it is unclear whether the standard model itself makes any sense beyond perturbation theory. 

One could argue that `quantum gravity already exists' in the form of an effective field theory. From this perspective, non-renormalizability, is just the statement that we do not understand the high-energy behaviour of the theory. Accordingly the problem of `quantum gravity' is to be taken in a broader sense: this is a way to point to the fact that GR and QFT are partly inconsistent. In particular, both singularities in classical GR and the non-renormalisability of its QFT version signal that there is \emph{something} we are missing at a fundamental level.

In fact, nearly as old as GR was the task to quantize gravity: Einstein himself indeed thought in his paper on gravitational waves \cite{Einstein:1918btx} that gravity should be modified by quantum effects. The history of quantum gravity \cite{Rovelli:2000aw, Carlip:2015asa} is long and painful for many technical problems were to be overcome. Many approaches have been developed (see \cite{Woodard:2009ns, Nicolai:2013sz} for reviews) with more or less partial success. 

\section*{Gravity, Quantum and a Matter of attitude}

Even though not directly tied up to quantum gravity the work presented in this thesis took its motivation from this problem and we thus wish to take some more time to consider the different possible attitudes towards it.

 One of the few essential aspects they all agree on is that something new should happen at a fundamental length scale, the so-called `Planck length' $l_p \simeq 10^{-30}m$. The relevance of the Planck length for quantum gravity can be motivated by an elementary argument: Let's consider a lump of matter of mass $m$. For any usual values of mass the Compton wavelength $\gl_{C} = \hbar /\left( c m \right)$ is much larger then the Schwarzschild radius $r_{S} = G m / c^2 $ and thus quantum mechanical effects will massively predominate on any gravitational ones.  As we increase the mass, however, the two lengths evolve towards another and eventually gravitational effects must end up to be of the same order of magnitude than the quantum ones. This is realised for the Planck mass $m_p = \sqrt{\hbar c/G}$. At this scale, both the Compton wavelength and the Schwarzschild radius equal the Planck length $l_P = \sqrt{G \hbar / c^3}$. Infinities proliferating both in GR and QFT are thought to be the result of our assumption that space-time is continuous and the related description of matter as point-like objects. The hope is therefore that they will disappear once the new physics popping in around this Planck scale is taken into account. 

The history of physics has seen many situations where different theoretical or empirical frameworks were contradicting each others and lead to conceptual revolutions. The solutions always came from a mixture of conservatism -aspects of the preceding theories where preserved and raised to the status of fundamental principles- and a revolutionary attitude -some aspects that were though to be carved in stones were dismissed as artefact given by a limiting procedure. 

Accordingly, special relativity was born from the tension between Galilean invariance and Maxwell equations. In the resulting theory both were reconciled but at the cost of loosing an absolute notion of time that had been unquestioned for centuries.  General relativity emerged from the contradiction between special relativity and Newton theory at the price of a definitive disappearance of the space-time fabric for a completely relational description of physics. Finally quantum mechanics appeared as a mean to solve the incompatibility of classical electrodynamics with the new theory of matter. What it took to overcome this contradiction, the introduction by Bohr of discrete orbits for the electron of the hydrogen atoms, cannot possibly be overstated. Not only, physicists, had to learn to deal with discreteness, but they also had to renounce to the, century old, notions of point particles and trajectories. Surprisingly, however, the framework of Hamiltonian and Lagrangian mechanics that were developed in the preceding century turned up to be crucial insights for developing the newborn theory. 

Any succeeding theory of quantum gravity is likely to see the same phenomenon appear. With some aspects of quantum theory and gravity to be preserved and other dismissed. It is enlightening to have a look at the different approach to quantum gravity in fashion today from this perspective.\\

In this respect the asymptotic safety scenario for gravity \cite{Weinberg:1980gg, Niedermaier:2006ns} certainly takes the most conservative approach. In this approach it is suggested that despite its non-renormalizability the QFT version of GR might be a consistent theory at all energies after all. Accordingly the renormalization group flow of perturbative GR could have a UV fixed point which would allow to make sense of the theory at any scale. What is more, if this space of `asymptotically safe' theories is small enough -and if one assumes that our world is described by such a theory- only a finite number of measurement might be enough to know the theory at any scale. In fact this is probably the only approach to quantum gravity were essentially nothing particular happens at the Planck scale. In some sense the hidden radical implication here is that all high energy physics is already contained at low energy. This might sounds a bit disappointing but it also has its own charm.

Next in line, as far as conservatism is concerned, is Loop Quantum Gravity \cite{Rovelli:2004tv,Thiemann:2007zz}. Here, only the perturbative framework tied up with QFT is rejected. It is indeed advocated that trying to quantize gravity as a perturbation around Minkowski space is inherently in contradiction with its relational nature and that this is responsible for the non-renormalisability of the theory. Apart from this, both the emphasise on general covariance and on canonical quantization are in line with the fundamentals of GR and QM and makes it a `narrow path to quantum gravity'. That it is so constrained is probably one of the most appealing feature of this approach.

String theory \cite{Green:1987sp,Polchinski:1998rq}, which trades point particles for extended objects, is in some sense the smallest possible modification of QFT that makes it UV finite. The motivation for this modification is that, just like the non-renomalizability of Fermi's theory of the weak interaction pointed towards the Weinberg-Salam model of electroweak interaction, the non-renormalizability of GR is a sign that it is not a fundamental theory. The beauty of the string theory lies in the fact that such a small modification -considering QFT of extended objects- has tremendous and unexpected implications: gravity seems to be built-in and, in order for the theory to make sense, is inevitably tied up with other interactions and particles (via super-symmetry) and higher dimensional geometry. In string theory the fundamental aspects pertaining to gravity (its field equations, diffeomorphism invariance but also the mere idea of space-time) are taken as secondary or `emergent' just as the Navier-Stokes equation can be derived from a microscopic description of fluids. In that respect String theory definitely sides with QM against GR. 

A third attitude, most vividly propounded py Penrose \cite{Penrose:1999cw,Penrose:2014nha} sides with GR against QM. This is made clear by the slogan `gravitization of quantum mechanics', versus the more usual `quantization of gravity'. It is here advocated that quantum mechanics is not a complete theory and needs to be modified in order to fit with gravity. This was the original aim of twistor theory to provide a description of gravity that would suggest how to do this.

Each of these approaches\footnote{There are certainly more approaches to quantum gravity than those we discussed here e.g Causal Sets, Dynamical Triangulation, Non-Commutative geometry, Relative Locality etc. We are obviously biased by our taste and work. This is however an enjoyable (but polemical as it turned out) game to try to decide for any strategy towards quantum gravity what is taken as fundamental principles both in GR and QM and what is played down. We urge the reader to play this game with his favourite theory.} faces its own problems. Penrose's twistor program could not overcome the `googly problem' i.e describing the two polarisations of gravity in twistorial terms; one still doesn't know if there exists a limit where loop quantum gravity can describe ordinary space-time; and this is still unclear whether or not string theory can describe our world (the standard model) nor what the theory really is beyond many of its perturbative versions -related by various non-perturbative dualities. Finally the asymptotic safety of gravity is still an open -interesting- problem.  As for now -and once again- `Quantum gravity' is therefore less the name of a theory to come than the name of a problem.

\section*{Following Penrose and Hitchin: Geometry as a guiding line}

This thesis is not about quantum gravity. Rather, its most obvious unifying theme is the description of classical\footnote{i.e non-quantum} gravity in terms of unusual geometrical structures. The motivation for looking for and studying such alternative descriptions of gravity however takes its root in the `Penrose's approach' to quantum gravity presented above. The hope is that, in the same spirit that the Hamiltonian formulation of classical mechanics first appeared as a technical trick but turned out to give significant insights in developing the quantum theory, the change of perspective on gravity suggested in this thesis might contribute to the collective task of quantizing gravity (or maybe gravitazing quantum mechanics). We however make no claim that significant steps have been achieved in this respect.

Another encouragement for considering geometrical reformulation of gravity is the example of twistor theory itself. Twistor theory proposed a radical change of perspective on space-time and, even thought it felt far apart from its original aim of providing a complete framework for quantizing gravity, proved to be very fruitful both on the physical and mathematical sides (see \cite{Atiyah:2017erd} for an overview).

In the Lorentzian signature context and for flat space-time, the twistor space $\T$ can be thought as the space of null lines in Minkowski space $M$. On the one hand, lines in Minkowski space corresponds to points in twistor space. On the other hand, a point in space-time can be equivalently described by its null cone. In the twistor space, the set of null lines forming this cone will trace a (projective complex) line. This is the essence of the twistor correspondence: lines correspond to points and reciprocally. Another crucial fact here is that the flat twistor space has a natural complex structure $\T \simeq \C^4$.

 In broader terms, twistor theory relates physical objects on some space-time $M$ to geometrical holomorphic structure on its associated twistor space $\T(M)$. As far as gravity is concerned, the main result of the theory is the \emph{non-linear-graviton theorem} \cite{Penrose:1976js, Ward:1980am}. This theorem demonstrates an equivalence between self-dual Einstein space-times and integrable complex structures on the associated twistor space. This is a deep theorem as it describe complicated solutions to differential equations on space-time as essentially free data on the twistor-space. However, the only Lorentzian self-dual Einstein space-times is Minkowski space so that, as far as physics is involved, this does not really helps. The problem of describing full gravity in twistorial term turned to be one of the main stumbling block for twistor theory and is commonly referred to as the `googly problem'.

A second unifying theme of this thesis is the emphasis on special geometry appearing in low dimension. Accordingly this thesis is made up of three parts respectively dealing with dimension four, six and seven. \\

Four dimensional space-times (we here consider complexified space-time for more generality) are specials in many respects. First this is the lowest dimension for which Einstein equations do not completely constrain the local form of the metric but rather leaves propagating degrees of freedom. Second in four dimension the local isometry group is not simple
 \begin{equation}\label{Introduction: SO4 = SO3 x SO3}
\SO(4,\C) = \left(\SL(2,\C) \times \SL(2,\C)\right)/\Z^2.
\end{equation} This is the only dimension where such a phenomenon appear and it has some interesting applications.  The main one is the existence of \emph{chiral} formulations of four-dimensional gravity, i.e formulation making use of only one half of \eqref{Introduction: SO4 = SO3 x SO3}. In the physics community, these formulations are usually associated to Ashtekar's `new variables' \cite{Ashtekar:1986yd}. Associated with these chiral formulations is a natural family of chiral deformations of GR, that were first studied by Bensgtsson (see e.g \cite{Bengtsson:1991bq}) and dubbed `neighbours of GR'. We prefer the term `chiral deformations of gravity' to emphasis the following facts: all those theories describe spin-2 particles with only 2 propagating degrees of freedom \cite{Krasnov:2008zz} and they differ from GR in a chiral way. All `chiral deformations of GR' indeed share the anti-self-dual sector of gravity (i.e , very schematically, the sector where the first half of the decomposition \eqref{Introduction: SO4 = SO3 x SO3} is turned off) but they differ when incorporating the self-dual sector. Both chiral formulations of gravity and their associated chiral deformations are reviewed in the first chapter of this thesis and will serve as a life-line all along this thesis.

 Now, the identity \eqref{Introduction: SO4 = SO3 x SO3} is just one of the many exceptional isomorphism of Lie group that happens in four dimensions. Another interesting one is
\begin{equation}
Conf(4,\C) = \SO(6,\C) = \SL(4,\C).
\end{equation}
The existence of this isomorphism is the starting point of twistor theory that associates with the conformal compactification of Minkowski space-time $\overline{\Mbb}$ an auxiliary space, the twistor space $\T \simeq \C^4$, on which the (complexified) conformal group, $\SL(4,\C)$, acts linearly. In Lorentzian signature, the representation of the conformal group on the twistor space means that null lines are sent to null lines under conformal transformations which, by itself, is a beautiful consequence of the correspondence. In Euclidean signature, the null line interpretation disappears and the twistor correspondence is less drastic (the twistor space is then a fibre bundle over space-time). However the above isomorphism of Lie group persist in an Euclideanized version. What's more this isomorphism then has a beautiful interpretation in terms of quaternions. This makes it all clear that the twistor correspondence have to do with the magic of the geometry in low dimensions.

The twistor correspondence and non-linear graviton theorem are well-known and rightfully celebrated results. It is however not always realised that twistor theory has a very nice interplay with chiral formulations of gravity. The main point of Part \ref{Part: Chiral Formulations of 4D Gravity and Twistors} of this thesis is to clarify this fact. In particular we present a new-proof of the non-linear graviton theorem with a strong `chiral' flavour. In some sense this version of the theorem allows to think of twistor theory as a far reaching consequence of chiral formulations of gravity.

The material of this Part is mainly taken from \cite{Herfray:2016qvg} as well as from \cite{Herfray:2015fpa,Herfray:2015rja,Fine:2015hef}.\\

Six and seven dimensions are also special but for a very different reason. In any even-number of dimensions there is a well known notion of non-degeneracy for two-forms. A global choice of such two-form then restrict the structure group of the manifold to $Sp(2n,\R)$. In dimensions six and seven one can define a similar notion for three-forms. Such three-forms are then called \emph{stable} after Hitchin \cite{Hitchin:2000sk,Hitchin:2001rw}. Just as for two-forms, a global choice of stable 3-forms reduces the structure group: to $\SL(3,\C) \times \SL(3,\C)$ in 6D and the special group $G_2$ in 7D. In \cite{Hitchin:2000sk}, Hitchin proposed variational principle for these three-forms in six and seven dimension.

In Euclidean signature, the twistor space of a Riemannian manifold is a two-sphere bundle. That self-dual gravity is so nicely described in these terms, by Penrose's theorem, suggests that also the full GR might be encoded into fields on the total space of bundles over space-time. This is our motivations for considering dimensional reduction of Hitchin theory from six to three and seven to four dimension. Accordingly we will consider fibre bundle over 3D and 4D manifold such that the total space is a 6D or 7D manifold. 

In the second part of this thesis we show that the dimensional reduction of 6D Hitchin theory  from 6D to 3D is 3D gravity(coupled with a constant scalar field). We also propose new variational principles for two and three forms in 6D and show that two of them are topological theories. These results were originally described in \cite{Herfray:2016std,Herfray:2017imd}.

In the third part of this thesis we turn to the 7D case. We first show that solutions to a certain chiral deformation of GR in 4D are associated with $G_2$ holonomy manifold in 7D. This is however not a dimensional reduction but rather a lift of some 4D theory to a 7D one. When we consider the dimensional reduction \emph{per se} we are less successful: Starting with a particular theory of 3-forms in 7D the 4D resulting theory turns out to be  some sort of scalar-tensor theory based on one of the chiral deformations of gravity - rather than usual GR. The original material of this part can be found in \cite{Herfray:2016azk} and  \cite{Krasnov:2016wvc,Krasnov:2017uam}. 

Finally, a striking example of geometrical structures that only exist in low dimension are \emph{division algebra} i.e complex numbers, quaternions and octonions, see \cite{Baez:2001dm} for a beautiful review. They will follow us all along this thesis and form another, somewhat hidden, unifying theme. Complex numbers and quaternion are indeed closely tied up with the Euclidean version of twistor theory. Three forms in six dimensions manifold defines almost complex structure and the $G_2$ structure appearing in seven dimension is best thought as the group of automorphism of octonions.

\counterwithin{equation}{section}		\setcounter{equation}{0}
\counterwithin{thmcnter}{chapter}		\setcounter{thmcnter}{0}

\end{Intro}

\begin{PartI}
\part[\\Chiral Formulations of 4D Gravity and Twistors]{Chiral Formulations of 4D Gravity \\and Twistors}\label{Part: Chiral Formulations of 4D Gravity and Twistors}

\section*{Introduction to Part 1:\\ \hspace*{1cm}Chiral and Twistor Formulations \\ \hspace*{2cm} of Four Dimensional Gravity}
\counterwithout{equation}{section}	\setcounter{equation}{0}
\counterwithout{thmcnter}{chapter}	\setcounter{thmcnter}{0}
\addstarredchapter{Introduction to Part 1}\markboth{}{Introduction to Part 1}

It is well known that gravity can be given `chiral formulations' \footnote{For the most striking ones see \cite{Plebanski:1977zz}, \cite{Jacobson:1988yy}, \cite{Krasnov:2011pp}. See also section \ref{ssection : Variational Principles} for a review of chiral Lagrangians for gravity}, i.e formulations where the full local isometry group \begin{equation}\label{Intro partII: SO4 decomposition}
\SO(4,\C)= \SL(2,\C) \times \SL(2,\C)\big/ \Z^2 
\end{equation}
loses its central role for one of the `chiral' (left or right) subgroup $\SL(2,\C)$. Note that, for generality, we here consider complexified gravity: in Lorentzian signature the two $\SL(2,\C)$ groups are complex conjugated while in Euclidean signature they are replaced by two independent $\SU(2)$ groups.

This shift in the local symmetries corresponds to a shift in the hierarchy of fields: in `chiral formulations' of GR the role of the metric is usually played down for other alternative variables with natural $\SL(2,\C)$ internal symmetries. Typically, the metric appears as a derived object and its associated local isometry group $\SO(4,\C)$ comes as an `auxiliary symmetry' that was somewhat hidden in the first place.

It's probably safe to say that the interest of the physics community for such reformulations started with Ashtekar `new' variables \cite{Ashtekar:1986yd} and the appealing form of the related diffeomorphism constraints.
In subsequent works \cite{Jacobson:1988yy}, \cite{Capovilla:1991qb} it was understood that the (ten year older!) Plebanski's action \cite{Plebanski:1977zz} gave a covariant description of Ashtekar variables. In Plebanski's pioneering work the metric completely disappears for $\SL(2,\C)$-valued fields. In both points of view, canonical and covariant, $\SL(2,\C)$-connections play a crucial role.

That $\SL(2,\C)$-connections appear is no surprise: In the more traditional metric perspective, the Levi-Civita connection comes as an $\SO(4,\C)$-connection. The decomposition of Lie group \eqref{Intro partII: SO4 decomposition} then corresponds to a splitting of the Levi-Civita connection into Left(or self-dual) and Right(or anti-self-dual) $\SL(2,\C)$-connections, which are in some sense the most natural `chiral' objects one can construct from the metric. 
The `chiral formulations' of GR essentially reverse this construction: they take $\SL(2,\C)$ fields (e.g connections) as a building block for the metric. This culminates in the so called `pure connection of GR' pursued in \cite{Capovilla:1991kx} and finally achieved in \cite{Krasnov:2011pp} where the only field that appears in the Lagrangian is an $\SL(2,\C)$-connection. Chiral formulations of GR will be reviewed from a general perspective in Chapter \ref{Chapter: Chiral formulation of gravity}.

On the other hand, it is not always apparent that twistor theory, at least in its original Penrose's program directed towards gravity \cite{Penrose:1999cw}, has a nice interplay with these reformulations and is in fact part of `chiral formulations' of GR in a broad sense. This is more clearly seen by taking a closer look at the main result of twistor theory on the gravity side, the `non-linear graviton theorem' \cite{Penrose:1976js}, \cite{Ward:1980am}.

The `non-linear graviton theorem' takes as a starting point an eight dimensional real manifold (the twistor space) equipped with an almost complex structure. This reduces the group of local symmetries to $\SL(4,\C)$ which is also the 4d (complex)conformal group $\SO(6,\C)  \simeq  \SL(4,\C)/\Z^2$ and indeed the first half of the non-linear graviton theorem asserts that, under some generic conditions, integrability of this almost complex structure is equivalent to a 4d complexified conformal anti-self-dual space-time (i.e such that self-dual part of Weyl curvature vanishes). That this theorem only describes anti-self-dual space-times clearly points in the direction of the intrinsic chirality of twistor theory, but there is more. 

The second half of the theorem requires additional data on twistor space in the form of a complex one-form up to scale, usually denoted as $\gt$. This is essentially equivalent to a 4d real distribution\footnote{In order to describe space-times with non-zero cosmological constant, which is our main concerned in this paper, this distribution should also be non-integrable $\gt\W d\gt \neq 0$.} at every point (the kernel of $\gt$). This one-form is taken to be `compatible' with the almost complex structure so that its kernel is in turn almost complex and identifies with $\C^2$. The restriction of the symmetry group $\SL(4,\C)$ to this distribution thus brings us down to the `chiral' group $\SL(2,\C)$: In fact such a one-form is naturally associated with a `chiral' $\SL(2,\C)$-connection on space-time (This is especially clear in the Euclidean context and we will come back to this in what follows). As connections are not conformally invariant, it fixes a scale in the conformal space-time. The second part of the Non-Linear-Graviton Theorem then essentially asserts that this scale is such that the resulting metric is anti-self-dual Einstein if one is given a `good enough' (holomorphic) one-form.

The usual approach to twistor theory generally emphasizes the metric aspect of the theorem and tend to overlook the fact that this one-form, which crucially fixes the scaling to give Einstein equations, is directly related to a $\SL(2,\C)$-chiral connection thus putting twistor theory in the general framework of `chiral formulations of gravity'. In Chapter \ref{Chapter: Twistors} we will review the basics of the curved twistor construction with an emphasis on the relation between the chiral connection and the $\Oc(2)$-valued one-form on twistor space $\gt$.

It is in fact well known to specialists that \emph{there is} an interplay between, for example, Plebanski formulation of GR and twistor theory as can be seen from the introduction of twistor variables in some recent spin-foam models \cite{Livine:2011vk,Speziale:2012nu} or in the conjoint use of Plebanski action and twistor theory \cite{Mason:2008jy} to investigate the structure of MHV gravity amplitudes. However, it is possible that not all consequences have been drawn from this overlapping.

Now, one of the `most radical' chiral formulations of GR is the pure connection formulation where only a $\SL(2,\C)$-connection is considered to be a fundamental field, the metric being a derived object. In this context Einstein equations take the form of second order field equations on the connection.

In the first part of this thesis we wish to emphasize the change of perspective on twistor theory that this extreme chiral reformulation of gravity suggests: We already stated that, on twistor space, the equivalent of this chiral connection on space-time is a complex one-form, $\gt$. In usual twistor theory this is just taken to be some additional data that complements the almost complex structure, the latter being fundamental. However the pure connection formulation of GR suggests that it is the one-form $\gt$ ( loosely related again to the chiral connection) that should be taken as the starting point, with the almost complex structure (related the conformal structure) arising as a derived objects.

We demonstrate in chapter \ref{Chapter: Twistors} that, at least in the Euclidean signature context, it is a valuable point of view and that it allows to reproduce nicely the results from the non-linear-graviton theorem while putting twistor theory firmly into the `chiral formulations' framework of gravity:

For a Riemannian manifold $M$ (i.e equipped with a metric of Euclidean signature) the associated twistor space $\T(M)$ is simply taken to be the 2-spinor bundle\footnote{$\left(\p_{A'},x\right)$ will be coordinates adapted to the fibre bundle structure $\C^2 \inj \T(M) \to M$.}. Then an $SU(2)$-connection, $A^{A'}{}_{B'} \in \su(2)$, allows to define the one-form on $\T(M)$:
\[ 
\gt = \p_{A'}\left(d\p^{A'} + A(x){}^{A'}{}_{B'} \p^{B'} \right)
\]
related to the preceding discussion.

We first show that this is enough to construct a Hermitian structure on $\PT(M)$, thus making contact with usual Euclidean twistor theory:
\begin{Proposition}{\emph{Almost Hermitian structure on $\PT(M)$}}\label{intro: prop AHS on PT}\mbox{} \\
	If $A$ is a definite connection \emph{(see below for a clarification of this notion)} then $\PT(M)$ can be given an almost Hermitian structure, i.e a compatible triplet $\left(\J_A , \go_A , g_A \right)$ of almost complex structure, two-form, and a Riemannian metric. 
	
	In general this triplet is neither Hermitian ($\J_A$ is not integrable) nor almost Kähler ($\go_A$ is non degenerate but generically not closed). In fact integrability of $\J_A$ is equivalent to the statement that $A$ is the self-dual connection of a self-dual Einstein metric with non zero cosmological constant. The metric on twistor space can be made Kähler if and only if $A$ is the self-dual connection of a self-dual Einstein metric with positive cosmological constant (ie if the definite connection is of `positive sign' ).
	
	Further more, the integrability condition is equivalent to $\gt\W d\gt\W d\gt=0$.
\end{Proposition}
The main difference with the traditional results from \cite{Atiyah:1978wi} is that integrability is not only related to anti-self-duality but is irremediably linked to Einstein's equations. This is because in the construction described in \cite{Atiyah:1978wi} one is only interested in a conformal class of metrics while here the use of connections automatically fixes the `right scaling' that gives Einstein equations.

The fact that the connection needs to be `definite' refers to a natural non-degeneracy condition. Such connection can be assigned a sign. This terminology first appeared in \cite{Fine:2008} and we will review it in chapter \ref{Chapter: Chiral formulation of gravity}.  
The possibility of associating a symplectic structure on $\PT(M)$ with a definite $SU(2)$-connection on $M$ was already pointed out in \cite{Fine:2008}. However, only in the integrable case does the symplectic structure described in this reference coincides with our $\go_A$. $\SL(2,\C)$-connections which are the self-dual connection of a self-dual Einstein metric with non zero cosmological constant were called `perfect' in \cite{Fine:2011} and are the one such that their curvature verify $F^i\W F^j \propto \gd^{ij}$. This well known (see e.g \cite{Capovilla:1990qi}) description of Einstein anti-self-dual metric in terms of connection will also be reviewed in chapter \ref{Chapter: Chiral formulation of gravity}.
\\

On the other hand, starting with a certain 6D manifold $\PTc$, the projective twistor space, together with a one-form valued in a certain line bundle $\gt$, we have a variant of the non linear graviton theorem:
\begin{Proposition}{\emph{Pure connection Non-Linear Graviton Theorem}} \label{intro: prop NLG} \mbox{}\\
	If $\gt$ is a definite one-form then $\PTc$ can be given an almost complex structure $\J_{\gt}$.
	
	Together with some compatible conjugation operation on $\PTc$ this is enough to give $\PTc$ the structure of a fibre bundle over a 4d manifold $M$: $\CP^1 \inj \PTc \to M$.
	
	Integrability of $\J_{\gt}$ is then equivalent to the possibility of writing $\gt$ as
	\[ 
	\gt = \p_{A'}\left(d\p^{A'} + A^{A'}{}_{B'} \p^{B'} \right)	
	\]
	with $A$ the self-dual connection of a Einstein anti-Self-Dual metric on $M$ with non zero cosmological constant.
	
	What is more the integrability condition reads $ \gt\W d\gt \W d\gt =0$.
\end{Proposition}

Bits and pieces of this last proposition were already known and developed in \cite{Mason05},\cite{Wolf:2007tx} and \cite{Adamo:2013tja} as part of a strategy to obtain twistor actions for conformal gravity, anti-self-dual gravity and gravity (the latter being still missing). However, we here give a new proof that emphasises the role of the connection as a fundamental object and we hope that by framing them in the general perspective of chiral approaches to gravity they will appear in a new light, i.e as more than just clever trick to construct twistor action.  In particular we hope to make it clear that one can effectively think of the (euclidean)non-linear-graviton theorem as a far reaching generalisation of the description of Einstein anti-self-dual metric in terms of connections. 

Our long term view in developing what could be called a `connection approach' to twistor theory, with the one-form $\gt$ being the main field instead of the almost complex structure, was to open new strategies to construct twistor action for gravity. However one faces difficulties that we could not overcome. We briefly explain in section \ref{section: Discussion on the would be `Twistor action for Einstein gravity'} our work in this direction and why it does not seem to offer a way to a twistor action for gravity.\\

In this whole part we stick to the Euclidean signature. This is for coherence with our results concerning twistor theory which only apply to this signature. \\

This part is organised as follows: In the beginning of chapter \ref{Chapter: Chiral formulation of gravity}, see section \ref{section: Chiral Formulations of GR - Fundations}, we review chiral formulations of gravity with an emphasis on the general geometric setting underlying any formulation of this type rather than on a particular Lagrangian. In section \ref{section: Definite Connections and Gravity}, we especially stress how to write equations for self-dual gravity (i.e Einstein anti-self-dual metric) in this framework and review the pure connection field equations for Einstein metrics. This will serve as a model for our `connection version' of the non-linear-graviton theorem. At the end of this chapter, see section \ref{section: Chiral Deformations of Gravity} we take some time to discuss `chiral deformations of GR'. This is an infinite family of spin-two theories with only two propagating degrees of freedom which is naturally related with chiral formulations of gravity. This lies somewhat out of the main line of development of this part but will be useful for Part \ref{Part: Variations on Hitchin Theory in Seven Dimensions}. We also describe variational principles associated with these `chiral deformations' and their relation with GR. This is in part a review of the existent literature but we also discuss some original results. 

In chapter \ref{Chapter: Twistors} we come to twistors. We first review the construction of the flat twistor space, its `curved' version and some essential results of the theory. This is done in Euclidean signature (see \cite{Atiyah:1978wi}, \cite{Woodhouse85} for Euclidean twistor theory) for coherence with the rest of the part but also because most physics literature logically emphasises the Lorentzian case and we believed the Euclidean twistor construction, with its beautiful relation to quaternions, deserves more attention. 
In section \ref{section: Twistor theory revisited} we come back on the `curved' twistor theory, again in Euclidean signature but from an unusual connection perspective, i.e we take a $\SU(2)$-connection to be the main field instead of a metric. From this data only we show how to construct very natural structures on twistor space, namely the one-form $\gt$, some associated connection on $\Oc(n)$ bundle and the triplet $\left(\J, \go, g\right)$ of compatible almost complex structure, two-form and Euclidean metric on twistor space of Prop \ref{intro: prop AHS on PT}. We also review, from \cite{Fine:2008}, some symplectic structure that is naturally constructed from the connection. Finally we investigate the condition for integrability of the almost complex structure as well as the condition for which the triplet $\left(\J, \go, g\right)$ is Kähler. These cases turn out to be given by the self-dual-gravity equations and therefore make contact with the usual Kähler structure on twistor space constructed from an instanton (i.e an anti-self-dual Einstein metric).

We then state and give a new proof for the non linear graviton theorem from a pure connection point of view (cf Prop \ref{intro: prop NLG}). 

Finally in section \ref{section: Discussion on the would be `Twistor action for Einstein gravity'} we explain how ideas from the previous sections suggest new ansätze for constructing twistor action for gravity. However this section will remain inconclusive and ideas described there should be seen as a few more elements on the chase (cf \cite{Mason05}, \cite{Adamo:2013tja}, \cite{Mason&Wolf09}, and \cite{Adamo:2013cra}) for this elusive (if existing) action.

\counterwithin{equation}{section}		\setcounter{equation}{0}
\counterwithin{thmcnter}{chapter}		\setcounter{thmcnter}{0}

\changetext{1cm}{}{}{}{} 
\parttoc
\changetext{-1cm}{}{}{}{} 

\chapter{Chiral Gravity} \label{Chapter: Chiral formulation of gravity}

In this chapter we review `chiral formulations of gravity'. In section \ref{section: Chiral Formulations of GR - Fundations}, we tried to adopt a broad perspective and describe the conceptual elements that are common to all these formulations rather than describe a particular Lagrangian. We also tried to avoid hiding the simplicity of the geometrical concepts involved under a debauchery of indices. We however provide appendix \ref{section: Appdx Decomposition of the Curvature} for the reader interested in the massive display of tensor indices that are sometimes required for precise calculations and proofs.

 The main objective of this chapter is to introduce in a natural way the description of self-dual gravity in terms of $\SU(2)$-connections and the related notion of definite connections. This is achieved at the end of section \ref{section: Definite Connections and Gravity}. We also review how to write Einstein equations in terms of $\SU(2)$-connections only.

 Naturally associated with these `chiral formulations' there is an infinite family of `chiral deformations of GR', which are spin-two theories with only two propagating degrees of freedom. Even thought they are not directly relevant for the rest of this part, they are so much interwoven with the `chiral formulations' that we thought it was best to describe them here, see section \ref{section: Chiral Deformations of Gravity}. These chiral deformations will be a central theme of part \ref{Part: Variations on Hitchin Theory in Seven Dimensions}.

\section{Chiral Formulations of Gravity : Geometrical Foundations}\label{section: Chiral Formulations of GR - Fundations}

Chiral formulations of gravity exploit the fact that Einstein equations can be stated using only `one half' of the decomposition $\SO(4,\C) = \SL(2,\C) \times \SL(2,\C)/\Z^2$. We here briefly review why this is possible. 

The whole discussion in this section could be treated in complexified terms but for clarity and coherence with the other sections we will restrict to the real form $\SO(4,\R) = \SU(2) \times \SU(2)/\Z^2$, i.e Euclidean signature.

\subsection{Chiral decomposition of the curvature tensor}
Let us consider a Riemannian manifold $\left(M, g\right)$. We note $\left\{e^I\right\}_{I\in 0..3}$ an orthonormal frame and $\left\{e_I\right\}_{I\in 0..3}$ a dual co-frame, they are defined up to $\SO(4)$ transformations. In order to see that Einstein equations can be stated using only one half of the decomposition $\SO(4)=\SU(2) \times \SU(2)/\Z^2$, the quickest way is to split the Riemann curvature tensor into self-dual/anti-self-dual pieces. This is classically done in spinor notation (see e.g \cite{Penrose_vol1}) or more directly as in \cite{Atiyah:1978wi}. We here make a presentation along the line of the second reference with an emphasise on the necessity of using a torsion-free connection in order for chiral formulations of gravity to be possible. 
See also \cite{Capovilla:1991qb,Krasnov:2009pu} for pedagogical expositions.

As a starting, point let us consider a 2n-dimensional manifold. A crucial remark is that the hodge duality $* \from \gO^k(M) \to \gO^{2n-k}(M)$ sends n-forms on n-forms. Self-dual (resp anti-self-dual) n-forms are then eigenvectors with eigenvalues $+1$ (resp $-1$) for the hodge duality\footnote{In fact for a generic dimension and signature the eigenvalues are either $\pm1$ or $\pm i$. Here we already have in mind the application to Euclidean four dimension.}. The case $2n=4$ it thus the only situation where two-forms can be decomposed in self-dual $\gO^+$ and anti-seld-dual $\gO^+$ two-forms. This is a happy accident that has several implications.

By using the metric, two-forms at a point $x \in M$ can be identified with anti-self-adjoint transformations of $\gO^1(M)_x$ and thus with $\so(2n)$:
\begin{equation}
b_{IJ} \frac{e^I \W e^J}{2} \in \gO^2(M)_x \quad \simeq \quad b_J{}^I \;e_I \otimes e^J\in End\left(\gO^1\right) \quad \simeq \quad \boldsymbol{b}\in \so(2n) 
\end{equation}
Where $b_{IJ} = b_I{}^K g_{KJ} $.\\

The `accidental' split of Lie algebra 
\begin{equation}\label{Chiral Formulations of GR - Fundations: Lie algebra split}
\so(4)= \su(2)\oplus \su(2)
\end{equation}
 then directly corresponds to the decomposition of two-forms into self-dual and anti-self-dual two-forms,
  \begin{equation}\label{Chiral Formulations of GR - Fundations: two-form split}
\gO^2 = \gO^2_+ \oplus \gO^2_-.
\end{equation}
Another accident is that \emph{curvature} forms are two-forms. In four dimensions curvatures can thus be decomposed into smaller elementary bits. This is useful both for Yang-Mills type theories but also for gravity. We now turn to the decomposition of the Riemann tensor:

Consider a connection $\nabla$ on the tangent bundle compatible with the metric, this is a $\SO(4)$-connection (Note that, at this stage, we do not assume that the torsion of this connection vanishes). It splits into two $\SU(2)$-connections $D$ and $\widetilde{D}$,
 \begin{equation}\label{Chiral Formulations of GR - Fundations: connection split}
\nabla = D + \widetilde{D}. 
\end{equation} 
 They naturally act as connections on the bundle of self-dual two-forms and anti-self-dual two-forms respectively. 

As a consequence of \eqref{Chiral Formulations of GR - Fundations: connection split} the curvature $\N^2$ two-form can be rewritten
\begin{equation}
\N^2 = D^2 + \Dt^2.
\end{equation}
At this point we only made use of the first `accident', the Lie algebra split \eqref{Chiral Formulations of GR - Fundations: Lie algebra split}. We can now make use of the second `accident' the two-form decomposition \eqref{Chiral Formulations of GR - Fundations: two-form split}: As we already pointed out, the curvature-forms $D^2$, $\Dt^2$ are indeed $\su(2)$-valued two-forms or equivalently $\gO^+$ (resp $\gO^-$) -valued two-forms,
\begin{equation*}
D^2 \in \gO^2\left(M, \su(2)\right) \simeq \gO^2\left(M, \gO^2_+\right), \qquad \widetilde{D}^2 \in \gO^2\left(M, \su(2)\right) \simeq \gO^2\left( M,\gO^2_-\right).
\end{equation*}

It follows from the decomposition, $\gO^2 = \gO^2_+ \oplus \gO^2_-$, that we can write them as bloc matrices:
\begin{equation}\label{Chiral Formulations of GR - Fundations: F, Ft decomposition}
D^2 = \Big( F , G \Big),\qquad \widetilde{D}^2= \left( \Gt , \Ft \right)
\end{equation}
where $F \in End\left(\gO^2_+\right)$, $G \in Hom\left(\gO^2_+, \gO^2_-\right)$, $\Ft \in End\left(\gO^2_-\right)$, $\Gt \in Hom\left(\gO^2_-, \gO^2_+\right)$.

Putting this altogether, the curvature of $\nabla$ can be written as a bloc matrix:
\begin{equation}\label{Chiral Formulations of GR - Fundations: Riemann decomposition 1}
\nabla^2 =
\left(\begin{array}{ll}
F & G \\ \Gt & \Ft
\end{array}\right) \quad \in \gO^2\left(M,\so(4)\right)\simeq End\left(\gO^2\right)
\end{equation}
It is also convenient to introduce the self-dual and anti-self-dual part of the Weyl tensor: 
\begin{equation}
\Psi = F -\frac{1}{3} tr F \; \Id, \qquad \Psit = \Ft -\frac{1}{3} tr \Ft \; \Id.
\end{equation}

Without any further assumptions this is as far as we can get. However, in the special case of the Levi-Civita connection, i.e if one assumes that the connection is torsion-free, we get a simpler picture: $\nabla^2\from \gO^2 \to \gO^2$ is then the usual Riemann curvature tensor and has some further symmetries. The torsion-free condition indeed implies that $\N^2$ has to be symmetric, i.e \begin{equation}\label{Chiral Formulations of GR - Fundations: Riemann symmetries}
G{}^t = \Gt,\qquad \Psi^t= \Psi,\qquad \Psit^t = \Psit.
\end{equation}
What is more, for the torsion-free connection $tr F = tr \Ft$. Using coordinates, one can indeed immediately see that this last identity is equivalent to the first Bianchi identity. See Appendix \ref{section: Appdx Decomposition of the Curvature} for more details.

From these considerations, we obtain the celebrated decomposition of the Riemann tensor into irreducible components:
\begin{equation}\label{Chiral Formulations of GR - Fundations: Riemann decomposition 2}
\nabla^2 = \underbrace{tr F \; \Mtx{ \Id & 0 \\ 0 & \Id}}_{\text{Scalar Curvature}}+ \underbrace{\Mtx{
0 & G \\ G^t & 0
}}_{\text{Ricci Traceless}} +\underbrace{\Mtx{
\Psi & 0 \\ 0 & \Psit
}}_{\text{Weyl Curvature}}.
\end{equation}

Let us now turn to Einstein equations. From the above decomposition it stems that,
\begin{equation}\label{Chiral Formulations of GR - Fundations: Chiral Einstein Equations}
\text{g is Einstein if and only if} \quad G=0
\end{equation}
and then the scalar curvature is $4\gL = 4 tr F$.

In particular one sees from $D^2 = F + G$ that Einstein equations can be stated in term of $D$ only: The metric is Einstein if and only if $D$ is a self-dual gauge connection, i.e if $D^2$ is a self-dual $\su(2)$-valued two-form.

Note however that, in order to achieve this `chiral formulation', the symmetries \eqref{Chiral Formulations of GR - Fundations: Riemann symmetries} were crucial. In case one does not assume the connection to be torsion-free the Riemann tensor does not enjoy the symmetries \eqref{Chiral Formulations of GR - Fundations: Riemann symmetries} and Einstein equations look much more complicated:
\begin{equation}\label{Chiral Formulations of GR - Fundations: Non-chiral Einstein equations}
G = -\Gt^t, \quad \Psi^t = \Psi, \quad \Psit^t = \Psit, \quad \text{and}\quad \frac{tr F + tr \Ft }{2}= cst =\gL.
\end{equation}
See Appendix \ref{section: Appdx Decomposition of the Curvature} for a derivation in coordinates.

As opposed to \eqref{Chiral Formulations of GR - Fundations: Chiral Einstein Equations} this last set of equations involve the whole of the Riemann tensor and are therefore not `chiral' at all. One easily checks however that equations \eqref{Chiral Formulations of GR - Fundations: Non-chiral Einstein equations} together with the symmetries \eqref{Chiral Formulations of GR - Fundations: Riemann symmetries} give back the chiral formulation of Einstein equations \eqref{Chiral Formulations of GR - Fundations: Chiral Einstein Equations}, as it should.

From this presentation we hope to make it clear that the general phenomenon allowing for chiral formulations of Einstein equations stems from the internal symmetries of the Riemann tensor, related to torsion-freeness, and not from a particular choice of signature.

\subsection{Urbantke metric}\label{ssection: Urbantke metric}
In the above, we explained how Einstein equations can be stated in an essentially chiral way, i.e in terms of $\su(2)$-valued fields. This general principle underlies any chiral formulation of gravity. However this was still very classical in spirit as we considered the metric as the fundamental field. We now describe an essential observation due to Urbantke \cite{Urbantke:1984eb} that allows to obtain a metric as a derived object from chiral (i.e $su(2)$-valued) fields. 

Suppose that we have a $\su(2)$-valued two-forms $\bdB$, using a basis of $\su(2)$, $\left(\gs_i\right)_{i\in 1,2,3}$, this gives us a triplet of real two-form $\left(B^i\right)_{i \in 1,2,3}$, such that $\bdB= B^i\gs_i$

Now, given such a triplet of two-forms $\left(B^i\right)_{i \in 1,2,3}$, there is a unique conformal structure that makes the triplet $\left(B^1,B^2,B^3\right)$ self-dual. We will refer to this conformal structure as the Urbantke metric associated with $\bdB$ and write it as $\tilde{g}_{\text{\tiny (B)}}$. There is even a way to make this conformal structure explicit through Urbantke formula \cite{Urbantke:1984eb},
\begin{equation}\label{Chiral Formulations of GR - Fundations: Urbantke metric}
\text{Urbanke metric}\text{:}\qquad \tilde{g}_{\text{\tiny (B)}}{}_{\mu\nu}\; = -\; \frac{1}{12}\epst^{\alpha\beta\gamma\delta}\eps_{ijk}B^i{}_{\mu\alpha}B^j{}_{\nu\beta}B^k{}_{\gamma\delta}.
\end{equation} 
Obviously, if the $B$'s do not span a 3 dimensional vector-space this cannot hold. In fact the `metric' \eqref{Chiral Formulations of GR - Fundations: Urbantke metric} will then be degenerated in the sense that it will not be invertible. A more precise statement, again from \cite{Urbantke:1984eb}, is the following:
given the triplet of two forms $\left(B^1,B^2,B^3\right)$, defines the conformal `internal metric' $\tilde{X}^{ij} = B^i \wedge B^j/ d^4x $ then Urbantke metric $\tilde{g}_{\text{\tiny (B)}}$ is invertible if and only if $\tilde{X}$ is. When Urbantke metric is invertible $\tilde{X}$ is just the metric on the space of self-dual two-forms given by wedge product.\footnote{As a side remark, on self-dual two-forms the `wedge' internal product $B^i \W B^j/d^4x$ coincide to the `metric' internal product $B^i\W *B^j /d^4x$}.

As we started with a triplet $\left(B^i\right)_{i\in 1,2,3}$ of \emph{real} two-forms, the associated Urbantke metric \eqref{Chiral Formulations of GR - Fundations: Urbantke metric} is also real. One the other hand, \emph{its signature is undefined}: self-dual two-forms in Lorentzian signature are complex so this signature is excluded but without further restriction it can still be either Euclidean or Kleinian. The signature of the internal metric $\tilde{X}$ however is enough information to fix this ambiguity: for an Euclidean conformal metric $\tilde{g}$ the metric $\tilde{X}$ on self-dual two-forms given by wedge product is Euclidean while for a Kleinian signature it would be Lorentzian.

Thus if we start with a triplet $\left(B^i\right)_{i\in 1..3}$ of real two-form such that the internal metric $\tilde{X}^{ij} = B^i\W B^j \big/ d^4x$ is definite, we are then assured that the associated Urbantke metric, $\tilde{g}_{\text{\tiny (B)}}$ is non degenerate (invertible) and of Euclidean signature. This suggests to introduce the following definition:
\begin{Definition}{\emph{Definite Triplet of two-forms}}\label{Chiral Formulations of GR - Fundations: Def: definite triplet} \mbox{} \\
	A triplet $\left(B^1, B^2, B^3\right)$ of real two-forms is called definite if the conformal metric constructed from the wedge product $\tilde{X}^{ij} = B^i\W B^j\big/ d^4x$ is definite.
\end{Definition}
As we just explained this is a useful definition because of the following:
\begin{Proposition}{\emph{Urbantke metric}} \mbox{} \label{Proposition: Urbantke metric}\\
	The Urbantke metric \eqref{Chiral Formulations of GR - Fundations: Urbantke metric} associated with a definite triplet of two-forms is non degenerate and of Euclidean signature.
\end{Proposition}
\begin{proof}
The equivalence between the non-degeneracy of $\gti_{\text{\tiny (B)}}{}_{\mu\nu}$ and the non-degeneracy of $\Xt^{ij}$ can be found in \cite{Urbantke:1984eb}. This reference also shows that the $B$'s are self-dual for $\gti_{\text{\tiny (B)}}{}_{\mu\nu}$. As already discussed above the 3D wedge product metric on self-dual two-forms $\Xt$ can only be real definite for a (conformal) Euclidean four dimensional metric.
\end{proof}
In this section we made two distinct but complementary observations, first Einstein equations can be stated in a chiral way (cf equation \eqref{Chiral Formulations of GR - Fundations: Chiral Einstein Equations}) ie using $\su(2)$-valued fields, second a (definite) $\su(2)$-valued two-form is enough to define a metric. Lagrangians that realise `chiral formulations' of GR all rely on some mixture of these two facts each with its own flavour and fields. See section \ref{ssection : Variational Principles} for some explicit variational principles.

However, for the most of this part, we won't be interested by a particular action but rather by how the general framework that we just describe intersects with twistor theory. Our main guide will be the description of anti-self-dual Einstein metric in terms of connections. 

Before we come to this it is useful to introduce two new tensors.

\subsection{Two useful tensors: the sigma two-forms} \label{ssection: Two useful tensors}

We already made the remark that a metric allows to identify the Lie algebra $\so(4)$ with the space of two-forms $\gO^2$. We denote by  \begin{equation}
\Phi \from \so(4)= \su(2)\oplus\su(2) \to \gO^2=\gO^2_+\oplus \gO^2_-
\end{equation} this isomorphism.

We choose a basis $\left( \,\gs^i , \gst^i \right)_{i\in 1,2,3}$ of $\so(4)= \su(2)\oplus\su(2)$ adapted to the decomposition and such that 
\begin{equation}
\left[\gs^i, \gs^j\right] = \eps^{ijk}\,\gs^k  ,\quad  \left[\gst^i, \gst^j\right] = \eps^{ijk}\gst^k ,\quad \left[\gs^i, \gst^j\right] = 0.
\end{equation}
Then one can define the sigma two-forms:
\begin{equation}\label{Chiral Formulations of GR - Fundations: Sigma def (geometric)}
\frac{1}{2} \gS^i = \Phi \left( \gs^i \right), \qquad \frac{1}{2} \gSt^i = \Phi \left( \gst^i \right).
\end{equation}
Thus $\Big(\gS^i \Big)_{i\in 1,2,3}$ (resp $\left(\gSt^i \right)_{i\in 1,2,3}$) form a basis of self-dual two-forms $\gO^2_+$ (resp anti-self-dual two-forms $\gO^2_-$). This basis is also defined (up to $SU(2)$ transformations) by the orthogonality relations
\begin{equation}
 \gS^i \W \gS^j = -\gSt^i \W \gSt^j = 2 \gd^{ij} Vol_g, \qquad \gS^i \W \gSt^j = 0.
\end{equation}

The awkward factor of one half in the definition is there for it to fit with the definition in terms of a tetrad that frequently appears in the literature:
\begin{align}\label{Chiral Formulations of GR - Fundations: Sigma def (tetrad)}
&\left(\gS^i= -e^0 \W e^i - \frac{\eps^{ijk}}{2} e^j \W e^k\right)_{i\in 1,2,3}, \qquad  
&\left(\gSt^i = e^0 \W e^i - \frac{\eps^{ijk}}{2} e^j \W e^k\right)_{i\in 1,2,3}.
\end{align}

The sigma two-forms are naturally $\su(2)^*$-valued two-forms or, using the Killing metric on $\su(2)$, $\su(2)$-valued two-forms:\footnote{In this thesis bold notation will indicate $\su(2)$-valued objects.}
\begin{equation}
\bdgS = \gS^i \,\gs^i \in \gO^2_+\left(M,\su(2)\right), \qquad \bdgSt =\gSt^i \,\gst^i \in \gO^2_-\left(M,\su(2)\right).
\end{equation}
Importantly they are compatible with the connections $D= d + \bdA$ , $\Dt = d + \bdAt$, in the following sense:
\begin{Proposition}\mbox{}\\
	Let $\bdgS$ (resp $\bdgSt$) be the $\su(2)$-valued self-dual (resp anti-self-dual) two-forms constructed from a metric as \eqref{Chiral Formulations of GR - Fundations: Sigma def (tetrad)}, let $D= d + \bdA$ be the self-dual part of the Levi-Civita connection associated with this metric. Then $D$ is the $\SU(2)$ connection satisfying
\begin{equation}\label{Chiral Formulations of GR - Fundations: Sigma/D compatibility}
d_{\bdA} \Big(\bdgS \Big)= d\bdgS + [\bdA , \bdgS ]  =0, \qquad d_{\bdAt} \left(\bdgSt \right)= d\bdgSt + [\bdAt , \bdgSt ]   =0.
\end{equation}
\end{Proposition}
\begin{proof}
See Appendix \ref{section: Appdx Decomposition of the Curvature} for a direct proof in coordinates.
\end{proof}
 This compatibility relations are important as they can be used as alternative definition for the chiral connection $D$ and $\Dt$.

Finally we can write the Einstein equations in terms of those two-forms. If $D=d + \bdA$ is the `left' or `self-dual' connection and $D^2 = \bdF$ its curvature, then we can rewrite the first half of the bloc decomposition \eqref{Chiral Formulations of GR - Fundations: Riemann decomposition 2} as
\begin{equation}\label{Chiral Formulations of GR - Fundations: F, Ft decomposition 2}
D^2 = F^i\,\gs^i = \left( F^{ij} \gS^j + G^{ij}\gSt^j \right)\gs^i.
\end{equation}
Then, as we already discussed, the self-dual part of Weyl curvature is \begin{equation}
\Psi^{ij} =F^{ij} - \frac{1}{3}tr F \gd^{ij},
\end{equation} the scalar curvature is $4 \gL= 4tr F$ and
\begin{equation}\label{Chiral Formulations of GR - Fundations: Chiral Einstein Equations 2}
\text{g is Einstein if and only if} \quad D^2 = M^{ij} \gS^j \,\gs^i.
\end{equation}

\section{Definite Connections and Gravity} \label{section: Definite Connections and Gravity}

We review here how to write equations for Einstein-anti-self-dual metric in terms of connections. This is a well known construction (cf  \cite{Capovilla:1990qi}) and we here use the terminology of \cite{Fine:2008},\cite{Fine:2011}. We also briefly recall how to write equations for full Einstein gravity in terms of connections from \cite{Krasnov:2011pp},\cite{Fine:2013qta}.

We now take $\bdA = A^i \,\gs^i$ to be the potential in a trivialisation of a $SU(2)$-connection $D =d + \bdA$ and $D^2 =\bdF = F^i \,\gs^i $ its curvature.

\subsection{Definite Connections}\label{ssection: Definite Connections}

We mainly consider definite connections, i.e connections such that the curvature two-form is a definite triplet:
\begin{Definition}{\emph{Definite Connections}}\label{Definite Connections and Gravity: Def: Definite Connections}\mbox{} \\
	A $SU(2)$-connection $D = d + A^i \,\gs^i$, is called \emph{definite} if the conformal metric, $\tilde{X}^{ij} = F^i\W F^j \big/d^4x$, constructed from its curvature, $D^2=F^i \,\gs^i$, is definite. 
\end{Definition}
For any $SU(2)$-connection with potential $A^i \,\gs^i$, there is a unique conformal class of metric $\tilde{g}_{\text{\tiny(F)}}$ such that the curvature $F^i \,\gs^i$ is self-dual. The definiteness of the connection then ensures that this conformal metric is invertible and of Euclidean signature (cf Def \ref{Chiral Formulations of GR - Fundations: Def: definite triplet} and Prop \ref{Proposition: Urbantke metric} ). Thus definite connections are associated with a `good' metric.

A definite connections also defines a notion of orientation. It is done by restricting to volume form $\mu_+$ such that $\tilde{X}^{ij} = F^i \W F^j \big /\mu_+ $ is \emph{positive} definite. In the following whenever there is a need for an orientation, we will always take this one.

We can also assign a \emph{sign} to a connection as follows: We consider co-frame $\left(e^I\right)_{I\in 0..3}$, orthonormal with respect to the Urbantke metric and oriented with the convention that we just described. They are defined up to Lorentz transformations and rescaling by a \emph{positive} function. From this tetrad we can construct a basis of self-dual two-form $\left(\gS^i \right)_{i\in 1,2,3}$ through the relation \eqref{Chiral Formulations of GR - Fundations: Sigma def (tetrad)}. Again $\left(\gS^i\right)_{i\in 1,2,3}$ is defined up to $SU(2)$ transformations and rescalings by \emph{positive} functions.
By construction, the curvature $D^2 =F^i \,\gs^i$ is self-dual for the associated Urbantke metric and we can thus write
\begin{equation}
D^2 = F^i \,\gs^i = M^{ij} \gS^j \,\gs^i.
\end{equation}
The sign of the connection is then defined as $s = \text{sign}\left(det\left(M\right)\right)$.
Note that this notion of sign makes sense as a result of $\left\{F^i\right\}_{i\in1,2,3}$ being defined up to $SU(2)$ transformations and $\left\{\gS^i\right\}_{i\in1,2,3}$ being defined up to $SU(2)$ transformations and positive rescaling.

We now have two $SU(2)$ transformations independently acting on $\left(F^i\right)_{i\in1,2,3}$ and $\left(\gS^i\right)_{i\in1,2,3}$, the first as a result of changing the trivialisation of the $SU(2)$ principal bundle of whom $D= d+ \bdA$ is a connection, the second as a result of changing the trivialisation of the bundle of self-dual two-forms associated with the Urbantke metric. Those two bundles can be identified (at least locally) by requiring $M^{ij}$ to be a definite symmetric matrix. Finally we also have two scaling transformations, one acting on $\Xt$ and the other one on $\gS$, we identify them by requiring that $ F^i \W F^j = \Xt^{ij} \; \frac{1}{3} \gS ^k \W \gS^k$.

In what follows these identifications will always be assumed unless we explicitly specify otherwise.

As a result of $\Xt^{ij}$ being definite we can make sense of its square root. In fact there is a slight ambiguity in this definition: we fix it by requiring $\sqrt{X}$ to be positive definite, i.e we take the positive square root.

With these choices of square root and identifications, we have
\begin{equation}
F^i= s\, \sqrt{X}^{ij}\gS^j \qquad \Leftrightarrow \qquad \gS^i = s \, \sqrt{X}^{-1}{}^{ij} F^j, \qquad \forall i \in 1,2,3.
\end{equation}

\subsection{Anti-self-dual gravity and Perfect Connections}\label{ssection: Self-dual gravity and Perfect Connections}

A metric is said to be `anti-self-dual' if the self-dual part of its Weyl curvature vanishes ie, if $W_+=0$ in \eqref{Chiral Formulations of GR - Fundations: Riemann decomposition 2}. 
As Weyl curvature is conformally invariant, this is a property of the conformal class of the metric rather than from the metric itself. 

A metric is Einstein-anti-self-dual if it is Einstein and anti-self-dual, ie if $W_+=0$ , $G=0$ in \eqref{Chiral Formulations of GR - Fundations: Riemann decomposition 2}. Alternatively, using \eqref{Chiral Formulations of GR - Fundations: F, Ft decomposition 2}, if
\begin{equation}\label{Definite Connections and Gravity: ASD metric def}
F^i = \frac{\gL}{3} \gS^i , \qquad \forall i \in 1,2,3.
\end{equation}
(then $4\gL$ is the scalar curvature). Note in particular that for $\gL\neq0$, $F^i\W F^j \big/ d^4x \propto \gd^{ij}$.

This motivates the following definition,
\begin{Definition}{\emph{Perfect Connections}}\mbox{}\\
	A definite connection is perfect if $F^i\W F^j = \gd^{ij} \frac{F^k \W F^k}{3}$.
\end{Definition}

The relevance of this definition comes from the following:
\begin{Proposition}\label{Proposition: Perfect connection}\mbox{}\\
	The Urbantke conformal metric associated with a perfect connection is anti-self-dual. What is more the representative with volume form $ \frac{3}{2\gL^2} F^k \W F^k$ is anti-self-dual-Einstein with cosmological constant $s \,|\gL|$, where $s$ is the sign of the connection.
\end{Proposition}

\begin{proof}\mbox{} \\
	Consider the Urbantke metric with volume form $\mu = \frac{3}{2 \gL^2} F^k \W F^k$. It is associated with a orthonormal basis of two-form $\left\{\gS^i\right\}_{i\in1,2,3}$ as in \eqref{Chiral Formulations of GR - Fundations: Sigma def (tetrad)}. By construction, they are such that $\gS^i \W \gS^j = 2 \gd^{ij} \; \mu$. 
	Together with our identification of the scaling transformations, $ F^i\W F^j = \Xt^{ij} \frac{1}{3} \gS^k \W \gS^k$, it gives
	\[ F^i\W F^j = 2\Xt^{ij} \mu.\]
	Now by hypotheses,
	\[  F^i \W F^j = \frac{\gd^{ij}}{3} F^k \W F^k = 2\gd^{ij} \frac{\gL^2}{9} \mu, \]
	from which we read $X^{ij} = \frac{\gL^2}{9} \gd^{ij}$ and
	\begin{equation}\label{Definite Connections and Gravity: Proof, SD gr eq1}
	F^i = s \, \sqrt{X}^{ij}\gS^j = s \, \frac{|\gL|}{3} \gS^i.
	\end{equation}
	From this last relation we see that Bianchi identity, $d_{\bdA} \bdF=0$, is now equivalent to $d_{\bdA} \bdgS= d\bdgS + [\bdA, \bdgS] =0$ which is the defining equation \eqref{Chiral Formulations of GR - Fundations: Sigma/D compatibility} of the self-dual connection. It follows that $D= d + \bdA$ is the self-dual connection of the Urbantke metric with volume form $\mu =\frac{3}{2 \gL^2} F^k \W F^k$. With this observation \eqref{Definite Connections and Gravity: Proof, SD gr eq1} are just the field equations for Einstein anti-self-dual gravity \eqref{Definite Connections and Gravity: ASD metric def} with cosmological constant $s\,|\gL|$.
\end{proof}

\subsection{Pure connection formulation of Einstein equations}\label{ssection: Pure connection formulation of Einstein equations} 
At this point it is hard to resist writing down the pure connection formulation of Einstein equations.\\

Consider a definite $SU(2)$-connection $D=d + \bdA$ with curvature $\bdF=F^i\,\gs^i$. As already explained, it is associated with an orientation, a sign $s$ and conformal class of metric $\tilde{g}_{\text{\tiny(F)}}$. We again denote $F^i\W F^j = \tilde{X}^{ij} d^4x$ and define the following volume form,
\begin{equation}
\frac{1}{2\gL^2}\left(Tr\sqrt{F\W F}\right)^2 \coloneqq \frac{1}{2\gL^2}\left(Tr\sqrt{\tilde{X}}\right)^2 d^4x .
\end{equation}
This is a well defined expression as a result of the following facts: the definiteness of the connection together with the orientation make $\tilde{X}^{ij}$ positive definite and thus we can take its square root, what is more $\left(Tr\sqrt{\tilde{X}}\right)^2$ being homogeneous degree one in $\tilde{X}$ the overall expression does not depends on the representative of the density $\tilde{X}$.

However there are signs ambiguity in this choice of square root. They amount to the choice of signature of the conformal metric $\sqrt{X}^{ij}$. We will always take this choice of square root such that $det \left(\sqrt{X}\right) >0$, then the only signatures that remains are $(+,+,+)$ and $(+,-,-)$. We thus need to make a choice for our definition of square root once and for all: either we stick with the `definite square root' or with the `indefinite square root'. 

\begin{Definition}{\emph{Einstein Connections}}\label{Chiral Formulations of GR - Fundations: Definition- Einstein connection} \mbox{}\\
	If $A^i$ is a definite connection, define $X^{ij}$ by the relation
	\begin{equation}\label{Definite Connections and Gravity: Proof, pure connection eq1}
	F^i \W F^j = 2 X^{ij}\frac{1}{2\gL^2}\left(Tr\sqrt{F\W F}\right)^2 .
	\end{equation}
	Then we will call it Einstein if
	\begin{equation}\label{Definite Connections and Gravity: Pure connection Einstein equations}
	d_A \left( \left(\sqrt{X}\right)^{-1}{}^{ij} F^j \right)=0.
	\end{equation}
\end{Definition}
Again, the two square roots in this definition need to be taken with the same convention, ie such that the resulting matrices have the same signature: either $(+,+,+)$ ( `definite square root') or indefinite $(+,-,-)$ (`indefinite square root'). Note that for perfect connections, $X^{ij} = \gd^{ij}\frac{\gL^2}{9}$, as a result of which perfect connections are special case of Einstein connections with the `definite square root' convention (note that perfect connections are \emph{not} Einstein connections for the `indefinite square root' as $d_A \left( \left(\sqrt{X}\right)^{-1}{}^{ij} F^j \right) \neq 0$ for $\sqrt{X}= diag\left(1,-1,-1\right)$).

The Definition \ref{Chiral Formulations of GR - Fundations: Definition- Einstein connection} is motivated by the following,
\begin{Proposition}{\emph{ Krasnov \cite{Krasnov:2011pp}}}\label{Proposition: Pure connection equation}\mbox{}\\
	For an Einstein connection, the Urbantke metric with volume form $\frac{1}{2\gL^2}\left(Tr\sqrt{F \W F}\right)^2$ is Einstein with cosmological constant $|\gL| \text{Sign}\left(s \, Tr\sqrt{F \W F} \right) $. What is more such a connection coincides with the self-dual Levi-Civita connection of the metric.
\end{Proposition}

\begin{proof} \mbox{}\\
	The metric in Urbantke conformal class with volume form $\nu=\frac{1}{2\gL^2}\left(Tr\sqrt{F \W F}\right)^2$ is associated with an orthonormal basis of self-dual two-form $\left\{\gS^i\right\}_{i\in1,2,3}$, $\gS^i\W \gS^j = 2 \gd^{ij}\nu$. It is defined up to $SU(2)$ transformation. By definition, $\left\{F^i\right\}_{i\in1,2,3}$ is a basis of self-dual two-forms for Urbantke metric and $F^i = M^{ij} \gS^j \; \forall i\in \{1,2,3\}$.
	
	As was already pointed out, \emph{a priori} $F$ and $\gS$ are valued in two different associated $SU(2)$ bundle: $D=d+A$ is a $SU(2)$ connection on a $SU(2)$ principal bundle P and the curvature naturally takes value in the adjoint bundle $ P \times_{SU(2)} \su(2)  $,  on the other hand $\left\{\gS^i\right\}_{i\in1,2,3}$ is a trivialisation of the bundle of self-dual two-forms associated with the Urbantke metric.
	
	We now come again to the subtle question of identifying the two: this can be done (at least locally) by requiring $M^{ij}$ to be a symmetric matrix. Once this is done, however there is still the possibility of acting with the diagonal transformation $\left(\gS^1, \gS^2, \gS^3 \right) \to \left(\gS^1, -\gS^2, -\gS^3 \right)$ and we thus have two possible identifications. We call them the `definite identification' and the `indefinite identification' depending whether or not the resulting matrix $M^{ij}$ is definite or not.
	
	As a rule, we now take the identification corresponding to the square root that we chose, ie if one chooses the `definite square root', we take the `definite identification'; on the other hand, if one takes the `indefinite square root' one should use the `indefinite identification'.
	
	Finally, just as in the case of perfect connections, we identify rescaling of $\Xt$ and rescaling of $\gS$ by imposing that $F^i\W F^j =  \Xt^{ij} \;\frac{1}{3} \gS^k \W \gS^k$. Together with the choice of volume form, $\gS^i \W \gS^j = 2 \gd^{ij} v$, this completely fixes all the scaling freedom: $F^i \W F^j = 2 X^{ij} \nu$. Note that this gives the same result as in definition \ref{Chiral Formulations of GR - Fundations: Definition- Einstein connection}.
	
	As a consequence of these different choices we have
	\begin{equation}
	F^i= s\, \sqrt{X}^{ij}\gS^j \qquad \Leftrightarrow \qquad \gS^i =  s\, \sqrt{X}^{-1}{}^{ij} F^j.
	\end{equation}
		The field equations \eqref{Definite Connections and Gravity: Pure connection Einstein equations} now read $d_{\bdA} \bdgS=0$ which are just the the defining equations \eqref{Chiral Formulations of GR - Fundations: Sigma/D compatibility} of the self-dual connection. It follows that $D=d + \bdA$ is the self-dual connection of the Urbantke metric with volume form $\nu$. Having this in mind, $F^i= s\, \sqrt{X}^{ij}\gS^j$, are Einstein equations \eqref{Chiral Formulations of GR - Fundations: Chiral Einstein Equations 2} with cosmological constant $s\,Tr\left(\sqrt{X}\right)$ . Finally, from \eqref{Definite Connections and Gravity: Proof, pure connection eq1}, one gets $|Tr\left(\sqrt{X}\right)| = |\gL|$.
\end{proof}

Note that one of the weakness of this formulation is that a particular choices of square root (ie `definite' or `indefinite') can only describe a particular subspace of Einstein metric, those such that the self-dual Weyl curvature $F^{ij}$ is respectively definite or indefinite. 

Interestingly, the integral of the volume form $\frac{1}{2\gL^2}\left(Tr\sqrt{F \W F}\right)^2$ also gives the correct variational principle for Einstein connections. This is the \emph{pure connection action} for GR \cite{Krasnov:2011pp}. It can be obtained by integrating fields successively from the Plebanski action, see also next section.

 \section{Chiral Deformations of Gravity} \label{section: Chiral Deformations of Gravity}
 
 While chiral formulations of GR described above are certainly known to differential geometers specialising in Einstein manifolds, the related `chiral deformations' of the Einstein theory are almost completely unknown to the community. It is however an interesting fact that the four-dimensional Einstein condition can be non-trivially deformed in a chiral way.
 
 On the one hand, it is well-known that GR can be modified, the simplest example of such a modification being the $R^2$ gravity, of relevance, e.g., as a good model of inflation \cite{Starobinsky:1980te, Ade:2015lrj}. However, this model is equivalent to GR coupled to an additional scalar field, and so propagates not just the two polarisations of the graviton, as in GR, but also a scalar. One can then consider more involved modifications of GR with higher powers of the curvature added to the Lagrangian. However, one can quickly convince oneself that, because of the higher derivatives present in these modified theories, they all propagate more degrees of freedom than does GR. Following this logic, if one insists on second-order field equations then GR is the unique theory of metric, at least in four dimensions. This is the content of several GR uniqueness theorems available in the literature. 
 
Consequently, it comes as a big surprise that it is indeed possible to modify GR without adding extra degrees of freedom if one starts from one of its chiral descriptions. The resulting chiral deformations of GR continue to have second-order field equations and a count of the number of degrees of freedom by the Hamiltonian analysis shows that they just describe the two propagating polarisations of the graviton. What is more and as will be reviewed below, there is  an \emph{infinite-parametric} class of such chiral modified gravity theories, in which GR is just a special member. The reason why it does not contradict the above discussion is that these theories, when rewritten in metric terms, exhibit an infinite number of higher derivative terms with precise coefficients. Each of these terms taken individually would lead to extra degrees of freedom, but taken altogether they `conspire' to forbid these extra propagating modes. 
 
 We should also stress that the type of modifications of gravity we are interested in here is unique in the following sense: One can inspect the proofs of the GR uniqueness, notably the modern proofs that deal with the scattering amplitudes, and note the particular assumptions in those proofs that are violated by these chiral deformations. Removing those assumptions, one can see that there results a new `uniqueness' theorem stating that these chiral modifications are the only ones that describe propagating gravitons with second-order field equations; see \cite{Krasnov:2014eza}.
 
 In this section, we first review the description of this infinite family of gravity theories in 4D can be described in terms of $\SU(2)$-connections. This description will be useful in Part \ref{Part: Variations on Hitchin Theory in Six Dimensions} of this thesis. The material discussed here is mainly from \cite{Krasnov:2011up, Krasnov:2011pp}, see also \cite{Fine:2013qta} for a more mathematical exposition. We then describe alternative action principles for these chiral deformations. Some of them appeared in the literature in the last ten years \cite{Krasnov:2008fm,Krasnov:2009iy,Krasnov:2011up} but some of them are related to more recent work, see \cite{Herfray:2015rja,Herfray:2015fpa} and \cite{Herfray:2016qvg}.
 
 \subsection{Chiral Deformations of Gravity}\label{ssection : Chiral Deformations of Gravity}
 
 We here describe chiral deformations of gravity in their most concise form, the pure connection formulation. It generalises the pure connection formulation of Einstein equations described above (see \ref{ssection: Pure connection formulation of Einstein equations}).

\subsubsection{Pure connection formulation of the Chiral Deformations of GR}
 
Let again $\bdA = A^i \gs^i$ be a definite $\SU(2)$-connection for a $\SU(2)$ principal bundle over a 4-dimensional manifold $M$,
\begin{equation}
\SU(2) \inj P \to M
\end{equation}
and  $\bdF = F^i \gs^i$ be its curvature two-form. The definiteness of the connection, as defined in \ref{ssection: Definite Connections}, amounts to the definiteness of $\Xt^{ij} \in M_{3}\left(\R \right)$ defined by
\begin{equation}\label{Chiral Deformations of Gravity: X def}
F^i \W F^j = \Xt^{ij} d^4x \,.
\end{equation}

A choice of chiral deformation of GR now amounts to a choice of $\SU(2)$-invariant function
 \begin{equation}
f \from  M_{3}\left(\R \right) \times \gO^4(M)  \to \gO^4(M).
\end{equation}
We can indeed evaluate such function on \eqref{Chiral Deformations of Gravity: X def} to get a volume form
\begin{equation}\label{Chiral Deformations of Gravity: Vol_f def}
\bdF \W \bdF \mapsto f( \bdF \W \bdF ) \gO^4(M),
\end{equation}
integrating this form against our manifold we obtain a functional:
\begin{equation}\label{Chiral Deformations of Gravity: S_f def}
S_f\left[\bdA \right] \coloneqq \int_{M} f(\bdF \W \bdF ).
\end{equation}

Let us call the above functions `deformation function', for practical purpose the following (equivalent) definition is more convenient:
\begin{Definition}{Deformation function}\label{Chiral Deformations of Gravity: f def}
A \emph{deformation function} is a $\SU(2)$-invariant function $f$ from three by three symmetric matrices to real numbers satisfying the following properties:
\begin{enumerate}
	\item gauge invariance $f(OXO^T)=f(X)$, where $O\in \SO(3)$,
	
	\item homogeneity of degree one, $f(\ga X) = \ga f(X)$ for any $\ga \in \R$ .
\end{enumerate}
\end{Definition}

By construction any deformation function gives a functional for connections, see  \eqref{Chiral Deformations of Gravity: S_f def}: the two above property ensure that it is gauge invariant and well defined.

Clearly, there are many deformation functions. As a count one can diagonalise the matrix $X$, deformations functions are then homogeneity degree one function of the eigenvalues. There are as many such functions as functions of two variables. We already saw that GR is given by
\begin{equation}\label{Chiral Deformations of Gravity: S_GR def}
S_{GR}\left[\bdA \right] = \int_{M} \left(\Tr \left( \sqrt{ \bdF \W \bdF } \right)\right)^2
\end{equation}
see however \ref{ssection: Pure connection formulation of Einstein equations} for a discussion on the precise choice involved in taking the square root.

Critical points of \eqref{Chiral Deformations of Gravity: S_f def} are $\SU(2)$-connections satisfying the following second order PDE's
\begin{equation}\label{Chiral Deformations of Gravity: feqs}
d_A \left( \frac{\partial f}{\partial \Xt^{ij}} F^j \right) = 0 \, .
\end{equation}
Note that the matrix of derivatives of the function $f$ with respect to $\Xt$ is homogeneity degree zero in $\Xt$, and is hence well-defined even though $\Xt$ is really volume-form valued.

\subsubsection{Discussion}

\paragraph{The anti-self-dual sector: Instanton solutions}

In general, solutions to \eqref{Chiral Deformations of Gravity: feqs} strongly depend on the theory. There is however a sector which is shared by all Chiral deformations, mainly the anti-self-dual sector of gravity i.e anti-self dual Einstein metrics:

As we already discussed in \ref{ssection: Self-dual gravity and Perfect Connections} perfect connection i.e satisfying $X^{ij}\sim \gd^{ij}$ give rise to metrics that are self-dual Einstein. 

It can be checked that, for any $f$, these connections are extremal points of \eqref{Chiral Deformations of Gravity: S_f def}, i.e satisfy \eqref{Chiral Deformations of Gravity: feqs}. For perfect connection the field equations \eqref{Chiral Deformations of Gravity: feqs} indeed reduce to the Bianchi identity for the curvature and are thus automatically satisfied.

As a result anti-self-dual Einstein metrics, which correspond to perfect connections, are solutions of \eqref{Chiral Deformations of Gravity: feqs} for any $f$. 

Accordingly all this theories coincide on the anti-self-dual sector of gravity and only differ when the self-dual sector is `turned on'. Thus the name `chiral deformations of GR'.  In particular the De Sitter solution, which is arguably the simplest anti-self-dual Einstein solution, is shared by all the chiral deformations. 

\paragraph{Metric interpretation.}

As already discussed in Proposition \ref{Proposition: Urbantke metric}, a $\SU(2)$-connection that satisfies the rather weak requirement of being definite defines a conformal Riemannian metric on $M$.  As already discussed, the triple of curvature two-forms is anti-self-dual  with respect to this (conformal) metric and this property defines it uniquely.

A choice of deformation function \eqref{Chiral Deformations of Gravity: f def} defines a volume form \eqref{Chiral Deformations of Gravity: Vol_f def}. We can make use of this volume to fix the conformal freedom of Urbantke metric. When the connection satisfies \eqref{Chiral Deformations of Gravity: feqs}, the metric defined by $\bdA$ is constrained. We already saw Einstein connections can be obtained by a proper choice of $f$ cf section \ref{ssection: Pure connection formulation of Einstein equations} and in particular Proposition \ref{Proposition: Pure connection equation}. In this case the Urbantke metric with volume form \eqref{Chiral Deformations of Gravity: Vol_f def} is Einstein and its self-dual connection is $\bdA$.

In general however the metric interpretation of \eqref{Chiral Deformations of Gravity: feqs} is unclear. In particular, $\bdA$ has no reasons to be the self-dual connection of Urbantke metric and nothing forces us to take \eqref{Chiral Deformations of Gravity: Vol_f def} as the volume of the Urbantke metric. There are indeed many other volume forms at hand e.g $\Tr \bdF \W \bdF$. 

Note that on the `anti-self-dual' sector, 
\begin{equation}
f\left(\bdF \W \bdF \right) = f\left(\bdgd \; \frac{\Tr\left(\bdF \W \bdF\right)}{3} \right) = f\left(\bdgd \right) \; \frac{\Tr \left(\bdF \W \bdF\right)}{3}
\end{equation}
so that it does not really matter which volume form we choose we consider.

\paragraph{More general solutions and propagating degrees of freedom.}

Even though we are far from understanding all Einstein metrics on 4-manifolds, some intuition as to how many solutions exist comes from the Lorentzian version of the theory. Indeed, GR with Lorentzian signature is a theory with local degrees of freedom, and so the space of solutions is infinite-dimensional. For example, solutions can be obtained by evolving initial data.

A similar description is also possible in the Riemannian context, in particular in the setting of asymptotically hyperbolic metrics. As is well known from \cite{FeffermanG}, one can indeed solve for asymptotically hyperbolic Einstein metrics in the form of an expansion in powers of the `radial' coordinate. The free data for this expansion are a conformal class of metric on the boundary (modulo boundary diffeomorphisms), together with a symmetric traceless transverse tensor. This second piece appears as free data in some higher order of the expansion. There are $2+2$ free functions on the boundary as free data, and this is the Riemannian analogue of the statement that GR has 2 propagating degrees of freedom.

We developed a similar expansion in the language of connections in \cite{Fine:2015hef}. One outcome of the analysis of this paper is that the expansion is universal for the whole class of theories \eqref{Chiral Deformations of Gravity: S_f def} i.e whatever the function $f$ is. Only the details of the expansion at sufficiently high order in the radial coordinate start to depend on $f$. In the first few orders, the expansion is completely independent of $f$. In particular, the count of free data that seeds the expansion is $f$-independent. This means that the free data to be prescribed to get an asymptotically hyperbolic solution of  theory \eqref{Chiral Deformations of Gravity: S_f def} (locally near the boundary) are $2+2$ free functions on the 3-dimensional asymptotic boundary. This illustrates the statement that the theory \eqref{Chiral Deformations of Gravity: S_f def} has as many solutions as GR.

\paragraph{Lorentzian signature and the physical significance of these deformations.}

 Once GR gets embedded into an infinitely large class of gravity theories all with similar properties, one is forced to ask a number of questions: What makes GR unique as compared to all these other theories? In fact, as all the chiral deformations of GR approximately look the same around DeSitter space could it be that the world we live is only approximately described by GR? The very fact that such chiral modified gravity theories exist forces us to understand them.
 
 Isotropic cosmological solutions are anti-self dual Einstein solutions and, as such shared by all the chiral deformations of GR. In the other hand, anisotropic cosmological solutions will depend on the particular theories considered. In \cite{Herfray:2015fpa} we considered the influence of simple modification terms on the cosmological singularity of `Kasner' cosmological model. This is interesting because such solutions are believed to encode the behaviour of a generic space-like singularity. We considered a simple type of deformation with the following property: as long as the Weyl curvature is small in Planck units solutions essentially behave like GR. Typically, there is then a regime `far away in time' from the would-be singularity where the solutions all look the same and approximate the Kasner solutions.  As the Weyl curvature increase, the modification start to show up and the deformed solutions run further and further away from GR. In simple case it is easy to choose these modifications in such a way that there is no singularity at all. Then `far away in time' before the would-be singularity the solution again approximate another Kasner solution.
 
 In general, however, one faces the following difficulty. In the Euclidean case, the use of definite $\SU(2)$-connections ensured the existence of an Euclidean Urbantke metric. On the other hand, when dealing with Lorentzian metric one should rather consider $\SL(2,\C)$-valued connection, i.e complex valued field. One then needs appropriate `reality conditions' to ensure the reality of the associated Urbantke metric. When considering GR, this procedure is (reasonably) well understood. In the case of chiral modifications of GR this is however not understood how to modify these `reality conditions'. In a sense, since the connection is the central field in these formulations and the metric interpretation of the field equation is generically obscure, one could take the position that whether or not the derived metric is real does not matter. Weather or not this attitude makes sense can only be decided by coupling those deformed theories with matter, which, as far as we are aware as never been seriously probed.
 
 Whatever the attitude one takes, this is clear that finding reality conditions for $\SL(2,\C)$-connection that adapted to a generic chiral deformation as such that the resulting Urbantke metric is real with Lorentzian signature is an open question.
 
 Having said that, we should also remark that there are many situations where the chiral deformations behave perfectly sensibly and admit the usual interpretation in terms of a real-valued space-time Lorentzian metric. This is typically the case when one considered solutions with particular symmetries, see for spherically symmetric solutions \cite{Krasnov:2007ky} and \cite{Herfray:2015fpa} for anisotropic cosmological solutions.
 In these situations the physical effects of the modification can be studied unambiguously, and have been studied, in particular in a paper \cite{Herfray:2015fpa} including the author of this thesis. We refrained from describing the results of this paper here because it would take us too far from the main line of development of this thesis

\subsection{Variational Principles}\label{ssection : Variational Principles}

In the above we gave a pure connection formulation of the Chiral deformation of GR \eqref{Chiral Deformations of Gravity: S_f def}. This deformations where parametrised by a choice of deformation-function $f$, GR itself being given by a particular representative \eqref{Chiral Deformations of Gravity: S_GR def}. We chose this presentation as it is the most compact one but many other descriptions of these chiral deformation exists. In fact, there are just as many way of describing these deformations as there are chiral formulations of GR. We now take some time to describe these formulations. 

These different actions are easily seen to be equivalent to the above pure connection action by integrating the relevant fields. As for for the GR ones, having the tools from section \ref{section: Chiral Formulations of GR - Fundations} and section \ref{section: Definite Connections and Gravity} in hand it should be easy to the reader to convince himself that they indeed describe GR metrics. For more details, see however the reference given below.

\subsubsection{Plebanski-like actions, $S[\bdA,\Psi, \bdB]$: }\label{section Appdx, ss: The Plebanski Action}

The most basic way to describe these chiral deformations is in terms of the following Lagragian
\begin{equation}\label{Chiral Deformations of Gravity: S[A,B,Psi]}
S[\bdA,\bdB,\Psi] = \int B^i \W F^i - \left(\Psi^{ij} + \frac{\gL\left(\Psi\right)}{3}\gd^{ij}\right) \frac{1}{2}B^i \W B^j.
\end{equation}
It is not a very economical action as is contains a lot of fields: a $\SU(2)$-connection $\bdA$ (which does not need to be a definite connection at this stage), a $\su(2)$-valued two-form $\bdB$ (again we do not need to require this triplet of two-forms to be a definite triplet) and a symmetric traceless field: $\Psi^{(ij)} = \Psi^{ij}$.

Here $\gL\left(\Psi \right)$ is any function of $\Psi$, it parametrise the chiral deformations. 

In this form GR is simply the special case where $\gL = cst$. One then obtain the Plebanski action for General Relativity see \cite{Plebanski:1977zz}, \cite{Capovilla:1991qb} for the original references. 

Despite the large number of fields involved, this formulation of the chiral deformation of GR is however interesting for the intuitive picture it gives of these otherwise unintuitive theories: For solutions to the Plebanski's action, i.e $\gL=cst$, $\Psi^{ij}$ is just the self-dual part of the curvature of the Einstein metric and $\gL$ its cosmological constant. Accordingly one can think of chiral deformations as theories where the cosmological constant in not a constant any more but rather a function of the Weyl curvature. This also makes it clear that any of these `chiral deformations' will behave as GR in the regime of not too high Weyl curvatures. In this regime one can indeed expand the function lambda of psi and keep only the constant part. This shows that the chiral deformations are UV modifications of GR that keep the number of its propagating DOF unchanged.

\subsubsection{Intermediate actions of the type $S[A,\Psi]$ : }\label{section Appdx, ss: Intermediate action (A, Psi)} 

Starting from the Plebanski-like action \eqref{Chiral Deformations of Gravity: S[A,B,Psi]}, the most direct way to see the equivalence with the pure connection action \eqref{Chiral Deformations of Gravity: S_f def} is to integrat out $\bdB$ and $\Psi$ in this order. The resulting intermediate action is
\begin{equation}\label{Chiral Deformations of Gravity: S[A,Psi]}
S[\bdA,\Psi] = \frac{1}{2}\int \left(\left(\Psi +\frac{\gL\left(\Psi\right)}{3} \gd \right)^{-1}\right)^{ij} F^i\W F^j
\end{equation}

Here again $\gL\left(\Psi \right)$ parametrise the deformation and $GR$ is recovered for $\gL= cst$.

As an interesting variant, that was first described in \cite{Herfray:2015fpa}, one can use a Lagrange multiplier $\mu \in \gO^4(M)$: 
\begin{equation}
S[\bdA,\mu, M] = \int M^{ij} F^i\W F^j + \mu \; g\left( M^{ij}\right)
\end{equation}
Each constraint $g\left(M^{ij}\right) =0$ will give a different theory. In particular, one can recover GR by considering
 \begin{equation}
g\left(\Psi\right) = Tr\left(M^{-1} \right) - \gL.
\end{equation}
Where $\gL$ is a constant. Interestingly one can easily describe `anti-self-dual gravity' in this formulation. Taking as a constraint $g\left(g\right) = Tr M$, one indeed finds that the field equation for $M$ forces the connection to be perfect.

At this point, the equations obtained by varying $M$ and $\mu$ can formally be solved by taking $M = f\left(\bdF \W \bdF \right)$ and $\mu = f'\left(\bdF \W \bdF \right)$ where $f$ and $f'$ are some functions that can in principle be computed from $g\left(M\right)$. The resulting volume is of the form $f''\left(\bdF \W \bdF \right)$ where $f''$ is some function that can in principle be computed from the above action and parametrizes the family of chiral deformations.

\subsubsection{Intermediate actions of the type $S[A,B]$}

There is a longer, but just as interesting, way to obtain the pure connection formulation of chiral deformations from \eqref{Chiral Deformations of Gravity: S[A,B,Psi]}. Instead of integrating out $\bdB$ one can indeed try to integrate out $\Psi$. The equations obtained by varying $\Psi$ can indeed be formally solved as $\Psi = f\left(\bdB \W \bdB \right)$, where $f$ is some function constructed from $\gL\left(\psi\right)$ that we leave implicit. The resulting action then is of the form
\begin{equation}\label{Chiral Deformations of Gravity: S[A,B]}
S\left[\bdA, \bdB\right] = \int \Tr\left(\bdB \W \bdF \right) + V\left( \bdB \W \bdB \right).
\end{equation}
This action reads like a `BF plus potential' type of action. Once again the `potential' $V\left(\bdB \W \bdB \right)$ is a free function parametrising the possible chiral deformations of GR. This action was first discussed in \cite{Krasnov:2009iy}.

 The particular potential necessary to describe GR is however not so easy to derive from the Plebanski-like action and was rather guessed and first discussed in \cite{Herfray:2015rja}(see however \cite{Celada:2016iah} for a derivation from a more complicated Lagrangian):
 \begin{equation}\label{Chiral Deformations of Gravity: S_GR[A,B]}
S_{GR}\left[\bdA, \bdB\right] = \int B^i \W F^i + \frac{\eps - \gL}{3}\; \frac{1}{2}\left( Tr \sqrt{B \W B} \right)^2 - \frac{\eps}{2} B^i\W B^i
\end{equation}

For generic $\eps$ and $\gL$ this action describes Einstein metrics with scalar curvature $4 \gL$. As described in \cite{Herfray:2015rja}, this action also has the following nice property: for $\eps=0$ one recovers again anti-self-dual gravity. Thus full gravity can be obtained from self-dual gravity by the addition of the simple $\bdB \W \bdB$ term.

By formally integrating out the $\bdB$ field one recovers the pure connection formulation of chiral deformations of GR. There is however an interesting intermediary step that is worth considering.

It is indeed convenient to parametrise the $\bdB$ field in terms of the tetrad $(e^I)_{I \in 0..3}$ associated with its Urbantke metric $\gt_{\bdB}$ and symmetric traceless square-matrix $\Mt = \Mt^{ij} \; \gs_i \otimes \gs_j$ such that
\begin{equation}
\bdB = \left(\Mt^{ij} + \gd^{ij}\right)  \gS^j\left(e \right)\; \gs_i
\end{equation}
where
\begin{equation}
\left(\gS^i(e) = -e^0 \W e^i - \frac{\eps^{ijk}}{2} e^j \W e^k\right)_{i\in 1,2,3}
\end{equation}
Note that there was an overall scaling freedom in the choice of the metric that we fixed by taking the factor in front of the delta function to be one. (One can always choose $\sqrt{\Mt}$ to be symmetric as it amounts to a particular identification of the $\SU(2)$-bundle with the bundle of self-dual two-forms for $\gti_{\bdB}$).

With this parametrisation, the above action reads,
\begin{equation}\label{Chiral Deformations of Gravity: S[A,e,M]}
S\left[\bdA, \Mt, e \right] = \int \gS^i(e) \W F^i +  \Mt^{ij} \gS^i \W F^j + V\left(\Mt\right) Vol(e).
\end{equation}

\subsubsection{The self-dual Palatini-like action $S[A,e]$}

In order to integrate out $\Mt$ from \eqref{Chiral Deformations of Gravity: S[A,e,M]}, it is convenient to introduce the following parametrisation for the curvature $\bdF(\bdA)$,
\begin{equation}
\bdF = \left(\sqrt{\Xt}^{ij} \gS^i + G^{ij} \gSt^j \right) \gs_i
\end{equation}
Where $\Xt$ is again a square-matrix $\Xt = \Xt^{ij} \gs_i \otimes \gs_j$. Inserting this definition into \eqref{Chiral Deformations of Gravity: S[A,e,M]} we obtain,
\begin{equation}
S\left[\bdA, \Mt, e \right] = \int \gS^i(e) \W F^i+ \Tr \left(\left(\Mt\sqrt{\Xt}\right) + V\left(\Mt\right)  \right)\; Vol(e).
\end{equation}
Varying with respect to $\Mt$, one obtains a set of equations that can be formally solved as $\Mt = f \left(\Xt \right)\; Vol(e)$, where again $f$ is some function left implicit. The resulting action is of the general form
\begin{equation}\label{Chiral Deformations of Gravity: S[A,e]}
S[\bdA,e] = \int \gS^i(e) \W F^i + \gL\left(\Xt\right)\; Vol(e)
\end{equation}
where $\gL(\Xt)$ is the free function parametrising the chiral deformations of GR. For $\gL =cst$ one recovers the self-dual Palatini action for GR with cosmological constant $\gL$. The self-dual Palatini action (or Ashtekar action) really is the covariant side of the canonical description of gravity in terms of Ashtekar variables (see \cite{Jacobson:1988yy}, \cite{Peldan:1993hi} for a precise derivation of the constraints). 

From this formulation, one again recovers the interpretation that chiral deformation of GR morally correspond to allowing the cosmological constant to be a function of the (self-dual) weyl curvature.

Integrating out the tetrad gives the pure connection formulation. One the other hand, integrating out $\bdA$ from \eqref{Chiral Deformations of Gravity: S[A,e]} gives a pure metric formulation of the chiral deformations. This last formulation was discussed in \cite{Krasnov:2009ik}. In the case of GR this is however easy and one obtains the usual Einstein Hilbert action.

\subsubsection{Intermediate action of type $S[A,M]$}

Starting back at \eqref{Chiral Deformations of Gravity: S[A,e,M]} one can instead integrate out the metric. The resulting field equations say that $e$ is a tetrad for the Urbantke metric $\gti_{\bdF}$ constructed from the curvature and with volume form $\left(\Tr \sqrt{ \bdF \W \bdF} \right)^2$:
\begin{equation}
S\left[\bdA, \Mt \right] = \int \Mt^{ij}\; \gS_A^i \W F^j + V\left( \Mt\right) \;\left(\Tr \sqrt{ \bdF \W \bdF} \right)^2  .
\end{equation}
 GR can be obtained as
\begin{equation}\label{Chiral Deformations of Gravity: S_{GR}[A,M]}
S[\bdA, \Mt] = \int \Mt^{ij}\; \gS_A^i\W F^j  - \frac{\eps}{2}\; \Tr (\Mt^2) \;\left(\Tr \sqrt{ \bdF \W \bdF} \right)^2.
\end{equation}
For $\eps\neq0$ this action describes gravity. In the case where $\eps=0$ this action describes anti-self-dual gravity. See section \ref{ssection:  new action for Gravity as a background invariant generalisation of the Chalmers-Siegel action.} for a direct proof that this action indeed describe gravity.

\chapter{Twistors}\label{Chapter: Twistors}

In this chapter, we wish to clarify that twistor theory is closely related to the above chiral formulations of GR. In particular, the description of self-dual Einstein metrics in terms of $\SU(2)$-connections has a very nice twistor counterpart in the form of the Non-Linear-Graviton theorem. This perspective leads to a new proof on the Euclidean version of this theorem. In turn this suggests new (though unsuccessful) approaches to constructing a twistor action for GR. We first review the Euclidean Twistor space of a Riemannian four dimensional manifold. This is done from a rather traditional `metric' perspective. We then come back on these results and reinterpret them from a `connection' point of view. Finally we close this chapter with a discussion on the still on-going chase for a twistor action for full GR.

The twistor/space-time correspondence is usually presented in complexified terms, we here decided to stick with the Euclidean signature. One the one hand this is coherent with the preceding chapter where we saw that chiral description of GR accommodate most easily this signature: then the relatively simple notion of \emph{definite connection} (see definition \ref{Definite Connections and Gravity: Def: Definite Connections}) was enough to construct Euclidean metric. On the other hand, the Euclidean Twistor space naturally has the structure of fibre bundle with structure group $\SU(2)$ and thus is most suited to a description in terms of $\SU(2)$ connections.

\section{Euclidean Twistor Space: Traditional Approach} \label{section: Euclidean metric Twistor Space}

We now review the geometry of the Twistor space $\T(M)$ associated with a Riemannian manifold $\left(M, g\right)$. In this Euclidean signature setting, this is just the primed spinor bundle over $M$:
 \begin{equation}
 \p_{A'} \inj \T(M) \xto{P} M.
 \end{equation}
 The correspondence between space-time points and twistor-space points is then just a projection: To any space-time $x$ point corresponds a complex line $P^{-1}(x) \simeq \C^2$ (parametrised by a spinor $\p_{A'}$) but a twistor-space point $z$ is sent by the projection on a unique space-time point $x=P(z)$. This is as opposed to the complexified case where a point in twistor-space was associated to a two-dimensional surface in (complexified)space-time. In some sense the Euclidean perspective weaken the non-locality of the twistor construction. This is unsatisfactory as non-locality was the most prominent feature of the theory. On the other hand the twistor-space of a general complexified space-time only exists for anti-self-dual metric while Euclidean twistor space always makes sense. It is then natural to start building up one's intuition with the Euclidean case with the hope that new insight obtained here can be generalised to other signatures.
 
 What is more there is some beauty of its own to twistor theory in the Euclidean setting (see e.g \cite{Atiyah:1978wi}, \cite{Woodhouse85}, \cite{LeBrun04} for nice expositions). First in the (conformally) flat case, it is tightly tied up to the geometry of quaternions via the Hopf bundle construction,
 \begin{equation}
 \Hbb \to \Hbb^2 \simeq \T \to \HP^1 \simeq \S^4.
 \end{equation}
Second, the Euclidean twistor space can be understood as the bundle of all possible (metric compatible) almost complex structures on $M$. This leads to the possibility of constructing a complex structure on $\T(M)$.
We now briefly review how this works.

\subsection{The Flat Case from Quaternions} \label{ssection : The Flat Case from Quaternions}
We first describe the \emph{flat twistor space} i.e the twistor space $\T$ associated with the conformal compactification $\S^4$ of the four dimensional Euclidean space. In the flat case, the twistor space has a beautiful interpretation in terms of quaternion geometry that we now review.

Our presentation will be non-standard in the following sense:
In the usual (complexified) approach to twistor theory (see e.g \cite{Huggett:1986fs}, \cite{Penrose_vol2}, \cite{Ward:1990vs}, \cite{Mason:1991rf}) one usually goes as follows: starting with a four dimensional complex vector space $\T\simeq \C^4$, `the twistor space', one construct the compactified (complex) space-time $\M_{\C}$ as the space of two-planes in $\T$. This gives the twistor correspondence where points in $\M_{\C}$ are planes in $\T$ and points in $\T$ are $\ga$ planes in $\M_{\C}$. Only then does one usually introduce a reality structure that picks up a particular signature: in the Euclidean case this is an anti-involution $\circonf \from \T \to \T$ with no fixed points. Euclidean space-time points $\Ebb$ are then taken to be planes in $\T$ that are left invariant by this anti-involution. It turns out that through any point $z$ in $\T$ passes a unique such plane (this is the plane going through the origine, $z$ an $\zh$) which therefore gives a projection $P \from \T \to \Ebb$.

As opposed to this approach, we here take $\left(\T, \circonf \right)$ as our starting point and interpret $\circonf$ as a quaternionic structure on $\T$. Practically it allows to identify $\T$ with $\Hbb^2$, then the four-sphere is directly constructed by taking a quotient $\S^4 \simeq \HP^1$. This approach culminates in the explicit realisation of the exceptional isomorphisms
\begin{equation}
Conf\left(\S^4\right) \simeq \PSL\left(2,\Hbb \right), \quad Isom\left(\S^4\right) \simeq \Sp(2)/\left\{\pm \Id\right\}.
\end{equation}

Even though the material here is known from experts, the presentation is somewhat original in the sense that we are not aware that it appears anywhere as such in the traditional literature on the subject (e.g \cite{Atiyah:1978wi}, \cite{Woodhouse85}). See however \cite{Baez:2001dm} for a general discussion on the division algebras and their relation to exceptional isomorphisms.

\subsubsection{Quaternion Geometry}
They are many descriptions of a quaternions $\mathbb{H}$ (also called Hamilton's numbers) : in terms of matrix, spinors, or in more abstract terms. The most common starting point is to describe quaternions as hyper-complex numbers:
\begin{equation}\label{Euclidean metric Twistor Space: q}
q = q_0 + \iq\;q_1 + \jq\;q_2 + \kq\;q_3 \in \Hbb
\end{equation}
Here $\left(q_0,q_1,q_2,q_3\right) \in \R^4$ while $\jq,\kq$ are generalisations of the unit imaginary number $\iq$, satisfying the algebra $\jq^2=\kq^2=-1$, $\iq \jq \kq=-1$ and anti-commuting which each others. The quaternion algebra is synthesized in Table \ref{Euclidean metric Twistor Space: Quaternion table} and Figure \ref{Euclidean metric Twistor Space: Quaternion circle}.

\begin{figure}[h]
\centering
\arrayrulecolor{white}
\arrayrulewidth=1pt
\begin{tabular}{| >{\columncolor{blue!40!white}}c| >{\columncolor{blue!20!white}}c| >{\columncolor{blue!20!white}}c| >{\columncolor{blue!20!white}}c| >{\columncolor{blue!20!white}}c| >{\columncolor{blue!20!white}}c|}
	\rowcolor{blue!40!white}
	 &	1&	$\iq$&	$\jq$&	$\kq$\\
	  \hline
	1&	1&	$\iq$&	$\jq$&	$\kq$\\
	$\iq$&	$\iq$& -1& $\kq$& -$\jq$\\
	$\jq$&	$\jq$& -$\kq$& -1&	$\iq$\\
	$\kq$&	$\kq$&	$\jq$& -$\iq$& -1\\	
	\hline
\end{tabular}
\captionof{table}{Multiplication Rules for Quaternions. (This table reads from left to right. e.g: $\iq \jq = \kq$.)}
\label{Euclidean metric Twistor Space: Quaternion table}

	\includegraphics[width=0.25\textwidth]{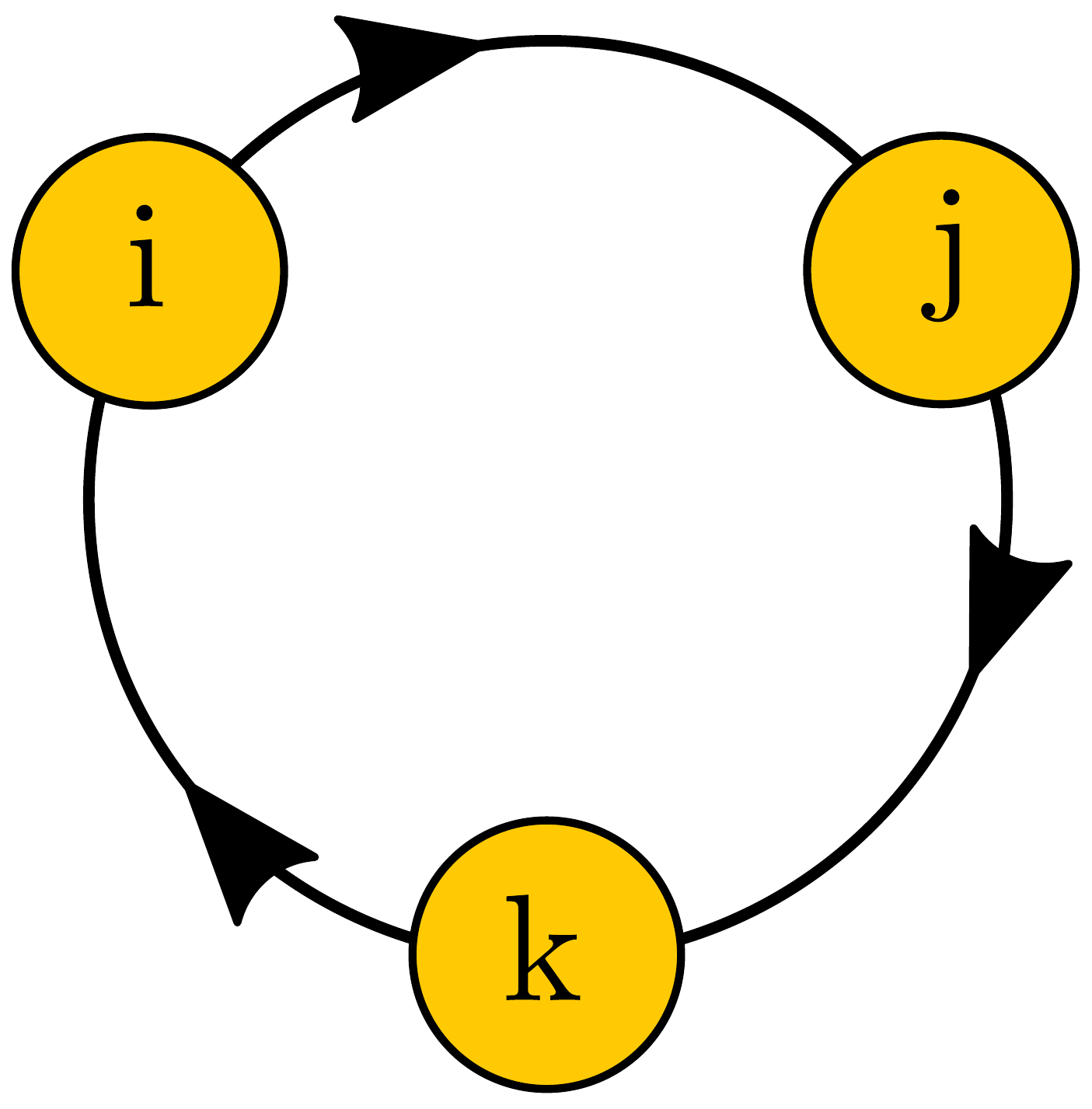}
\captionof{figure}{A picture mnemonic for the multiplication of unit quaternions. E.g $\iq \jq = \kq$, $\jq \kq = \iq$, etc.}
\label{Euclidean metric Twistor Space: Quaternion circle}
\end{figure}

An alternative useful point of view is to think of quaternions $\Hbb$ as a complexification of complex numbers $\C$ (by the same procedure octonions $\Obb$ can be obtained as complexification of $\Hbb$). A practical way of doing the identification $\Hbb \simeq \C^2 $ is to rewrite \eqref{Euclidean metric Twistor Space: q} as :
\begin{equation}\label{Euclidean metric Twistor Space: q = a+b}
q = \ga + \jq\; \gb, \qquad \left(\ga, \gb\right) =  \left(q_0 + \iq\; q_ 1, q_2 - \iq \;q_3\right) \in \C^2.
\end{equation}

One can define a conjugation operation (i.e an involution on $\Hbb$) as
\begin{equation}\label{Euclidean metric Twistor Space: quaternions conjugation}
\bar{q} = q_0 - \iq\; q_1  - \jq\; q_2 - \kq\; q_3 = \ga^* -\jq\; \gb.
\end{equation}
For any $p,q \in \Hbb$, it satisfies $\overline{pq} = \bar{q} \bar{p} $ and gives a metric structure on quaternions through\footnote{In particular, \begin{equation}
	|q|^2 = q_0^2 + q_1^2 + q_2^2 +q_3^2 = |\ga|^2 + |\gb|^2.
	\end{equation} and it follows that $q^{-1} = \qb / |q|^2= \frac{\left( \ga^* - \jq \gb \right)}{|\ga|^2 + |\gb|^2} $}
\begin{equation}\label{Euclidean metric Twistor Space: quaternion metric}
\bra p, q \ket \coloneqq \frac{1}{2}\left(\qb p + \pb q\right) = p_0 q_0 + p_1 q_1 + p_2 q_2 + p_3 q_3.
\end{equation}

As opposed to complex numbers, quaternions are non-commutative. They, however preserve associativity (contrary to octonions) this allows for a matrix representation as the matrix group $U(2,\C)$: In terms of the notation \eqref{Euclidean metric Twistor Space: q = a+b}, this is easily done as
\begin{equation}\label{Euclidean metric Twistor Space: quaternion matrix}
q \simeq  \begin{pmatrix}
\ga & -\gb^* \\
\gb & \ga^*
\end{pmatrix} \in U(2,\C) .
\end{equation}
What is more, the isomorphism \eqref{Euclidean metric Twistor Space: quaternion matrix} also identifies unit quaternions $U(1,\Hbb)$ (i.e such that $ q \bar{q} = 1$) with $SU(2,\C)$ matrices,
\begin{equation}
U(1,\Hbb) \simeq SU(2,\C).
\end{equation}

It is also easy to see that the action of $SU(2) \times SU(2) \simeq U(1,\Hbb)\times U(1,\Hbb)$ on quaternions defined by
\begin{equation}\label{Euclidean metric Twistor Space: quaternions SO(4) action}
(u,v).q = u q \bar{v}, \quad \left(u,v\right) \in U(1,\Hbb)\times U(1,\Hbb) 
\end{equation}
preserves the metric \eqref{Euclidean metric Twistor Space: quaternion metric} thus giving a concrete realisation of the isomorphism $SO(4) \simeq \left(SU(2)\times SU(2) \right)/{\pm \Id}$

One the other hand, if one restricts oneself to the left $SU(2)$ action only, quaternions can be identified with 2-spinors and their natural $SU(2)$ action:
\begin{equation}\label{Euclidean metric Twistor Space: spinor representation of quaternions}
q \simeq \begin{pmatrix} \ga \\ \gb \end{pmatrix} \in \C^2.
\end{equation}
This is then clear that the `hat operation', defined by
\begin{equation}\label{Euclidean metric Twistor Space: ^ on quaternions}
\hat{q} = q \;\jqb= \gb^* - \jq\; \ga^* 
\end{equation}
is preserved by this action. Clearly the hat operation is an anti-involution. 

Because we restricted ourselves to left $\SU(2)$ transformations there must be an invariant hermitian metric. This is constructed as follows. First we define a skew-symmetric `dot product' on quaternions
\begin{equation}\
\begin{array}{ccc}
\Hbb^2 & \to & \C \\
\left(x,y\right)& \mapsto & x.y
\end{array}\end{equation}
 by the relation
\begin{equation}
\xb y \coloneqq  \xh.y + \;\jq \; x.y  \qquad \forall x,y \in \Hbb.
\end{equation}
In practical terms, if
\begin{equation}
x = \ga + \jq \;\gb \quad\text{and}\quad y = a + \jq\; b
\end{equation}
then
\begin{align}
x.y = -y.x =  \ga b -\gb a.
\end{align}
Finally, combining the dot product with the hat operation we obtain the Hermitian metric on quaternions
\begin{equation}
\xh.y =  \ga^* a + \gb^* b.
\end{equation}

\subsubsection{Quaternionic Structure}

In most situations we will not be working directly with quaternions but rather with vector spaces equipped with a quaternion structure.

A quaternionic structure on a \emph{complex} vector space $V \simeq \C^{2n}$ is given by an operator $J$ on $V$ which is both anti-linear, $J(\gl v) = \gl^* J(v), \forall \gl \in \C$ and an anti-involution $J^2=-1$. This is also common, and essentially equivalent, to describe a quaternionic structure on a \emph{real} vector space $V \simeq R^{4n}$ in terms of two almost complex structures $I$ and $J$, $I^2=J^2=-1$, which anti-commute $IJ=-JI$. There is then a whole 2-sphere of almost complex structure as any tensor of the form
\[ 
x I + y J + z JI, \;\text{such that}\; x^2+y^2+z^2=1
\]
indeed squares to minus identity. We will however here stick to the first point of view. 

One can always choose a basis on $V \simeq \C^{2n}$ such that the action of $J$ is
\begin{equation}
\begin{array}{ccc}
X=\begin{pmatrix}
\ga_{1} \\ \gb_{1} \\ ...\\ \ga_{n} \\ \gb_{n}
\end{pmatrix} \in \C^{2n}
& \mapsto &
J(X)=\begin{pmatrix}
\gb^*_{1} \\ -\ga^*_{1} \\ ...\\ \gb^*_{n} \\ -\ga^*_{n}
\end{pmatrix} \in \C^{2n}.
\end{array}
\end{equation}
(compare with \eqref{Euclidean metric Twistor Space: ^ on quaternions})
In effect it realises the identification $V \simeq \C^{2n}\simeq \Hbb^n$,
\begin{equation}
\begin{array}{ccc}
X = \begin{pmatrix}
\ga_{1} \\ \gb_{1} \\ ...\\ \ga_{n} \\ \gb_{n}
\end{pmatrix} \in \C^{2n}
& \mapsto &\begin{pmatrix}
x_1 \\ ...\\ x_n
\end{pmatrix}=
\begin{pmatrix}
\ga_{1} + j \gb_{1} \\ ...\\ \ga_{n} + j \gb_{n}
\end{pmatrix} \in \Hbb^{n}
\end{array}
\end{equation}
(compare with \eqref{Euclidean metric Twistor Space: q = a+b}). This identification really is defined up to the subgroup of $GL(2n,\C)$ preserving $J$ and is clearly isomorphic to $GL(n,\Hbb)$. The action of $J$ in the $\Hbb^{n}$ representation is just a left multiplication by $\jqb$, $JX \simeq X \jqb$.

As a result a complex vector space $V \simeq \C^{2n}$ equipped with a quaternionic structure has both a $\GL(n,\Hbb)$-action on the left and a $\Hbb$-action on the right.

Let us now consider more structure in the form of a compatible skew-symmetric complex-bilinear form on $V \simeq \C^{2n}$, $\go \in \gL^2(V)$. This bilinear product is said to be compatible with the quaternionic structure $J$ if \begin{equation}\label{Euclidean metric Twistor Space: compatibility of 2forms with quaternionic structure}
\go(J(.),J(.)) = \left(\go(.,.)\right)^*.
\end{equation}
This is a useful condition because then
 \begin{equation}
 g(.,.) = \go(J(.), .)
 \end{equation}
  is a Hermitian product on $V \simeq \C^{2n}$. In general it will not be definite. Through the identification $V \simeq \C^{2n}\simeq \Hbb^n$ the compatible metric $g$ then becomes an Hermitian structure on $\Hbb^{n}$. By a proper choice of basis one can always put this metric in the canonical form $diag(s_1 ... s_n), \; s_i\in\{-1,0,1\}$. Practically, in this basis \footnote{Here, \[ \begin{array}{ccc}
 		Y = \Mtx{
 			a_{1} \\ b_{1} \\ ...\\ a_{n} \\ b_{n}
 	} \in \C^{2n}
 		& \simeq &\Mtx{
 			y_1 \\ ...\\ y_n
 		}=
 		\Mtx{
 			a_{1} + \jq\; b_{1} \\ ...\\ a_{n} + \jq\; b_{n}
 		} \in \Hbb^{n}
 	\end{array} \]},
\begin{align}
g\left(X,Y\right) = \sum_{i=1}^{i=n} s_i\;\left( \ga^*_{i} a_{i} + \gb^*_{i} b_{i}\right) \in \C
\end{align} then
\begin{equation}
 \go\left(X,Y\right) = \sum_{i=1}^{i=n} s_i\;\left( \ga_{i} b_{i} - \gb_{i} a_{i}\right) \in \C
\end{equation}
and altogether
\begin{equation}
g\left(X,Y\right) + \jq\; \go\left(X,Y\right) = \sum_{i=1}^{i=n} s_i\;\bar{x}_i y_{i} \in \Hbb.
\end{equation}
 The subgroup of $GL(2n,\C)$ preserving both a two-form and a hermitian metric is $USp(2n,\C) = U(2n,\C) \cap Sp(2n,\C)$. From the discussion above it is clearly isomorphic to $U(n,\Hbb)$. This group is more commonly referred to as $Sp(n)$:
 \[ 
 Sp(n) \simeq U(n,\Hbb) \simeq  U(2n,\C) \cap Sp(2n,\C).
  \]

\paragraph{Example: Euclidean Spinors} The simplest example of quaternion structure is that of a two dimensional complex vector space $S \simeq \C^2$ together with an anti-linear, anti-involutive operator $J$. It is best to think of $S$ as the space of spinors $\go^{A}\in S$. In this context we will write $J =  \circonf $. Choosing an adapted basis identifies $S$ with $\Hbb$ as
\begin{equation}
\begin{array}{ccc}
S\simeq C^2 & \to &\Hbb\\
\go^{A}=\Mtx{
\ga \\ \gb
} & \mapsto & \go = \ga + j\gb.
\end{array}
\end{equation}
Then the hat operator is just quaternionic multiplication by $\jqb$ on the right,
\begin{equation}
\begin{array}{ccc}
S & \to &\Hbb\\
\goh^{A}=\begin{pmatrix}
\gb^* \\ -\ga^*
\end{pmatrix} & \mapsto & \goh = \go \jqb.
\end{array}
\end{equation}
Transformations preserving $J$ form the group $GL(1,\Hbb)\simeq \C \times SU(2,\C)$. 

 If one makes a choice of compatible two-form $\eps$ \footnote{We raise and lower spinor indices according to the usual rules $\go_{A}= \go^{B} \eps_{BA}$, $\go^{A}= \eps^{AB}\go_{B}$.} we obtain a hermitian product:
\begin{equation}
\eps\left(\goh, \go \right)  = \goh_{A}\,\go^{A} =  \ga^* \ga + \gb^* \gb= \bar{\go}\,\go.
\end{equation}
Here the compatibility condition \eqref{Euclidean metric Twistor Space: compatibility of 2forms with quaternionic structure} reads
\begin{equation}
\eps_{AB} \in M_{2}\left(\R \right)
\end{equation}
i.e is equivalent to $\eps$ having real coordinates.

As already discussed the subgroup of $GL(1,\Hbb)$ preserving the hermitian metric is $Sp(1) \simeq U(1,\Hbb) \simeq SU(2,\C)$.

We will come back rapidly to the next simplest case $\C^4 \simeq \Hbb^2$ as it is just the structure of the twistor space of the four-sphere $\S^4$. Before that, it is best to introduce the last piece of the puzzle and make a short detour to describe Hopf bundles.

\subsubsection{Hopf Bundles}

The Hopf bundles are fibre bundles made up of spheres only i.e of the form $\S^p \inj \S^q \to \S^r$. It turns out that the only possible cases are
\begin{align*}
\S^1 &\inj \S^3 \to \S^2 \\
\S^3 &\inj \S^7 \to \S^4 \\
\S^7 &\inj \S^{15} \to \S^8. \\
\end{align*}
If one defines $\S^0$ to be the set of two points $\{-1, 1\}$ one could also add to this list \begin{equation*}
\S^0 \inj \S^1 \to \S^1
\end{equation*}
but this is a bit of a singular case and we won't consider it here.

The first of these Hopf bundles, the  fibration of $\S^3$ by circles $\S^1$ certainly is the most famous and is pictured in Figure \ref{Euclidean metric Twistor Space: Hopf Fibration}.

\begin{figure}[h]
	\centering
	\includegraphics[width=0.8\textwidth]{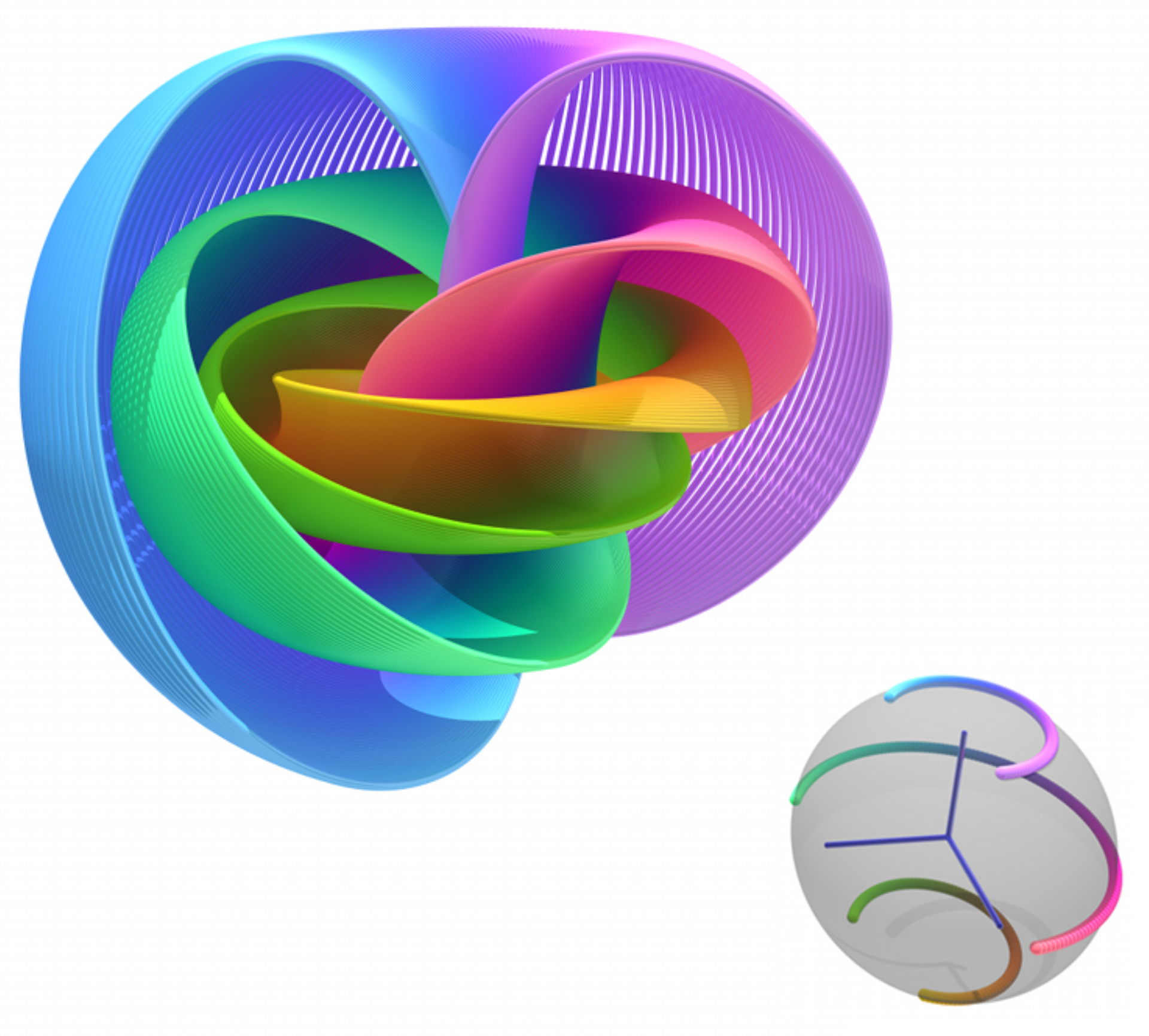}
	
	\caption[]{A pictural representation\footnotemark of the simplest Hopf fibration: the three-sphere $\S^3$ is thought as a three-ball where all the points on the boundary are identified (i.e this is essentially a variant of the stereographic projection). It is fibred by circles of different colours. The two-sphere in the bottom-right corner represents the base space.}\label{Euclidean metric Twistor Space: Hopf Fibration}
\end{figure}
\footnotetext{This nice piece of art is due to Niles Johnson and can be found on the Wikipedia page untitled "Hopf fibration".}

That these Hopf bundles exist in finite numbers and for very special dimensions might seem miraculous at first sight. The miracle will seem however less surprising once one realises that it is directly related to the existence of complex number $\C$, quaternions $\Hbb$ and octonions $\mathbb{O}$ respectively. Indeed Hopf bundles are easily seen to be consequences of the `tautological bundles' above the following projective spaces
\begin{align*}
\C &\inj \C^2 \to \CP^1 \\
\Hbb &\inj \Hbb^2 \to \HP^1 \\
\Obb &\inj \Obb^2 \to \OP^1. \\
\end{align*}
So that there is only one miracle, the existence of division algebras. The existence of the preceding bundles is also responsible for the existence of numerous `accidental isomorphisms' of Lie groups in low dimensions. Of particular importance for us are:
\begin{table}[h]
	\centering
	\arrayrulecolor{white}
	\arrayrulewidth=1pt
	\begin{tabular}{ >{\columncolor{blue!40!white}}l >{\columncolor{blue!20!white}}c >{\columncolor{blue!20!white}}c}
		\rowcolor{blue!40!white}
	
Action on $\S^n$&  n= 2 & n=4 \\
Conformal group & $SO(3,1) \simeq PSL(2,\C)$ & $SO(5,1) \simeq PSL(2,\Hbb)$ \\
Isometry group   &  $SO(3) = SU(2,\C)/_{\pm\Id}$ & $SO(5) \simeq Sp(2)/_{\pm\Id} = U(2,\Hbb)/_{\pm\Id}$
\end{tabular}
\arrayrulecolor{black}
\end{table}

The first and most famous Hopf bundle
\begin{equation*}
\S^1 \inj \S^3 \to \S^2
\end{equation*}
 is thus related to complex numbers and its geometry can be nicely described in spinor notation as we will soon recall. This is the simplest example and it is often illuminating to keep it in mind when considering its higher dimensional counterpart. The second Hopf bundle
 \begin{equation*}
 \S^3 \inj \S^7 \to \S^4
 \end{equation*}
  is essentially the Euclidean twistor space over the four-sphere and is tied up with quaternions and will be our main interest. We will not touch to the last Hopf bundle (over the eight-sphere).

\paragraph{Example: Euclidean Spinors and the Riemann Sphere} As a baby example, let us briefly consider the simplest case $\C \inj \C^2 \to \CP^1$. The idea is to construct a metric (resp a conformal metric) on the quotient space $\CP^1$ such that the action of the isometry group $\SU(2,\C)$ (resp the conformal group $\PSL(2,\C)$) acts linearly on the total space $\C^2$. It will serve as a warming up exercise as essentially the same methods will be used in the slightly less trivial twistor construction.

Here again we take $S\simeq \C^2$ to be a two dimensional complex vector space that we think of as the space of unprimed spinors. Thus let $\go^{A}$ be coordinates on $S$. The particular dimension of $S$ allows to use a skew-symmetric tensor $\eps_{AB}$ (defined up to a global factor) to raise and lower indices. We can then construct the following metric (defined up to a global scaling).\footnote{Here and thereafter $A \odot B= A \otimes B + B\otimes A$. }
\begin{equation}\label{Euclidean metric Twistor Space: CP^1 conformal metric}
\gti_{\CP^1} =  \frac{1}{2}\go^*_{A}\;d\go^*{}^{A} \odot \go_{B}\;d\go^{B}.
\end{equation}
This metric vanishes in the Euler directions (that generates complex rescaling)
\begin{equation}\label{Euclidean metric Twistor Space: CP^1 Euler direction on CP^1}
E = \go^{A}\frac{\pa}{\pa \go^{A}}, \qquad  E^* = \left( \go^{A}\frac{\pa}{\pa \go^{A}}\right)^*,
\end{equation}
and what is more the Lie derivative along this directions just rescale the metric. One therefore obtains a conformal metric on $\CP^1$, in fact it is conformally equivalent to the round metric, see below. The subgroup of $\GL(2,\C)$ preserving $\eps_{AB}$ is $\SL(2,\C) \simeq \Sp(2,\C)$. As a result, the action of $\SL(2,\C)$ on spinors preserves the metric \eqref{Euclidean metric Twistor Space: CP^1 conformal metric} and thus realises the isomorphism $\PSL(2,\C) \simeq Conf(\S^2) \simeq \SO(3,1)$.

We now suppose that $S$ is equipped with a quaternionic structure $\circonf \from S \to S$ and take $\eps_{AB}$ to be real so that the two structures are compatible. We then have a hermitian product on $S$, $\bra \go,\go \ket = \goh_{A} \go^{A} \coloneqq \goh.\go$. Now consider
\begin{equation}\label{Euclidean metric Twistor Space: FS metric on CP^1}
g_{\CP^1} =  4 R^2 \;\frac{\goh_{A}\;d\goh{}^{A} \odot \go_{B}\;d\go^{B}}{ 2\left(\goh . \go\right)^2}.
\end{equation}
As compare to \eqref{Euclidean metric Twistor Space: CP^1 conformal metric}, the Lie derivative of this metric in the Euler directions \eqref{Euclidean metric Twistor Space: CP^1 Euler direction on CP^1} vanishes and it therefore descends to the round metric on $\CP^1$. Note that it does not depend on the precise scaling of $\eps$. The precise numerical factor in front of this expression has been chosen so that the radius of $\S^2$ is $R$.
\begin{ExtraComputation}
	\begin{framed}
 (For the n-sphere of radius $R$ the scalar curvature $S$ is $S=n(n-1)/R^2$. Accordingly the scalar curvature is here $2/R^2$).
 \end{framed}
\end{ExtraComputation}

  Again $Sp(1) = U(1,\Hbb) \simeq SU(2)$ preserves the hermitian structure on $S \simeq \Hbb$. This thus realises 
  \begin{equation}
  Sp(1)/_{\pm \Id} \simeq SO(3).
  \end{equation} 

One could also have directly started with the metric on $S\simeq \Hbb$
\begin{equation}
g_{S} = \frac{1}{2} d\gob \odot d\go = \frac{1}{2} d\goh_{A} \odot d\go^{A}.
\end{equation}
playing with indices\footnote{All one needs is the identity $\gd^{A}{}_{B} \;\goh.\go = \go^{A}\goh_{B} - \goh^{A}\go_{B}$.} this can be rewritten in a form that fit with the bunlde structure $\C \inj \C^2 \to \CP^1$
\begin{equation}
g_{S} = -\frac{\goh_{A} d\go^{A} \odot \go_{B} d\goh^{B}}{2 \;\goh . \go} + \frac{\goh_{A} d\goh^{A} \odot \go_{B} d\go^{B}}{2\;\goh . \go}
\end{equation}
i.e the first term is a metric along the fibre while the second is proportional to Fubini-Study metric \eqref{Euclidean metric Twistor Space: FS metric on CP^1} on $\CP^1$.

\begin{ExtraComputation}
	\begin{framed}
Alternatively, making use of the identification $\C^2 \simeq \Hbb$ a useful set of coordinates on $\C^2$ is
\begin{equation}
 \go^{A} \simeq \go = r g \in \Hbb, \quad\text{with}\; r \in \R\quad \bar{g}g=1.
\end{equation}
Where $g \in U(1 \Hbb)$ is a unit quaternion and is best thought as a coordinates on $SU(2) \simeq S^3$, e.g the Maurer-Cartan form on $SU(2)$ is $g^{-1}dg$.
Then the quaternionic metric reads
\begin{equation}\label{Euclidean metric Twistor Space: Quaternion- metric2}
g_{S} = \frac{1}{2}d\bar{\go} \odot d \go = dr^2 + r^2\left(dg\right)^2.
\end{equation}

The two formulas are related by
\begin{equation}
r^2= \goh.\go,\qquad \left(g^{-1} dg\right)^{AB} = \frac{1}{\goh.\go}\left(\go^{(A}d\goh^{B)}- \goh^{(A}d\go^{B)}\right).
\end{equation}
	\end{framed}
\end{ExtraComputation}

\subsubsection{The Twistor space of $\S^4$}
We now come to the twistor space of the four-sphere $\S^4$. This is just the total space of the Hopf bundle 
 \begin{equation}
\Hbb \inj \Hbb^2 \to S^4 \simeq \HP^1.
\end{equation}
Just as in the above example we want to construct a metric (resp a conformal metric) on $\HP^1$ such that the isometry group $\Sp(2)$ (resp the conformal group $\SL(2,\Hbb)$) acts linearly.

As a starting point, we take the `flat twistor space' $\T \simeq \C^4$ to be a four dimensional complex vector space. What is more, we take $\T$ to be equipped with a quaternionic structure in the form of an anti-linear, anti-involutive, `hat' operator 
\begin{equation}
\circonf \from \T \to \T.
\end{equation}
As previously described, it allows us to identifies $\T$ with $\Hbb^2$ up to an action of $GL(2,\Hbb)$ on the left. We thus introduce quaternionic coordinates on $\T$ as $(\p , \go) \in \Hbb^2$. We will move freely from complex coordinates $\Z^{\ga} \in \T \simeq \C^4$ to quaternionic ones $(\p , \go) \in \T \simeq \Hbb^2$.  The quaternionic structure also allows us to define a quaternion multiplication on the right: \begin{equation}\label{Euclidean metric Twistor Space: Quaternions- right H action}
\forall \; q=\ga + j \gb \in \Hbb,\; \forall \; Z^{\ga}\in \T, \qquad  Z^{\ga}.q^{-1} \coloneqq \frac{Z^{\ga}\; \ga^* +\Zh^{\ga} \gb}{|\ga|^2+|\gb|^2}  = \left(\p q^{-1} , \go q^{-1} \right).
\end{equation}
Quotienting by this action we obtain points on $S^4 \simeq \HP^1$. They are best written in terms of homogeneous coordinates $[Z^{\ga}] = \left[\pi , \go \right] $.

\paragraph{Conformal Structure on $\S^4$} The four-sphere can now be given a conformally flat metric as follows. Making use of the four dimensional skew-symmetric tensor $\eps_{\ga\gb,\gc\gd}$ (defined up to a scale), we introduce a metric (also defined up to a global rescaling) on $\T$ as
\begin{equation}\label{Euclidean metric Twistor Space: Flat Twistor- Conformal metric}
\tilde{g}_{S^4} = \eps_{\ga\gb\gc\gd} \Zh^{\ga} Z^{\gb} d\Zh^{\gc} \odot dZ^{\gd}.
\end{equation}
Now this metric has four degenerate directions
\begin{equation}\label{Euclidean metric Twistor Space: Flat Twistor- rescaling directions}
Z^{\ga} \frac{\pa}{\pa Z^{\ga}},\quad \Zh^{\ga} \frac{\pa}{\pa \Zh^{\ga}}, \quad  \Zh^{\ga} \frac{\pa}{\pa Z^{\ga}}, \quad Z^{\ga} \frac{\pa}{\pa \Zh^{\ga}}.
\end{equation}
These vector fields generate the action on the right of $\Hbb$ on $\T$ (cf  eq \eqref{Euclidean metric Twistor Space: Quaternions- right H action}). What is more it can be checked that the Lie derivative of \eqref{Euclidean metric Twistor Space: Flat Twistor- Conformal metric} along the two first vector fields of \eqref{Euclidean metric Twistor Space: Flat Twistor- rescaling directions} just rescale the metric while the Lie derivative along the last two is zero.  As a result this metric descends to a non degenerate conformal metric on $S^4 \simeq \HP^1$. 

This is clear that the $GL(2,\Hbb)$ action on $\Hbb^2$ preserves this conformal metric. We will see shortly that it is in fact the usual conformally flat metric on $S^4$. It follows that the action of $PSL(2,\Hbb)$ on $\HP^1 \simeq S^4$ is just the action of the conformal group of the four sphere $Conf(4) \simeq SO(5,1)$, thus realizing the exceptional isomorphisms
 \begin{equation}
PSL(2,\Hbb) \simeq SO(5,1).
\end{equation}

It is now convenient to introduce stereographic coordinates on $S^4 \simeq \HP^1$. Let us choose a point $I^{\ga} \in \HP^1$, and take a properly adapted quaternionic coordinate system such that $I^{\ga} = \left[0 , 1 \right]$. We can then write any point of $\HP^1\!\!\smallsetminus\!\! \{I\}$ as $\left[\pi , \go \right] = \left[1 ,X \right]$. Here $X \in \Hbb$ are the stereographic coordinates on $\HP^1\!\!\smallsetminus\!\! \{I\} \simeq \R^4$. By construction, we have the euclidean version of the \emph{incidence relation}:
\begin{equation}
\go = X \pi.
\end{equation}
The incidence relation relates point `space-time' with points of the twistor space. Making use of this coordinates and restricting \eqref{Euclidean metric Twistor Space: Flat Twistor- Conformal metric} to $\p = cst$, one indeed obtains the conformally flat metric on $S^4$,
\begin{equation}
ds^2 \propto d\bar{X} \odot dX.
\end{equation}

\paragraph{Infinity Twistor}
If, on top of the quaternionic structure, one is given an `infinity twistor' i.e a compatible complex-bilinear form on $\T$, $I \in \gL^2(\T)$ such that $I(\hat{Y},\hat{Z}) = \left(I(Y,Z)\right)^*$ one obtains an hermitian structure $g(X,Y) = I(\hat{X},Y)$ on $\T$. In a suitable basis,
\begin{equation}\label{Euclidean metric Twistor Space: Flat Twistor- quaternionic metric}
g(Z,Z)= I_{\ga\gb} \Zh^{\ga} Z^{\gb} = \gpb\p + \gL\;\gob\go.
\end{equation}
Here $\gL \in \R$ is a parameter whose sign characterises the signature of the metric.

As a general rule we will write the contraction by the infinity twistor with a dot product:
\begin{equation}\label{Euclidean metric Twistor Space: Flat Twistor- quaternionic metric2}
\Zh.W \coloneqq \Zh^{\ga}W^{\gb} I_{\ga\gb}, \qquad \forall Z,W \in \T.
\end{equation}

We can use the infinity twistor to fix the conformal scaling in \eqref{Euclidean metric Twistor Space: Flat Twistor- Conformal metric}
\begin{equation}\label{Euclidean metric Twistor Space: Flat Twistor- S^4 metric}
g_{\S^4} = 12\;\frac{\eps_{\ga\gb\gc\gd} \Zh^{\ga} Z^{\gb} d\Zh^{\gc} \odot dZ^{\gd}}{\left( \Zh.Z\right)^2}
\end{equation}
This is indeed straightforward to check that the Lie derivative of this metric along the four directions \eqref{Euclidean metric Twistor Space: Flat Twistor- rescaling directions} vanishes. In other terms this metric is invariant under quaternionic multiplication and therefore gives a proper metric on $\HP^1$. In stereographic coordinates,
\begin{equation}\label{Euclidean metric Twistor Space: Flat Twistor- S^4 metric2}
g_{\S^4} = 12 \;\frac{ d\bar{X} \odot dX}{2\left(1+ \gL \; |X|^2 \right)^2}.
\end{equation}
Here the precise numerical factor has been chosen such that the scalar curvature is constant with value $4\gL$. 
\begin{ExtraComputation}
	\begin{framed}
In $n$-dimension, the metric
 \begin{equation}
g = 4(n-1)\;\frac{ dX^I \otimes dX^I}{\left(1+ \gL X^2\right)^2}
\end{equation}
has scalar curvature $S = n\; \gL$, and therefore cosmological constant $\gL$. For $\gL>0$ this is the $n$-sphere with radius $R^2 =(n-1)/\gL$.
	\end{framed}
\end{ExtraComputation}

In general the metric \eqref{Euclidean metric Twistor Space: Flat Twistor- S^4 metric} is only well defined where $\Zh.Z \neq0$. Points on $\HP^1$ such that $\Zh.Z =0$ are then `at infinity', thus the name of the infinity twistor. There are three different possibilities:\\

For $\gL=0$, the infinity twistor is degenerate and therefore factorises $I_{\ga\gb} = \hat{I}_{[\ga}I_{\gb]}$. There is a unique point at infinity written $[I^{\ga}]$ and the metric \eqref{Euclidean metric Twistor Space: Flat Twistor- S^4 metric}  is the flat metric on $\R^4 \simeq \HP^1\!\smallsetminus\!\! \{I\}$.

For $\gL <0$, the infinity twistor has signature $\left(1,1\right)$ and there are "null directions". The metric \eqref{Euclidean metric Twistor Space: Flat Twistor- S^4 metric} is the hyperbolic metric on the four-dimensionnal Poincarré `four-ball'. Then the action of $\U\left(1,1;\Hbb \right)$ on $\Hbb^2$ preserves this metric and it follows that $\U\left(1,1,\Hbb \right)/_{\pm \Id}$ is the group of isometry of the hyperbolic space $\bdH^4$, i.e $\U\left(1,1,\Hbb \right)/_{\pm \Id} \simeq \SO(4,1)$.

For $\gL>0$, \eqref{Euclidean metric Twistor Space: Flat Twistor- quaternionic metric} is the usual hermitian metric on $\Hbb^2$ and the metric \eqref{Euclidean metric Twistor Space: Flat Twistor- S^4 metric} is the round metric on the four-sphere with radius $R^2 = 3/\gL$.  The action of $\Sp(2) = \U(2,\Hbb)$ on $\T$ preserves this structure and thus the metric \eqref{Euclidean metric Twistor Space: Flat Twistor- S^4 metric}. It follows that the action of $\Sp(2)/_{\pm \Id}$ on $\HP^1\simeq \S^4$ are isometries of $\S^4$, thus realising the exceptional isomorphism 
\begin{equation}
\Sp(2)/_{\pm \Id}\simeq \SO(5).
\end{equation}
Note the very nice relationship between the sign of the curvature of the base space metric \eqref{Euclidean metric Twistor Space: Flat Twistor- S^4 metric2} and the signature of the quaternionic metric on the total space \eqref{Euclidean metric Twistor Space: Flat Twistor- quaternionic metric}. Interestingly this says that the twistor space above $\bdH^4$ naturally comes with a metric of split signature $(4,4)$.

Finally the `infinity twistor' allows to define a metric on $\T$,
\begin{equation}\label{Euclidean metric Twistor Space: Flat Twistor- quaternionic metric3}
g_{\T}= I_{\ga\gb}\; d\Zh^{\ga} \odot dZ^{\gb} = d\gpb \odot d\p + \gL\;d\gob \odot d\go.
\end{equation} 
In order to make sense of this metric, let us consider the patch
\begin{equation}
\Hbb \inj \T\!\smallsetminus\!\! \{ Z \;\text{st}\; Z= \left(0, \go \right) \} \to \HP^1\!\!\smallsetminus\!\! \{I\}
\end{equation}
together with the following trivialisation
\begin{equation}\label{Euclidean metric Twistor Space: Flat Twistor- convenient trivilisation}
\phi \left|
\begin{array}{cccc}
\T\!\smallsetminus\!\! \{ Z \;\text{st}\; Z= \left(0, \go \right) \}& \to & \Hbb \times  \HP^1\!\!\smallsetminus\!\! \{I\} \\ \\

Z^{\ga} = \frac{1}{\sqrt{1+\gL |X|^2}} \Mtx{ \p \\ X \p} & \mapsto & \left(\; \p \;,\; X\; \right)
\end{array}\right.
\end{equation}
Then we can put the above metric in a form adapted to the fibre bundle  structure,\footnote{Here $D \p = d \p + \iq A \p$ where $A = \frac{\gL}{2 \iq} \frac{dX \Xb -Xd\Xb}{1+\gL |X|^2}$ is the self-dual part of the Levi-Civita connection.}
\begin{equation}\label{Euclidean metric Twistor Space: Flat Twistor- quaternionic metric4}
g_{\T} = \frac{1}{2} D\gpb \odot D \p + \frac {\gL \;d\Xb \odot dX}{2\left(1+\gL |X|^2 \right)^2}.
\end{equation}
Accordingly, the first term is just the flat metric on the fibre and the second term is the conformally flat metric on the base with scalar curvature $sgn\left(\gL\right) \;4\times12$. When $\gL>0$ this is the round metric with radius $R^2=1/4$.\footnote{In general, the radius $R^2$ of the $n$-sphere is related to its scalar curvature through $S = n(n-1) /R^2 $.} Note that the signature of \eqref{Euclidean metric Twistor Space: Flat Twistor- quaternionic metric4} is coherent with the signature of \eqref{Euclidean metric Twistor Space: Flat Twistor- quaternionic metric3} as it should.

\subsubsection{Projective Twistor Space}
We just described the essential geometry of the \emph{twistor space} $\T \simeq \Hbb^2$ of $\S^4$. One crucial feature here is that $\T \simeq \C^4$ naturally is a four dimensional complex manifold. Even more interesting is the \emph{projective Twistor space} of $\S^4$, the space of complex lines of $\T\simeq \C^4$ i.e $\PT \simeq \CP^3$.

 The Fubini-Study metric $g_{\PT}$ on $\CP^3$ is defined from \eqref{Euclidean metric Twistor Space: Flat Twistor- quaternionic metric3} as follows:
\begin{equation}\label{Euclidean metric Twistor Space: Flat Twistor- Fubini-Study}
g_{\T} = \frac{Z.d\Zh \odot \Zh.dZ}{Z.\Zh} + \left(Z.\Zh \right) g_{\PT}
\end{equation}
This expression has the following interpretation, the twistor space $\T$ is the total space of a complex line bundle over $\PT$,
\begin{equation}
\C \inj \T \to \PT.
\end{equation}
The metric on $\T$ \eqref{Euclidean metric Twistor Space: Flat Twistor- Fubini-Study} accordingly split into a fibre metric and a base part.
 
Because we have a quaternionic structure, $\PT \simeq \CP^3$ itself is fibre bundle
\begin{equation}
\CP^1 \inj \CP^3 \to \S^4
\end{equation}
and the Fubiny-Study metric on $\PT$ can be refined into 
\begin{equation}
 g_{\PT} =\frac{Z dZ \odot \Zh d\Zh }{2\left(\Zh.Z\right)^2} +\frac{\gL}{12} g_{\S^4}.
\end{equation}
With the second term given by \eqref{Euclidean metric Twistor Space: Flat Twistor- S^4 metric}. In stereographic coordinates:
\begin{equation}\label{Euclidean metric Twistor Space: Flat Twistor- PT metric}
g_{\PT} =  \frac{\p .D \p \odot \ph. D\ph}{2\left(\pp\right)^2} + \frac{\gL \; d \Xb \odot dX}{2\left(1+\gL |X|^2\right)^2}.
\end{equation}

In the following, this metric will serve as a model for our curved twistor space constructions.

\subsection{4D Euclidean Space-Time, Complex structure and Spinor Conventions}\label{ssection: 4D Euclidean Space-Time, Complex structure and Spinor Conventions}

We now consider a general 4D Riemannian manifold $\left(M,g\right)$. Let $\left(e^{I}\right)_{I\in 0,1,2,3}$ be a orthonormal co-frame and $\left(e_{I}\right)_{ I\in 0,1,2,3}$ be a dual orthonormal frame.
Everywhere Latin indices are raised and lowered with the Euclidean metric $\eta_{IJ} = diag\left(1,1,1,1\right)$.

One here wishes to establish the following point: A choice of spinor $\p_{A'}$ at a point $x$ on $M$ amounts to a choice of almost complex structure on the tangent space at this point. This will serve as a motivation for introducing, in the next section the twistor space $\T(M)$ associated with $M$ as the total space of the primed spinor bundle $\p_{A'} \inj \T(M) \to M$. Indeed the twistor space can then be understood as the bundle of all possible almost complex structure. 

In our way to prove this (rather elementary) fact we will review the spinor notation and elements of complex geometry, essentially with a view of fixing our conventions. See for example \cite{Penrose_vol1} for more on spinors in four dimensions and \cite{Huygbrechts, Wells} for the geometry of complex manifolds.
 
\subsubsection{Spinor notation}

We already described in the previous sections how the isomorphism $\SO\left(4\right) \simeq \SU(2) \times \SU(2) /_{\pm \Id}$ can be made explicit by using identification $\Hbb \simeq \R^4$. Then $\SU(2)$ actions are just multiplications by unitary quaternions on the right and left respectively. It is sometimes convenient to use this isomorphism in the most explicit way possible i.e in a tensorial notation. This is the role of spinors.

Let $V$ be a vector field on $M$
\begin{equation}
V= V^{I} e_{I}.
\end{equation}
In order to convert space-time indices into spinor ones let us introduce the tensor $ e_I^{AA'}$ defined by the convention:
\begin{align} \label{Euclidean metric Twistor Space: Twistor- V^AA' def}
V^I e_I^{AA'} = \frac{1}{i\sqrt{2}}\Mtx{-i V^0 + V^3 & V^1-iV^2 \\  V^1+iV^2 & -iV^0-V^3}.
\end{align}
As we already discussed, the action of the orthogonal group on vector indices $V^I$ amounts to an action of $\SU(2)\times \SU(2)/_{\pm \Id}$ respectively acting on primed and unprimed indices.

Spinors indices are lowered and raised with the skew symmetric tensor
\begin{equation}
\eps^{AB} = \eps_{AB} = \eps^{A'B'} = \eps_{A'B'} = \Mtx{0&1 \\ -1 & 0}
\end{equation}
according to the usual convention:
\begin{equation}
\ga_{A} = \ga^{B} \eps_{BA},\quad \ga^{A} =  \eps^{AB} \ga_{B}.
\end{equation}
(with similar convention for primed spinors) 

Altogether, this is such that
\begin{equation}
V^2 = V^I V_{I} = V^{AA'} V_{AA'}.
\end{equation}
Contraction of primed and unprimed spinors are treated on an equal footing,
\begin{equation}
\ga.\gb \coloneqq \ga_{A'}\gb^{A'},\qquad \ga.\gb \coloneqq \ga_{A}\gb^{A}.
\end{equation}
\begin{align}
\bra\ga, \gb\ket  \coloneqq \gah_{A'}\gb^{A'}  = \gah.\gb \geq 0, \qquad 
\bra\ga, \gb\ket  \coloneqq \gah_{A}\gb^{A}   = \gah.\gb \geq 0 .
\end{align}

Now, let $V^{AA'}$ be in $M_4\left(\C \right)$. In general,
\begin{equation}
V = V^{AA'} e_{AA'} \in T_{\C}M
\end{equation}
will describes a vector of the \emph{complexified} tangent bundle. It will be of the form \eqref{Euclidean metric Twistor Space: Twistor- V^AA' def} and thus describe a \emph{real} tangent vector if and only if for any $\p^{A'} \in S'$ there exists $\go^{A} \in S$ such that
\begin{equation}
V^{AA'} = \frac{1}{\pp}\left(\go^{A}\ph^{A'} -\goh^{A}\p^{A'}\right).
\end{equation}
Accordingly a choice of primed spinor $\p^{A'}\in S'$ at a point $x\in M$ defines an almost complex structure on $TM_x$ in the form of an identification of $TM_x$ with $S\simeq \C^2$:
\begin{equation}\label{Euclidean metric Twistor Space: almost complex structure on R^4 (1)}
\begin{array}{ccc}
TM_x & \to & S \\
V & \mapsto & \go^{A} = V^{AA'} \p_{A'}
\end{array}
\end{equation}

\subsubsection{Almost Complex Structure}

We here very briefly recap some standard results of complex geometry that we will need in the following.

\begin{Definition} An \emph{almost complex structure } on a differentiable manifold M is a differentiable endomorphism in the tangent bundle,
 $ \J: TM\mapsto TM$, such that $ \J^2=-Id  $.
 
A differentiable manifold with some fixed almost complex structure is called an almost complex manifold.
\end{Definition}

We can extend $J$ to the complexification of $TM$, $T_{\C}M$ and define the holomorphic tangent bundle and anti-holomorphic tangent bundle as the eingenspaces of eigenvalues $+i$ and $-i$:
\[\begin{array}{llr}
T^{1,0}=\{V\in T_{\C}M \;|\; JV=iV\}  &\quad,\qquad & T^{0,1}=\{V\in T_{\C}M \;|\;  JV=-iV\}.
\end{array} \]
An almost complex structure is thus equivalent to a decomposition:
\begin{equation}
T_{\C}M = T^{1,0} \oplus T^{0,1}.
\end{equation}
It also induces a decomposition of k-forms on M : $\gO_{\C}^k M = \underaccent{p+q=k}{\bigoplus} \gO^{p,q}M$.

We will note $\pi^{p,q}$ the projection $\gO_{\C}^k M \mapsto \gO^{p,q}M$ (here $p+q=k$), most of the time we will simply write this projection with a bar e.g if $\ga \in  \gO^2(M)$, \begin{equation}
\ga\big|_{0,2} = \pi^{0,2} \left(\ga\right).
\end{equation} This is useful to define the \emph{Dolbeault operators}:

\begin{Definition}{Dolbeault operators}
\[ 
\begin{array}{llr}
\pa \coloneqq \pi^{p+1,q} \circ d : \gO^{p,q}M \mapsto \gO^{p+1,q}M
&\quad \text{and} \quad &
\pab \coloneqq \pi^{p,q+1} \circ d : \gO^{p,q}M \mapsto \gO^{p,q+1}M.
\end{array}
\]
\end{Definition}

What is essential here is that we only used the almost complex structure to make sense of these definitions. In particular we did not suppose that M has the structure of a complex manifold. There are \emph{a priori} no holomorphic coordinates.  Of course if $M$ is a complex manifold with holomorphic coordinates $\{z^I\}$ then
\[ T^{1,0}M=\left\{\frac{\pa}{\pa z^I}\right\},\quad T^{0,1}M=\left\{\frac{\pa}{\pa \overline{z}^I}\right\},\quad \gO^{1,0}\left(M\right)=\left\{dz^I\right\},\quad \gO^{0,1}\left(M\right)=\left\{d\overline{z}^I\right\} \] but not every almost complex manifold is a complex manifold.

\begin{Definition}
An almost complex manifold is called \emph{integrable} if it is induced by a complex structure.
\end{Definition}

The \emph{Newlander-Nirenberg} theorem tells us when an almost complex structure is integrable:

\begin{Theorem}{Newlander-Nirenberg} \\
	The following statements are equivalent:
	\begin{itemize}
		\item The almost complex manifold M is integrable
		\item $[T^{0,1}M,T^{0,1}M] \subset  T^{0,1}M$, i.e $T^{0,1}M$ is integrable as a distribution
		\item $\pab^2=0$
		\item $d\alpha= \pa\alpha+\pab \alpha$ ,  $\forall\;\alpha\in \Lambda^k , k\neq0 $
		\item $d\alpha= \pa\alpha+\pab \alpha$ , $\forall\;\alpha\in \Lambda^{0,1} $
		\item $\pi^{0,2} \circ d\alpha=0$ , $\forall\;\alpha\in \Lambda^{0,1} $
	\end{itemize}
\end{Theorem}
One easily shows that the last five points are equivalent (cf e.g \cite{Huygbrechts}), but the highly non trivial part is to prove that they imply the existence of a complex structure.

\subsubsection{Almost complex structure on a Riemannian manifold}

We are now interested in the almost complex structures on a Riemannian manifold that are compatible with the metric structure.

\begin{Definition}
An almost complex structure $\J$ is said to be compatible with a metric $g$ if for any vector fields $X,Y$
 \begin{equation}\label{Euclidean metric Twistor Space:: Complex geometry- compatibility condition}
g\left( \J(X) , \J(Y) \right)= g\left( X , Y \right).
\end{equation}
\end{Definition}
As $\J^2=-Id$, this is equivalent to 
\begin{equation}\label{Euclidean metric Twistor Space: Complex geometry- go def}
\go\left(X,Y\right) \coloneqq g\left(\J\left(X\right), Y \right)
\end{equation}
being a two-form, $g\left( \J(X) , Y \right) = - \left( \J(Y) , X \right)$.

In this thesis we will use the following terminology. It is essentially standard but there seem to be fluctuations.
\begin{Definition}\mbox{}
	\begin{itemize}
\item A differentiable manifold with a compatible triplet $\left(\J, g, \go \right)$ of almost complex structure, metric and two-forms will be called and \emph{almost Hermitian manifold}.
\item An almost hermitian manifold with integrable complex structure will be called \emph{Hermitian}.
 \item An almost hermitian manifold with a closed (symplectic) two-forms $\go$ will be called \emph{almost Kähler}.
  \item  Finally a \emph{Kähler manifold} is an almost hermitian manifold with both integrable complex structure and a closed two-form. 
   \end{itemize}
\end{Definition}

As any two-form in four dimension, $\go$ can be decomposed in terms of the self-dual and anti-self-dual basis (see Appendix eq \eqref{Appdx: two-form decomposition}):
\begin{equation}
 \go= \ga^i\gS^i + \gb^i\gSt^i. 
\end{equation}
It just takes a direct calculation using the algebra of the sigma matrices (again given in Appendix see eq\eqref{Appdx: Sigma algebra}) to see that $\J^2=-Id$ is equivalent to
\begin{equation}\
\begin{array}{llll}
\go=\ga^i\gS^i \quad\text{with}\quad \ga^i\ga^i=1&
\text{or}& \go=\gb^i\gSt^i \quad\text{with}\quad \gb^i\gb^i=1.
\end{array}
\end{equation}
We will call the first a `self-dual almost complex structures' and the  second `anti-self-dual almost complex structures'.

We now concentrate on self-dual almost complex structures, these are the one which preserve the orientation. The above shows that the space $Z$ of almost complex structures compatible with both the metric and orientation is isomorphic to the two-sphere $\S^2$:
\begin{IEEEeqnarray*}{ccccc}
	\S^2  &\qquad \mapsto\qquad & \gL^+ &\qquad \mapsto\qquad & Z \\
	\p^i &\mapsto & \p^i \gS^i &\mapsto & \p^i \gS^i{}^J{}_{I} \;e^{I} \otimes e_J
\end{IEEEeqnarray*}
Alternatively, $Z \simeq S^2 \simeq \CP^1$. We can make this even more explicit: Using spinor notation every \emph{euclidean real} two-form $\go$ can be written as:\footnote{In particular in our conventions
	 \begin{equation}
	\gS^{A'B'} = -\frac{e{}_{C}{}^{A'} \W e^{CB'}}{2}.
	\end{equation}
	This is such that it coincides with our convention for the $\gS$ matrices of the first chapter (see eq \eqref{Appdx: Sigma def (tetrad)}) when one converts spinor indices according to the rule given in appendix (see eq \eqref{Appdx: su(2)/Spinor indices conversion}). This convention also naturally allows to interpret self-dual two-forms as representation of $\su(2)$ see in the appendix eq \eqref{Appdx: self-dual representation of su2} and below.}
\begin{align}
\go &= -\left(\ga_{(A'} \gah_{A')}\;\eps_{AB} + \gb_{(A} \gbh_{A)}\;\eps_{A'B'}\right)\;\frac{e^{AA'} \W e^{BB'}}{2} \\ \\
&= \ga_{A'}\gah_{B'} \;\gS^{A'B'} + \gb_{A}\gbh_{B} \;\gS^{AB}
\end{align}
So that the isomorphism $\CP^1 \simeq Z$ can be rewritten as
\begin{IEEEeqnarray*}{ccccc}\label{Euclidean metric Twistor Space: Isomorphism CP1/J}
	\CP^1 &\quad \mapsto\quad & \gL^+ &\quad\mapsto\quad& Z \yesnumber\\
	\left[\p_{A'}\right]&\mapsto &   2i\;\frac{\p_{A'}\ph_{B'}}{\pp}\; \gS^{A'B'}  &\mapsto &  i\frac{\p_{A'}\ph^{B'}}{\pp}\; e^{AA'} \otimes e_{AB'} + i\frac{\ph_{A'}\p^{B'}}{\pp}\; e^{AA'} \otimes e_{AB'}.
\end{IEEEeqnarray*} 

It is then clear from \eqref{Euclidean metric Twistor Space: Isomorphism CP1/J} that the set of $(1,0)$-vectors associated with a point $\p \in \CP^1$ is: $\left(\ph_{A'}e^{A'A}\right)_{A\in 0,1}$.

This is coherent with what we already saw above, see eq \eqref{Euclidean metric Twistor Space: almost complex structure on R^4 (1)}), a choice of primed spinors $\p^{A'}$ at a point $x\in M$ decomposes the tangent space\footnote{This comes again from the identity $\gd^{A'}{}_{B'}= \frac{1}{\pp}\left(\ph^{A'}\p_{B'} - \p^{A'} \phi_{B'}\right)$.} :
\begin{equation}\label{Euclidean metric Twistor Space: almost complex structure}
\begin{array}{ccccc}
T_{\C}M_x & = & T^{1,0}M_x & \oplus & T^{0,1}M_x \\ \\
V^{AA'} e_{AA'} & =  &  V^{AA'}\p_{A'}\; \frac{\ph^{B'}}{\pp} e_{AB'} &-&V^{AA'}\ph_{A'}\; \frac{\p^{B'}}{\pp} e_{AB'}.
\end{array}
\end{equation}

The bundle over $(M,g)$ of almost complex structures compatible both with the metric and the orientation is called the projective Twistor bundle and its total space $\PT(M)$ is called the projective Twistor space of $M$. As we just saw it is isomorphic to the bundle of projective primed spinors.

\subsection{The Twistor Space  of a General Riemannian 4-manifold; essential results}

We are now in a position to describe the essential structure of the twistor space of a Riemannian manifold. See \cite{Woodhouse85}, \cite{Atiyah:1978wi} and reference therein for the original results. We however presents these results in a way that is suitable for our `pure connection' generalisation.

\subsubsection{The Twistor Space of a Riemannian manifold}

Given a Riemannian manifold $\left(M, g\right)$ the \emph{Twistor space of M}, $\T(M)$, is the total space of the primed bundle, i.e locally $\T(M) \simeq S' \times M$. The associated \emph{projective Twistor space} $\PT(M)$ is just $\T(M)$ with projectivised fibres, locally $\PT(M) \simeq \CP^1 \times M$.

The discussion from the previous section gives a more geometrical interpretation of the projective twistor space of a Riemannian manifold as the bundle of the self-dual almost complex structures over $M$. In particular a section of $\PT(M)$ is the same as a choice of almost complex structure on $M$. Somehow working with $\PT(M)$ means that we are not choosing and that we are considering all the possible almost complex structures on $M$ at the same time. 

Because the space of almost complex structure at a point is itself a complex manifold,  $\CP^1$, it is no surprise that $\T(M)$  can be given an almost complex structure. Let us see schematically how it can be done:

 The self-dual part of the Levi-Civita connection, being a $SU(2)$-connection, induces a connection on twistor space. Suppose we are at a point $Z=(x,\p_{A'})\in \T$, the Levi-Civita connection then splits the tangent space $T_Z\T$ in it's vertical part $V_Z$, naturally isomorphic to $T_\p S'\simeq\C^2$, and its horizontal part $H_Z$, isomorphic to $T_x M$:
\begin{equation}
T_Z \T = V_Z \oplus H_Z \simeq T_\p S' \oplus T_x M
\end{equation}

We can now choose for $T_\p S'\simeq\C^2$ its canonical almost complex structure and for $T_xM$ the almost complex structure associated with $\p$ \eqref{Euclidean metric Twistor Space: almost complex structure}, all in all it defines an almost complex on $\T(M)$. $\PT(M)$ naturally inherits this almost complex structure.

This almost complex structure turns out to be conformally invariant. What is more it is integrable if and only if the base manifold is a anti-self-dual i.e if the self-dual part of the Weyl tensor vanishes: $\Psi_{A'B'C'D'}=0$ (or, in the notations of the first section, $\Psi^{ij}=0$).

Before we come back to this construction in some more details, it is good to take some time to describe the $\SU(2)$-geometry of the twistor space. 

\subsubsection{$\SU(2)$-Connection and the Geometry of the Twistor Space}

We here emphasise the geometry induced on $\T(M)$ by a $\SU(2)$-connection only. It will serve as a starting point for our `connection approach' to Twistor theory.

Accordingly, we now take `space-time' to be a $SU(2)$-principal bundle
\begin{equation}
\SU(2) \inj P \to M
\end{equation}
over a four dimensional manifold $M$ equipped with a $SU(2)$-connection \begin{equation}
D=d+\bdA.
\end{equation} We will describe this connection by its potential in a trivialisation, $\bdA= A^i \,\gs^i$. 

The associated `twistor space' $\T(M)$ is simply the spinor bundle over $M$, this is an associated vector bundle for our $\SU(2)$-principal bundle: \begin{equation}
\C^2 \inj \T(M) \to M.
\end{equation} 
We will use adapted local coordinates $\left(x^\mu , \p_{A'} \right)$ to describe this bundle. As always, we raise and lower spinor indices with the anti-symmetric tensor $\eps_{A'B'}$ (Here this is simply the metric volume form preserved by the $SU(2)$ action). Having $SU(2)$ structure group, the $\C^2$ fibres of this bundle come equipped with a hermitian metric we represent by an anti-linear, anti-involutive map,
\begin{equation}
  \circonf \colon \left\{\begin{array}{ccc}
\C^2 &\to &\C^2 \\ \p_{A'} &\mapsto &\ph_{A'}
\end{array} \right. 
\end{equation}
such that
\begin{equation}
\ga ,\gb \in \C^2, \qquad \langle \ga , \gb \rangle \coloneqq \gah_{A'}\gb^{A'}.
\end{equation}
We already discussed the interpretation of the hat operator as a quaternionic structure in the previous sections. 

Making use of the fundamental representation of $SU(2)$, the $SU(2)$-connection $D= d + \bdA$ naturally acts as a connection on twistor space :\\
if 
\begin{equation}
 s \left\{ \begin{array}{ccc} M &\to &\T(M) \\ x &\mapsto &\p_{A'}\left(x\right) \end{array} \right. 
\end{equation}
is a section of $\T(M)$ then its covariant derivative with respect to $A$ is
\begin{equation}\label{Euclidean metric Twistor Space: Covariant derivativ of section}
\nabla \p_{A'} = d\p_{A'} - A^{B'}{}_{A'} \;\p_{B'}, \qquad  A^{A'}{}_{B'} \in \su(2).
\end{equation} 
Now we can also re-interpret this last equality in terms of forms: We define the one-forms $D\p_{A'} \in \gO^1\left(\T(M)\right)$ on the full space of the bundle $\T(M)$ as
\begin{equation}\label{Euclidean metric Twistor Space: Dp definition}
D\p_{A'} = d\p_{A'} - A^{B'}{}_{A'} \; \p_{B'} \quad \in \gO^1\left(\T(M)\right).
\end{equation}
These are in fact the coordinates of a projection operator, the projection operator on the vertical tangent space to $\T(M)$:
\begin{equation}\label{ Proj on V of T(M)}
Proj = D\p_{A'} \otimes \frac{\pa}{\pa \p_{A'}} \quad \in End\left(T \T(M) \right) .
\end{equation}
The kernel of this operator is the \emph{horizontal distribution} associated with the connection $D=d+\bdA$. Thus \eqref{Euclidean metric Twistor Space: Covariant derivativ of section}, \eqref{Euclidean metric Twistor Space: Dp definition} corresponds to the usual dual points of view on connections: either as a differential operators acting on sections or as a horizontal distribution on the total space of the bundle.

The associated `projective twistor space' $\PT(M)$ is the projectivised version of $\T(M)$, with fibres isomorphic to $\CP^1$: 
\begin{equation}
\CP^1 \inj \PT(M) \to M.
\end{equation} 
We will most frequently use homogeneous coordinates $\left(x^{\mu}, \left[\p_{A'} \right] \right)$ to describe this bundle. The main advantage with this notation is that section of $\Oc(n,m)$-bundle over $\CP^1$ (and by extension over $\PT(M)$) are equivalent to functions $f(x,\p_{A'})$ with homogeneity $n$ in $\p_{A'}$ and $m$ in $\ph_{A'}$ .\footnote{The $\Oc\left(n,m\right)$ bundles are `natural' complex line bundle over $\CP^1$.  Here one can take as a definition representations of their sections in terms of functions $f\left(\p, \ph\right)$ on $\C^2$ with homogeneity $n$ and $m$ respectively in $\p$ and $\ph$. See however appendix \ref{section: Appdx Geometry of the Riemann sphere} for more details.}

Similarly k-forms on $\PT(M)$ with values in $\Oc(n,m)$ are uniquely represented by k-forms on $\PT(M)$ with homogeneity $n$ in $\p_{A'}$, $m$ in $\ph_{A'}$ which vanishes on $E = \p_{A'} \frac{\pa}{\pa \p_{A'}}$, $\Eb = \ph_{A'} \frac{\pa}{\pa \ph_{A'}}$: 
\begin{equation}
\begin{array}{ll}
&\ga' \in \gO^k\left(\PT, \Oc(n,m)\right) \\
\Leftrightarrow \quad & \\
&\ga \in \gO^k\left(\T\right) \quad \text{st} \quad E\id \ga = 0, \quad \Eb \id \ga=0, \quad \Ld_{E} \ga = n \ga, \quad \Ld_{\Eb} \ga = m \ga. 
\end{array}
\end{equation}
Here $E$ and $\Eb$ the `Euler vectors', they generate the vertical tangent space of the complex line bundle $\C \inj \T(M) \to \PT(M)$.

As a concrete example,
\begin{equation}
\gt \coloneqq \p_{A'} D \p^{A'}
\end{equation}
represents a $\Oc(2)$-valued one-form on $\PT(M)$ but $\ph_{A'} D \p^{A'}$ does not represent a well defined object on $\PT(M)$ as it does not vanish on $Span\left(E,\Eb \right)$.

We can use this fact to define a connection on the $\Oc(n,m)$ bundles.
For suppose $f(x ,\p_{A'})$ represents a section of the $\Oc(n,m)$ bundle, $\Ld_{E}f=n f$, $\Ld_{\Eb} = m f$. Then we can define its covariant derivative as
 \begin{equation}
d_{(n,m)} f \coloneqq df + n\;\frac{\ph_{A'} D \p^{A'}}{\pp}\;f - m\;\frac{\p_{A'} D \ph^{A'}}{\pp}\;f
\end{equation}
It is a simple exercise to verify that $E \id d_{(n,m)} f=0$, $\Eb \id d_{(n,m)} f =0$, $\Ld_{E} d_{(n,m)} f = n \;d_{(n,m)} f$ , $\Ld_{\Eb} d_{(n,m)} f = m \; d_{(n,m)} f $ and thus that $d_{(n,m)} f$ indeed represents a $\Oc(n,m)$-valued one-form on $\PT(M)$.

This connection also preserves the following Hermitian metric on the $\Oc(n,m)$-bundles:
\begin{equation}\label{Euclidean metric Twistor Space: Hermitian metric on O(n)}
\ga ,\gb \in \Oc(n,m), \qquad \langle \ga , \gb \rangle = \overline{\ga} \;  \gb \; \left(\pp \right)^{-n-m}.
\end{equation}
A simple calculation indeed shows that $d_{(n,n)} \left(\pp\right)^{n} =0$. In particular, when restricted to each $\CP^1$ this connection is the natural Chern-connection on $\Oc(n)$ bundle induced by the Kähler structure. 

This connection on $\Oc(n,m)$ bundle over $\PT(M)$ extends to a connection on $\Oc(n,m)$-valued k-forms in the usual way. It is for example instructive to check that,
\begin{equation}\label{d tau}
d_{(2)} \gt = F^{A'}{}_{B'}\;\p_{A'}\p^{B'}.
\end{equation}

We thus see that the $SU(2)$-connection that we started with induces two natural geometric objects on $\PT(M)$: a $\Oc(2)$-valued one-form $\gt = \p_{A'}D\p^{A'}$ and a covariant derivative $d_{(n,m)}$ on the $\Oc(n,m)$-bundle over $\PT(M)$.

\subsubsection{The Almost Complex Structure on $\T(M)$}

We now come back to a metric context. We take $\left(M, g\right)$ to be a Riemannian manifold and $\T(M)$ the associated twistor space. As we already explained the self-dual part of the Levi-Civita connection gives a $\Oc(2)$-valued one-form on $\PT(M)$ and a connection on the $\Oc(n,m)$-bundle over $\PT(M)$.

The almost complex structure on $\PT(M)$ can then be defined by the $\Oc(4)$-valued $\left(3,0\right)$-form
\begin{equation}\label{Euclidean metric Twistor Space: integrable ACS}
\gO^{3,0} = \p_{A'}D\p^{A'} \W e^0{}^{B'}\p_{B'}\W e^1{}^{C'}\p_{C'}
\end{equation}
In fact there is another natural almost complex structures on $\T(M)$.
\begin{equation}
\gO^{3,0} =\frac{1}{\left(\pp\right)^2} \left(\ph_{A'}D\ph^{A'} \W e^0{}^{B'}\p_{B'}\W e^1{}^{C'}\p_{C'}\right)
\end{equation}

 We will respectively refer to these almost complex structures as `the integrable almost complex structure of $\T(M)$' and `the non-integrable almost complex structure of $\T(M)$' because the first one can be integrable under certain conditions (cf Proposition \ref{Proposition: Metric Twistor, integrability}) while the second never is. In this section we will only consider the first one, see however the end of section \ref{section: Hermitian structure and NKM} for a discussion on the `non-integrable' one.

As we already explained, in the presence of almost complex structure we can define the projection $\gO^r \to \gO^{(p,q)}$ (here $p+q=r$). For practical purpose, we will write this map as $\ga \mapsto \ga\big|_{(p,q)}$. For example,
\begin{equation}\label{Euclidean metric Twistor Space: Sigma Restrictions}
\gS^{A'B'}\big|_{(2,0)} = \gS\p\p\, \frac{\ph^{A'}\ph^{B'}}{(\pp)^2}, \qquad  \gS^{A'B'}\big|_{(0,2)} = \gS\ph\ph\, \frac{\p^{A'}\p^{B'}}{(\pp)^2},
\end{equation}
\begin{equation*}
\gS^{A'B'}\big|_{(1,1)} = -2\gS\p\ph\, \frac{\p^{(A'}\ph^{B')}}{(\pp)^2},
\end{equation*}
(In order to lighten notations here and thereafter $\gS\pi\pi$ stands for $\gS^{A'B'}\p_{A'}\p_{B'}$ etc).

Let us now turn to integrability. A direct calculation gives
\begin{equation}
d \left(e^{AA'}\p_{A'}\right) \big|_{0,2} = 0
\end{equation}
\begin{equation}
d \gt \big|_{0,2} = F\p\p \big|_{0,2} = \Psi\p\p\p\p \; \frac{\gS\ph\ph}{(\pp)^2},
\end{equation}
so that we have the following
\begin{Proposition}\label{Proposition: Metric Twistor, integrability}
 The almost complex strucutre defined by the $\left(3,0\right)$-form \eqref{Euclidean metric Twistor Space: integrable ACS} is integrable if and only if the metric on the base manifold is anti-self-dual i.e is the self dual part of the Weyl tensor vanishes $\Psi^{A'B'C'D'}=0$.
\end{Proposition}

It turns out that this almost complex structure is conformally invariant. Because the Weyl tensor is a conformal invariant the integrability condition of Prop \eqref{Proposition: Metric Twistor, integrability} also is, which is reassuring.

\subsubsection{The Contact Structure on $\T(M)$}

In the context of an almost complex manifold, it is natural to introduce Dolbeault operators on the space $\gO^{p,q}\left[n,m\right]$ of $\Oc\left(n,m\right)$-valued $(p,q)$-forms as
\begin{equation}\label{Definition: Dolbeault Operators}
\begin{array}{lll}
\pa : \left|
\begin{array}{ccc}
\gO^{p,q}\left[n,m\right] &\to & \gO^{p+1,q}\left[n,m\right] \\ \\
\ga & \mapsto & \left(d_{(n,m)}\ga \right) \big|_{(p+1,q)}
\end{array} \right.
,\;
\pab : \left|
\begin{array}{ccc}
\gO^{p,q}\left[n,m\right] &\to & \gO^{p,q+1}\left[n,m\right] \\ \\
\ga & \mapsto & \left(d_{(n,m)}\ga \right) \big|_{(p,q+1)}
\end{array} \right.
\end{array}. 
\end{equation}

Then
\begin{equation}
\pab \gt = -2\; \Psi\p\p\p\ph \;\frac{\gS\p\ph}{(\pp)^2} + \p\p G_{AB} \;\gSt^{AB}
\end{equation}
so that we have the following
\begin{Proposition}\label{Proposition: Metric Twistor, holomorphic gt}
	The $\Oc(2)$-valued  $\left(1,0\right)$-form $\gt = \p_{A'}D\p^{A'}$ is holomorphic if and only if the metric on the base is anti-self-dual Einstein. 
\end{Proposition}

Note that while the almost complex structure on $\PT(M)$ was conformally invariant, $\gt$ which is directly related to the self-dual part of the Levi-Civita connection is not. This is in line with the fact that the Einstein condition on metric is not conformally invariant. 

\subsubsection{The Kähler Structure on $\T(M)$}

Let us now consider the following hermitian structure on $\PT(M)$ (compare with the flat case \eqref{Euclidean metric Twistor Space: Flat Twistor- quaternionic metric4})
\begin{align}\label{Euclidean metric Twistor Space: Kahler strcture, g}
g &= 4R^2\frac{\p_{A'}D\p^{A'} \odot \ph_{B'}D\ph^{B'}}{2\left(\pp \right)^2} + \frac{1}{2}e^{AA'}\odot e_{AA'} \nonumber\\
&= 4R^2\frac{\p_{A'}D\p^{A'} \odot \ph_{B'}D\ph^{B'}}{2\left(\pp \right)^2} - \frac{e^{AB'}\p_{B'}\odot e_{A}{}^{C'}\ph_{C'}}{\pp}.
\end{align}
and
\begin{align}\label{Euclidean metric Twistor Space: Kahler strcture, go}
\go &= 4iR^2\;\frac{\p_{A'}D\p^{A'} \W \ph_{B'}D\ph^{B'}}{2\left(\pp \right)^2} - i\frac{e^{AB'}\p_{B'}\W e_{A}{}^{C'}\ph_{C'}}{\pp} \nonumber \\
&= 2iR^2 \left(\frac{\p_{A'}D\p^{A'} \W \ph_{B'}D\ph^{B'}}{\left(\pp \right)^2} -\frac{1}{R^2}\frac{\gS^{B'C'}\p_{B'}\ph_{C'}}{\pp}\right)
\end{align}

A direct calculation shows the following 
\begin{Proposition}\label{Proposition: metric Twistor, Kähler condition}
The Kähler form $\go$ on $\PT(M)$ is closed if and only if the metric on the base is anti-self-dual Einstein with cosmological constant $\frac{3}{R^2}$.
\end{Proposition}

\paragraph{Example: $\PT(\S^4) \simeq \CP^3$} As we already saw, in the flat case the projective twistor space is the projective space $\CP^3$. The above proposition just says that the Fubini-Study metric \eqref{Euclidean metric Twistor Space: Flat Twistor- PT metric} is Kähler. We here recall the form of this metric for convenience
\begin{equation}
g_{\PT} = \frac{\p.D \ph \odot \ph.D\p}{2\left(\pp \right)^2} + \frac{dX^{AA'}\odot dX_{AA'}}{2\left(1 + |X|^2\right)^2}.
\end{equation}
 Here the radii of both the 3-sphere (the fibre) and the 4-sphere (the base) is $R = 1/2$. Accordingly, the scalar curvature of the base manifold has value $4 \times 12$.

\subsection{The Non-Linear Graviton Theorem}

In its original form, see \cite{Penrose:1999cw}, the aim of twistor theory was to realise solutions of complicated differential equations on space-time in terms of simpler, essentially free, geometrical data on the associated Twistor space. The key insight was holomorphicity. The original success of twistor theory takes the form of three theorems, each of these being an equivalence, between
\begin{itemize}
	\item solutions to the zero rest mass equations on Minkowski space and some cohomology group on the associated twistor space (this is the `Penrose Transform', see \cite{Penrose:1969ae})
	\item solutions to self-dual Yang Mills equations on Minkowski space and holomorphic fibre bundle on twistor space (this is  the `Ward transform', see \cite{Ward:1977ta})
	\item solutions to self-dual Einstein equations and deformations of the complex structure of the twistor space (this is the `non-linear graviton theorem', see \cite{Penrose:1976js} and \cite{Ward:1980am}).
\end{itemize}	
In its original form, this program could not overcome the `googly problem' i.e the difficulty of describing anti-self-dual fields. However, taken in a broader sense, twistor theory have proved a very fruitful framework, both for physicist and mathematicians, see \cite{Atiyah:2017erd} for an overview of its achievement over the last fifty years.

In this thesis we will be mainly concerned with the non-linear gravitons theorem for anti-self-dual space-time with non-zero scalar curvature.
\begin{Theorem}{Non-Linear Graviton Theorem \cite{Penrose:1976js},\cite{Ward:1980am}}\label{Theorem: original non-linear graviton theorem}
	\begin{itemize}
\item There is a natural one-to-one correspondence between holomorphic conformal structures $[g]$ on some four-dimensional (complex) manifold $M$ whose self-dual Weyl curvature vanishes, and three-dimensional complex manifolds $\PT$ (the twistor space) containing a rational curve (a $\CP^1$) with normal bundle $N = \Oc(1) \oplus \Oc(1)$.
\item The existence of a conformal scale for which the trace-free Ricci tensor vanishes, but for which the scalar curvature is non-vanishing, is equivalent to $\PT$ admitting a holomorphic $\Oc(2)$-valued one-form $\gt$ such that $\gt \W d \gt \neq 0$.

\item What is more, there is a real Euclidean slice in $M$ if and only if there exists an anti-holomorphic involution $\circonf \from \PT \to \PT$ with no fixed points.
\end{itemize}
\end{Theorem}
The topological requirements of the first point can typically be realised by small (but finite) deformations of the complex structure of the flat twistor space.

\newpage
\section{Euclidean Twistor Theory Revisited: a Connection Point of View}  \label{section: Twistor theory revisited}

We now come back on some of the preceding results but from an unusual `connection point of view', the presentation and results from this section are taken from \cite{Herfray:2016qvg}.

 Accordingly, we now take `space-time' to be a $SU(2)$-principal bundle
\begin{equation}
\SU(2) \inj P \to M
\end{equation}
over a four dimensional manifold $M$ equipped with a $SU(2)$-connection \begin{equation}
D=d+\bdA.
\end{equation} We will describe this connection by its potential in a trivialisation, $\bdA= A^i \,\gs^i$. 

The associated `twistor space' $\T(M)$ is simply the spinor bundle over $M$.
In the preceding section we already saw that this is enough to define both a $\Oc(2)$-valued one-form $\gt= \p_{A'}D\p^{A'}$ and a connection $d_{m,n}$ on the $\Oc\left(n,m\right)$ line bundle.

\subsection{Symplectic and Almost Hermitian Structure on $\PT(M)$ from a Definite Connection}

We now restrict ourselves to the case of definite connections \eqref{Definite Connections and Gravity: Def: Definite Connections}, ie the case where $\tilde{X}^{ij} = F^i\W F^j /_{d^4x}$ is a definite 3x3 conformal metric. This is in fact equivalent to the requirement that no real 3-vector $\left(v^i\right)_{i\in 1,2,3}$ is such that $v^i\;F^i$ is a simple two-form:
\begin{equation*}
A\; \text{is a definite connection} \qquad \Leftrightarrow \qquad  \forall v^i \in \R^3 , \; v^i F^i \W v^j F^j = v^i v^j \tilde{X}^{ij} d^4x \neq 0.
\end{equation*}
 
A definite connection on $\PT(M)$ naturally gives a symplectic structure:
\begin{Proposition}{ \emph{Symplectic structure on $\PT(M)$  } (Fine and Panov \cite{Fine:2008})} \label{Proposition: symplectic structure} \mbox{} \\
	If $A$ is a definite connection then $\go_s = \left(n-m\right)^{-1} \left( d_{(n,m)} \right)^2 $ , $n \neq m$, is a symplectic structure on $\PT(M)$.
\end{Proposition}
\begin{proof}\mbox{}\\
	As $d_{(n,m)}$ is a covariant derivative on a line bundle, its curvature two-form \[  \go_s = \left(n-m\right)^{-1} \left( d_{(n,m)} \right)^2 = \left(n-m\right)^{-1} d \left(\frac{n\;\ph_{A'} D\p^{A'}-m\;\p_{A'} D\ph^{A'}}{\pp}  \right) \] is automatically closed. A direct computation shows that,
	\begin{equation}\label{Twistor theory revisited: PT sympleptic structure}
	\go_s = \frac{\p_{A'}D\p^{A'} \W \ph_{B'}D\ph^{B'}}{\left(\pp \right)^2} - F^{A'B'} \frac{\p_{A'} \ph_{B'}}{\pp}
	\end{equation} and therefore $\go_s$ is independent of $n$ and $m$. From this last expression one also sees that non degeneracy is equivalent to the definiteness of the connection.
\end{proof}

We also have an almost Hermitian structure obtained by a modification of the classical one described in the previous section and due to \cite{Atiyah:1978wi},\cite{Woodhouse85}:

\begin{Proposition}{\emph{Almost Hermitian structure on $\PT(M)$}} \label{Proposition: Almost Hermitian structure} \mbox{} \\
	If $A$ is a definite connection then $\PT(M)$ can be given an almost Hermitian structure, i.e a compatible triplet $\left(\J_A , \go_A , g_A \right)$ of almost complex structure, two-form, and a Riemannian metric.  In general this triplet is neither Hermitian ($\J_A$ is not integrable) nor almost Kähler ($\go_A$ is non degenerate but generically not closed). 
\end{Proposition}

\begin{proof}\mbox{}\\
	We first describe how to construct the almost complex structure $\J_A$ on $\PT(M)$ from a definite connection:
	Because the connection is definite, one can make sense of the square root (we take the positive square root) and inverse  of X. Define $\gS^i = X^{-\frac{1}{2}}{}^{ij} F^j$. By construction $\gS^{i} \W \gS^j \propto \gd^{ij}$. It implies that $\gS_{\pi} = \gS^{A'B'} \p_{A'}\p_{B'}$ is simple, $\gS_{\pi} \W \gS_{\pi}=0$. We now define the almost complex structure by the requirement that $\gO_A = \gt \W \gS_{\pi}$ be a $(3,0)$-form. It makes sense as its kernel,  $\{X\; st\; X\id \gO_A=0\}$, is 3 dimensional and thus can be identified with the $(0,1)$-distribution:
	\begin{equation}\label{Twistor theory revisited: Hermitian strct: J_A on PT}
	X \in T^{0,1}\PT(M) \quad \Leftrightarrow \quad X\id\; \gt \W \gS_{\pi} =0.
	\end{equation}
	
	This construction has a simple metric interpretation: We already explained how to construct a conformal, non degenerate, Euclidean metric from a definite connection. We will note $e^{AA'}$ the associated null tetrad. It is then easy to see that the construction leading to $\gS^i$ is in fact just an alternative way of constructing $\gS^{A'B'}= \frac{1}{2}e^{A'C}\W e_C{}^{B'}$ ( or equivalently \eqref{Chiral Formulations of GR - Fundations: Sigma def (tetrad)}, see Appendix \ref{Section : Appdx Spinor conventions} for our conventions on spinors). Then the $(3,0)$-form
	\begin{equation}\label{Twistor theory revisited: Hermitian strct: gO_A on PT}
	\gO^{3,0}_A = \gt \W \gS^{A'B'}\p_{A'}\p_{B'} = \p_{A'}D\p^{A'} \W e^0{}^{A'}\p_{A'}\W e^1{}^{B'}\p_{B'}
	\end{equation}
	is of the same form as in the metric case \eqref{Euclidean metric Twistor Space: integrable ACS}.

	One now comes to the compatible metric on $\PT(M)$. From the definite connection we have a conformal metric. One fixes the scaling freedom by requiring the volume form to be $\frac{3}{2 \gL^2} F^k\W F^k$. We will note $e^{AA'}$ the associated null tetrad.
	This gives a metric on the horizontal tangent space (as defined by $A$), on the other hand the vertical tangent space comes equipped with a metric and altogether this gives the following metric on $\PT(M)$ \footnote{Here $A \odot B = A \otimes B + B \otimes A$}:
	\begin{align}\label{Twistor theory revisited: Hermitian strct: g_A on PT}
	g_A &= 4R^2\frac{\p_{A'}D\p^{A'} \odot \ph_{B'}D\ph^{B'}}{2\left(\pp \right)^2} + \frac{1}{2}e^{AA'}\odot e_{AA'} \nonumber\\
	 &= 4R^2\frac{\p_{A'}D\p^{A'} \odot \ph_{B'}D\ph^{B'}}{2\left(\pp \right)^2} - \frac{e^{AB'}\p_{B'}\odot e_{A}{}^{C'}\ph_{C'}}{\pp}.
	\end{align}
	We leave R, the radius of the fibres, as a parameter but we will see that the Kähler condition will relate it uniquely with $\gL$.
	
	From this, one readily sees that the two-form,
	\begin{align}\label{Twistor theory revisited: Hermitian strct: go_A on PT}
	\go_A 
	&= 2iR^2 \left(\frac{\p_{A'}D\p^{A'} \W \ph_{B'}D\ph^{B'}}{\left(\pp \right)^2} -\frac{1}{R^2}\frac{\gS^{B'C'}\p_{B'}\ph_{C'}}{\pp}\right) \nonumber \\
	&= 4iR^2\;\frac{\p_{A'}D\p^{A'} \W \ph_{B'}D\ph^{B'}}{2\left(\pp \right)^2} - i\frac{e^{AB'}\p_{B'}\W e_{A}{}^{C'}\ph_{C'}}{\pp}
	\end{align}
the metric \eqref{Twistor theory revisited: Hermitian strct: g_A on PT} and the almost complex structure \eqref{Twistor theory revisited: Hermitian strct: J_A on PT} are compatible. 
	
	This is clear as \eqref{Twistor theory revisited: Hermitian strct: gO_A on PT},\eqref{Twistor theory revisited: Hermitian strct: g_A on PT},\eqref{Twistor theory revisited: Hermitian strct: go_A on PT} are already in the canonical form
\begin{equation}
	\gO_A^{3,0}= dz^1\W dz^2 \W dz^3, \qquad g_A = h_{i\bar{j}}\; dz^i \odot d\bar{z}^{\bar{j}}, \qquad \go_A = i\; h_{i\bar{j}}\; dz^i \W d\bar{z}^{\bar{j}}.
\end{equation}
	
\end{proof}

Essentially this construction is a variation of the one described in the previous section. As compared to the classical construction from \cite{Atiyah:1978wi} there are however small differences:\\ 
First the conformal structure is obtained from the connection. \\
Second one does not use the notion of horizontality associated with the (Levi-Civita connection of the) conformal structure but the one given by \emph{our original} $SU(2)$-connection. In general those two connections differ. The special case where they coincide in fact corresponds to the Einstein case, i.e the base metric is Einstein.

For clarity, we expand a little bit on this last point even though this is more related to the pure connection formulation of Einstein equations (that we reviewed in \eqref{Proposition: Pure connection equation}) and somewhat lies out of the main line of development: Suppose that, $\bdA$, the $SU(2)$-connection that we took as starting point coincide with $\bdA_g$, the (Left-chiral or `self-dual' part of the) Levi-Civita connection, then their curvature also coincide: $\bdF =\bdF_g$. Now by construction Urbantke metric is such that it makes $\bdF$ self-dual. Therefore $\bdF_g$ is self-dual and this is just the chiral way of stating Einstein equations (Cf first part of section 2).

A natural question is then to ask when the almost Hermitian structure introduced in Prop \ref{Proposition: Almost Hermitian structure} is Hermitian, i.e $\J_A$ is integrable. 
\begin{Proposition}\label{Proposition : Integrability of A} \mbox{}\\
	$\Leftrightarrow$ $\J_A$ is integrable \\
	$\Leftrightarrow$ $\pab \gt =0 $ \\
	$\Leftrightarrow$  $d_{(n)}$ is compatible with $\J_A$, ie $\left(d_{(n)}\right)^2 \big|_{(0,2)} =0$ \\
	$\Leftrightarrow$ $\gt \W d\gt \W d\gt=0$ \\
	$\Leftrightarrow$ $A^i$ is perfect : $F^i \W F^j \propto \gd^{ij} d^4x$.\\
	
	It follows that under this conditions the $\Oc(n)$-bundles are holomorphic with Hermitian metric \eqref{Euclidean metric Twistor Space: Hermitian metric on O(n)} and $d_{(n)}$ is the associated Chern connection.
\end{Proposition}
We recall that by proposition \eqref{Proposition: Perfect connection} a perfect connections is the connection of an anti-self-dual Einstein metric. In particular, under the assumption of proposition \ref{Proposition : Integrability of A}: \[ \text{\emph{The Urbantke metric with volume form $\frac{3}{2 \gL^2} F^k \W F^k$ is anti-self-dual Einstein}}. \]

\begin{proof}\mbox{}\\
	
	We now prove that each point taken separately is equivalent to perfectness of the connection.
	
	It is easy to check that $\gt \W d\gt \W d\gt = \gt \W F^{A'B'} \W F^{C'D'} \p_{A'}\p_{B'}\p_{C'}\p_{D'}$ and thus $\gt \W d\gt \W d\gt=0$ is directly equivalent to the perfectness of the connection.
	
	In the previous chapter, we saw that $F^i = s \, \sqrt{X}^{ij} \gS^j$. Thus we can write
	\begin{equation}\label{Twistor theory revisited: Proof: F decomposition}
	F^{A'B'} = \Psi^{A'B'}{}_{C'D'} \gS^{C'D'} + \gl(x)\gS^{A'B'}\quad \text{with} \quad \Psi^{A'B'C'D'} = \Psi^{(A'B'C'D')}.
	\end{equation}
	It was also explained in the previous chapter that the self-dual Einstein equations, i.e perfectness of the connection, are equivalent to $\Psi=0$. Then our choice of volume form $\mu = 2\gS^i \W \gS^i = \frac{3}{2 \gL^2} F^k \W F^k$ gives $ \gl(x) = s |\gL|$.
	
	Taking \eqref{d tau} and \eqref{Twistor theory revisited: PT sympleptic structure} together with \eqref{Euclidean metric Twistor Space: Sigma Restrictions}, \eqref{Definition: Dolbeault Operators} and \eqref{Twistor theory revisited: Proof: F decomposition} one easily shows that 
	\[ \left(d_{(2)}\gt \right)\big|_{0,2} = \Psi\p\p\p\p \; \frac{\gS \ph\ph}{(\pp)^2},\qquad \pab \gt = \Psi\p\p\p\ph \; \frac{\gS \p \ph}{(\pp)^2},\]
	and
	\[
	\frac{1}{n}\left(d_{(n)}\right)^2 \big|_{(0,2)}= \Psi\p\p\p\ph \frac{\gS \ph\ph}{(\pp)^2}.\]
		Therefore $d_{(2)}\gt \big|_{(0,2)} =0$,  $\pab \gt = 0$ and $\left(d_{(n)}\right)^2 \big|_{(0,2)}=0$  are separately equivalent to $\Psi =0$, i.e to the perfectness of the connection.
	
	Finally, all is left to show is that integrability of $\J_A$ is equivalent to the perfectness of the connection. However integrability is equivalent to having both $d\gt \big|_{(0,2)} =0$ and $d\left(e^{AA'}\p_{A'}\right)\big|_{(0,2)}$ and thus imply perfectness of the connection. On the other hand, if the connection is perfect then \[ d\left(e^{BB'}\p_{B'}\right)\big|_{(0,2)} = \left(d_A e^{BB'} \right)\big|_{(0,2)} \p_{B'} + e^{BB'} \W D\p_{B'}\big|_{(0,2)} = \left(d_A e^{BB'} \right)\big|_{(0,2)} \p_{B'}  \] holds identically as a result of $\bdA$ being the self-dual connection associated with the tetrad.
\end{proof}

The main difference with the traditional results from \cite{Atiyah:1978wi} (that we described in the previous section cf Prop \ref{Proposition: Metric Twistor, integrability} and Prop \ref{Proposition: Metric Twistor, holomorphic gt}) is that integrability is not only related to the anti-self-duality but is irremediably linked to Einstein equations. This is because in the construction described in \cite{Atiyah:1978wi} one is only interested in a conformal class of metric while here the use of the connection automatically fixes the `right scaling' that gives Einstein equations.

It is also natural to ask under which condition the almost hermitian structure is almost Kähler and Kähler. In fact those two situations necessarily come together but depends on the sign of the connection: 

\begin{Proposition} \label{Proposition: Kahler condition}\mbox{}\\
	$g_A$ is Kähler \\
	$\Leftrightarrow$ $\J_A$ is integrable, $s=1$ and $R^2 = \frac{3}{|\gL|}$\\
	$\Leftrightarrow$ $\go_A$ is closed (an thus sympleptic) \\
	$\Leftrightarrow$ $\go_A = 2i R^2 \;\go_s$ 
\end{Proposition}
\begin{proof} \mbox{}\\
	$\go_A = 2i R^2 \;\go_s$ is easily shown to be equivalent to $F^i = \frac{1}{R^2} \gS^i$. This is only possible if the connection is perfect with positive sign. The same is true for the closeness of $\go_A$, ie direct computations shows the equivalence of the last two point with $F^i = \frac{1}{R^2} \gS^i$.
	
	Now, from proposition \ref{Proposition : Integrability of A} perfectness of the connection is equivalent to integrability. Incidentally one sees from $F^i = \frac{1}{R^2} \gS^i$ that the metric associated with the connection is self-dual Einstein with cosmological constant $\gL = \frac{3}{R^2}$.
\end{proof} 

When the connection is perfect, the Hermitian structure that we described restrict to the usual Hermitian structure on twistor space constructed from an Instanton cf Prop \ref{Proposition: metric Twistor, Kähler condition}. The discussion on the sign of the connection parallel the well known fact that this Hermitian structure can be made Kähler only if the cosmological constant is positive.

\subsection{The Mason-Wolf Action for Self-Dual Gravity}\label{ssection: Mason Wolf action}

In \cite{Mason&Wolf09}, L.Mason and M.Wolf described a twistor action for self-dual gravity. It is an action for an $\Oc(2)$-valued one-form $\gt$ and a $\Oc(-6)$-valued one-form $b$ on some 6d real manifold, the  `projective twistor space'. It essentially used a new version of the non linear graviton theorem relying on the equation $\gt\W d\gt\W d\gt=0$. This equation was understood as a sufficient condition for the integrability of a certain almost complex structure and thus, relying on Penrose-Ward Non-Linear-Graviton theorem \cite{Penrose:1976js} \cite{Ward:1980am}, as describing some Einstein anti-self-dual space-time. The Mason-Wolf action implements this constraint with a Lagrange multiplier:
\begin{equation}\label{Twistor theory revisited: Action: Mason-Wolf}
S\left[\gt, b\right] = \int_{\PT} b\W \gt\W d\gt\W d\gt .
\end{equation}
Even thought the logic that lead to this Lagrangian was somehow different, in retrospect one sees that this Lagrangian could have been guessed from the description of self-dual gravity in terms of perfect connections. Indeed, as already explained in section 2, in terms of $SU(2)$-connections the equations for self-dual gravity read $F^{(A'B'} \W F^{C'D')}=0$ and therefore we can easily obtain an action for self-dual gravity by implementing this constraint by a Lagrange multiplier:
\begin{equation}\label{Twistor theory revisited: Action : Space-time SD gravity FF}
S\left[B, A\right] = \int_{M} B_{A'B'C'D'} F^{A'B'}\W F^{C'D'},
\end{equation}
where the $B$ field is completely symmetric, $B^{A'B'C'D'} = B^{(A'B'C'D')}$.

Now, as discussed before the natural `Penrose transform' of a $SU(2)$-connection is the $\Oc(2)$-valued one-form on $\PT(M)$,  \begin{equation}
\gt = \p_{A'}\left(d\p^{A'} + A^{A'}{}_{B'} \p^{B'}\right). 
\end{equation} We also take the Penrose transform of $B$ to be  \begin{equation}
B^{A'B'C'D'} = \int_{\CP^1} \p^{A'}\p^{B'}\p^{C'}\p^{D'}\; b \W \gt
\end{equation} with $b$ a $\Oc(-6)$ valued (0,1)-form on $\PT(M)$. This is just the usual Penrose transform for massless fields, see eg \cite{Woodhouse85}.  We recall from the previous discussion that \begin{equation}
\gt \W d\gt \W d\gt = \gt \W F^{A'B'}\p^{A'}\p^{B'}\W F^{C'D'}\p^{C'}\p^{D'}.
\end{equation}

From this one readily sees that the Mason-Wolf action \eqref{Twistor theory revisited: Action: Mason-Wolf} is the immediate generalisation of \eqref{Twistor theory revisited: Action : Space-time SD gravity FF}. 

The aim of the next sub-section is to make the relation between the Mason-Wolf action and the connection description of Einstein self-dual connections even more precise by giving a new proof of the non-linear graviton theorem that emphasises this relation.

Our proof will indeed make it clear that this version of the non linear graviton theorem has a strong `connection' flavour. It might therefore suggest new types of generalisation to full gravity. We discuss some of those in the next section where we will described strategies towards a twistor action for full gravity. 

\subsection{The Non-Linear-Graviton Theorem Revisited} \label{ssection: the NLG revisited}

Up to now we constructed different geometrical structure on $\PT(M)$ from a definite connection. In particular we saw that $\PT(M)$ can be given a Kähler structure when $\gt \W d\gt \W d\gt =0$, with $\gt = \p_{A'} \left( d\p^{A'} + A^{A'}{}_{B'}\p^{B'}\right)$. 

We are now interested in the reverse problem: We take `projective twistor space' $\PTc$ to be an oriented manifold diffeomorphic to $\R^4\times S^2$ together with a one-form $\gt \in \gO^1_{\C}\left[\PTc, L \right]$ with values in a line bundle $L$ over $\PTc$. We suppose this line bundle to be such that its restriction to each $S^2$ has Chern class 2. This is enough to define an almost complex structure $\J_{\gt}$ on $\PTc$ as we now describe.\footnote{We thank L.Mason for important discussions and suggestions that greatly contributed to this presentation.}

\paragraph{The almost complex structure $\J_{\gt}$}\mbox{}\\
We first introduce the 4-dimensional  `horizontal distribution' $ H\subset T_{\R}\PTc $ defined as the kernel of $\gt$, $H = Ker\left(\gt\right) $. 

We then determine $\gl\in C^{\infty}(\PTc)$ as
\begin{equation}\label{Twistor theory revisited: lambda def}
\gt \W \gtb \W \left( \gl d\gt + d\gtb \right)^2 =0.
\end{equation}
This is a quadratic equation for $\gl$. We then construct $a$, a connection on the $L$ bundle, defined modulo the addition of multiple of $\gt$ and $\gtb$ by requiring, 
\begin{equation}\label{Twistor theory revisited: a def}
\gtb \W \left( d\gtb + \gl \left(d\gt + a \W \gt  \right)\right)^2 =0.
\end{equation}
This is in fact linear in $a$ and has the right number of components to determine $a$ modulo $\gt$ and $\gtb$.
From all this we define the complex three-form,
\begin{equation}\label{Twistor theory revisited: J from tau def}
\gOb = \gtb \W \left( d\gtb + \gl\,d_a\gt \right).
\end{equation}
This three-form in turn defines an almost complex structure, $\J_{\gt}$: We just define the holomorphic tangent space to be the kernel of $\gOb$,
\[ 
X \in T^{1,0}\PTc \quad \overset{def}{\Longleftrightarrow} \quad X \id \gOb =0.
\]
From \eqref{Twistor theory revisited: a def} we see that the Kernel of $\gO$ indeed is 3-dimensional as required for an almost complex structure. Note that this definition of $\J_{\gt}$ is equivalent to requiring that $\gOb$ is $(0,3)$. In particular its complex conjugate $\gO$ is (3,0).

\paragraph{Spacetime from $\J_{\gt}$} \mbox{}\\
Having constructed an almost complex structure, $\J_{\gt}$ on $\PTc$ we are now in a similar situation as in \cite{Mason05} where the almost complex structure is taken as a starting point. 

Following the same steps as in this reference \emph{we can construct a Euclidean `space-time' $M$ from $\left(\PTc, \J_{\gt}\right)$}. Then $\PTc$ has the structure of a fibre bundle over $M$ : $\CP^{1} \inj \PTc \to M$. Twistor space $\mathcal{T}$ is taken as the total space of a special line bundle over $\PTc$. We here recall how this works for completeness.\\

We first introduce a conjugation $\circonf \from \PTc \to \PTc $, $\circonf^2 = 1$, that reverses $\J_{\gt}$, i.e $\circonf^* \J_{\gt} = - \J_{\gt}$. We also assume that this conjugation has no fixed points. This is a common in twistor theory and will lead to a Euclidean Space-time, the other alternative (existence of fixed points for $\circonf$) would lead to Lorentzian signature.

We now take as \emph{`complexified space-time'} $\Mc$ the moduli space of pseudo-holomorphic rational curves in $\PTc$, ie the space of embedded $S^2$ in $\PTc$ in the same topological class as the $S^2$ factors in $\PTc \simeq \R^4 \times S^2$ such that $\J_{\gt}$ leaves the tangent space invariant and thus inducing a complex structure on these embedded two-spheres. Theorems in McDuff and Salamon \cite{McDuff&Salamon2004} imply that $\Mc$ exists and is 8-dimensional if $\J_{\gt}$ is close to the standard complex structure on a neighbourhood of a line in $\CP^3$. We assume this condition to be satisfied. This can be done by requiring that our one-form $\gt$ is close to the standard holomorphic one-form with values in $\Oc(2)$ on $\CP^3$.

The conjugation $\circonf$ induces a conjugation on $\Mc$, $\circonf \from \PTc \to \PTc$ and we define our \emph{Euclidean space-time} $M$ as the 4-dimensional fixed point set of $\circonf$ on $\Mc$. There is then a natural projection $P \from \PTc \to M$ as a consequence of the fact that from our assumption that there will be a unique rational curves in $\PTc$ through $Z$ and $\hat{Z}$. By construction our projective twistor space $\PTc$ now is the total space of a fibre bundle over $M$ with fibre $\CP^1$: $\CP^1 \inj \PTc \xto{P} M$.

We will also assume that $\J_{\gt}$ is such that the canonical bundle $\gO^{3,0}$ has Chern class $-4$ on each $S^2$ in $\PTc$. This will be the case if we construct $\gt$ by a small deformation of the standard holomorphic one-form with values in $\Oc(2)$ on $\CP^3$.

We then define the associated twistor space $\Tc$ to be the fourth root of the canonical bundle. It is thus a complex line bundle over $\PTc$, $\C \inj \Tc \xto{\Pi} \PTc$. We denote the complex line bundle $\left(\gO^{3,0}\right)^{-\frac{n}{4}}$ by $\Oc(n)$. When restricted to each $\CP^1$ fibres in $\PTc$, these bundles will restrict to the usual $\Oc(n)$ holomorphic bundle on $\CP^1$ and thus the notation is coherent. We can now think of $\Tc$ as a complex rank two vector bundle over $M$ with structure group $SU(2)$, $\C^2 \inj \Tc \xto{P'} M$.

\paragraph{A non linear graviton theorem} \mbox{}\\
We now give a new proof of the (euclidean) non-linear-graviton theorem. As explained in introduction, the essential result of this theorem already appeared in \cite{Mason&Wolf09} but the presentation that we make here is original.\\

Introduce coordinates that form a trivialisation of $\mathcal{T}$, $\{x^{\mu}, \p_{A'} \}$.  $\p_{A'} \; \text{with} \; \text{\footnotesize A'} \in\{0,1\}$ are linear coordinates on the fibres of $\C^2 \inj \Tc \xto{P'} M$ and $x^{\mu}$ are local space-time coordinates on the base.\\ Then,
\begin{Proposition}\label{Proposition: Non Linear graviton} \mbox{}\\
	(i)  \quad 	$\J_{\gt}$ is integrable\\
	(ii) \quad $\Leftrightarrow$ $\gt \W d\gt \W d\gt =0$ \\
	(iii)\quad $\Leftrightarrow$ $\gt = \gc \left(\p_{A'}d\p^{A'} + A(x){}^{A'}{}_{B'}\p^{B'}\p_{A'} \right)$ with $\gc \in \C^{\infty}\left(\PTc \right)$ and $A^{A'}{}_{B'}$ a perfect connection on $M$.
\end{Proposition}

\begin{proof} \mbox{} \\
	We now prove $(i) \Rightarrow (ii) \Rightarrow (iii) \Rightarrow (i)$.\\
	$(i) \Rightarrow (ii)$\\
	By construction, $\gt \W \left(d\gt + \bar{\gl} d_{\bar{a}} \gtb \right)^2 = \gO \W \left( d\gt + \bar{\gl} d_{\bar{a}} \gtb \right)=0$, integrability means that $\gO \W d\gt =0$ and thus $\bar{\gl} \;  \gO \W d_{\bar{a}}\gtb = 0$. It follows that either $\gl=0$ or $d_{\bar{a}}\gtb \big|_{(0,2)} =0$.
	If $\gl=0$ then $\gt\W d\gt \W d\gt =0$. Suppose $d_{\bar{a}}\gtb \big|_{(0,2)} =0$, integrability implies that $ d_{\bar{a}} \gtb \big|_{(2,0)} = 0$ and therefore $d_{\bar{a}} \gtb \in \gO^{1,1}$. It follows that both $d_{\bar{a}} \gtb \in \gO^{1,1}$ and $d_{a} \gt \in \gO^{1,1}$. However this is in contradiction with $\gt \W \left( d_{a}\gt + \gl d_{\bar{a}} \gtb \right) \in \gO^{3,0}$. \\
	$(ii) \Rightarrow (iii)$\\
	If $\gt \W d\gt \W d\gt =0$ then by construction $\gt \W d\gt \in \gO^{3,0}$. We now take $\gz$ to be coordinates on $\CP^1$, $\pa_{\gzb}$ is the anti-holomorphic vertical tangent vector. It follows that $\pa_{\gzb} \id d\gt \propto \gt$. Using coordinates adapted to the trivialisation we can write $\gt =\gl\left( d\gz + A_{\mu} dx^{\mu}\right)$ and $\gt\W \gzb \id d\gt =0$ implies $\pa_{\gzb}A_{\mu}=0$. $A_{\mu}$ being $\Oc(2)$ valued this last relations implies, by a generalisation of Liouville's theorem, $A = A(x){}^{A'}{}_{B'}\p^{B'}\p_{A'}$. Now  \ref{Proposition : Integrability of A} implies that $A^{A'}{}_{B'}$ is perfect.\\
	$(iii) \Rightarrow (i)$\\
	Starting with a perfect connection $A$, we have from Proposition \ref{Proposition : Integrability of A} that $\gt \W d\gt \W d\gt =0$. From the definition of the almost complex structure, this implies $\gt \W d\gt = \gt \W F\p\p \in \gO^{(3,0) }$. For a perfect connection, $F^i = s\, \frac{\gL}{3} \gS^i $, and the almost complex structure is therefore the same as in proposition \ref{Proposition: Almost Hermitian structure}. Finally, by proposition \ref{Proposition : Integrability of A} perfectness of the connection implies integrability.
\end{proof}

From this point of view, the `non-linear graviton theorem', with central equation $\gt\W d\gt\W d\gt=0$, can therefore be understood as a deep generalisation of the description of self-dual gravity in terms of perfect connections $F^i\W F^j = \frac{\gd^{ij}}{3} F^k\W F^k$, cf \eqref{Proposition: Perfect connection}. As we already reviewed, full gravity can be described in terms of $SU(2)$-connection only (cf Prop \eqref{Proposition: Pure connection equation}) and this is therefore suggestive of a twistor description of full gravity in terms of the one-form $\gt$ only. 

\section{Discussion on the would be `Twistor action for Einstein gravity'} \label{section: Discussion on the would be `Twistor action for Einstein gravity'}

In \cite{Mason05} two new variational principles for Yang-Mills theory and conformal gravity based on fields living on twistor space were presented. The fact that the fields which appear in this action live on a 6d manifold (`projective twistor space') is compensated by new symmetries of the action and the propagating degrees of freedom thus remain the same as in the usual Yang-Mills or Conformal gravity action as expected. 
One of the nice features of these actions is that they give a natural explanation for why there is a MHV formalism (see for the original paper \cite{Cachazo:2004kj}) for Yang-Mills and Conformal gravity. Because of the extra symmetries that these actions enjoys (as compared to the space-time actions) they allow to choose a special gauge ( referred to as CSW gauge) that makes a MHV formalism manifest (cf \cite{Jiang08}, \cite{Adamo:2013cra}, \cite{Adamo:2011pv} and \cite{Adamo:2013tja}).
If such an action existed for GR one could expect the same phenomenon and it could serve as a proof for the existence (or the obstruction to the existence) of a MHV formalism for GR.

As already described in section \ref{ssection: Mason Wolf action} a twistor action for self-dual gravity was presented in \cite{Mason&Wolf09}. Then, in \cite{Adamo:2013tja} a conjectured twistor action for full gravity was proposed. However, in spite of the many interesting features of this conjectured twistor action, some geometrical understanding is lacking, mainly because it is formulated around a fixed background, and this makes it unclear whether it actually describe GR or not.

Both the twistor action for Yang-Mills and Conformal gravity where obtained by a generalisation of the respective space-time action. We very briefly sketch how this works for Yang-Mills, but refer the reader to \cite{Mason05} for details on the construction. This is of interest for us because we will see that, together with the description of metric in terms of connection described in section \ref{section: Chiral Formulations of GR - Fundations} it has an immediate generalisation to gravity. The discussion given here already appeared in \cite{Herfray:2016qvg}.

\subsection{The Twistor action for Yang-Mills from the Chalmer-Siegel action}\label{ssection: Chalmer Siegel action}

In \cite{Mason05}, the Chalmer-Siegel \cite{Chalmers:1996rq} action for Yang-Mills was taken as a starting point on the way to a twistor action:
\begin{equation}
S\left[\bdA, \bdB\right] = \int_M Tr\left( \bdB \W \bdF - \frac{\eps}{2} \bdB \W \bdB \right)
\end{equation}
where $B$ is taken to be a lie algebra valued \emph{self-dual} two-form, ie $\bdB = \bdB_{A'B'} \gS^{A'B'}$. Here $\gS^{A'B'}$ is a basis of self-dual two-forms associated with a fixed background flat metric and constructed as in \eqref{Chiral Formulations of GR - Fundations: Sigma def (tetrad)}. As already described, the Euclidean twistor space is the total space of the primed spinor bundle over $M$, the almost complex structure on $\PT$ is given by taking the $(3,0)$-form on $\PT$ to be $\gO = \gt \W \gS^{A'B'}\p_{A'}\p_{B'} = \p_{C'}d\p^{C'}\W \gS^{A'B'}\p_{A'}\p_{B'}$. For coherence with the previous sections, we also introduce the $\Oc(2)$-valued $\left(1,0\right)$-form $\gt= \p_{A'}d\p^{A'}$ on $\PT$.

An interesting feature of this action is that for $\eps=0$ we are left with an action for self-dual Yang-Mills. This is a key point to make contact with twistor theory as self-dual Yang-Mills solutions are fully understood in terms of geometry of the twistor space through the Ward transform \cite{Ward:1977ta}.

If we take the Penrose transform of $\bdB_{A'B'}$ to be\footnote{In order to make sense of the integral of a lie algebra valued form this is here implicit that one should take an holomorphic frame for the associated bundle. By Liouville theorem, holomorphicity on a compact manifold ensure that this choice of frame is unique up to a global `rotation'. }
\begin{equation}
\bdB_{A'B'} = \int_{\CP^1} \p_{A'}\p_{B'} \;\bdb\W \gt
\end{equation}
(where $\bdb$ is a Lie algebra valued $(0,1)$-form on $\PT$ with values in $\Oc(-4)$) and plug it into the action, we see that it is suggestive of the twistor action for Yang-Mills \cite{Mason05}:
\begin{equation}
S\left[\bda, \bdb\right] = \int_{\PT} Tr\left( \bdb \W \bdf \W \gO \right) - \frac{\eps}{2} \int_{M\times \CP^1 \times \CP^1} Tr\left(\bdb_1 \W \gt_1 \W \bdb_2 \W \gt_2 \left(\p_{1}{}_{A'} \p_{2}{}^{A'}\right)^2 \right).
\end{equation}
Where now $\bda$ is taken to be a (0,1) $SU(N)$-connection of a Yang-Mills bundle \emph{over $\PT$} and $\bdf \in \gO^{0,2}\left(\PT \right)$ its curvature. For $\eps=0$ varying the action with respect to $\bdb$ gives $\bdf=0$ and thus gives this bundle the structure of a holomorphic bundle over $\PT$. By a theorem from Ward \cite{Ward:1977ta}, this is equivalent to self-dual Yang-Mills equations see also \cite{Woodhouse85} for Euclidean methods in twistor theory. What is more, it turns out that this action describes full Yang-Mills for $\eps\neq 0$. Again, the aim here is just to sketch how this action is constructed and refer to \cite{Mason05} for a proper discussion.

\subsection{A first twistor ansatz... and why it fails}

We now would like to take as starting point the following space-time action 
\begin{equation}
	S[\bdA,\Psi] = \frac{1}{2}\int \left(\left(\Psi +\frac{\gL}{3} \gd \right)^{-1}\right)^{ij} F^i\W F^j.
\end{equation}
This is an action for gravity and can be obtained from Plebanski's action by integrating out the $\bdB$ field (see section \ref{ssection : Variational Principles}). It is interesting to consider what happens if one expand the first term in series. Then, up to a global rescaling and a topological term we obtain
\begin{equation}
S[A, \Psi] = \int \Psi^{ij}\; F^i\W F^j + \sum_{k=0}^{\infty} \left(\frac{3}{\gL}\right)^{k+1} \; \left(\Psi^{k+2}\right)^{ij} F^i \W F^j.
\end{equation}

This action, which does not seem to have attracted much attention up to now, has the following interesting interpretation: in the limit where $\gL$ goes to infinity we recover an action for anti-self-dual gravity. For a finite $\gL$ however this action describe full GR as an interacting theory around the anti-self-dual background with the cosmological constant playing the role of coupling constant. This parallels the Chalmers-Siegel action for Yang-Mills.

It suggests the following twistor ansatz,

\begin{align}
S\left[\gt, \psi \right] &= \int_{\PTc} \psi \W \gt \W d\gt \W d\gt \\
& \hspace{1cm}  + \sum_{k=0}^{\infty} \left(\frac{3}{\gL}\right)^{k+1}\;\smashoperator{\int_{\hspace{1.75cm} M \times \CP^1 \times \CP^1}} \psi_1 \W \gt_1 \W d\gt_1 \W \psi_2 \W \gt_2 \W d\gt_2  \left(\Psi\right)^k{}^{A'B'}{}_{C'D'}\;\p_1{}^{C'}\p_1{}^{D'}\p_2{}_{A'}\p_2{}_{B'} \nonumber
\end{align} 
where  $\gt$ is a $\Oc(2)$-valued one-form on $\PTc$. As we already described in \ref{ssection: the NLG revisited} such a one-form is enough to construct an almost complex structure $\J_{\gt}$ on $\PTc$ and to give it a fibre bundle structure over some space-time $M$, $\CP^1 \inj \PTc \to M$.

This action also contains $\psi\in \gO^1_{\C} \otimes \Oc\left(-6\right)$ a one-form on $\PTc$ with values in $\Oc(-6)$, its Penrose transform is $\Psi(x){}^{A'B'}{}_{C'D'}$:
\begin{equation}
\Psi\left(x\right){}^{A'B'}{}_{C'D'} = \int_{\CP^1_x} \psi\W \gt \;\p{}^{A'}\p{}^{B'}\p{}_{C'}\p{}_{D'}
\end{equation}
where $\CP^1_x = P^{-1}(x)$ is the fibre above $x\in M$.

Interestingly, when $\gL$ goes to infinity one recovers the Mason-Wolf action for self-dual gravity described in section \ref{ssection: Mason Wolf action}. What is more, truncating the infinite sum to the first term one recovers an action that looks like a background independent version of the twistor action conjecture in \cite{Adamo:2013tja}.

Unfortunately, despite those encouraging features, one cannot prove that this twistor action is related to gravity. To do so one would hope that varying this action with respect to $\psi$ would give enough field equations to recover a $SU(2)$-connection from $\gt$ as was the case in our proof of the non-linear graviton theorem  \ref{ssection: the NLG revisited}. However this does not seem to be the case here. We are thus unable to make contact with a space-time counter part of this action and the interpretation of the fields equations remains obscure.

\subsection{A new action for Gravity as a background invariant generalisation of the Chalmers-Siegel action.}\label{ssection:  new action for Gravity as a background invariant generalisation of the Chalmers-Siegel action.}

Let us now come back to the Chalmer-Siegel action and consider the special case of a $SU(2)$-connection:
\begin{equation}\label{Action: Chalmer-Siegel}
S\left[A, B\right] = \int_M B^i_{A'B'} \gS^{A'B'}\W F^i - \frac{\eps}{2} B^i_{A'B'}B^i{}_{C'D'} \gS^{A'B'}\W \gS^{C'D'}
\end{equation}
Here $\gS^{A'B'}$ is a basis of self-dual two-forms associated with a fixed background flat metric and constructed as in \eqref{Chiral Formulations of GR - Fundations: Sigma def (tetrad)}. This action has obviously nothing to do with an action for gravity as a background metric is present.

However, as explained in section \ref{section: Chiral Formulations of GR - Fundations}, a definite $SU(2)$-connection is enough to define a conformal class of metric. If we now choose a representative in the conformal class, and use it to parametrise the $\gS$'s, $\gS_A = \gS(g_A) = \gS(A) $, the action \eqref{Action: Chalmer-Siegel} becomes background independent:
\[ 
S[A^i, B^{ij}] = \int B^{ij} \gS_A^i\W F^j  - \frac{\eps}{2} B^{ij}B_{ik}\; \gS_A^j\W \gS_A^k
\]
If we take $B^{ij}$ to be unconstrained, the action ends up to be topological. However if we take $B^{ij}$ to be traceless with the good choice of volume form then the action happens to describe gravity in the pure connection formulation (in fact, we already encounter this action at the very end of section \ref{ssection : Variational Principles} see \eqref{Chiral Deformations of Gravity: S_{GR}[A,M]}):
\begin{Proposition}\label{Action: generalised CS}\mbox{}\\
	The action
	\begin{equation}
	S[A^i, B^{ij}] = \int B^{ij} \gS_A^i\W F^j  - \frac{\eps}{2} B^{ij}B_{ik}\; \gS_A^j\W \gS_A^k,
	\end{equation}
	with $\gS_A^i$ the basis of orthonormal self-dual two-form associated with Urbantke metric with volume $\frac{1}{2\gL^2}\left(tr \sqrt{F\W F}\right)^2$ and $B^{ij}$ a traceless matrix,  describes the vacuum solution of Einstein equations with non zero cosmological constant. What is more for $\eps=0$ this action describes anti-self-dual gravity.
\end{Proposition}
\begin{proof}
	
	By construction the $\gS_A$'s are such that,
	\begin{equation}
	 F^i = M^{ij} \gS_A^j.
	\end{equation}
	Our choice of volume form,
	\begin{equation}
	\frac{1}{3}\gS^i \W \gS^i = \frac{1}{\gL^2}\left(tr \sqrt{F\W F}\right)^2,
	\end{equation}
	is such that $Tr M = \gL$ is a constant.
	
	Now, varying the action with respect to $B$, we get 
	\begin{equation}
	M^{ij}\big|_{trace-free} = \eps B^{ij} 
	\end{equation}
	which is equivalent to
	\begin{equation}
	F^i =  \left(\eps B^{ij}+ \frac{\gL}{3} \gd^{ij}\right)\gS^j.
	\end{equation}
		For $\eps=0$, these are the equations for self-dual gravity in terms of connection.
	For $\eps\neq 0$ we can solve for $B$, plugging this back into the action we obtain
	 \begin{equation}
	S\left[A\right] = \frac{1}{\eps} \int \frac{1}{2} F^i\W F^i - \frac{1}{6} \left(tr \sqrt{F\W F}\right)^2.
	\end{equation}
	
	Up to a topological term, this is just the pure connection action for gravity \cite{Krasnov:2011pp}. 
\end{proof}

\subsection{Discussion on a second ansatz}

The action in Proposition \eqref{Action: generalised CS} looks like a promising starting point to construct ansatz for twistor action for gravity. It indeed has many appealing features. First it explicitly separates the self-dual sector ($\eps=0$) of the theory from the full theory ($\eps \neq 0$). Second it superficially looks like the space-time counterpart of the twistor action conjectured in \cite{Adamo:2013tja}. Finally, as explained in the previous section the $SU(2)$-connection on $M$ can naturally be lifted as a one-form $\gt$ on $\PTc$. Starting with an action of this type would again allow to use the machinery described in section \ref{section: Twistor theory revisited}.

However as to now, despite many attempts from the author of this paper, none of the ansatz that are suggested by this action seem to lead to an interesting gravity action. We describe here one attempt that seemed at some point the most promising to the author. We will see that it can indeed eventually lead to a certain variational principle in twistor space but at the expense both of technical complications and the addition of an unnatural constraint. Thus the result seems both too complicated to be directly useful (let say for computing scattering amplitudes) and to anaesthetic to be otherwise appealing. However, on the way the interested reader should get some glimpses on the type of difficulties that one faces when one tries to construct such a variational principle in twistor space.

The essential idea here is it that we now would like to construct an action of the type $S\left[\gt, \psi \right]$ on $\PTc$, some real 6d manifold. First using the results from section \ref{section: Twistor theory revisited}, one can construct an almost complex structure $\J_{\gt}$ which in turn allows to construct some space time $M$, giving $\PTc$ a fibre bundle structure $\CP^1 \inj \PTc \xto{\pi} M $. A look at the action \ref{Action: generalised CS} then suggests the following `twistor ansatz':
\begin{align}\label{twistor ansatz 2}
S\left[\gt, \psi \right]&= \int_{\PT} \psi \W \gt \W d\gt \W \gS_{\gt} + \frac{\eps}{2} B^{A'B'} \W B_{A'B'}
\end{align}
where $\gS_{\gt}$ should be constructed from $\gt$ only,
\begin{equation}
\psi \in \gO^{1}_{\C}\otimes \Oc(-6)
\end{equation}
and 
\begin{equation}
B(x)^{A'B'}= \int_{\CP_X^1} \p^{A'}\p^{B'} \; \psi \W \gt \W \gS_{\gt}.
\end{equation}

Where in this last line one should integrate over $\pi^{-1}(x) \simeq \CP^1$.

An appealing feature of actions of this type is that, linearising around a given background (let say describing flat space-time) we obtain $\gd\psi \in H^{0,1}\left(\PTc, \Oc(-6)\right)$ and $\gd \gt \in H^{0,1}\left(\PTc, \Oc(2)\right)$ which are then naturally interpreted as the Penrose transform of a propagating self-dual $\Psi_{A'B'C'D'}$ and anti-self-dual $\Psi_{ABCD}$ gravitons.

The difficult part now is to make sense of $\gS_{\gt}$. We propose the following.
In section \ref{section: Twistor theory revisited} we partly defined a connection $a$ on $\Oc(2)$,  through the relation
\begin{equation*}
\gtb \W \left( d\gtb + d_a \gt\right)^2 =0
\end{equation*}
however as for now it is only defined up to multiple of $\gt$, $\gtb$. If we require as some non-degeneracy condition that $\gt$ does not vanish on the $\CP^1$ that fibres $\PTc$, we can then completely fix $a$ by requiring that the $\Oc(2)$-connection that is induces on each $\CP^1$ fibres is the Levi-Civita connection of the Kahler metric on $\CP^1$. Now that $a$ is completely defined we have access to its curvature $\left( d_a\right)^2$.

Consider the following triple of two-forms: $\left(d_a\gt,d_a\gtb, \left( d_a\right)^2  \right)$ . Generically it spans a 3d subspace of the two-forms of each horizontal space and thus allows us to defines a conformal metric (the associated Urbantke metric cf section \ref{section: Chiral Formulations of GR - Fundations}) \emph{on each horizontal tangent spaces}. Let us see how it works explicitly:\\ Define
 \begin{equation}
B^{A'B'} \coloneqq \frac{\p^{A'}\p^{B'}}{\left( \pp\right)^2} d_a\gtb + \frac{\ph^{A'}\ph^{B'}}{\left( \pp\right)^2} d_a\gt + \frac{\p^{(A'}\ph^{B')}}{\pp} \left( d_a\right)^2
\end{equation}
and
\begin{equation}
B^i = \gs^i_{A'B'} \;B^{A'B'}.
\end{equation}

This last object should be understood as an $\su2$-valued two-form. From this we can follow the same procedure as in the first section and construct $\gS$:
\begin{equation}
\gS^i\left( x,\gz\right)= X^{-\frac{1}{2}}{}^{ij}B^j \qquad \gS^{A'B'}=\gs^{A'B'}_i \gS^i
\end{equation}
such that $\gS^i \W \gS^j \propto \gd^{ij}$.
It is associated with a conformal class of metric on each horizontal space $e{}^{AA'}\left( x,\p\right)$, $\gS^{A'B'}= e^{A'}{}_{A}\W e^{B'A}$.

Importantly at this point the tetrad on the horizontal tangent space $e^{AA'}$ varies along the fibre $e^{AA'} = e^{AA'}(x,\p)$.

In the end we define the $\Oc(2)$-valued two-form on $\PT$:
 \begin{equation}
\gS_{\gt} (x, \pi) = \gS^{A'B'} (x, \p) \p_{A'}\p_{B'} = \theta^0\W \theta^1, \quad \text{with} \; \theta^{A}= e^{AA'}\p_{A'}
\end{equation}

This construction is not as arbitrary as it might seem at first sight: in the particular case where there is an underlying $SU(2)$-connection such that $\gt = \p_{A'}\left(d\pi^{A'} + A^{A'}{}_{B'}\pi^{B'} \right)$, it precisely coincides with the construction from proposition \ref{Proposition: Almost Hermitian structure}. The connection $a$ on $\Oc(2)$ then coincides with the connection described at the beginning of section \ref{section: Twistor theory revisited} and the restriction of the triplet  \begin{equation}
\left(d_a\gt,d_a\gtb, \left( d_a\right)^2  \right)
\end{equation}  to the horizontal tangent space then indeed is just  \begin{equation}
\left(F^{A'B'}\p_{A'}\p_{B'},F^{A'B'}\ph_{A'}\ph_{B'} , F^{A'B'}\p_{A'}\ph_{B'} \right). 
\end{equation} Note that we did not need to assume the connection to be perfect. It can then be checked that, under such conditions, the twistor action \eqref{twistor ansatz 2} coincides with the original  space-time action from proposition \eqref{Action: generalised CS}.

Therefore we could hope that with this definition for $\gS_{\gt}$, the action \eqref{twistor ansatz 2} would describe gravity: all we need are the field equations for $\psi$ to imply the existence of an $SU(2)$ connection such that $\gt = \p_{A'}\left(d\pi^{A'} + A^{A'}{}_{B'}\pi^{B'} \right)$. 

At this point however, it seems that we are out of luck. Varying \eqref{twistor ansatz 2} with respect to $\psi$ we obtain
\begin{equation}
\gt \W d\gt \W \gS_{\gt} + \eps \gt \W \gS_{\gt} \W B^{A'B'} \p_{A'}\p_{B'}=0.
\end{equation}
For simplicity let us consider the case $\eps=0$. Then the action in proposition \ref{Action: generalised CS} is an action for self-dual gravity and we thus would like to interpret, 
\begin{equation}\label{Twistor self dual field equations}
\gt \W d\gt \W \gS_{\gt}=0
\end{equation}
as implying the integrability of some almost complex structure and/or as the perfectness of some $SU(2)$ connection arising on the way.

However, on the one hand, due to the important non linearities involved in constructing $\gS_{\gt}$, it seems very complicated to interpret this field equations in terms of the almost complex structure from section \ref{ssection: the NLG revisited}. On the other hand, one could be tempted to consider as another almost complex structure: the one that makes $\gt \W \gS_{\gt}$ a $(3,0)$-form, then the field equations \eqref{Twistor self dual field equations} just read $d\gt \big|_{0,2}=0$.

At this point, to obtain self-dual gravity, it would be enough to be able to conclude that there exists a $SU(2)$ connection such that $\gt = \p_{A'} \left(d\p^{A'} + A^{A'}{}_{B'} \p^{B'}\right)$.

This is however not the case: generically $\gt$ can be written $\gt = d\gz + A_{\mu} dx^{\mu} $ with $A_{\mu} \in \Gamma(\Oc(2))$. From this is follows that
\begin{equation}
 \pab_{\gzb} \id d\gt =0 \qquad \Leftrightarrow \qquad \pab_{\gzb}\left(A_{\mu}\right)dx^{\mu} = 0
\end{equation}
would indeed imply that $A_{\mu}= A^{A'B'}\p_{A'}\p_{B'}$. On the other hand 
 \begin{equation}
\pab_{\gzb} \id \left(d\gt \big|_{0,2}\right) =0 \qquad \Leftrightarrow \qquad \pab_{\gzb}\left(A_{\mu}\right) dx^{\mu}\big|_{0,1} = 0
\end{equation}
are just not enough field equations to conclude that $\pab_{\gzb}\left(A_{\mu}\right)=0$. In this last case one indeed misses one half of the necessary field equations:
\begin{equation}
 \pab_{\gzb} \id \left(d\gt \big|_{1,1}\right) =0. 
\end{equation}

In principle, this missing set of equations could be implemented as a constraint in our twistor ansatz \eqref{twistor ansatz 2}: this would at last gives a twistor action for gravity. However, on top of definitely spoiling any remaining geometric aesthetics, it would also add another layer of complexity to our already complicated construction, making it more than unlikely to be useful.

\end{PartI}

\begin{PartII}
\part[\\Variations on Hitchin Theory in Six Dimensions]{Variations on Hitchin Theory \\in Six Dimensions}\label{Part: Variations on Hitchin Theory in Six Dimensions}

\section*{Introduction to Part 2:\\ \hspace*{1cm}Hitchin Theory and Six Dimensions}
\counterwithout{equation}{section}		\setcounter{equation}{0}
\counterwithout{thmcnter}{chapter}		\setcounter{thmcnter}{0}
\addstarredchapter{Introduction to Part 2}\markboth{}{Introduction to Part 2}

In the preceding part we saw that, solutions of self-dual gravity are naturally described in the following terms. Start with a $\SU(2)$-principal bundle
\begin{equation}\label{Intro partII: SU2 bundle}
\SU(2) \inj \P^7 \to \M^4
\end{equation}
together with a connection $\bdA$. Then, construct a certain almost complex structure $\J_{\bdA}$ on the associated bundle 
\begin{equation}\label{Intro partII: twistor bundle}
\S^2 \inj \PT(M) \;\to \M^4, \qquad \PT(M)\coloneqq \;\left(\S^2 \times \P^7 \right)/\SU(2).
\end{equation}
Solutions to self-dual gravity are then equivalent to integrability of $\J_{\bdA}$, see Proposition \ref{Proposition : Integrability of A}. 

When we turn to full GR things gets more complicated: On the one hand we know that Einstein equations can be stated in terms of $\bdA$ only, see Proposition \eqref{Proposition: Pure connection equation}. On the other hand, writing up a twistor formalism for Einstein equations seems nearly as hard as in the metric formalism. We face the `googly problem' once again. This is despite the fact that our connection approach suggests some natural new ansatz, see section \ref{section: Discussion on the would be `Twistor action for Einstein gravity'}. 

It might however be that sticking to complex geometry in six dimension is too restrictive. In fact, from our connection perspective, the total space of \eqref{Intro partII: SU2 bundle} seems just as natural as the associated bundle \eqref{Intro partII: twistor bundle}.

This is our motivation for investigating the relationship between certain theories of differential forms due to Hitchin, hereafter `Hitchin's theories' (see \cite{Hitchin:2000sk,Hitchin:2001rw,Hitchin:2002ea}), and gravity.

Hitchin's theories are action functionals for stable forms. Both the notion of stability for differential forms and Hitchin theories will be reviewed in general terms, i.e independently of the choice of a particular dimension, in Chapter \ref{Chapter: Hitchin theory, geometrical basis}.

For two-forms there is a well known notion of non-degeneracy. In \cite{Hitchin:2000sk}, Hitchin defined \emph{stable} k-forms in dimension $n$ as k-forms such that their orbit under the action of $\GL(n)$ is open. Stable two-forms correspond to the non-degenerate ones and therefore `stability' generalise `non-degeneracy' to any form. 

 `Stability' of differential forms is in fact is quite a stringent condition and the only interesting cases for three-forms arise in dimension six and seven. In fact stable k-forms for k bigger than two are scarce see below.

Apart for the first chapter of this part (Chapter \ref{Chapter: Hitchin theory, geometrical basis} of this thesis), we will restrict ourselves to the six dimensional case, leaving the seven dimensions case for part \ref{Part: Variations on Hitchin Theory in Seven Dimensions}. 

The first aspect of this part is to consider `variations' around Hitchin theories. In particular we propose new diffeomorphism invariant actions for two and three forms in six dimensions and demonstrate that some of them are topological. In this thesis a `topological theory' means a theory with no propagating degrees of freedom. In particular the phase space of a diffeomorphism-invariant topological theory (on a compact manifold) is at most finite dimensional. Let us briefly describe these action functionals.

We take as a starting point for our action functionals for differential forms the kinematical term
\begin{equation}\label{Intro partII: schwarz actions}
C_p \W dC_{n-p-1},
\end{equation}
where $C_p \in \gO^p\left(M^n\right)$ are $p$-forms on a n-dimensional manifold. Theories of the form \eqref{Intro partII: schwarz actions} are very well-known see \cite{Schwarz:1978cn}, \cite{Schwarz:1979ae} and we will refer to these as `Schwarz type' theories. These are obviously diffeomorphism invariant and they are known to be topological. The partition function of Schwarz type theories is a variant of Ray-Singer analytic torsion of a manifold.

 Accordingly, in our six dimensional context, we consider a theory of two- and three-forms 
\begin{equation}
B \in \gO^2(M^6), \qquad C \in \gO^3(M^6)
\end{equation}
with kinematical term,
\begin{equation}\label{Intro partII: BC schwarz actions}
B \W d C .
\end{equation}
We then consider the Hitchin volumes - written $\Phi$ and to be reviewed in the rest of this part-  for two- and three-forms, 
\begin{equation}\label{Intro partII: BC Hitchin action}
\Phi\left[B\right], \qquad \Phi\left[C\right],
\end{equation}
as potential terms that can be added to \eqref{Intro partII: BC schwarz actions}.  With these potential terms the resulting theories are  diffeomorphism invariant and we demonstrate that they remain topological. 

The reason why this is not so surprising is that these theories are closely related to Hitchin's. In Hitchin's original perspective \cite{Hitchin:2000sk,Hitchin:2001rw}, each of the functionals \eqref{Intro partII: BC Hitchin action} where meant to be evaluated on closed forms and varied inside a cohomology class. The reason for this prescription is that \eqref{Intro partII: BC Hitchin action} are constructed algebraically from $B$ or $C$ and therefore need some other constraints to give interesting differential equations. 
In the theories we consider, the kinematical term \eqref{Intro partII: BC schwarz actions} effectively implements the constraint that one of the form must be closed. Therefore they only differ from Hitchin's prescription because the forms do not have to stay in a given cohomology class. As Hitchin's theories are believed to be topological, this is not surprising that these other theories are topological as well. Here these can however be explicitly checked. 

An interesting question is to consider what happens if one adds both terms \eqref{Intro partII: BC Hitchin action} at the same time to \eqref{Intro partII: BC schwarz actions}. The resulting theory turns out to describe so-called \emph{nearly Kähler structure} on our six dimensional manifold. These structure are more natural in the context of holonomy of a Riemannian manifold. See section \ref{section: Holonomy} for more. This is another nice `variation' on Hitchin theories in six dimensions as nearly Kähler manifold had already been obtained in \cite{Hitchin:2002ea} but from a different functional. This theory is however most likely not topological.

The end chapter (Chapter \ref{Chapter: Three Dimensional Gravity as a Dimensional Reduction of Hitchin Theory in Six Dimensions}) of this part describes our main results as far as six dimensions is concerned: We demonstrate that the $\SU(2)$ reduction of Hitchin theory in six dimensions is just three dimensional gravity together with a constant scalar field. More precisely
\begin{Proposition}
Let $\SU(2)$ acts freely on a six dimensional manifold $P^6$. In particular $P^6$ has the structure of a principal bundle
\begin{equation}
\SU(2) \inj P^6 \to M^3
\end{equation}
Let us consider the Hitchin volume $\Phi[C]$ evaluated on a closed $\SU(2)$-invariant three-form $C$. The resulting theory on $M^3$ is (Euclidean) 3D gravity with non-zero cosmological constant coupled to a (constant) scalar field.

 In particular, the Hitchin functional is then simply related to the pure connection action of 3D gravity as
\begin{equation}
\int_{p^6} \phi\left[C\right] \propto\int_{M^3} \left(1+ \gL \gr^2 \right) v\left[\bdw \right]
\end{equation}
Here $\bdw$ and $\gr$ respectively are a $\SU(2)$ connection and a scalar field that parametrize the closed, $\SU(2)$-invariant, three-form $\gO$. Varying with respect to the scalar field one obtains $\gr=0$ and the resulting action
\begin{equation}
\int_{M^3} v\left[\bdw \right]
\end{equation}
 is the pure connection action for 3D gravity.

 What is more the sign of the cosmological constant corresponds to the sign of the orbit of the corresponding stable three-form.  
\end{Proposition}

The `sign of the orbit' appearing in this proposition refers to the following: In six dimensions, real stable three-forms can only live in one of the two possible orbits of $\GL(6,\R)$, these orbits being characterised by a sign.

We will try to convey here why the above proposition can happen, see chapter \ref{Chapter: Three Dimensional Gravity as a Dimensional Reduction of Hitchin Theory in Six Dimensions} for a proof. For a three-forms $\gO$ living in the positive orbit, its stabiliser $Stab_{\gO} \subset \GL(6)$ is $\SL(3,\R) \times \SL(3,\R)$ while in the negative case this is $\SL(3,\C)$. A choice of such a stable three-forms accordingly reduces the structure group to one of the above. In particular, in the first case it defines two three-dimensional \emph{real} distributions on $P^6$, $D$ and $\Dt$ such that
\begin{equation}
D^3 \oplus \Dt^3 \simeq T\P^6.
\end{equation}
at every point of the six dimensional manifold $P^6$. In the negative case, it defines two complex-conjugated three-dimensional \emph{complexified} distributions
\begin{equation}
D^3 \oplus \Db^3 \simeq T_{\C}\P^6.
\end{equation}
When $P^6$ is taken to be a $\SU(2)$ principal bundle, it turns out that the above tangent space decomposition can be identified with the following Lie algebras
\begin{equation}
\su(2) \oplus \su(2) \simeq T\P^6
\end{equation}
and 
\begin{equation}
\sl(2,\C) \simeq T\P^6.
\end{equation}
So that \emph{at the infinitesimal level}, a three-form $\gO$ allows to identify $P^6$ with the Lie group $\SU(2) \times \SU(2)$ or $\SL(2,\C)$. When $\gO$ is taken to be $\SU(2)$-invariant this identification `fits' with the $\SU(2)$ action in such a way that 
\begin{equation}\label{Intro partII: infinitesimal 3d-Hitchin1}
TP^6 / V \simeq \su(2) \oplus \su(2) /\su(2) \quad \text{or} \quad \sl(2,\C)/\su(2)
\end{equation}
where $V \simeq \su(2)$ is the vertical tangent space as defined by the structure of $\SU(2)$-principal bundle.

The field equations of Hitchin theory then make this identification local\footnote{By local, we mean that for any point $p\in P^6$ this identification holds on an open subspace containing $p$.} i.e
\begin{equation}
P^6/ \SU(2) \simeq \SU(2) \times \SU(2) /\SU(2) \quad \text{or} \quad \SL(2,\C)/\SU(2).
\end{equation}
Which is just the statement that the base manifold $M^3 = P^6 / \SU(2)$ has the structure of Euclidean 3D gravity with positive or negative cosmological constant. 
Again, this is only the essential idea of the proof that will be detailed in Chapter \ref{Chapter: Three Dimensional Gravity as a Dimensional Reduction of Hitchin Theory in Six Dimensions}.\\
 
 This part is organised as follows. In chapter \ref{Chapter: Hitchin theory, geometrical basis} we review the concept of stable forms and their associated Hitchin functionals in a way that is agnostic of a particular dimension. We then discuss different action functionals for differential forms in six dimensions. This first chapter is intended to provide the minimum set of tools and concepts to navigate in the rest of the thesis.
 
  In the second chapter (chapter \ref{Chapter: Hitchin Theories in Six Dimensions}) we specialise to six dimensions. We first review in details the geometry of stable three-forms. In particular we describe how to construct explicitly the functionals for two- forms and three-forms presented in general terms in the preceding  chapter.  We then come to the situation where two- forms and three-forms interact to give hermitian structure on a six dimensional manifold and present a new functional for nearly Kähler structure. We make a brief review of nearly-Kähler manifold. As a bridge with Part \ref{Part: Chiral Formulations of 4D Gravity and Twistors}, we also describe the nearly-Kähler structure on the Twistor space of a self-dual Riemannian four manifold.

 Finally, in chapter \ref{Chapter: Three Dimensional Gravity as a Dimensional Reduction of Hitchin Theory in Six Dimensions}, we show that the $\SU(2)$ reduction of Hitchin theory in six dimensions gives 3D gravity. This is done in two steps.  First we show that 3D gravity can be naturally embedded into this theory, in particular we review the pure connection formulation of 3D gravity and show that it coincides with the Hitchin functional. Then, we consider the $\SU(2)$-reduction of the theory.

\counterwithin{equation}{section}		\setcounter{equation}{0}
\counterwithin{thmcnter}{chapter}		\setcounter{thmcnter}{0}

\parttoc

\chapter{Hitchin Theory: An Overall Picture}\label{Chapter: Hitchin theory, geometrical basis}

The aim of this chapter is two-fold. One the one hand we review after \cite{Hitchin:2000sk,Hitchin:2001rw, Hitchin:2002ea} the notion of stability for differential forms and the related geometrical construction, in particular Hitchin's volume and functional. On the other hand we make use of these notions to introduce action functionals for differential forms in six dimensions that will be further discussed in the next sections.

\section{Geometry of Stable Forms}\label{section: Geometry of Stable Forms}

\subsection{Stable Forms}

The notion of `stability' of a skew-linear form is, in essence, a purely algebraic concept. It is thus simpler to start at a linear algebra level before considering generalisation to differential geometry. Accordingly, everywhere in this section, let $E$ be a n-dimensional vector space. Stability of skew linear form is a generalised version of the notion of `non-degeneracy' of two-forms that we first recall.

\subsubsection{Non-degenerate two-forms}

For any skew-bilinear forms $\ga \in \W^2 E^*$ there is a well-known notion of \emph{non-degeneracy}. Consider the endomorphism $\gi \ga \from E \to E^*$ defined by the interior product, \begin{equation}
\gi_{X} \ga \coloneqq \ga(X,.).
\end{equation}
One says then that $\ga$ is non-degenerate (or pre-symplectic) if the kernel of $\gi \ga$ is trivial. However, from linear algebra considerations, the rank of a skew-linear endomorphism must always be even. It follows that a two-forms in odd dimension cannot be non-degenerate. Thus, it is convenient to introduce the following definition
\begin{Definition}
	A 2 form $\ga \in \W^2 E^*$ in dimension $n$ is said to be \emph{maximally non-degenerate} if the rank of $\gi \ga \from E \to E^*$ is maximal i.e $rank(\gi \ga) = n$ if n is even,  $rank(\gi \ga) = n-1$ if n is odd.
\end{Definition}

\subsubsection{Action of the General Linear Group on Forms}

Let us now consider the group of linear isomorphisms $End\left(E\right) \simeq GL(n)$ on $E$. Its natural action on $E$ extends to k-skew-linear form: Let $\ga \in \W^k E^* $,  then for any $g \in End\left(E\right) \simeq GL(n)$ we define
\begin{equation}\label{Geometry of Stable Forms: Gl(n) action}
g.\ga\left(X_1, ..., X_k\right) \coloneqq \ga\left(g(X_1), ..., g(X_k)\right), \qquad \forall \left(X_1,..., X_n\right) \in E^k.
\end{equation}

For any $\ga \in \W^k E^* $, we can define its \emph{orbit} $\Oc_\ga \subset \W^k E^* $ under $\GL(n)$ as
\begin{Definition}
	\begin{equation}\label{Geometry of Stable Forms: orbit}
	\Oc_{\ga} \coloneqq \left(\gb \in \W^k E^* \;| \; \exists\; g \;\in End\left(E\right)  \;\text{s.t}\; \gb = g.\ga  \right)
	\end{equation}
\end{Definition}
Finally we can also define the \emph{stabiliser of $\ga$} as the subset of $End(E)$ which leaves $\ga$ invariant:
\begin{Definition}
	\begin{equation}\label{Geometry of Stable Forms: stabiliser}
	Stab_{\ga} \coloneqq \left( g \in End(E)\;| \; \exists\; g.\ga = \ga  \right)
	\end{equation}
\end{Definition}

One can in particular consider the action \eqref{Geometry of Stable Forms: Gl(n) action} of $\GL(n)$ on 2-skew-linear forms. When $n$ is even, the subgroup of $GL(n)$ that stabilise a given non-degenerate 2-skew-linear forms $\ga$ is the `classical group' $Stab_{\ga} \simeq Sp(n, \R)$. A two-form on an even-dimensional differential manifold which is non-degenerate has at every point is `pre-symplectic'\footnote{In fact this is not a completely fixed terminology. For certain author, pre-symplectic is reserved to closed (but not necessarily non-degenerate) two-forms. }, it is 'symplectic' if it closed everywhere. 

\subsubsection{Stable forms}
One might wonder if there is a way of generalising these notions to any k-form.
In \cite{Hitchin:2000sk},  \cite{Hitchin:2002ea} Hitchin considered the following
\begin{Definition}
A k-skew-linear form of $E$ is said to be \emph{stable} if and only if $\Oc_{\ga}$ is an open subset of $\W^{k} E^*$
\end{Definition}
 Stable 2-skew-linear forms coincide with the maximally non-degenerate ones. Therefore stability of k-forms extends the notion of "degeneracy" of forms. Practically, starting with a stable k-form $\ga \in \W^k E^*$ we can reach any other `nearby' k-form by the action of $End(E)$. What is more
 \begin{equation}
 \Oc_{\ga} \simeq End(E) \big / Stab_{\ga}.
 \end{equation}
This implies, however that stable forms are in fact very rare. The space of endomorphism of $E$ should indeed be big enough to satisfy the following inequality:
 \begin{equation}\label{Geometry of Stable Forms: inequality}
 dim\left(\W^k E^*\right)  \leq  dim\left(End(E)\right)
 \end{equation}
But the dimension of $End(E)$ grows quadratically $dim(End(E)) = n^2$ while the dimension of $\W^k E^*$ grows in general much faster $dim(\W^k E^*) = \frac{n!}{(n-k)! k!} \simeq n^k$. It follows that, in general, the inequality \eqref{Geometry of Stable Forms: inequality} cannot be satisfied for large $n$ and that, apart for the peculiar case $k=2$ stable forms can only exist in low dimensions. 

In fact, in n dimensions, the only interesting case of  ` stable k-forms ' are the following:
\begin{table}[h]\label{Geometry of Stable Forms: list}
	\arrayrulecolor{black}
	\begin{tabular}{cccc}
 n & k & Stabiliser (over $\C$) & Stabiliser (over $\R$)\\ 
 \hline
 for all n &  1 & $Gl(\C)$ & $Gl(\R)$ \\
n=2m, n=2m+1 & 2 & $Sp(2m, \C)$ & $Sp(2m,\R)$ \\
n=6 & 3  & $SL(3,\C)\times SL(3,\C)$ & $Sl(3,\C)$ or $Sl(3,\R)\times Sl(3,\R)$ \\
n=7 & 3 & $G_2(\C)$ & one of the real form of $G_2$ \\
n=8 & 3 & $PSL(3,\C)$ & $SU(3)$ or $SU(2,1)$ or $SL(3,\R)$
	\end{tabular}
\end{table}

Importantly, because in n dimensions any non-zero n-bivector $\W^n E$ is stable, it follows, by the isomorphism $\W^k E^* \otimes \W^n E \simeq \W^{n-k} E^*$, that if $k$-forms can be stable in n dimensions, then $(n-k)$-forms can also be stable. E.g in $n=2m$ dimensions there is a notion of stable $(2m-2)$-form but it really is just dual to stable two-forms and this is the reason why it is not included above. Practically, a $(2m-2)$-form $\ga$ in $2m$ dimensions is stable if and only if there exists a stable $2$-form $\go$ such that $\ga = \go^{m-1}$.

\subsection{Hitchin Functionals}

\subsubsection{Definition}

We now take $\left(k,n\right) \in \Nbb^2$ such that $k$-forms exist in $n$ dimensions. Let $\gr$ be any stable $k$-form and take $\Oc_{\gr}$ its orbit. By definition, this is an open subset $\Oc_{\gr} \subset \W^k E^*$. Looking at the list \eqref{Geometry of Stable Forms: list} one sees that all stabilisers are subgroups of $\SL(n)$ and thus preserve a volume form. For any $\gr$ there is thus an associated volume form $\Phi(\gr)$ constructed  from $\gr$: The `Hitchin volumes' $\Phi$ are then the $GL(n)$-equivariant maps:
\begin{equation}
\Phi : \W^k E^* \to \W^n E^*.
\end{equation} 
In the following we will see in concrete situation how to construct these volumes explicitly. For now it is enough to see that they are generalisations of Liouville form  for pre-symplectic two-form in $2m$ dimensions:

\paragraph{Hitchin functional for stable 2 form}\mbox{}
Let $\go \in \W^2E^*$ be a stable form in dimension n= 2m. Then
 \begin{equation}\label{Geometry of Stable Forms: Liouville form}
 \Phi(\go) = \frac{1}{m!}\go^m.
\end{equation}

As a slightly less trivial example, one can look at the dual case i.e $( 2m-2)$-forms:
\paragraph{Hitchin functional for stable $2m-2$-forms}\mbox{}
Let $\gr\in \W^{2m-2}E^*$ be stable. As already discussed this implies that $\gr = \frac{1}{(m-1)!}\go^{m-1}$ with $\go$ a stable 2 form. Then Hitchin functional is again the Liouville form for $\go$ \eqref{Geometry of Stable Forms: Liouville form}.

This can be constructed directly from $\gr$ as follows. First we make use of the isomorphism $\W^k E^* \simeq \W^{n-k} E \otimes \W^n E^*$ to obtain $\grt \W^{2} E \otimes \W^{2m} E^*$, see appendix \ref{Section : Appdx Density }.
Then, taking $m$ times the wedge product of $\grt$ one obtains a tensor of the type $\W^{2m} E \otimes \left(\W^{2m} E^*\right)^{m}$ and contracting this form with itself and taking its $(m-1)$-square-root one obtains the Hitchin functional for stable $2m-2$ forms:
\begin{equation}
\Phi\left[\gr \right]=  \left(\grt^{a_1 a_2}...\grt^{a_{2m-1} a_{2m}} \epsut_{a_1 a_2 ... a_{2m-1} a_{2m}}\right)^{\frac{1}{m-1}}
\end{equation}
One can check directly that, for $\gr = \frac{1}{(m-1)!}\go^{m-1}$, it coincides with \eqref{Geometry of Stable Forms: Liouville form}.

\subsubsection{The `hat' operator on stable forms}

The differential of $\Phi$ at $\gr$ is 
\begin{equation}
\gd \Phi \in \left(\W^k E^* \right)^*\otimes \W^n E^* \simeq \W^k E \otimes \W^n E^*
\end{equation}

 Using again the canonical isomorphisms  $\W^k E \otimes \W^n E^* \simeq \W^{n-k} E^* $, one can thus define an operator 
 \begin{equation}
 \circonf \from \W^k E^* \to \W^{n-k} E^* 
 \end{equation}
  such that
 \begin{equation}\label{Geometry of Stable Forms: hat operator def}
\gd\Phi[\gr] = \grh \W \gd \gr.
\end{equation}
 By equivariance of $\Phi$, acting with scalar matrices $\gl \Id_n \in Gl(n)$ on $\Phi[\gr]$ gives $\Phi\left[(\gl \Id_n).\gr \right] = (\gl \Id_k). \Phi\left[\gr \right] $ or in other term $\Phi \left[\gl^k \gr \right] = \gl^n \Phi \left[\gr \right]$, i.e  the Hitchin volumes are homogeneous degree $n/k$:
\begin{align}\label{Geometry of Stable Forms: homogeneity}
\Phi \left( \gl \gr \right) = \gl^{n/k} \Phi \left( \gr \right) 
\end{align}
Differentiating \eqref{Geometry of Stable Forms: homogeneity}, making use of \eqref{Geometry of Stable Forms: hat operator def} and taking $\gl=1$ one obtains 
\begin{equation}\label{Geometry of Stable Forms: Psi = k/n gOh W gO}
\Phi(\gr) = \frac{k}{n}\grh \W \gr.
\end{equation}

For two-form in $n=2m$ dimensions, varying the Liouville action directly gives $\goh = \frac{1}{(m-1)!}\go^{m-1}$. On the other hand, starting with a stable $(2m-2)$-form $\gr = \frac{1}{(m-1)!} \go^{m-1}$ gives $\grh = \go$.

\subsubsection{Critical Points}

Let us now consider a n dimensional manifold $M^n$. We say that $\gr \in \gO^k(M^n)$ is a stable k-form on $M^n$ if it is globally defined and stable at each point. In particular, it reduces the structure group of the tangent space to the $Stab_{\gr} \subset GL(n)$. 

We can now make use of Hitchin volumes to set up a variational principle for the following functional on stable forms:
\begin{equation}
\gr \mapsto S(\gr) = \int_{M^n} \Phi\left[\gr\right] \in \R
\end{equation}
  It is well defined because stable forms have open orbits under $Gl(n)$. 
  
   In principle we could directly look at critical points of Hitchin functionals, however the resulting field equation would be $\grh =0$ which is much to strong. Hitchin volumes are indeed \emph{algebraic} functions of $\gr$. The strategy followed in \cite{Hitchin:2000sk} to obtain interesting field equations is to take $\gr$ to be closed and restrict variations to a given cohomology class. 
\begin{Theorem} (Hitchin \cite{Hitchin:2001rw} p7) \label{Theorem: Hithin Theory}
	Let $\gr \in \gO^k(M^n)$ be a stable k-form such that $d \gr = 0$ (closedness condition). It is a critical point of $\Phi\left[\gr\right]$ in its cohomology class if and only if  $d\grh =0$ (field equations).
\end{Theorem}
 This theorem comes directly from the remark that $\gd \Phi = \grh \W \gd \gr$ together with the fact that, because we restrict variation to a given cohomology class, $\gd \go = d \ga$.
 
For a closed stable $2$-forms $\go$ in dimension $2m$ the equations $d \goh = d \left(\go^{m-1}\right) =0$ are automatically satisfied by closeness. On the other hand, if one start with a closed stable 2m-2 form, $\gr = \frac{1}{(m-1)!} \go^{m-1}$ then $d\grh=0$ implies $d\go =0$.

In the following, we will refer to this variational principle has `Hitchin Theory'. 

\subsection{On the `Background independence' of Hitchin Theories}

A subtle point in `Hitchin theory' is the prescription of varying \emph{inside a cohomology class}. It means that one must first \emph{choose} a cohomology class $\left[\gr_0\right]$ and only then vary the form. Practically this often implies to write the form as
\begin{equation}
\gr = \gr_0+ dB.
\end{equation}
Then $\gr_0$ will effectively play the role of a background: That is it reflects the fact that \emph{we must make a choice of cohomology class} in order to make sense of the theory. 

From a physicists point of view this looks quite unsatisfactory to be forced to make such a choice beforehand. On the other hand, in the context of dimensional reduction -that we will consider in the next sections- such an attitude turns out to be very painful as the background fields of the $\gr_0$ type proliferate and tend to make the geometrical interpretation obscure. See for example \cite{Herfray:2016std}.

Accordingly we will favour in this thesis a slight variation on Hitchin theory:  Instead of considering the Hitchin functional
\begin{equation}
\gr \mapsto \int \Phi\left(\gr \right)
\end{equation}
together with the prescription of `varying inside a cohomology class' we simply constrain the form $\gr$ to be closed. This approach, even tough closely related to Hitchin's, has the advantage of avoiding making any choice of `background' cohomology class. 

From now-one we will refer to this approach as \emph{Background Independent Hitchin Theory}.

It can also be given an explicit Lagrangian formulation by implementing the closeness constraint with a Lagrange multiplier.
\begin{equation}\label{New Actions in 6d: background invariant Hitchin theory}
\int db \W \gr - \Phi\left(\gr\right).
\end{equation}
The first term then effectively plays the role of a kinematic term. In general, by assembling this kinematical term together with Hitchin volumes one can obtain different diffeomorphism invariant theories of differential forms, see below.

Varying \eqref{New Actions in 6d: background invariant Hitchin theory} with respect to $\gr$ gives 
\begin{equation}
\grh = db.
\end{equation}
So that  altogether the field equations for the background independent Hitchin theory are
\begin{equation}\label{New Actions in 6d: background invariant Hitchin field eqs}
d\gr=0, \qquad \grh =db.
\end{equation}

By going from one theory to the other we slightly weaken the constraint on $\gr$, allowing to vary between different cohomology class. As a result for the `background independent' version we obtain slightly stronger field equations than for Hitchin theory \eqref{Theorem: Hithin Theory}. Indeed here not only should $\grh$ be closed but also exact. 

\section{New Actions for 2- and 3-Forms in 6d}\label{section: New Actions in 6d}

\subsection{`Background Independent' Hitchin theories in Six Dimensions } 

In the setting of a 6-dimensional manifold let's consider the following topological theory: 
\begin{equation}\label{New Actions in 6d: Swharz type Lagrangian}
S\left[B,C\right] =\int_{M^6} \;B \W dC\qquad B\in\gO^2(M^6),\quad C\in\gO^3(M^6).
\end{equation}
As was already pointed out in the introduction to this part, this is a particular example of Schwarz theory \cite{Schwarz:1978cn}, \cite{Schwarz:1979ae} that are genealogically of the type 
\begin{equation}
S\left[C_p,C_{n-p-1}\right] = \int_{M^n} \;C_p \W d C_{n-p-1}
\end{equation}
where $C_p\in\gO^p(M^n)$ are $p$-forms on an $n$-dimensional manifold $M^n$. These theories are obviously diffeomorphism invariant and what is more are known to be topological field theories: Their partition function is a variant of Ray-Singer analytic torsion of the manifold. 

 In \cite{Herfray:2017imd} we pointed out that this obviously topological theory admits modifications that keep its topological character unchanged. They are just `background independent' Hitchin theories discussed in the above.

In six dimensions, the `background independent' Hitchin theory for two-forms is 
\begin{equation}\label{New Actions in 6d: action-symp}
S\left[B,C\right] =\int_{P^6} BdC + \Phi\left(B\right)
\end{equation}

Indeed, the simplest `potential' for the kinematical term \eqref{New Actions in 6d: Swharz type Lagrangian} is the Liouville form for $B$, which is also the Hitchin volume for two-forms in 6D,
\begin{equation}
\Phi_{B} = \frac{1}{3}B^3.
\end{equation}
Assuming that $B$ is stable (non-degenerate), this top form is non-zero. The Euler-Lagrange equations that follow by extremising this action functional are
\begin{equation}
dB=0, \qquad dC=- B\W B.
\end{equation}

The three-form field $C$ plays the role of the Lagrange multiplier imposing the condition that $B$ is closed. Critical points of this theory are therefore symplectic manifolds, with the additional constraint that $B\W B$ is exact. 

We note that the numerical coefficient that could have been put in front of the second term in \eqref{New Actions in 6d: action-symp} can be absorbed by the simultaneous redefinition of the $B$ and $C$ fields.

Thus the `background independent' Hitchin theory for two-forms \eqref{New Actions in 6d: action-symp} just amounts to adding $B^3$ to \eqref{New Actions in 6d: Swharz type Lagrangian}. It is natural to wonder if we can repeat the same construction with $C$, i.e adding a top form constructed from $C$. The most straightforward attempt does not however give anything as the wedge product $C\W C$ vanishes. Nevertheless, we can turn to the `background independent' Hitchin theory for three-forms
\begin{equation}\label{New Actions in 6d: action-complex}
S\left[B,C\right] =\int_{M^6} BdC + \Phi\left(C\right)
\end{equation}
 Just as for \eqref{New Actions in 6d: action-symp}, any parameter that may have been put in front of the second term can be absorbed by a field rescaling. 
The Hitchin volumes $\Phi\left[C\right]$ for three-forms in six dimensions is a highly non-linear algebraic function of $C$. We will discuss this functional it in more detail in the next chapter. We however already see that it can always be written as
\begin{equation}
\Phi_{C} = \frac{1}{2} \Ch \W C.
\end{equation}
Where $\Ch$ is a three-form constructed from $C$ in an algebraic but non-trivial way.

This is schematically constructed as follows. As will be explicitly described in the next chapter (see in particular section\ref{section: Stable form in 6d}), a stable three-forms in six dimensions defines a linear operator on the tangent space $J_C \from TM^6 \to TM^6$ such that $J_C^2 = \pm \Id$. The exact sign in this expression depends on the three-forms considered. Whatever the sign, $\Ch$ is obtained by acting on the three indices of $C$ with $J_C$,
\begin{equation}
\Ch\left(.,.,.\right) \coloneqq C\left(J_C(.),J_C(.),J_C(.)\right).
\end{equation}

The Euler-Lagrange equations that describe extrema of this functional are
\begin{equation}\label{New Actions in 6d: BIHT C field equations}
dC=0,  \qquad dB = \hat{C}.
\end{equation}
In particular the second equation implies that 
\begin{equation}
d\hat{C}=0,
\end{equation}
and thus the three-form is closed and "co-closed" in the sense of the Hitchin story. These are the field equations for the `background independent' Hitchin theory.  Once again, these equations are stronger than those of Hitchin theory because the second equation in \eqref{New Actions in 6d: BIHT C field equations} says that the three-form $\hat{C}$ is exact, not just closed.

For three-forms in the `negative orbit of $\GL(6)$' we have $J_C^2 = - \Id$ and therefore such forms define an almost complex structure. The above field equations then are equivalent to the integrability of this almost complex structure together with the closeness of a certain $\left(3,0\right)$-form. See next chapter for more on this.

In \cite{Herfray:2017imd} we were mostly interested by the following question: does the theory \eqref{New Actions in 6d: Swharz type Lagrangian} starts to have any local degrees of freedom after deformation by a given `potential term'? We proved for both theories \eqref{New Actions in 6d: action-symp} and \eqref{New Actions in 6d: action-complex} that this is not the case. Both theories are therefore diffeomorphism invariant topological theories i.e. there are no propagating degrees of freedom. Naively a good reason for this is that these really are the `background independent' versions of Hitchin theory for two-forms in six dimensions and that Hitchin theory itself is believed to be topological.

 The `background independent' Hitchin theory for three-forms is particularly interesting because its dimensional reduction to three dimensions gives 3D gravity, see below. This is coherent with the above statement as a 6D diffeomorphism invariant topological gets reduced to a 3D diffeomorphism invariant topological theory.

\subsection{A New Action Functional for Nearly-Kähler Manifolds}

In the above we successively added to the kinetic term \eqref{New Actions in 6d: Swharz type Lagrangian} the Hitchin volume for two-forms and the Hitchin volume for three-forms. We obtained diffeomorphism invariant topological theories that we coined `background invariant' Hitchin Theories.

At this point it is natural to put the two Hitchin volumes together and consider
\begin{equation}\label{New Actions in 6d: action-nK}
S[B,C] = \int B\W dC + \frac{1}{2} C\W \hat{C} + \frac{1}{3} B\W B\W B.
\end{equation}
The numerical coefficients that could have been put in front of the second and third terms can be absorbed by a field redefinition of $B$ and $C$ up to multiplying the action by an overall constant. Consequently, the only parameter in the above theory is the coefficient in front of the action, or, in physics terminology, the Planck constant. This parameter only matters in the quantum theory, where the partition function of the theory will depend on it. The Euler-Lagrange  equations that describe the extrema of the functional \eqref{New Actions in 6d: action-nK} are 
\begin{equation}\label{New Actions in 6d: nK feq}
dC = - B\W B, \qquad dB = \hat{C}.
\end{equation}
Which are just the field equations of \emph{nearly Kähler manifolds}, see next chapter for a geometrical discussion. See also section \ref{section: Holonomy} where this structure is put in its natural context which is holonomy on Riemannian manifolds. In particular neither the two-form or the three-form are closed. 

On the one hand, stable two-forms are the non-degenerate two-forms, on the other hand (negative)stable three-forms define almost complex structure. It turns out that the above field equations make those structure compatible, again see section \ref{section: Hermitian structure and NKM} for a proof. As a result we obtain an almost hermitian structure $\left(\J_C,B, g \right)$ on $M^6$, where $\J_C$, $B$ and $g$ are respectively an almost complex structure, a two-form and a metric which are compatible with each others, i.e such that
\begin{equation}
g\left(X, Y\right)= B\left( \J_C(X) , Y\right).
\end{equation}

For the field equation \eqref{New Actions in 6d: nK feq}, this metric, whose construction uses both $B$ and $C$ then gives $M^6$ the structure of a nearly-Kähler manifolds.

Nearly Kähler manifolds, which is a concept that only exists in six dimensions, have special properties. They are Einstein manifolds of positive scalar curvature. They also admit a spin structure and admit real Killing spinors. In fact the Einstein property follows directly from the fact that such manifolds admit Killing spinors. A useful reference on nearly Kähler manifolds (and much more) is \cite{Alexandrov:2004cp}, see in particular section 4.2 and references therein. See also section \ref{section: Holonomy} where they appear in the context of holonomy of a Riemannian manifold.

Therefore, the field equations for the theory \eqref{New Actions in 6d: action-nK} imply that the metric constructed from $B,C$ is Einstein, and in this sense it can be viewed as a gravity theory. This theory is however unlikely to be topological, see \cite{Herfray:2017imd}. 

\paragraph{Remarks} \mbox{}

In \cite{Hitchin:2004ut} Hitchin described a generalisation of the volume functional $C\W \hat{C}$ to all odd or even polyforms in 6D. There is thus a generalisation of all 3 theories \eqref{New Actions in 6d: action-symp}, \eqref{New Actions in 6d: action-complex} and \eqref{New Actions in 6d: action-nK} to polyforms, necessarily involving forms of all degree. It would be interesting to study these theories, and characterise them in terms of the degrees of freedom they propagate as well as their dimensional reduction. We leave this to future work.

\chapter{Hitchin Theories in Six Dimensions}\label{Chapter: Hitchin Theories in Six Dimensions}

In this chapter we look into the geometrical details of different constructions related to stable forms in six dimensions. The aim is to allow to understand the content of the theories described at the end of the preceding chapter and give a precise sense to the different statements that were made. Accordingly we first review the geometry of stable three-forms in six dimensions. Then we discuss how to obtain $\SU(3)$(or almost hermitian) structures in six dimensions from stable forms and the equations for nearly Kähler manifolds.

\section{Geometry of Stable 3-Forms in Six Dimensions}
\label{section: Stable form in 6d}

This section is a review of the geometry of stable forms in six dimensions. In particular we recall how to explicitly construct Hitchin volumes. See \cite{Hitchin:2000sk,Hitchin:2001rw} for the modern references.

\subsection{Stable three-forms in Six Dimensions}

\subsubsection{Stable three-forms over $\C$}

A stable three-form in six dimensions is a form that lies in an open $\GL(6)$ orbit. For complex three-forms, there is a single open orbit. On the other hand for real three-forms there are two distinct orbits that can be distinguished by a sign. It is thus easier to start with the description of the situation over $\C$ and later specialise to the real case.

For a stable three-forms over complex numbers, the component of the stabiliser connected to the identity is the group 
\begin{equation}
\SL(3,\C)\times\SL(3,\C) \subset \GL(6,\C).
\end{equation}
Accordingly, if $M$ is a six-dimensional manifold, a choice of stable complex three-forms $C$ on $M$ will define two three-dimensional complex distributions $D \subset T_{\C}M$ and $\Dt \subset T_{\C}M$ on $M$. Each of the factor of the stabiliser groups $\SL(3,\C)\times\SL(3,\C)$ naturally acts on each of this distribution.

Practically, we have the following proposition:
\begin{Proposition}{Hitchin \cite{Hitchin:2000sk}}\label{Proposition: Hitchin's decomposition in six d} \mbox{}\\
 A three-form $C \in \gO^3\left(M^6\right)$ is stable if and only if it is the sum of two decomposable three-forms, 
\begin{equation}\label{Stable forms in 6d: om-canonical}
C=\ga^1\W \ga^2\W \ga^3 + \gb^1\W \gb^2\W \gb^3, \qquad \ga^1\W \ga^2\W \ga^3 \W \gb^1\W \gb^2\W \gb^3 \neq 0.
\end{equation}
\end{Proposition}

While each of the sets of forms $\left(\ga^i \right)_{i\in \{1,2,3\}}$, $\left(\gb^i \right)_{i\in \{1,2,3\}}$ are only defined up to an action of the stabiliser group  $\SL(3,\C) \times \SL(3,\C)$ -with the obvious action of each $\SL(3,\C)$ on each family of forms- the following distributions are invariantly defined:
\begin{equation}\label{Stable forms in 6d: D def}
D = \textrm{Ker}\left(\ga^1 \right) \cap \textrm{Ker}\left(\ga^2 \right) \cap \textrm{Ker}\left(\ga^3 \right),\;  \Dt = \textrm{Ker}\left(\gb^1 \right) \cap \textrm{Ker}\left(\gb^2 \right) \cap \textrm{Ker}\left(\gb^3 \right).
\end{equation}

The condition $ \ga^1\W \ga^2\W \ga^3 \W \gb^1\W \gb^2\W \gb^3 \neq 0$ means that the set of one-forms $\left(\ga^i, \gb^i \right)_{ i\in \{1,2,3\}}$ is a basis of $T^*_{\C}M$. As a consequence, the distributions they define span the whole tangent space: 
\begin{equation}
T_{\C}M = D \oplus \Dt.
\end{equation}

The decomposable three-forms $\ga^1\W \ga^2\W \ga^3$ and $\gb^1\W \gb^2\W \gb^3$ are unique but in general we cannot distinguish between the two: in what follows $\ga$'s and $\gb$'s will always play symmetric roles.

Let us now describe how to concretely obtain the distributions \eqref{Stable forms in 6d: D def} in terms of $C$. This will be done by constructing at every point $x \in M$ an endomorphism `up to scale' \begin{equation}
\Kt_{C} \from T_\C M \to T_\C M \otimes \gO_\C^n(M)
\end{equation} and noticing that $D \subset T_\C M$ and $\Dt \subset T_\C M_x$ are eingenspaces of $\Kt_{C}$ of opposite eigenvalues.

Let $\gx \in T_\C M_x$ be a vector field. One first produce a five-form by using a proper combination of interior derivative and wedge product
\begin{equation}
\gi_{\gx} C \W C \in \gO_{\C}^5(M),
\end{equation}
and then uses the isomorphism $\gO^5_{\C}(M) \simeq T_{\C}M \otimes \gO^6_{\C}(M)$ (cf eq\eqref{Appendix: Weight def} of Appendix \ref{Section : Appdx Density }) to obtain \begin{equation}
\Kt_{C}(\gx) \in T_\C M \otimes \gO_\C^n(M).
\end{equation}

In tensorial notation, taking $\left(e_a\right)_{a \in \{1...6\}}$ a frame on $TM$ and $\left(e^a\right)_{a \in \{1...6\}}$ a co-frame,
\begin{equation}
\Kt_{C} = \frac{1}{12} C_{\ga \mu_1 \mu_2} C_{\mu_3 \mu_4 \mu_5} \epst^{\mu_1 \mu_2 \mu_3 \mu_4 \mu_5 \gb} \; e_{\ga}\otimes e^{\gb} \otimes \left(e^1\W...\W e^6\right).
\end{equation}

This last expression might look reassuring (or not), but is essentially useless in most situations. Rather, making use of the decomposition \eqref{Stable forms in 6d: om-canonical} and taking a basis of tangent vector $\left(\gx_i,\gz_i \right)_{i\in\{1,2,3\}}$ dual to $\left(\ga^i,\gb^i \right)_{i\in\{1,2,3\}}$ this endomorphism can be rewritten
\begin{equation} \label{Stable forms in 6d: Kt alpha beta formulation}
\Kt_{C} = \left( \ga^i \otimes \gx_i - \gb^i \otimes \gz_i \right) \otimes\; \ga^1\W\ga^2\W \ga^3 \W \gb^1\W \gb^2\W \gb^3
\end{equation}
So that, in general, $D$ and $\Dt$ can be characterised as the eingenspaces of $\Kt_{C}$ with opposite eigenvalues:
\begin{align}\label{Stable forms in 6d: eigen values}
\xi \in D \quad &\Leftrightarrow \quad \Kt_{C}(\xi) = \;\xi\; \otimes\; \Phi_{C} \qquad   \Phi_{C} \in \gO^0\left(M\right) \\
\gz \in \Dt \quad &\Leftrightarrow \quad \Kt_{C}(\gz) = -\gz\; \otimes \; \Phi_{C} \nonumber
\end{align}

Now, clearly $\Kt_{C}^2$ is a multiple of the identity operator
\begin{equation}
\Kt_C^2 = \Id \otimes \left(\Phi_{C}\right)^2.
\end{equation}
Taking the trace, we have
\begin{equation}\label{Stable forms in 6d: Tr Kt}
\frac{1}{6} \Tr(\Kt_C^2) \in \left(\gO^6(M)\right)^2
\end{equation}
This defines a maps
\begin{equation}
\begin{array}{ccc}
\gO^3(M) &\to& \left(\gO^6(M) \right)^2 \\ \\
C &\mapsto&  \frac{1}{6} \Tr(\Kt_C^2) = \left(\Phi_{C}\right)^2
\end{array}
\end{equation}
which is clearly equivariant under the action of $\GL(6)$. For complex three-forms this object is such that it is everywhere non-zero if and only if $C$ is stable. This last statement is directly related to proposition \ref{Proposition: Hitchin's decomposition in six d}, see \cite{Hitchin:2000sk}.

Here, $\Phi_{C}$ cries out to become the Hitchin volume. However one needs to take a square root of \eqref{Stable forms in 6d: Tr Kt} which leads to a sign ambiguity at every points on $M$. A related problem is that in general there is no way to make a distinction between $D$ and $\Dt$, e.g eq \eqref{Stable forms in 6d: Kt alpha beta formulation} is symmetric under permutation of $\ga$'s and $\gb$'s. Therefore there is no way to choose between one of the two possible `square roots' $\Phi_C$ and $-\Phi_C$ in \eqref{Stable forms in 6d: eigen values}. In what follows we will however consider \emph{real} three-forms where this ambiguity simplifies to a choice of orientation.

Before coming to real three-forms, let us describe a constructive way of obtaining the decomposable three-forms $\ga^1\W \ga^2\W \ga^3$ and $\gb^1\W \gb^2\W \gb^3$. First, define the operator
\begin{equation}\label{Stable forms in 6d: K def c}
K_C := \frac{1}{\Phi_C} \Kt_C \in \gO(M) \otimes TM.
\end{equation}
This is a genuine operator (however only defined up to a sign, due to the above discussion) as the density weight\footnote{Here both $\Phi_C$ and $\Kt_C$ are volume-form-valued object so that dividing one by the other results in a proper scalar.} of $\Phi_C$ and $\Kt_C$ compensate each others. It has eigenvalues $\pm 1$ and thus squares to the identity. The distributions $D$ and $\Dt$ correspond to the eingenspaces of eigenvalues one and minus one respectively. Note that changing $\Phi_{C}$ by a sign exchange the role of $D$ and $\Dt$.  

We can then make use of \eqref{Stable forms in 6d: K def c} to construct the decomposable forms
\begin{equation} \label{Stable form in 6d: aaa in terms of gO c}
2 \ga^1\W \ga^2\W \ga^3  = C + \Ch, \qquad 
2 \gb^1\W \gb^2\W \gb^3 = C - \Ch,
\end{equation}
where $\Ch$ is the result of the action of $K_{C}$ on all three form-indices of $C$,
\begin{equation}\label{Stable forms in 6d: def Ch}
\hat{C}(.,.,.)\coloneqq C\left(K_{C}(.), K_{C}(.), K_{C}(.)\right) \quad \in \gO^3_{\C}(M^6).
\end{equation}

 Not surprisingly the overall sign ambiguity in the definition of $\Phi_{C}$ (and thus $K_{C}$) translates into an ambiguity between $\ga$'s and $\gb$'s.

A look at equations \eqref{Stable forms in 6d: eigen values} and \eqref{Stable forms in 6d: Kt alpha beta formulation} then shows that 
\begin{equation} \label{Stable 6 form in 6d: Hitchin functionnal c}
\Phi_C = \ga^1\W \ga^2\W \ga^3 \W \gb^1\W \gb^2\W \gb^3
\end{equation} (again exchanging the role of $\ga$ and $\gb$'s is equivalent to changing $\Phi_{C}$ by a sign). With this in hand and once the decomposable three-form factors \eqref{Stable form in 6d: aaa in terms of gO c} of $C$ have been obtained, one easily gets that
\begin{equation}\label{Stable 6 form in 6d: Hitchin functionnal d}
\Phi_C=\frac{1}{2} \Ch\W C.
\end{equation}
Finally one can check from \eqref{Stable 6 form in 6d: Hitchin functionnal c} that $\gd \Phi_C = \Ch \W \gd C$ so that the notation of this chapter is coherent with the previous one.

As we emphasised all along, in the complex case there is a systematic sign ambiguity everywhere that cannot be fixed. We now consider real three-forms where this ambiguity is less drastic.

\subsubsection{Stable forms over $\R$}

There are exactly two $\GL(6,\R)$ stable orbits of three-forms, characterised by the sign of \begin{equation}\label{Stable forms in 6d: tr Kt 2}
\frac{1}{6} \Tr(\Kt_C^2) \in \left(\gO^6(M)\right)^2.
\end{equation} Note that, despite the fact that \eqref{Stable forms in 6d: tr Kt 2} is not really a number but rather a section of $\left(\gO^6(M)\right)^2$, its sign is invariantly defined as follows. Take any volume form $v\in \gO^6(M)$ then the sign of \eqref{Stable forms in 6d: tr Kt 2} is the sign of $\gl$ as defined by the expression
\begin{equation}\label{Stable forms in 6d: gl def}
\frac{1}{6} \Tr(\Kt_C^2) = \gl \; \left(v\right)^2 .
\end{equation}
Clearly this sign does not depend on the choice of $v$ ( in particular it does not
depend on a choice of orientation).

In what follows three-forms $C$ such that $ \frac{1}{6} \Tr(\Kt_C^2) >0$ will be said to be \emph{in the positive orbit} while those with $ \frac{1}{6} \Tr(\Kt_C^2) <0$ will be said to be \emph{in the negative orbit}. When $ \frac{1}{6} \Tr(\Kt_C^2) =0$, $C$ is simply not stable.

\subsubsection{Stable forms over $\R$, case $\gl(C)>0$}
Let us first consider the `positive orbit'. This sign corresponds to stable forms with stabiliser
\begin{equation}
\SL(3,\R) \times\SL(3,\R)
\end{equation}
Then the three-form $C$ has the canonical form \eqref{Stable forms in 6d: om-canonical} with $\left(\ga^a\right)_{a\in \{1,2,3\}}$, $\left(\gb^a\right)_{a\in \{1,2,3\}}$ \emph{real} one-forms and the distributions they define (cf eq\eqref{Stable forms in 6d: D def}) are real.  

We now want to define the Hitchin volume as
\begin{equation}
\Phi_C \coloneqq \pm \sqrt{\frac{1}{6} \Tr(\Kt_C^2)}.
\end{equation}
As already discussed, because a square root is involved there is a sign ambiguity in this definition. However this is less problematic than in the complex case as fixing this ambiguity is now equivalent to picking an orientation on $M$. We suppose that such a choice has been made and choose $\Phi_{C}$ such that it is oriented accordingly.

Practically, if $v$ is an oriented volume form we can write
\begin{equation}
\Phi_{C} := \sqrt{\gl} \;v.
\end{equation}
Where $\gl$ is defined by \eqref{Stable forms in 6d: gl def}.

Integrating this volume form we get the Hitchin functional for the positive orbit
\begin{equation}
S[C] := \int_M \Phi_C \quad \in \R.
\end{equation}

The operator
\begin{equation}\label{Stable forms in 6d: K def +}
K_C := \frac{1}{\Phi_C} \Kt_C \in \gO^1(M) \otimes TM
\end{equation}
is now \emph{real} with eigenvalues $\pm 1$ and thus squares to the identity, \begin{equation}
K_C^2=\Id_{TM}.
\end{equation} This is sometimes called a \emph{para-complex structure}.

Just as in the complex case,
\begin{equation}
\hat{C}:= K_C(C) \quad \in\gO^3(M).
\end{equation}and the decomposable forms are obtained as
\begin{equation} \label{Stable form in 6d: aaa in terms of gO +}
2 \ga^1\W \ga^2\W \ga^3  = C + \hat{C} \quad \in\gO^3(M),\qquad 
2 \gb^1\W \gb^2\W \gb^3 = C - \hat{C} \quad \in\gO^3(M).
\end{equation}
The only difference is that all the forms are now real. 

Finally, the Hitchin volume can expressed as
\begin{equation} \label{Stable 6 form in 6d: Hitchin functionnal+}
\Phi_C =\frac{1}{2} \hat{C}\W C = \ga^1\W \ga^2\W \ga^3 \W \gb^1\W \gb^2\W \gb^3 \in \gO^6(M).
\end{equation}

\subsubsection{Stable forms over $\R$, case $\gl(C)<0$}

The negative orbit, \begin{equation}\label{Stable forms in 6d: tr kt 3}
\frac{1}{6} \Tr(\Kt_C^2)<0
\end{equation} is more interesting. In this case, the stabiliser group is 
\begin{equation}
\SL(3,\C)
\end{equation}
and the canonical form for $C$ is
\begin{equation}\label{Stable forms in 6d: om-complex}
C = \ga^1\W \ga 2 \W \ga^3 + \gab^1\W \gab^2\W \gab^3, \qquad \ga^1\W \ga^2 \W \ga^3 \W \gab^1\W \gab^2\W \gab^3 \neq 0,
\end{equation}
where $\left(\ga^{i}\right)_{i \in \{1,2,3\}}$ are now complex-valued one-forms, and $\left(\gab^{i}\right)_{i \in \{1,2,3\}}$ are the complex conjugate forms. The distributions $D$ and $\Db$ they define are therefore complex conjugated to each others. Here again there is an ambiguity between $\ga$'s and $\gab$'s. As in the positive sign case solving this ambiguity amounts to making a choice of orientation.

We now define the Hitchin volume as
\begin{equation}\label{Stable forms in 6d: Psi 2}
\Phi_{C} \coloneqq \pm \;\sqrt{-\frac{1}{6} \Tr(\Kt_C^2)} 
\end{equation}
 Integrating this form we obtain the Hitchin functional for the negative orbit
\begin{equation}
S[C] := \int_M \Phi_C \quad \in \R.
\end{equation}

\begin{ExtraComputation}
	\begin{framed}
Note that, with the definition \eqref{Stable forms in 6d: Psi 2} this is really "$i \Phi_C $" which is a square-root of \eqref{Stable forms in 6d: tr kt 3}. In particular, with the definition \eqref{Stable forms in 6d: Psi 2} $\Kt_{C}$ has eigenvalues $\pm i \; \Phi_{C}$. This is as compare with the previous definition \eqref{Stable forms in 6d: eigen values}.
	\end{framed}
\end{ExtraComputation}

As a result,
\begin{equation}\label{Stable forms in 6d: ACS}
\J_C := \frac{1}{\Phi_{C}} \Kt_C, \in \gO(M) \otimes TM
\end{equation}
is real with eigenvalues $\pm i$ and thus is an \emph{almost-complex structure} on $M$ \begin{equation}
\J_C^2 = -\Id_{TM}.
\end{equation} 
The distributions $D$, $\bar{D}$ are now invariantly defined as the eigenspaces of eigenvalues $i$ and $- i$ respectively.
Accordingly, we still have
\begin{equation}
\hat{C} := \J_C \left( C \right) \quad \in \gO^3\left(M\right).
\end{equation}
but now
\begin{equation}\label{Stable form in 6d: aaa in terms of gO -}
2\; \ga^1\W \ga^2 \W \ga^3 = C + i \hat{C} \quad \in \gO^3_{\C}\left(M\right)
\end{equation}
Finally,
\begin{equation} \label{Stable 6 form in 6d: Hitchin functionnal-}
\Phi_C = \frac{1}{2} \hat{C}\W C = -\rmi \ga^1\W \ga^2\W \ga^3 \W \gab^1\W \gab^2\W \gab^3.
\end{equation}

\subsection{Hitchin Theory for 3 forms in 6d}\label{ssection: Hitchin Theory for 3 forms in 6d}

\subsubsection{Hitchin Functional}

Let us consider a three-form $C$ on $M$. We formally write the Hitchin functional as
\begin{equation}\label{Stable 6 form in 6d: action-hitchin}
\Phi\from \left|\begin{array}{ccc}
\gO^3_{\C}(M) & \to & \C \\ \\
C & \mapsto & \int_M \Phi_C.
\end{array}\right.
\end{equation}
Because of the sign ambiguity for $\Phi_{C}$ at each points of the manifold this does not really make sense for complex forms. For real forms however and as already explained it is well defined once an orientation for $M$ is given. For now we just take \eqref{Stable 6 form in 6d: action-hitchin} as a formal way for treating both real orbits at the same time. 

\subsubsection{Critical Points}

We can now consider `Hitchin Theory' i.e variations of the three-form $C$ staying within a fixed cohomology class. As stated in Theorem \eqref{Theorem: Hithin Theory}, the resulting critical points are three-forms $C$ such that
\begin{equation}
dC = 0, \qquad d \Ch =0.
\end{equation}
Making use of \eqref{Stable form in 6d: aaa in terms of gO +} this is equivalent to
\begin{equation}
d\left(\ga^1\W \ga^2 \W \ga^3 \right)=0, \qquad d\left(\gb^1\W \gb^2 \W \gb^3 \right)=0.
\end{equation}
In turns this implies that each distribution $D$ and $\Dt$ are separately integrable. Taking the exterior derivative of
\begin{equation}
\ga^i \W \ga^1\W \ga^2 \W \ga^3 =0 \quad \forall i \in \{1,2,3\}
\end{equation}
one indeed obtains
\begin{equation}
d\ga^i \W  \ga^1\W \ga^2 \W \ga^3=0 \quad \forall i \in \{1,2,3\}
\end{equation}
which is just the integrability condition for the distribution defined by the kernel of the $\ga$'s. The same obviously holds for $\gb$'s.

In the particular case where $C$ is in the negative orbit, $C = 2\;\textrm{Re} \left( \ga^1\W \ga^2\W \ga^3\right)$ it defines an almost complex structure together with a globally defined $\left(3,0\right)$-form $\ga^1\W \ga^2\W \ga^3$. The critical points of Hitchin theory are therefore integrable almost complex structure with a holomorphic $\left(3,0\right)$-form. This result is synthesised by the following theorem.

\begin{Theorem}{Hitchin \cite{Hitchin:2000sk}}\label{Theorem: Hitchin Theory, critical points in 6d} \mbox{}\\ 
	Let M be a compact complex 3-manifold with trivial canonical bundle
	and $C$ the real part of a non-vanishing holomorphic three-form. Then $C$ is a critical point
	of the functional $\Phi$ restricted to the cohomology class $C \in H^3(M,\R)$.
	
	Conversely, if $C$ is a critical point of $\Phi$ on a cohomology class of an oriented closed 6-
	manifold M and $\Phi_{C} < 0$ everywhere, then it
	defines on M the structure of a complex manifold such that it is the real part of a non-vanishing holomorphic three-form.
\end{Theorem}

\section{Stable forms, Hermitian structure and Nearly Kähler Manifold}\label{section: Hermitian structure and NKM}

In the preceding section we saw how a negative stable three-form in six dimensions $C$ defines an almost complex structure $\J_C$. Recall, that stable two-forms $B$ are the non-degenerate ones. We now come to the case where the two structures interact to give an (almost) hermitian structure, i.e a compatible triplet $\left(\J_C, B , g\right)$  We then discuss variational principle for nearly Kähler manifold. This is again essentially reviewed from \cite{Hitchin:2001rw} and \cite{Hitchin:2002ea}.

\subsection{$\SU(3)$ Structure in six dimensions from Stable Forms}

In previous sections we saw that the only stable forms in 6D are two-forms and three-forms (In fact four-forms can also be stable but are dual to two-forms in six dimensions).

Let $C \in \gO^{3}\left(M^6\right)$ be a stable three-form and $B \in \gO^2\left(M^6\right)$ be a stable two-form.
 In this section we will everywhere assume that $C$ is in the negative orbit. Then $C$ and $B$ respectively define an almost complex structure $\J_{C}$ and a non-degenerate two-from. In other terms, they respectively reduce the structure group to $\SL\left(3,\C\right)$ and $\Sp\left(6,\R\right)$.

In order for $\left( \J_{C} , B\right)$ to be an \emph{almost hermitian structure} i.e reduce the structure group to $\SU(3) = \SL(3,\C) \cap  \Sp\left(6,\R\right)$, one needs a further compatibility condition:
\begin{equation}\label{Hermitian structure and NKM: SU(3) condtion1}
B\left(X,Y \right) = B\left(J_{C}\left(X\right),J_{C}\left(Y\right)\right).
\end{equation}
This is conveniently rewritten in terms of $C$ and $B$ as
\begin{equation}\label{Hermitian structure and NKM: SU(3) condtion2}
C \W B =0.
\end{equation}
The two conditions are in fact equivalent to $\go$ being a $\left(1,1\right)$-form for the almost complex structure defined by $C$. Indeed, recall that when $C$ is in the negative orbit it can be rewritten
\begin{equation}
C = \ga^1 \W \ga^2 \W \ga^3 + \gab^1 \W \gab^2 \W \gab^3
\end{equation}
where $\ga^1 \W \ga^2 \W \ga^3$ is a $(3,0)$-form for $\J_{C}$ (accordingly $\gab^1 \W \gab^2 \W \gab^3$ is a $(0,3)$-form). It follows that \eqref{Hermitian structure and NKM: SU(3) condtion2} is equivalent to 
\begin{equation}
\ga^1 \W \ga^2 \W \ga^3 \W B =0, \qquad \gab^1 \W \gab^2 \W \gab^3 \W B =0
\end{equation}
which in turn implies that $B$ is $(1,1)$. The equivalence with \eqref{Hermitian structure and NKM: SU(3) condtion1} is classical (see e.g \cite{Huygbrechts}) and can be obtained directly by expanding $B$ in terms of the $\ga$'s and $\gab$'s.

As a consequence \eqref{Hermitian structure and NKM: SU(3) condtion2} gives an almost hermitian structure, i.e a compatible triplet $\left(\J_{C}, B , g \right)$ of almost complex structure, pre-symplectic two-form and metric.

Generically however, the hermitian product defined as
\begin{equation}\label{Hermitian structure and NKM: hermitian metric}
g\left(X,Y\right) = B\left(\J_{C}\left(X\right), Y \right)
\end{equation}
does not have to be definite. In what follows we will always suppose that this is the case and that \eqref{Hermitian structure and NKM: hermitian metric} is a genuine Hermitian metric. This is obviously an open condition.

There is however more to the doublet $\left(C,B\right)$ than an hermitian structure $\left(\J_{C}, B, g\right)$: suppose that \eqref{Hermitian structure and NKM: SU(3) condtion2} is verified and take $\left(\ga^i\right)_{i \in 1,2,3}$ a basis of $(1,0)$-form adapted with the hermitian structure i.e\footnote{We recall that our notation it that $A \odot B\coloneqq A\otimes B+ B\otimes A$.}
\begin{equation}
B= i \; \gab^i \W \ga^i, \qquad g= \gab^i \odot \ga^i.
\end{equation}
This basis is unique up to a $\U(3)$ action. Then there exists $\gl \in \Cc^\infty\left(M^6, \C \right)$ such that
\begin{equation}
C = \gl \;\ga^1 \W \ga^2 \W \ga^3 + \glb\; \gab^1 \W \gab^2 \W \gab^3.
\end{equation}
Using the $\U(3)$ freedom we can take $\gl$ to be real positive, $\gl \in \Cc^\infty\left(\R^+, M^6 \right)$. This choice reduces the freedom in choosing the $\ga$'s to a $\SU(3)$ action. One sees that there is however a remaining scalar field so that in general the pair of stable forms $\left(C,B\right)$ together with the compatibility condition \eqref{Hermitian structure and NKM: SU(3) condtion2} is slightly more than just a hermitian structure $\left(\J_{C}, B , g\right)$. Accordingly without further constraints the two Hitchin functionals,
\begin{equation}
\Phi\left[B\right]= \frac{1}{3}B^3, \qquad \qquad 
\Phi\left[C\right] = \frac{1}{2} \Ch \W C,
\end{equation}
are distinct volume-forms:
\begin{align}
\Phi_{C} = \frac{|\gl|^2}{2}\; \Psi_{\go}
\end{align}
Therefore in order for $\left(C, B\right)$ to describe a hermitian structure only one should impose on top of \eqref{Hermitian structure and NKM: SU(3) condtion2} that the two Hitchin functionals are proportional
\begin{equation}\label{Hermitian structure and NKM: SU(3) condtion3}
\Psi\left[B\right] = c \Psi\left[C\right]
\end{equation}
where $c$ is a constant.

\paragraph{Calabi-Yau manifold}\mbox{}

Let $B$ and $C$ respectively be two-forms and three-forms satisfying the compatibility equations \eqref{Hermitian structure and NKM: SU(3) condtion2} and \eqref{Hermitian structure and NKM: SU(3) condtion3} and satisfying the necessary open conditions for
\begin{equation}\label{Hermitian structure and NKM: calabi-yau triplet}
\left(\J_{C}, B, g \right)
\end{equation}
to be a good almost hermitian structure. I.e such that $\J_{C}$ is an almost complex structure and $g$ is definite. 

Let us further suppose that they satisfy
\begin{equation}\label{Hermitian structure and NKM: Calabi-Yau equations}
d B = 0,\qquad dC = 0, \qquad d\Ch=0.
\end{equation}
The first equation means that $V$ is a symplectic form, making the triplet \eqref{Hermitian structure and NKM: calabi-yau triplet} an almost Kähler structure. As discussed in the previous chapter, the last two equations implies that the almost complex structure is integrable $\J_{C}$ and that the $(3,0)$-form $C + i \Ch$ is holomorphic. This respectively makes \eqref{Hermitian structure and NKM: calabi-yau triplet} Kähler and Calabi-Yau.

\subsection{Nearly Kähler Manifold}

At the end of chapter \ref{Chapter: Hitchin theory, geometrical basis} we considered the following variational principle
\begin{equation}\label{Hermitian structure and NKM: nk action}
s\left[C, B\right] = \int_{M^6} BdC + \Phi\left(B\right) + \Phi\left(C\right),\qquad  C\in \gO^3\left(M^6\right),\qquad B \in \gO^2\left(M^6\right).
\end{equation}
The resulting field equations are (after rescaling the fields)
\begin{equation}\label{Hermitian structure and NKM: Nearly kahler field equations}
d\Ch = B^2,\qquad dB = C.
\end{equation}

A crucial point is that these equations imply the compatibility equations \eqref{Hermitian structure and NKM: SU(3) condtion2} and \eqref{Hermitian structure and NKM: SU(3) condtion3} for
\begin{equation}
B \W C = B \W d B = d \left(B^2\right) =d^2 \Ch = 0.
\end{equation}
As already discussed, this implies that $B$ is $(1,1)$ for the almost complex structure defined by $C$. In particular $B \W \Ch=0$. This is useful to show that the two Hitchin functionals are proportional:
\begin{equation}
\begin{array}{l}
3\;\Phi\left(B\right)
 =B^3 = B \W d \Ch = d\left(B \W \Ch \right) - dB \W \Ch = \Ch \W \C
 = 2\;\Phi\left(C\right).
\end{array}
\end{equation}

The set of equations \eqref{Hermitian structure and NKM: Nearly kahler field equations} define a \emph{nearly Kähler manifold} (or \emph{weak holonomy $\SU(3)$}, a terminology that we shall not use here). 

\begin{ExtraComputation}
	\begin{framed}
It is known, see Hitchin \cite{Hitchin:2001rw} Theorem 6, that a cone over such a structure gives a $G_2$-structure in 7 dimensions that defines a manifold of holonomy $G_2$. Take $t$ to be coordinate on the radial direction to $\R \times M^6$, then the three-form in 7-dimensions is $C = t^2 dt \W B + t^3 \hat{C}$ is closed $dC=0$ and co-closed $d*C=0$ for ${}^*\Omega = t^3 dt \W C - (t^4/2) B\W B$. See last part of this thesis for more on $G_2$ structure.
	\end{framed}
\end{ExtraComputation}

\paragraph{Constrained Hitchin functional}

As a point of comparison, we briefly described how nearly Kähler manifold were obtained in \cite{Hitchin:2001rw}.

Let, $C$ be a stable \emph{exact} three-form, and let $\gr$ be a stable \emph{exact} 4-form
\begin{equation}
C = d \ga, \qquad \gr=B^2 = d \gb.
\end{equation}
In \cite{Hitchin:2001rw}, Hitchin considered a \emph{constrained} variational principle. Mainly he looked for the critical points (in a cohomology class) of
\begin{equation}
\int_{M^6} \Phi\left[\gr\right] + \Phi\left[C\right] = \int_{M^6} \frac{1}{3}B^3 + \frac{1}{3} \Ch C
\end{equation}
lying on the constraints surface
\begin{equation}\label{Hermitian structure and NKM: constraints}
\int_{M^6} \ga \W d \gb = cst.
\end{equation}
The resulting field equations are just \eqref{Hermitian structure and NKM: Nearly kahler field equations} (up to rescaling the fields).

Note however that, even though superficially similar, this is quite different from the variational principle \eqref{Hermitian structure and NKM: nk action} described above. In particular in \eqref{Hermitian structure and NKM: nk action} there are no extra constraint to be added to the variational principle.

\subsubsection{Example: The nearly Kähler structure of $\PT(M^4)$ on Instantons}

As a neat example of nearly Kähler structure, and in order to relate this part with the first one, we briefly describe here the nearly Kähler structure on the twistor space of an anti-self-dual Einstein metric.

In the first part of this thesis we considered a first almost hermitian structure on $\PT(M^4)$
\begin{equation}
\gO^{3,0} = \p_{A'}D\p^{A'} \W e^0{}^{B'}\p_{B'}\W e^1{}^{C'}\p_{C'}
\end{equation}
\begin{align}
\go &= 4iR^2\;\frac{\p_{A'}D\p^{A'} \W \ph_{B'}D\ph^{B'}}{2\left(\pp \right)^2} - i\frac{e^{AB'}\p_{B'}\W e_{A}{}^{C'}\ph_{C'}}{\pp} \nonumber \\
&= 2iR^2 \left(\frac{\p_{A'}D\p^{A'} \W \ph_{B'}D\ph^{B'}}{\left(\pp \right)^2} -\frac{1}{R^2}\frac{\gS^{B'C'}\p_{B'}\ph_{C'}}{\pp}\right)
\end{align}
see eqs \eqref{Euclidean metric Twistor Space: integrable ACS},\eqref{Euclidean metric Twistor Space: Kahler strcture, g},\eqref{Euclidean metric Twistor Space: Kahler strcture, go}.

We dubbed it the \emph{integrable} almost hermitian structure as it is indeed integrable on anti-self-dual manifolds see Proposition \ref{Proposition: Metric Twistor, integrability} and is in fact Kähler on anti-self-dual Einstein manifolds with cosmological constant $\gL=\frac{3}{R^2}$, see Proposition \ref{Proposition: metric Twistor, Kähler condition}.
In particular on $\CP^3$ this gives the Fubini-Study metric (which is indeed known to be Kähler-Einstein)

However we here wish to now consider the alternative `non-integrable' almost hermitian structure:
\begin{equation}
C = \gO + \gOb
\end{equation}
with
\begin{equation}
\gO = \frac{1}{\left(\pp\right)^2}\;\p_{A'}D\p^{A'} \W e^0{}^{B'}\ph_{B'}\W e^1{}^{C'}\ph_{C'}
\end{equation}
\begin{align}
B = 2iR^2 \left(\frac{\p_{A'}D\p^{A'} \W \ph_{B'}D\ph^{B'}}{\left(\pp \right)^2} +\frac{1}{R^2}\frac{\gS^{B'C'}\p_{B'}\ph_{C'}}{\pp}\right)
\end{align}

\begin{Proposition}\mbox{}\\
The almost Hermitian complex structure given by $\left(C, B\right)$ on $\PT(M)$ is almost nearly Kähler, i.e satisfies 
\begin{equation}
dB = k \; \Ch, \qquad d C = \kt\; B \W B,
\end{equation}
if and only if the base manifold is anti-self-dual Einstein with cosmological constant $\gL = \frac{3}{2 \;R^2}$. Then $k=3$ and $\kt =\frac{\gL}{3}$.
\end{Proposition}
\begin{proof}
	This proposition can be checked by a direct computation.
\end{proof}

In particular this gives another `squashed' Einstein metric on $\CP^3$: starting with the Fubini-Study metric on $\CP^3$ this other Einstein metric is just obtained by squashing radius $R$ of the two-sphere $\S^2$ of the twistor fibration $\S^2 \to \CP^3 \to \S^4$.

\chapter{Three Dimensional Gravity as a Dimensional Reduction of Hitchin Theory in Six Dimensions} \label{Chapter: Three Dimensional Gravity as a Dimensional Reduction of Hitchin Theory in Six Dimensions}

In this chapter, we want to establish that $\SU(2)$ reduction of the `background independent' version of Hitchin theory is 3D gravity coupled with a constant scalar field. We first show how 3D gravity can be naturally understood from a 6D point of view and, what is more, is a subset of solutions to Hitchin theory. Once this is done, we consider the $\SU(2)$ reduction of the full theory.

\section{3D Gravity in Terms of 3-Forms on the Principal Bundle of Frames}\label{section: 3D Gravity in Terms of 3-Forms}
There are several alternative ways of writing the canonical form of $C$. We already considered
\begin{equation}
C = \ga^1\W \ga^2 \W \ga^3 + \gb^1\W \gb^2 \W \gb^3.
\end{equation}
Which is given by the proposition \ref{Proposition: Hitchin's decomposition in six d}.

Yet another way of writing $C$ arises if we set
\begin{equation}
\ga^i = W^i+ \sqrt{\gL} \;E^i, \qquad \gb^i = W^i - \sqrt{\gL}\; E^i, \qquad \forall i \in\{1,2,3\}
\end{equation}
where $\left(W^i\right)_{i\in\{1,2,3\}}$ and  $\left(E^i\right)_{i\in\{1,2,3\}}$ are real one-forms and $\sqrt{\gL} \in \C$. We get
\begin{equation}
C = \frac{\eps^{ijk}}{3}\; W^i\W W^j\W W^k + \gL \; \eps^{ijk}\; W^i\W E^j\W E^k.
\end{equation}

This suggests the following interpretation. We take our six-dimensional manifold $P^6$ to be a $\SU(2)$ principal bundle over a 3D manifold $M^3$,
\begin{equation}
\SU(2) \inj P^6 \to M^3.
\end{equation}
Then if $M^3$ is a Riemannian manifold one can relate $\bdW$ and $\bdE$ respectively with the Levi-Civita connection and the frame field. We now recall how this is done.

\subsection{Lift from 3D gravity to 6D}

We now recall the relationship between the Riemannian geometry of a base manifold $M^3$ with that of the total space $P^6$ of the related $\SU(2)$ principal bundle. Our notations for 3D gravity are standard ones so we don't feel that it is necessary to detail them in the main body of this thesis. See however appendix \ref{Section : Appdx 3d gravity Conventions} for a brief review of these conventions. In particular we use \emph{Lie algebra valued} forms to describe the frame field $\bde$ and the potential of a $\SU(2)$-connection $\bdw$ (when the field equations are satisfied, of course this is just the potential of the Levi-Civita connection). The reader unfamiliar with this notation can again report to the appendix. As a convention, all Lie algebra valued field are from now-on written in bold notation.

\paragraph{Connection forms and forms on a principal $\SU(2)$ bundle}\mbox{}

Thus, we want to establish the classical relation between the potential one-form $\bdw$ of a $\SU(2)$ connection on $M^3$ and a more geometrical description in terms of Lie algebra valued one-forms on the total space $P^6$ of the principal bundle. 

Consider the total space $P^6$ of the principal $\SU(2)$ bundle over $M^3$,
\begin{equation}
\SU(2) \inj P^6 \xto{\pi} M^3.
\end{equation}
We work locally and choose a local trivialisation of this bundle so that every fiber is identified with a copy of the group. Let $p$ be a point in $P^6$ and $U_p \subset P^6$ be an open subset containing $p$. Let $x = \pi(p)$ and $V_{x} = \pi(U_p)$ be their respective projection on $M^3$. Then a local trivialisation $\phi_{U}$ is a map identifying $U_p$ with $SU(2) \times V_{x}$,
\begin{equation}
\phi_{U} \left|
\begin{array}{ccc}
U_p \subset P^6 & \to& SU(2) \times \left( V_{x} \subset M^3 \right) \\
 p &\mapsto& \left(g , x \right)
\end{array}\right.
\end{equation}
A local trivialisation thus gives local coordinates adapted to the bundle structure. We will use the notation $\simeq$ whenever we need to identify geometrical objects on $P^6$ with their expression in a trivialisation e.g \begin{equation}
p \simeq \left(g,x\right) \in SU(2)\times M^3.
\end{equation}

 A choice of trivialisation $\phi \from U_p \to SU(2) \times V_{x}$ is equivalent to choosing a section $s_{\phi} \from V_{x} \to U_p$ such that
\begin{equation}
\phi \circ s_{\phi} \left|
 \begin{array}{ccc}
 V_{x} \subset M^3 & \to& SU(2) \times V_{x} \\
x&\mapsto & s(x) \simeq \left(e , x \right)
\end{array}\right.
\end{equation}
Accordingly, if we choose another trivialisation $s_\psi$ the two are related by
\begin{equation}
\begin{array}{lccc}
\psi \circ s_{\phi} \from & V_{x} \subset M^3 & \to& SU(2) \times V_{x} \\
& x & \mapsto & s_{\psi}(x) \simeq \left(h^{-1}(x) , x \right)
\end{array}
\end{equation}
where $h$ is a section $h \in \gG\left[M^3, \SU(2) \right]$. Therefore, a change of trivialisation amounts to a change of coordinate described by the following diagram.
 \begin{center}
 	\begin{tikzpicture}
 	
 	\node(x1) at (0,0) 
 	{$\begin{array}{c}
 		U_p\\
 		p \in
 		\end{array}$};
 	 	
 	\node(y0) at (6,1.)
 	{$\begin{array}{c}
 		\SU(2) \times V_x\\
 		\left(g, x\right) \in
 		\end{array}$};
 	
 	\node(y2) at (6,-1.)
 	{$\begin{array}{c}
 		\SU(2) \times V_x\\ 
 		\left(h^{-1}(x).g, x\right) \in
 		\end{array}$};
  	
 	\draw [->] (x1) -- (y0.west) node[above, pos=0.5]{$\phi$}; 
 	\draw [->] (x1) --  (y2.west) node[above, pos= 0.5] {$\psi$};
 	\draw [->] (y0.south) --  (y2.north)node[right, pos= 0.5] {$\psi \circ \phi^{-1}$};
 	\end{tikzpicture}
 \end{center}
 The take-home message of this diagram is that \emph{change of coordinates associated with a change of trivialisation act on the left}.
 
  Related to this observation is that any principal bundle comes with a \emph{global right action}. In a given trivialisation, the action of any $h\in SU(2)$ on $p \in P^6$ reads,
  \begin{equation}
  p.h \simeq \left(g.h , x\right).
  \end{equation}
 Clearly this definition does not depend on the choice of trivialisation.
 
 Change of trivialisation are directly related to \emph{gauge transformations}, which are defined as the automorphisms\footnote{In particular this is a subgroup of the diffeomorphisms of $P^6$.} of $P^6$ that preserves each fibres: change of trivialisation and gauge transformations are the two faces (passive/active point of view) of the same coin. 
We here adopt a passive point of view so that for us `gauge transformation' and `gauge invariance' will simply mean `change of trivialisation' and `well-defined'. In fact this terminology is often misleading and we will refrain to talk about `gauge' in the following.

For example, the \emph{Maurer Cartan} frame associated with a trivialisation
 \begin{equation}
 \bdm:= g^{-1} dg
 \end{equation}
 is not a meaningful geometrical object because it is defined by a particular trivialisation \footnote{One could say that it is not `gauge invariant' but really this is more misleading than anything else, the unique meaningful point here is that this form is tied up with a trivialisation.} and will look completely different in any other.  
 
 The same is true for the potential of the $\SU(2)$-connection, $\bdw$. In geometrical terms $\bdw$ is a one-form on $M^3$ taking values in sections of the associated bundle $P^6 \times_{\SU(2)} \su(2)$ so that its exact representation depends on the trivialisation. Changes of trivialisation correspond to `gauge transformations' of $\bdw$.

However, while both $\bdm$ and $\bdw$ depend on the trivialisation chosen, the connection one-form\footnote{In this expression we abuse notation as $\bdw$ really stands for the pull-back of the potential by the projection operator $\pi^* \bdw$. We will systematically make this abuse of notation as this considerably lighten the notation and there is generally no ambiguity.} 
\begin{equation}\label{3D Gravity in Terms of 3-Forms: trivialis}
\bdW:= g^{-1} dg +  g^{-1} \bdw g \quad  \in\gO(P^6) \times \su(2).
\end{equation}
does not. Accordingly, this is a geometrically simple object on $P^6$ - a Lie-algebra valued one-form. 

In general terms a \emph{connection one-form} is a Lie-algebra valued one-form in the total space of the bundle, whose kernel defines the notion of horizontal vector fields. Importantly it \emph{reduces to the Maurer Cartan frame} when restricted to vertical vector fields and is \emph{equivariant} under the right action,
 \begin{equation}
R^*_{h} \bdW = Ad_{h^{-1}}\left(\bdW\right).
\end{equation} These are in fact the two defining properties for a connection form.

 An easy calculations gives the curvature of the connection 
\begin{align}
\bdF&= d\bdW+\bdW \W \bdW \\ &= g^{-1} (d\bdw+\bdw\W \bdw) g \coloneqq g^{-1} \bdf g ,
\end{align}
which is a $\su(2)$-valued \emph{basic} two-form, i.e it is zero on vertical vector fields. It is also equivariant $R^*_{h} \bdF = Ad_{h^{-1}}\left(\bdF\right) $. Again, these are the two defining properties for $\su(2)$-valued forms coming from forms on $M^3$ with values in sections of $P \times_{\SU(2)} \su(2)$.

We will also need the $\su(2)$-valued one-form on $P^6$ defined by the frame field $\bde$
\begin{equation}
\bdE \coloneqq g^{-1} \bde g \quad \in \gO(P^6) \times \su(2).
\end{equation}
One easily checked that it is basic and equivariant as well, as it should be.

\paragraph{Chern-Simons Connection forms and Hitchin Equations}\mbox{}

A standard approach to 3d gravity is in terms of the so called \emph{Chern-Simons formulation}. See appendix \ref{Section : Appdx 3d gravity Conventions} for conventions.

Let $\bda \coloneqq \bdw + \sqrt{\gL} \bdE$ and $\bdat \coloneqq \bdw - \sqrt{\gL} \bdE$ be the potential of the Chern-Simons connections. Here $\sqrt{\gL}$ is a mnemonic standing for $\sqrt{|\gL|}$ for $\gL>0$ and $i \sqrt{|\gL|}$ for $\gL<0$. In the first case (positive $\gL$) $\bda$ and $\bdat$ are two independent real-valued object, in the second situation (negative $\gL$) they are conjugated complex forms. Chern-Simons connections are useful because the flatness of both connections is equivalent to the field equations for 3D gravity.

We can now introduce the lift of the Chern-Simons connections as
\begin{equation}\label{3D Gravity in Terms of 3-Forms: CS form def}
\bdA \coloneqq \bdW + \sqrt{\gL} \bdE = g^{-1}dg + g^{-1} \; \bda \; g.
\end{equation}
\begin{equation*}
\bdAt \coloneqq \bdW - \sqrt{\gL} \bdE = g^{-1}dg + g^{-1} \; \bdat \; g.
\end{equation*}
This should be understood as follows. When $\gL>0$ then $\bdA$ and $\bdAt$ are two $\su(2)$-valued one-forms, while when $\gL<0$ then $\bdA$  and $\bdAt$ are two complex conjugated $\sl(2,\C)$-valued one-forms.
The proper geometrical interpretation of these objects is in term of \emph{Cartan connections}, see section \ref{ssection: A Very Brief Introduction to Cartan Geometry}, but we won't need it here. It will however be useful when we come to consider $\SU(2)$ reduction of Hitchin theory.

Instead of going in that direction let us consider the following three-form on $P^6$ as (matrix multiplication implied)
\begin{equation}\label{3D Gravity in Terms of 3-Forms: gO matrix form}
C = -2\;\frac{\Tr}{3}\left(\bdA \W \bdA \W \bdA + \bdAt \W \bdAt \W \bdAt \right).
\end{equation}
Introducing a coordinate notation rather than a matrix notation\footnote{See Appendix \eqref{Appdx: Spin 1/2 representation of su2} for our conventions on the basis $\gs_i$. In particular with these conventions
	 \begin{equation*}
-\frac{2}{3}\Tr \bdA \bdA \bdA = -\frac{2}{6}\Tr \bdA [\bdA \bdA ] = \frac{\eps^{ijk}}{6} A^i A^j A^k.	
	\end{equation*}} \begin{equation}
\bdA = A^{i} \gs^i, \qquad \bdAt = \At^{i} \gs^i
\end{equation}
this three-form can be rewritten as
\begin{equation}\label{3D Gravity in Terms of 3-Forms: gO form}
C = A^1\W A^2 \W A^3 + \At^1\W \At^2 \W \At^3.
\end{equation}
Which is just of the form \eqref{Stable forms in 6d: om-canonical} and thus $C$ is stable. What is more, the sign of its orbit corresponds to the sign of $\gL$!

  Indeed when $\gL$ is positive both families $\left(A^i\right)_{ \in 1,2,3}$ and $\left(\At^i\right)_{ \in 1,2,3}$ are real forms which implies that \eqref{3D Gravity in Terms of 3-Forms: gO form} is in the positive orbit. On the other hand, when $\gL$ is negative they are complex conjugated and \eqref{3D Gravity in Terms of 3-Forms: gO form} is then in the negative orbit.

In fact in order for the three-form \eqref{3D Gravity in Terms of 3-Forms: gO form} to be stable one also need to check that $\left(A^i , \At^i \right)_{i\in 1,2,3}$ is a basis of one-form. This obviously requires $\gL\neq 0$. When $\gL\neq0$, $\left(A^i , \At^i \right)_{i\in 1,2,3}$ is a basis if and only if $\left(W^i , E^i \right)_{i\in 1,2,3}$ is. The last three forms are basic (they vanish on vertical tangent vectors) while the first three are not so that one only need to check that separately each of the set of $W$'s and $E$' are made up of independent forms. Because the Maurer-Cartan frame $g^{-1}dg m^i \gs_i$ is a basis and \eqref{3D Gravity in Terms of 3-Forms: trivialis}, the one-forms $\left(W^i\right)_{i\in 1,2,3}$ are always independent. Therefore the stability of \eqref{3D Gravity in Terms of 3-Forms: gO form} is equivalent to the non-degeneracy of the triad $\left(e^i\right)_{i \in 1,2,3}$. We will always suppose that this is the case from now on.

Now we saw in section \ref{ssection: Hitchin Theory for 3 forms in 6d} that critical points of Hitchin theory are stable three-forms 
\begin{equation}
C=\ga^1\W \ga^2\W \ga^3 + \gb^1\W \gb^2\W \gb^3, 
\end{equation}
 such that
 \begin{equation}
 d C =0 \quad\text{and}\quad   d \Ch =0
 \end{equation}
 or equivalently
\begin{equation}
\left(d\ga^i\right)\W \ga^1\W \ga^2\W \ga^3 =0 \quad\text{and}\quad\left(d\gb^i\right)\W \gb^1\W \gb^2\W \gb^3 =0 \quad\forall i \in 1,2,3.
\end{equation}
This last equations implies that the distribution $D$ and $\Dt$ defined by the kernel of the $\ga$'s and $\gb$'s (cf \eqref{Stable forms in 6d: D def}) are integrable.

Now in our $\SU(2)$ principal bundle setting \eqref{3D Gravity in Terms of 3-Forms: gO form}, $\bdA$ and $\bdAt$ are connection forms and integrability of the distribution they define is equivalent to vanishing of their curvature:
\begin{equation}
D \coloneqq Ker\left(\bdA \right) \text{is integrable} \quad \Leftrightarrow \quad \bdF = d\bdA + \frac{1}{2} \left[\bdA \W \bdA\right] =0
\end{equation}
As a result Hitchin equations implies 3d gravity! In fact a closer look at the equations shows that this is an equivalence:
\begin{Proposition}\label{Prop: eg 3D grav in 6D} \mbox{}\\	
For three-forms on $P^6$ constructed from a frame field on $M^3$ and a $\SU(2)$-connection as in \eqref{3D Gravity in Terms of 3-Forms: gO matrix form} and \eqref{3D Gravity in Terms of 3-Forms: CS form def}, the following system of equations are equivalent
\begin{enumerate}
	\item $dC=0$,\; $d\Ch=0$ \qquad `Hitchin's Equations'
	\item $\bdF=0$, \;$\bdFt=0$ \qquad `Chern-Simons connections are flat'
	\item $d_{\bdW} \bdE =0 $, \; $\bdF_{\bdW} + \gL \bdE \W \bdE=0$ \qquad `3D gravity'.
\end{enumerate}
Note that we here both need $\bdE$ to be a non-degenerate frame and $\gL \neq0$ in order for $C$ to be stable.
\end{Proposition}

This is a nice interplay between the field equations of 3D gravity and Hitchin Theory but what about the variational principle? At this point it is indeed natural to wonder whether the Hitchin functional \eqref{Stable 6 form in 6d: Hitchin functionnal-} has any interpretation in terms of 3D gravity. In fact, there is. However, perhaps surprisingly the resulting variational principle is the \emph{pure connection} action for 3D gravity. This is a slightly non-standard description of 3D gravity and we will briefly review it before moving on to the Hitchin functional.

\subsection{The pure connection formulation of 3D gravity}
\label{ssection: The pure connection formulation (3D)}

In this section we review the pure connection description of 3D gravity. It seems that the pure connection formulation of 3D gravity was first worked out in \cite{Peldan:1991mh}, starting from the Hamiltonian point of view. A simpler description, directly at the level of the Lagrangian, appears in Section 3.4 of \cite{Peldan:1993hi}. We here only give the Lagrangian description. 

The essential idea is to start with the first-order Einstein-Cartan action,\begin{equation}\label{3D Gravity in Terms of 3-Forms: EC action}
S[\bde,\bdw] = - \int_M \Tr\left( \bde\W \bdf + \frac{\gL}{3} \bde\W\bde\W \bde\right),
\end{equation}
solve the equation of motion $\bdf+ \frac{\gL}{2}\left[\bde\W \bde\right] =0$ for $\bde$ as a function of $\bdf$ and finally substitute the result back into the action.  

\begin{ExtraComputation}
	\begin{framed}
		\begin{align}
&- \int_M \Tr\left( \bde\W \bdf + \frac{\gL}{3} \bde\W\bde\W \bde\right)	\\
&= - \int_M \Tr\left( -\gL\;\bde\W \bde \W \bde + \frac{\gL}{3} \bde\W\bde\W \bde\right)	\\
&=\frac{2\gL}{3}\int_M \Tr\left( \bde\W \bde \W \bde\right) \\
&=\frac{\gL}{3}\int_M \Tr\left( \bde\W \left[ \bde \W \bde\right]\right) \\
&= -\frac{\gL}{3}\int_M 6\; e^1\W e^2 \W e^3 \\
&= -\gL \int_M  (e)^3
		\end{align}
	\end{framed}
\end{ExtraComputation}

To describe the solution, we introduce the notion of definiteness and sign of a connection. For now they will only have 3D interpretation but we will soon see that these notions can in fact be related to those of stability and the sign of a certain\footnote{In fact the `Chern-Simons' three-form.} three-form on $P^6$!

\paragraph{Definite $\SU(2)$-connections in 3D}\mbox{}

For now let us consider again a $\SU(2)$-principal bundle $P^6$ on a 3-manifold $M^3$. Let $\bdw$ be the potential of a $\SU(2)$-connection. Let $\bdf=d\bdw+\bdw\W \bdw$ be its curvature two-form. It is convenient to see the curvature as a map from bi-vectors to $\su(2)$:
\begin{equation}
\bdf \from \gL^2\;TM \to \su(2)
\end{equation}
In three dimensions, bi-vectors form a three dimensional vector space. We will say that a connection $\bdw$ is \emph{definite} if its curvature is non-degenerate when understood as the map above. 
Now, the Lie algebra $\su(2)$ has a natural orientation given by the Lie group product, i.e a basis of $\su(2)$ $\left(a,b,c\right)$, is positively oriented if \footnote{Here we think of $\su(2)$ element as two by two hermitian matrix, see Appendix \ref{Appdx: Spin 1/2 representation of su2} for our convention.}
\begin{equation}
-2\;\Tr \left( a \left[b,c\right] \right) > 0 .
\end{equation}
 Bi-vectors also have a natural orientation: introducing a metric to raise and lower indices one can indeed define a bracket
 \begin{equation}
 \left[\ga , \gb\right]^{ab} \coloneqq \ga^{a}{}_{c}\gb^{cb} - \ga^{a}{}_{c}\gb^{cb}
 \end{equation}
 and take $\left(\ga, \gb , \gc \right)$ to be an oriented basis of bi-vectors if and only if
 \begin{equation}
 \left[\ga, \gb \right]^{ab} \gc_{ab} > 0 
 \end{equation}
Because any two metrics can be continuously deformed into another this choice of orientation does not depend on the choice of metric.
 
 An alternative way to know if a basis $\left(\ga, \gb, \gc \right)$ of bivectors is oriented according to this rule is as follows. We first use the isomorphism $\gO^2(M^3) \simeq T \left(M^3\right) \otimes \gO^3(M^3)$ (see appendix \ref{Section : Appdx Density })
   \begin{equation}
   \ga \in \gL^2T(M^3) \;\mapsto\; \gat \in  T \left(M^3\right) \otimes \gO^3(M^3).
   \end{equation}
  to construct a triplet $\left(\gat,\gbt, \gct \right)$ of volume-form-valued tangent vector.

Then taking the wedge products of these tangent vectors we obtain
 \begin{equation}\label{3D Gravity in Terms of 3-Forms: det bivector}
\gat \W \gbt \W \gct \in \gL^3T(M^3) \otimes \left(\gO^3\left(M^3\right)\right)^3 \simeq \left(\gO^3(M^3)\right)^2  \end{equation}
The sign of \eqref{3D Gravity in Terms of 3-Forms: det bivector} is well defined. For any volume form $\nu$,
\begin{equation}
\gat \W \gbt \W \gct = - \gl \left(\nu \right)^2, \quad \gl \in \R
\end{equation}
Then $\left(\ga, \gb, \gc \right)$ is oriented according to the above rules if and only if $\gl>0$. Note the minus sign necessary here. That the two orientations then coincide can be checked explicitly using coordinates and our conventions given in Appendix\ref{Section : Appdx Density } and the algebra of epsilon symbol given in Appendix\ref{Appdx: Spin 1 representation of su2} :
		\begin{align}
\ga_{ab}\left(\gb^{ac} \gc_c{}^b - \gc{}^{ac} \gb_c{}^b\right) &= \left(\eps^{ac}{}_i \eps_c{}^b{}_j - \eps^{ac}{}_i \eps_c{}^b{}_j \right) \gbt^i \gct^j \nonumber\\
&= \ga_{ab} \left( -\eps^{ijk} \eps^{ab}{}_k\right) \gbt^i \gct^j \nonumber\\
&= -2 \eps^{ijk} \gat^k \gbt^i \gct^j.\nonumber
		\end{align}

Therefore, if $\bdw$ is a definite connection their are two possibilities: either its curvature sends oriented basis of bi-vectors into oriented basis of $\su(2)$ or not. We call definite positive the first case and definite negative the second case. This defines the \emph{sign} of a definite 3D connection.

For practical purpose, this sign is directly computed as follows: We first apply the isomorphism $\gO^2(M^3) \simeq T \left(M^3\right) \otimes \gO^3(M^3)$ to $\bdf$
\begin{equation}
\bdf \in \gO^2\left(M^3\right) \otimes \su(2) \quad \mapsto \quad \bdft \in T\left(M^3\right) \otimes \gO^3\left(M^3\right) \otimes \su(2)
\end{equation}
\begin{ExtraComputation}
	\begin{framed}
\begin{equation}
\bdft^{\mu} = \frac{\epst^{\mu\nu\rho}}{2} \;\bdf_{\nu\rho} 
\end{equation}
	\end{framed}
\end{ExtraComputation}

Then the connection is definite if and only if
\begin{equation}\label{3D Gravity in Terms of 3-Forms: vol^2}
\frac{1}{3}\Tr \left( \bdft \left[\bdft, \bdft \right] \right) \in \left(\gO^3(M^3)\right)^2
\end{equation}
is non-zero and the sign of \eqref{3D Gravity in Terms of 3-Forms: vol^2} is the sign of the connection. Explicitly:
\begin{align*}
&\left(dx^3\right)^3 \;\frac{1}{3}\Tr \left( \bdft^{\mu} \left[\bdft^{\nu}, \bdft^{\gr} \right] \right) \;\pa_{\mu}\otimes \pa_{\nu} \otimes \pa_{\gr}\\	&= -\left(dx^3\right)^3\;\frac{\eps^{ijk}}{6} \ft_i^{\mu} \ft_j^{\nu} \ft_k^{\rho} \;\pa_{\mu}\otimes \pa_{\nu} \otimes \pa_{\gr}\\
&= -\left(dx^3\right)^3\; det\left(\ft_i^{\mu}\right) \frac{\epsut^{\mu\nu\rho}}{6} \;\pa_{\mu}\otimes \pa_{\nu} \otimes \pa_{\gr}\\
 &= -\left(dx^3\right)^2\; det\left(\ft_i^{\mu}\right). 
\end{align*}
So that the sign of the connection is the sign of $det\left(\ft_i^{\mu}\right)$.

If we choose an orientation on $M^3$, one can then take the square root of \eqref{3D Gravity in Terms of 3-Forms: vol^2} and obtain a volume form:
\begin{equation}\label{3D Gravity in Terms of 3-Forms: vol}
v_{\bdf} \coloneqq \sqrt{\frac{1}{3}\left|\Tr \left( \bdft \left[\bdft, \bdft \right] \right)\right|}
\end{equation}
or
\begin{equation}
v_{\bdf}= \sqrt{|det\left(\ft_i^{\mu}\right)|}\; d^3x
\end{equation}
where $d^3x$ is any oriented volume form.

\paragraph{The pure connection formulation}\mbox{}

Consider a definite connection $\bdw$ with sign $\gL=\pm1$.  The pure connection formulation gravity action is just the total volume
\begin{equation}\label{3D Gravity in Terms of 3-Forms: pure-conn-action}
S_{GR}[\bdw] = -\gL\; \int_{M^3}  \; v_\bdf.
\end{equation}

An interesting property of definite connections with sign $\gL = \pm1$ is that they define a frame field $\bde_\bdf \in \gO^1\left(M^3\right) \otimes \su(2)$ such that 
\begin{equation}\label{3D Gravity in Terms of 3-Forms: f-theta-eqn}
\bdf +\gL\; \frac{1}{2}\left[\bde_\bdf\W\bde_\bdf\right] =0.
\end{equation}

\begin{ExtraComputation}
	\begin{framed}
		\begin{equation}
		\bdf +\gL \bde_\bdf\W\bde_\bdf =\left(\bdf^i_{jk} + \gL \eps^{i}{}_{jk}\right)\; \frac{e^j\W e^k}{2}=0
		\end{equation}
		\begin{equation}
		\Rightarrow \qquad \bdft = -\left(\bde^3\right)\;\gL\; \gd^i_k e^k
		\end{equation}
		
		\begin{equation}
		\Rightarrow \qquad \frac{1}{3}\Tr \left( \bdft \left[\bdft, \bdft \right] \right)  = \left( \bde^3\right)^2  \left(\gL\right)^3
		\end{equation}	
	\end{framed}
\end{ExtraComputation}

The above triad is obtained via the following construction. One first construct the densitized triad,
\begin{equation}
\bdet_{\bdf} = \frac{1}{2}\left[\bdft \id \bdf \right] \in \gO^1(M^3) \otimes \gO^3(M^3) \otimes \su(2)
\end{equation}
where $\id$ is the interior derivative.	More explicitly,
\begin{equation}
\bdet = \frac{1}{2}	\eps^{ijk} \ft^{\mu j} f^k{}_{\mu\nu}  \;dx^{\nu} \otimes \left(d^3x\right) \otimes\; \gs_i .
\end{equation}
\begin{ExtraComputation}
	\begin{framed}
\begin{align}
\bdet &= \frac{1}{2}	\eps^{ijk} \ft^{\mu j} f^k{}_{\mu\nu}  \;dx^{\nu} \left(dx^3\right) \; \gs_i \\
&= \frac{1}{2}	\eps^{ijk} \gL^2 \;\gd^{\mu}{}_j \eps^k{}_{\mu\nu}  \; e^{\nu} \left(\bde^3\right) \;\gs_i \\
&= e^i \;\gL^2\; \left(\bde \right)^3 \; \gs_i
\end{align}
	\end{framed}
\end{ExtraComputation}

One then fixes the scaling by dividing by the volume form \eqref{3D Gravity in Terms of 3-Forms: vol}.
\begin{equation} \label{3D Gravity in Terms of 3-Forms: e def}
\bde_{\bdf} = \frac{1}{2}\left[\bdft \id \bdf \right] \big/ v_{\bdf}
\end{equation}
\begin{equation}
\bde = \frac{1}{2}\; \eps^{ijk} \ft^{\mu j} f^k{}_{\mu\nu}\; \left(\sqrt{|det\left(\ft_i^{\mu}\right)|}\right)^{-1}  \;dx^{\nu} \otimes\; \gs_i .
\end{equation}

The frame $\bde_\bdf$ defines the metric $ds^2_\bdf:= - 2\, \Tr(\bde_\bdf\otimes \bde_\bdf)$, which is of Riemannian signature. One can check that with these definitions, $\bde_\bdf$ is such that 
\begin{equation}\label{3D Gravity in Terms of 3-Forms: volume-f}
v_\bdf = -\frac{2}{3} \Tr(\bde_\bdf\W \bde_\bdf\W \bde_\bdf),
\end{equation}
i.e \eqref{3D Gravity in Terms of 3-Forms: vol} is the volume form of the metric. In particular this frame is non-degenerate if and only if $v_\bdf \neq0$ i.e if and only if $\bdw$ is definite (which is, of course, the whole point of the definition).

Note that the action \eqref{3D Gravity in Terms of 3-Forms: pure-conn-action} is just the value of the first-order action \eqref{3D Gravity in Terms of 3-Forms: EC action} on the solution \eqref{3D Gravity in Terms of 3-Forms: e def} of \eqref{3D Gravity in Terms of 3-Forms: f-theta-eqn}.

\paragraph{The first variation and Euler-Lagrange equations}\mbox{}

The expression \eqref{3D Gravity in Terms of 3-Forms: volume-f}  makes it clear that the first variation of the pure connection action \eqref{3D Gravity in Terms of 3-Forms: pure-conn-action} is given by
\begin{equation}\label{3D Gravity in Terms of 3-Forms: var-vf}
\gd S[\bdw]= - \gL \;\int \,\Tr(\gd \bde_\bdf \W \bde_\bdf\W \bde_\bdf) = - \gL\;
\int \Tr(\gd (\bde_\bdf \W \bde_\bdf) \W \bde_\bdf) =
\int \Tr(\gd \bdf \W \bde_\bdf).
\end{equation}
This shows that the critical points of the pure connection action are connections satisfying the following second-order PDE
\begin{align}\label{3D Gravity in Terms of 3-Forms: pure-conn-feqs}
d_{\bdw} \bde_\bdf &= d \bde_{\bdf}  + \left[\bdw , \bde_{\bdf} \right] \\
&= d \bde_{\bdf} + \bdw \W \bde_{\bdf} + \bde_{\bdf}\W \bdw \\  \nonumber
 &=0. \nonumber
\end{align}
 This equation says that the connection $\bdw$ is the unique torsion-free connection compatible with the frame $\bde_\bdf$. The equation \eqref{3D Gravity in Terms of 3-Forms: f-theta-eqn} that defines $\bde_\bdf$ then becomes the statement that the metric constructed from $\bde_\bdf$ is of constant negative curvature. This shows that \eqref{3D Gravity in Terms of 3-Forms: pure-conn-action} is indeed the pure connection formulation of 3D gravity (non -zero cosmological constant $\gL$).

\subsection{Hitchin Functional, the Chern-Simons three-form and the Pure Connection Formulation of 3D Gravity}\label{ssection: Hitchin Functional and the Chern-Simons three-form}

Let us now come back to the 6D notations. We consider again the three-form \eqref{3D Gravity in Terms of 3-Forms: gO matrix form} which we rewrite here for convenience.
\begin{equation}\label{3D Gravity in Terms of 3-Forms: gO matrix form2} 
C = -2\frac{\Tr}{3}\left(\bdA \W \bdA \W \bdA + \bdAt \W \bdAt \W \bdAt \right).
\end{equation}

Our initial question was \emph{What is `Hitchin Theory' for this particular three-form?} By `Hitchin Theory' we really mean a variational principle on stable \emph{closed} three-form in 6D.\footnote{That is what we referred above as `background independent' Hitchin theory.} The three-form \eqref{3D Gravity in Terms of 3-Forms: gO matrix form2} is generically (i.e for $\gL \neq 0$) stable but not closed. Therefore before considering Hitchin functional one needs to impose that the exterior derivative of \eqref{3D Gravity in Terms of 3-Forms: gO matrix form2} vanishes.

\paragraph{Closing $C$}\mbox{}

In order to get a geometric interpretation of the constraints $dC =0$ it is best to open up again the Chern-Simons connections in terms of connection and triad:
\begin{align}\label{3D Gravity in Terms of 3-Forms: ChernSimon conections}
\bdA &\coloneqq \bdW + \sqrt{\gL} \bdE = g^{-1}dg + g^{-1} \; \bda \; g \\
\bdAt &\coloneqq \bdW - \sqrt{\gL} \bdE = g^{-1}dg + g^{-1} \; \bdat \; g.
\end{align}

Rewriting \eqref{3D Gravity in Terms of 3-Forms: gO matrix form2} in terms of \eqref{3D Gravity in Terms of 3-Forms: ChernSimon conections} we get
\begin{equation}\label{3D Gravity in Terms of 3-Forms: gO matrix form EW}
C = -4\;\Tr\left(\frac{1}{3} \bdW \W \bdW \W \bdW +\gL \;\bdW \W \bdE \W \bdE \right).
\end{equation}
A direct computation shows that
\begin{align}\label{3D Gravity in Terms of 3-Forms: dgO}
dC= -4\;\Tr \left(\bdW \W \bdW \W \left(\bdF_{\bdW} +\gL \; \bdE \W \bdE \right)\right) -2\gL\;\Tr\;\left( \bdW \W d_{\bdW} \left[\bdE \W \bdE \right]\right). 
\end{align}
It follows that
\begin{equation}\label{3D Gravity in Terms of 3-Forms: closness of gO}
dC=0 \quad \Leftrightarrow \qquad \bdF_{\bdW} +\gL \bdE \W \bdE =0.
\end{equation}
This is indeed obtained by evaluating the above equation on two vertical vector field. Note that the second term in \eqref{3D Gravity in Terms of 3-Forms: dgO} does not give more equations as it vanishes identically once the first term does (due to Bianchi identity).

We thus obtain the `second half' of the equations for 3D gravity that we recall here for convenience:
\begin{equation}\label{3D Gravity in Terms of 3-Forms: closness of gO 3d gravity eq }
d_{\bdW}\bdE=0,\qquad \bdF_{\bdW} +\frac{\gL}{2} \left[\bdE \W \bdE\right] =0.
\end{equation}

 This result is not really surprising because we know from Proposition \eqref{Prop: eg 3D grav in 6D} that altogether $dC =0$ and $d\Ch=0$ are fully equivalent to 3D gravity. This was however not completely obvious in the first place. In \eqref{3D Gravity in Terms of 3-Forms: dgO} equations coming from the term proportional to $\bdW$
\begin{equation}\label{3D Gravity in Terms of 3-Forms: closness of gO other eq}
d_{\bdW} \left(\bdE \W \bdE \right)=0
\end{equation}
 could have been an independent equation but `miraculously' turns out to be a consequence of the equations coming from the term proportional to $\bdW \W \bdW$. In fact if instead of starting with the closeness of $C$ we had looked at $d\Ch=0$ we would \emph{not} have obtained the first half of \eqref{3D Gravity in Terms of 3-Forms: closness of gO 3d gravity eq } only. Rather we would have found some additional equations that are necessary condition of the second half of \eqref{3D Gravity in Terms of 3-Forms: closness of gO 3d gravity eq }  (i.e equation of the type \eqref{3D Gravity in Terms of 3-Forms: closness of gO other eq}).

The essential result following from \eqref{3D Gravity in Terms of 3-Forms: closness of gO} is that imposing the closeness of $C$ amounts to solving the triad $\bdE$ in terms of the connection $\bdW$. In the previous section we precisely saw how this can be done. Then the sign of the connection encodes the information about the sign of $\gL$. Once the constraints \eqref{3D Gravity in Terms of 3-Forms: closness of gO} are satisfied we  have a theory of a $\SU(2)$ connection only!

 Coming back to the three-forms \eqref{3D Gravity in Terms of 3-Forms: gO matrix form EW}, it can now be rewritten (just making use of \eqref{3D Gravity in Terms of 3-Forms: closness of gO})
\begin{equation}
C = 4\;\Tr\left(-\frac{1}{3} \bdW \W \bdW \W \bdW + \bdW \W \bdF_{\bdW} \right)
\end{equation}
or equivalently
\begin{equation}\label{3D Gravity in Terms of 3-Forms: chern simon three-form}
C = -2\;CS\left(\bdW\right) \coloneqq 4\;\Tr\left(\bdW \W d\bdW +\frac{2}{3} \bdW \W \bdW \W \bdW \right).
\end{equation}
Which is the \emph{Chern-Simon three-form} on $P^6$ associated with $\bdw$. Note that despite the fact that the usual Chern-Simon three-form on $M^3$
\begin{equation}
CS\left(\bdw\right) \coloneqq -2\;\Tr\left(\bdw \W d\bdw +\frac{2}{3} \bdw \W \bdw \W \bdw \right)
\end{equation}
 is not gauge invariant this three-form on $P^6$ is well defined i.e \eqref{3D Gravity in Terms of 3-Forms: chern simon three-form} makes sense independently of any choice of trivialisation.
 
 Now, on the one hand we have a theory of a $\SU(2)$ connection $\bdw$ and the associated notions of definiteness and sign of a connection. On the other hand we have theory of three-forms $C$ in 6D and the associated notions of stability and sign of an orbit. The three-form $C$ being related to the connection $\bdw$ by \eqref{3D Gravity in Terms of 3-Forms: chern simon three-form}.
 
 It is reassuring that the different notions involved are, in that particular case, equivalent:
 \begin{Proposition}\mbox{} \\
 Stability of the Chern-Simon three-form $CS\left(\bdW\right)$ is equivalent to the definiteness of $\bdw$.\\
 When this is satisfied, the sign of the orbit of $CS\left(\bdW \right)$ and the sign of $\bdw$ coincide. 
 \end{Proposition}
The proof of the right-to-left implication is more or less already contained in a scattered way in the previous sections. In section \ref{ssection: The pure connection formulation (3D)} we saw that the curvature of a $\SU(2)$ connection defines a frame-field $\bdE$ such that
\begin{equation}\label{3D Gravity in Terms of 3-Forms: prop constraint}
\bdF + \gL \bdE \W \bdE =0.
\end{equation}
Then the Chern-Simons three-form \eqref{3D Gravity in Terms of 3-Forms: chern simon three-form} can be rewritten in terms of $\bdW$ and $\bdE$ as \eqref{3D Gravity in Terms of 3-Forms: gO matrix form EW}. Then, as we already discussed (see the discussion around eq \eqref{3D Gravity in Terms of 3-Forms: gO form}), stability of the three-form is equivalent to the non-degeneracy of the frame field which is in turn equivalent to the definiteness of the connection. Finally introducing the Chern-Simons connections \eqref{3D Gravity in Terms of 3-Forms: CS form def} it can in turn be rewritten as \eqref{3D Gravity in Terms of 3-Forms: gO form} from which one directly sees that the sign corresponds. \qed

All that remains to be seen is that the Hitchin functional for the Chern-Simon three-form $CS\left(\bdW \right)$ coincide with the pure connection action $S_{PC}\left[\bdw \right]$.
 
\paragraph{Hitchin Functionnal and The Pure Connection Action}\mbox{}

We now compute the Hitchin's action \eqref{Stable 6 form in 6d: action-hitchin},\eqref{Stable 6 form in 6d: Hitchin functionnal d} on our three-form $\Omega$. We use the expressions \eqref{Stable form in 6d: aaa in terms of gO +},\eqref{Stable form in 6d: aaa in terms of gO -} and rewrite $\hat{\Omega}$
\begin{equation}
\Ch = -2\left(\sqrt{\gL}\right)^{3}\frac{\Tr}{3}\left(\bdA \W \bdA \W \bdA - \bdAt \W \bdAt \W \bdAt \right). 
\end{equation}
Note that here and thereafter $\gL \in \left( 1, -1\right)$ and $\sqrt{\gL} \in \left(1,i\right)$.

 In terms of $\bdW$ and $\bdE$, we have
\begin{equation}\label{3D Gravity in Terms of 3-Forms: hat-omega}
\Ch =  -4\; \Tr\left( \frac{\gL}{3} \bdE\W \bdE\W \bdE + \bdW \W \bdW \W \bdE)\right)
\end{equation}
and
\begin{equation}
C = -4\;\Tr\left(\frac{1}{3} \bdW \W \bdW \W \bdW +\gL \;\bdW \W \bdE \W \bdE \right).
\end{equation}
This gives
\begin{equation}
\frac{1}{2}\Ch \W C = -8 \gL \; \left[ \frac{1}{3} \Tr (\bdW^3) \frac{1}{3} \Tr(\bdE^3) +\Tr\left( \bdW \bdE^2\right) \Tr\left( \bdW^2 \bdE\right)\right],
\end{equation}
where we omitted the wedge product signs. We can now replace $\bdW$ here by the Maurer-Cartan form $\bdm=g^{-1} dg$, as the part of $\bdW$ that is a one-form on the base does not contribute. We can also rewrite the last term here as a multiple of the first, using some simple properties of the Lie algebra generators $\gs^i$
\begin{equation}\label{3D Gravity in Terms of 3-Forms: trace-ident}
\Tr\left( \bdW \bdE^2\right) \Tr\left( \bdW^2 \bdE\right)= \frac{1}{3} \Tr (\bdW^3) \Tr(\bdE^3).
\end{equation}
Thus, overall
\begin{align}\label{3D Gravity in Terms of 3-Forms: v-omega}
\Phi\left[C\right] = \frac{1}{2} \Ch \W C &= 4\left(\sqrt{\gL}\right)^{3} \; \frac{1}{3} \Tr \bdA \W \bdA \W \bdA \W \frac{1}{3} \Tr \bdAt \W \bdAt \W \bdAt \\
 &= -8\gL \left(-\frac{2}{3}\Tr (\bdm^3)\right)  \left(-\frac{2}{3}\Tr(\bde_\bdf^3)\right)\\
 &= 8\;\left(-\frac{2}{3} \Tr (\bdm^3)\right) \W \left(- \gL\, v_\bdf \right),
\end{align}
where the volume form on the base $v_\bdf$ is given by \eqref{3D Gravity in Terms of 3-Forms: vol}. 

We thus have the following proposition
\begin{Proposition}
Let $\SU(2) \in P^6 \to M^3$ be a $\SU(2)$ principal bundle over a 3D manifold $M^3$. Let $\bdw$ be the potential of the a $\SU(2)$ connection on $P^6$ and $\bdW$ the associated connection one-form.
	
The Hithin functional $S[C]$ (see \eqref{Stable 6 form in 6d: action-hitchin}) evaluated on the Chern-Simons three-form $C = CS\left(\bdW \right)$ (see \eqref{3D Gravity in Terms of 3-Forms: chern simon three-form} ) is proportional to the pure connection action to 3D Gravity $S_{GR}\left[\bdw \right]$ (see \eqref{3D Gravity in Terms of 3-Forms: pure-conn-action}):

\begin{equation}
\int_{P^6}\Phi\left[ CS\left(\bdW \right) \right] = 8\left(\int_{\SU(2)} -\frac{2}{3} \Tr (\bdm^3) \right) \times S_{GR}\left[\bdw \right]
\end{equation}

Equivalently, if $\bde$ is a $\su(2)$-valued one-form on $M^3$, the above action is obtained by evaluating the Hitchin functional on the three-form $C\left(\bde, \bdw \right)$ given by \eqref{3D Gravity in Terms of 3-Forms: CS form def} and \eqref{3D Gravity in Terms of 3-Forms: gO matrix form} together with the constraint $dC =0$. Solving this constraint then amounts to solving $\bde$ as a function of $\bdw$ (see \eqref{3D Gravity in Terms of 3-Forms: e def}).

\end{Proposition}

\section{$\SU(2)$ Reduction from 6d to 3d}\label{section: Hitchin6D reduction}

In this section we dimensionaly reduce Hitchin from 6D to 3D by an $SU(2)$ action. Practically we suppose that $\SU(2)$ acts freely on $P^6$ such that it has the structure of a principal bundle. We note $M^3$ the resulting quotient manifold.
\begin{equation}
	\SU(2) \inj P^6 \to P^6\big / \SU(2) \simeq M^3
\end{equation}
 We then consider Hitchin theory for three-forms $C$ on $P^6$ that are invariant under the $\SU(2)$ action
 \begin{equation}
 R^* C= C.
 \end{equation}

The resulting theory turns out to be 3d gravity coupled with a (constant) scalar field. This can be seen in different ways but we believe that the most convincing proof is by using concept from Cartan Geometry. Cartan geometry is a beautiful framework generalising both Klein geometry, i.e the geometry of homogeneous spaces and the essential idea of Riemannian geometry which is the make the geometry local. This is however slightly out of fashion today and not so well known, at least from the physicist community. Therefore before we conduct the reduction of Hitchin theory we review the basics of Cartan-Geometry. This presentation is essentially taken from the beautiful book \cite{Sharp}.

\subsection{A Very Brief Introduction to Cartan Geometry}\label{ssection: A Very Brief Introduction to Cartan Geometry}

Cartan geometry is a generalisation both of Riemannian geometry and Klein geometry.\\
In Riemannian geometry a d-dimensional manifold can be infinitesimally identified with $\R^d$. This is indeed the role of the metric. The Riemann curvature tensor is then the obstruction to make this identification local, we won't discussed global aspects here.

On the other hand Klein geometry is the geometry of homogeneous spaces. If $G$ is a $n$-dimensional Lie group and $H$ a subgroup of dimension $n-d$ then the homogeneous space $G/H$ is a d-dimensional manifold. In particular, associated with this homogeneous space we have a principal $H$ bundle 
\begin{equation}
H\inj G \to G/H.
\end{equation}

Cartan geometry applies the essential idea of Riemannian geometry, \emph{make the geometry infinitesimal} to the geometry of homogeneous space. Accordingly a Cartan geometry modelled on $\left(G,H\right)$ makes a infinitesimal identification, i.e an identification at the level of the tangent space, of a manifold with some fixed homogeneous space the \emph{model} $G/H$. Doing this identification is the role of the \emph{Cartan connection}, the curvature of the connection then is the obstruction to making this identification local (i.e in an open subset). Again, we won't really consider global problem here. The following results are crucial to understand how this is done.

\subsubsection{Non Abelian Generalisation of the Fundamental Theorem of Calculus}

Let $G$ be a lie group, $\frg$ its Lie algebra. We note $\bdm_G$ the Maurer-Cartan form on $G$. We take $M$ to be a smooth manifold.

If $f \from M \to G$ is a smooth map we can define its `Darboux derivative' $\bdgo_f$, a $\frg$-valued one-form on $M$, as
\begin{equation}
 \bdgo_f = f^* (\bdm_G). 
\end{equation}
Reciprocally we will refer to $f$ as the primitive of $\bdgo_f$.

Just as for the usual primitive of a function from $\R$ to $\R$, the primitive (if it exists) is defined up to an integration constant (cf \cite{Sharp} p115):\\
if $\bdgo_{f_1} = \bdgo_{f_2}$ then there must exists $C\in G$ the `integration constant' such that $f_1 = C.f_2$.

One can always  find a primitive (`integrate') a $\frg$-valued one-form, $\bdgo$, along a path $\gs$ in $M$. The result is a path $\gst$ in $G$, referred to as the `development of $\bdgo$ along $\gs$'. This procedure is essentially unique up to the choice of a "starting point" for $\gst$:
\begin{Theorem} "Development of $\bdgo$ along a path"\mbox{}\\
	Let $\bdgo$ be a $\frg$-valued one-form on M and $\gs\from [0,1] \to M$ is a smooth path in $M$. Then there exists a unique smooth map $\gst \from [0,1] \to G$ satisfying $\gst(0) = g$ and $\gs^*(\bdgo) = \gst^*(\bdm_G)$. This map is called the development of $\bdgo$ along $\gs$ starting at $g$.
\end{Theorem}
\begin{proof}
For a proof see \cite{Sharp} p120.
\end{proof}

However in general the endpoint of a development $\gst(1)$ will depend not just on the endpoints of the path $\gs(0)$ and $\gs(1)$ but on the details of the path. This is related to the fact that not all $\frg$-valued one-form are Darboux derivatives.

 From the definition, it is clear that a necessary condition for $\bdgo$ to be a Darboux derivative i.e $\bdgo = \bdgo_f$, is that it must satisfy the Maurer-Cartan (or "structural") equation:
\begin{align}
d\bdgo_f + \frac{1}{2} [\bdgo_f,\bdgo_f] =0
\end{align}

It turns out that locally this is the only obstruction:
\begin{Theorem}"Local generalisation of the fundamental theorem of calculus" (E.Cartan). \mbox{}\\
	Let $\bdgo$ be a $\frg$-valued one-form on $M$ satisfying the structual equation, $d\bdgo + \frac{1}{2} [\bdgo,\bdgo] =0$. Then, for each point $p\in M$, there is a neighborhood $U$ of p and a smooth map $f\from U \to G$ such that $\bdgo\big|_U = \bdgo_f$.
\end{Theorem}

A related theorem is 
\begin{Theorem} "Monodromy representation of $\bdgo "$ \mbox{}\\
	Let $p$ and $q$ be two point in $M$. Let $\gs$ be a path in M starting at p and ending at q. If $\bdgo$ verifies the structural equation then the endpoint $\gst(1)$ of its development starting at $g$ along $\gs$ only depends on the homotopy class of $\gs$. In particular this gives a well defined map 
\begin{equation*}
	\Phi_{\bdgo}\from \pi_1(M,p) \to G
\end{equation*}
	called the "monodromy representation of $\bdgo$".
\end{Theorem}
Practically the monodromy representation encodes the global obstruction for $\bdgo$ to have a primitive:
\begin{Theorem}"Generalisation of the fundamental theorem of calculus"\mbox{}\\
	Let $\bdgo$ be $\frg$-valued one-form on $M$ then \\
	
		$\bdgo$ is the Darboux derivative of some map $M \to G$ \\
		$\Leftrightarrow$ \\
		$\bdgo$ verifies the structural equation and 		the monodromy representation
		 \begin{equation*}
		\Phi_{\bdgo}\from \pi_1(M,p) \to G
		\end{equation*} is trivial.
\end{Theorem}
\begin{proof}
	For a proof of the theorems above, see \cite{Sharp} p116, p121 and p124.
\end{proof}

Let us briefly sum up what we've just learned. Any $\frg$-valued one-form on M identifies `infinitesimally' $M$ with $G$. Development realises this identification along a path. It can be made local, in an open subset, if and only if the one form satisfy the Maurer-Cartan equation. Thus in some sense $d\bdgo + \frac{1}{2}[\bdgo,\bdgo]$ plays the same role as the Riemannian tensor.
 
 This is a first step but this is only part of the road to Cartan geometry as the aim of Cartan geometry is to identify a manifold $M$ not with a Lie group $G$ but with an homogeneous space $G/H$. This will require a little bit more structure on the manifold.

\subsubsection{Cartan Geometry and the Tractor Connection}

Let $M$ be a d dimensional manifold and let $H \inj P \to M$ be a principal $H$ bundle. 

In general the topology of $M$ (respectively $P$) will be very different from the topology of $G/H$ (respectively G). However one can follow the philosophy of Riemannian geometry and try to make an infinitesimal identification of
\begin{equation}
H \inj P \to M
\end{equation}
with our model space
\begin{equation}
H \inj G \to G\big/H.
\end{equation}
 From the description we just gave of "the generalisation of the fundamental theorem of calculus" one can already infer a strategy to do just that: A $\frg$-valued one-form $\bdgo$ on $P$ indeed allows to "infinitesimally" identify $P$ with $G$.

 We also already encounter the only obstruction to make this identification local, i.e the structural equation. However generically this local identification $P\simeq G$ will not be compatible with the $H$-bundle structure. A related problem is that in general $\bdgo$ has no geometrical interpretation from the point of view of the base manifold $M = P/H$. One thus need to `tie up' $\bdgo$ with the $H$-action on P to fix these problems.

\paragraph{The Flat Case} \mbox{}

Before we get to the full fledge definition of Cartan geometry it is good to take some times to look at the flat model:
  \begin{equation}\label{Hitchin6D reduction: flat cartan model}
 H\inj P=G \to M=G/H.
 \end{equation} 
Accordingly $G$ is though of as the total space of a $H$ bundle over $G/H$.
 
 On $G$ we have the Maurer-Cartan one-form $\bdm_G$. It's fundamental geometrical meaning is to identify each tangent space to $G$ with the Lie algebra $\frg$. It is related to the $H$-bundle structure by the following properties
\begin{IEEEeqnarray}{lClr}\label{Hitchin6D reduction: H Maurer-Cartan properties}
	R^*_h \bdm_G &=& Ad(h^{-1}) \;\bdm_G &\qquad \forall h\in H \\
	\bdm_G\left(X_{\xi}\right) &=& \xi \;&\qquad \forall \xi\in \mathfrak{h}\nonumber
\end{IEEEeqnarray}
where $X_{\xi}$ is the vector field induced on $G$ by the differentiation of the right action of $H$. This is this property that we want to generalise to the curved case.

A second important remark is that one can use the left action of $H$ on $G$ to construct a principal \emph{$G$ bundle} over $G/H$, \begin{equation}\label{Hitchin6D reduction: flat cartan tractorB}
G \inj P' \to G/H.
\end{equation}
 Then \eqref{Hitchin6D reduction: flat cartan model}, seen as a principal bundle, is naturally included in \eqref{Hitchin6D reduction: flat cartan tractorB}. It turns out that there is a unique principal $G$-connection on $P'$, \emph{the tractor connection}, such that its connection one-form
 \begin{equation}
 \bdgot \in \gO^1\left(P', \frg\right)
 \end{equation}
 coincides with $\bdm_G$ when restricted to $G$ i.e \begin{equation}\label{Hitchin6D reduction: Tractor connection on G}
\bdgot\big|_{G} = \bdm_G.
\end{equation}
Therefore the Maurer-Cartan one-form on $G$ descends on $G/H$ as a principal $G$ connection. This connection is the `tractor connection on G/H'. 

\paragraph{General Cartan Geometry}\mbox{}

The generalisation to curved (non-flat) homogeneous space is now straightforward.\\

Let $M$ be a d-dimensional manifold. A Cartan geometry on $M$ modelled on $(G,H)$, where $G$ is a Lie group of dimension $n$ and $H$ a subgroup of dimension $n-d$, consists of the following data:

A principal $H$ bundle over $M$
\begin{equation}
H\inj P \to M,
\end{equation}
\noindent together with a $\frg$-valued one-form on $P$, that we will write 
 \begin{equation}
 \bdgo \in \gO^1\left(P, \frg \right)
 \end{equation}
and refer to as "the Cartan connection".

  What is more the Cartan connection must verify \eqref{Hitchin6D reduction: H Maurer-Cartan properties}, i.e
\begin{Definition}{Cartan Connection}\mbox{}\\
	A cartan connection on a principal bundle $H \inj P \to M$ is a $\frg$-valued one-form on $P$ such that
\begin{enumerate}
	\item $R^*_h \bdgo = Ad(h^{-1}) \;\bdgo \qquad \forall h\in H$
	\item $\bdgo\left(X_{\xi}\right) = \xi \qquad \forall \xi\in \mathfrak{h}$
	\item $\bdgo \from T_p P\to \frg$ is an isometry for any $p \in P$
\end{enumerate}
\end{Definition}
The two first properties are familiar. This is the third property of Cartan connections that makes them differ from more usual Ehresmann connections. An Ehresmann connection is not an isometry on the tangent bundle and indeed its kernel gives the horizontal tangent vectors. On the contrary the role of a Cartan connection is to infinitesimally identify each tangent space of $P$ with the Lie algebra $\frg$.

We now want to interpret the Cartan connection from the base manifold point of view. The resulting object is the \emph{tractor connection}. 

Just as in the flat case, one can use the left action of $H$ on $G$ to construct a $G$-bundle on $M$,
\begin{equation}
G \inj P' \to M.
\end{equation}
Then $H \inj P \to G/H $ is naturally included in $G \inj P' \to G/H$.

Cartan connection on $P$ are uniquely identified with principal $G$-connection on $P'$ whose kernel is not tangent to $P$ in $P'$:
\begin{Theorem} "Cartan connection and Tractor connection" \mbox{}\\
	Let $P$ and $P'$ be principal $H$ and $G$ bundles, respectively, over a manifold M. Assume $\emph{dim G = dim P}$ and that $\varphi\from P \to P'$ is an $H$ bundle map. Then the correspondence
	\begin{IEEEeqnarray*}{lCr}
		\left.\begin{array}{l}
			\text{\emph{Principal $G$ connection on $P'$}}\\ \text{\emph{whose kernel do not meet $\varphi_*(T(P))$}}
		\end{array}\right\}
		& \xto{\varphi^*} &
		\begin{array}{l}
			\text{\emph{Cartan connections on P}}
		\end{array}
	\end{IEEEeqnarray*}
	is a bijection.
\end{Theorem}
\begin{proof}For a proof see \cite{Sharp} p365.
\end{proof}
 Thus, associated with each Cartan connection on $P$ there is a unique principal G-connection on $P'$, we will refer to this connection as the `Tractor connection'. This is a `connection' in the usual sense (a \emph{Ehresmann} connection) and descends on $M$ as usual.

\subsubsection{Curvature}
One can now define the curvature of a Cartan connection as \begin{equation}
\bdgO = d\bdgo + \frac{1}{2} [\bdgo,\bdgo]
\end{equation} or equivalently as the curvature of the tractor connection. The two being related by restriction from $P'$ to $P$. The generalised version of the fundamental theorem of calculus asserts that when the curvature vanish, $P$ can locally be identified with $G$. Then, as we expect the conditions \eqref{Hitchin6D reduction: H Maurer-Cartan properties} are enough to locally identify $M$ with $G/H$.
\begin{Theorem}
	Let $M$ be a flat effective Cartan geometry modeled on $(G,H)$. Then each point of $M$ has a neighbourhood U which is canonically isomorphic as a Cartan geometry to an open subset of the model geometry $G/H$.
\end{Theorem}
\begin{proof}
Cf \cite{Sharp} p212.
\end{proof}
 Here "effective" means that $(G,H)$ must be such that $H$ contains no normal subgroup in G. Starting with any homogeneous space $G/H$, one can always make the geometry effective. Let $K$ be the largest subgroup of $H$ that is normal in $G$. Then $\left(G/K , H/K\right)$ is effective and $\left(G/K\right)/\left(H/K\right) \simeq G/H$. Cf \cite{Sharp} p151.

\subsubsection{A canonical example: 3d gravity}

Gravity in three dimensions has no propagating degrees of freedom and Einstein's equations are just the statement that locally $M^3$ is an homogeneous space. This is exactly what Cartan connections are good for! Depending on the sign of the cosmological constant all we need to do is find a Cartan connection associated with the model spaces $\left(SU(2)\times SU(2)\right)/SU(2)$ or $SL(2,\C)/SU(2)$ then 3d gravity is just the vanishing of its curvature. In fact this just lead to a reinterpretation of the "Chern-Simons" connections. In some sense it explains why this is such a fruitful point of view: this is really not a trick, this is something deep about the geometry of 3d gravity.

If $\bde$ is a frame on $M^3$ (here thought as an $\su(2)$-valued one-form) and $\bdw$ the potential of a $SU(2)$ connection on $M^3$ then let us introduce the following Tractor connection:
 \begin{equation}
\Ac = \left(\bda ,\bdat\right) = \left(\bdw + \sqrt{\gL} \bde \;,\; \bdw - \sqrt{\gL} \bde \right).
\end{equation}
This gives a Cartan geometry modelled on $\left(SU(2)\times SU(2)\;,\; SU(2)\right)$ or, depending on the sign of the cosmological constant, $\left(SL(2,\C) \;,\; SU(2)\right)$. Note that what makes this object different from a usual $SU(2)\times SU(2)$ (or $SL(2,\C)$ ) principal connection is that the gauge transformations are only $SU(2)$ gauge transforms. As discussed above this is the interplay between $G$-valued objects and $H$-actions that gives Cartan connections their particular flavour. In particular this implies that it can be lifted to the $\SU(2)$ principal bundle to obtain the Cartan connection form on $P^6$:
\begin{equation}
\bdgo = \left(\bdA ,\bdAt\right) = \left(\bdW + \sqrt{\gL} \bdE \;,\; \bdW - \sqrt{\gL} \bdE \right)
\end{equation}

Now, theorems from the preceding sections asserts that the field equations of 3d gravity are just the vanishing of the curvature of this Cartan connection,
\begin{equation}
\bdgO = \left(\bdF , \bdFt\right) =0
\end{equation}
Of course, this is in line with what we know from the Chern-Simons formulation of 3D gravity.

\subsection{3D Gravity as SU(2) Reduction of 6D Hitchin Theory}

We here would like to show that having the Cartan point of view on geometry in mind drastically simplifies the proof given in \cite{Herfray:2016std} that the general $SU(2)$ reduction of Hitchin theory is 3D gravity.

\subsubsection{6d Hitchin theory}

As we have already seen in section \ref{section: Stable form in 6d}, in six dimensions a general complex stable three-form $C \in \gG_{\C}^3(P^6)$ is equivalent to two independent triples of complex-valued one-forms $\left(A^i , \At^i \right)_{i\in 1,2,3}$ defined up to $SL(3,\C) \times SL(3,\C)$ gauge transformations, where each $SL(3,\C)$ transform acts on one of the two triples only. These two triples have to be independent in the sense that altogether the six one-forms form a basis. Then $C$ can be parametrized as
\begin{equation}
C =\frac{ \eps^{ijk}}{6}\left(A^i\W A^j \W A^k + \At^i \W \At^j \W \At^k \right).
\end{equation}

When $C$ is taken to be real, there are two distinct cases. In the `positive orbit' case, both $A$ and $\At$ are taken to be real and defined up to $SL(3,\R) \times SL(3,\R)$ transformations. In the `negative orbit' case, $\At$ is taken to be the complex conjugate of $A$ and both are defined up to the obvious $SL(3,\C)$ action.

\subsubsection{SU(2) reduction of 6d Hitchin field equations}
We now consider the case where we have a free $SU(2)$ action on $P^6$. We note $M^3$ the 3d quotient manifold $M = P^6/SU(2)$. In particular the infinitesimal version of this action gives us an identification of the Lie algebra $\su(2)$ with vertical right invariant vector fields:
\begin{IEEEeqnarray}{llll}
	R_* &: \su(2) & \to & \gG\left[TP^6\right].
\end{IEEEeqnarray}
Taking a canonical basis of $SU(2)$ $\left(\gs_i\right)_{i\in 1,2,3}$ such that $\left[\gs_i , \gs_j\right] = \eps_{ijk} \gs_k$ allows us to define a vertical basis of right-invariant vector fields:
\begin{equation}
X_i = R_* \left(\gs_i \right).
\end{equation}

As the basis $\left(\gs_i\right)_{i\in 1,2,3}$ is defined up to $Ad_{SU(2)}$ action only, $\left(X_i\right)_{i\in 1,2,3}$ is only defined up to a \emph{global} $SU(2)$ action.

We now fix the $SL(3,\C) \times SL(3,\C)$ freedom in the definition of the $A$'s and $\At$'s by requiring
\begin{equation}\label{Hitchin6D reduction: A gauge fixing}
A^i\left(X_j\right) \propto \gd^i{}_j, \qquad  \At^i\left(X_j\right) \propto \gd^i{}_j.
\end{equation}

This dramatically reduces the local $SL(3,\C) \times SL(3,\C)$ freedom to a \emph{global} $SU(2)$ action.
It is then convenient to rescale $A$'s and $\At$'s such that the preceding equations are equalities. To keep track of the remaining degrees of freedoms we introduce two complex scalars $\ga$ and $\gat$. Then 
\begin{equation}
C =  \frac{ \eps^{ijk}}{6}\left( \; \ga \;A^i\W A^j \W A^k + \;\gat\;\At^i \W \At^j \W \At^k \right).
\end{equation}

Once this gauge fixing is done this is legitimate to think of $A$ and $\At$ as lie algebra valued forms. For concreteness
\begin{equation}
\bdA = A^i \gs^i,\qquad \bdAt = \At^i \gs^i.
\end{equation}
Then
\begin{equation}
C = -2\frac{\Tr}{3}\left( \ga\; \bdA \W \bdA \W \bdA  + \gat\; \bdAt \W \bdAt \W \bdAt \right)
\end{equation}

With this new parametrisation the Hitchin field equations read,
\begin{align}\label{Hitchin6D reduction: Hitchin field equations}
\Tr\left( d\bdA \W \bdA \W \bdA  +\frac{d \ga}{3\ga} \W \bdA \W \bdA \W \bdA \right)&= 0,\\ \Tr\left( d\bdAt \W \bdAt \W \bdAt  +\frac{d \gat}{3\gat} \W \bdAt \W \bdAt \W \bdAt\right)&= 0.
\end{align}

Up to now we only used the principal bundle structure of $P^6$ to parametrize a \emph{general} stable three-form.

 One now restrict to invariant three-forms. The invariance of $C$ under the $SU(2)$ action implies that $A$ and $\At$ have to be equivariant \begin{equation}
 R^*\bdA = Ad\left(\bdA\right).
 \end{equation}
  This is because $\bdA$ can be completely defined in term of its kernel and its action on the $X$'s. Now, on the one hand, the action on $X$'s is fixed by the condition \eqref{Hitchin6D reduction: A gauge fixing}, what is more this is an SU(2)-equivariant condition. On the other hand if $C$ is $\SU(2)$ invariant, the kernel of the $A$'s has to be $SU(2)$ invariant because it can be defined as eingenspaces of $J_C$ which is. Finally requiring $SU(2)$ invariance of $C$ implies that $\ga$ and $\gat$ are constant along the fibres (and thus scalars functions on M). 

One cannot get any further in the complex case, however in the real cases we have a nice geometrical interpretation. In the real cases with positive type, we can now indeed think of $\bdgo = \left( \bdA , \bdAt \right)$ as a $\SU(2)$-equivariant $\su(2) \oplus \su(2) $-valued one-form. This makes
 \begin{equation}\label{Hitchin6D reduction: 3D cartan geometry}
\left(M^3 = P^6/SU(2), \bdgo \right)
\end{equation}
 a Cartan geometry modelled on $\left( \SU(2)\times \SU(2), \SU(2) \right)$. In the negative case  $\bdA$ is an $\SU(2)$-equivariant $\sl(2,\C)$-valued one-form and  \eqref{Hitchin6D reduction: 3D cartan geometry} is a Cartan geometry modelled on $ \left(\SL(2,\C) , \SU(2)\right)$.

The field equations \eqref{Hitchin6D reduction: Hitchin field equations} then implies that these Cartan geometries are flat,
\begin{equation}
\bdgO = \left(d\bdA + \frac{1}{2}\left[\bdA,\bdA\right] \;,\; d\bdAt + \frac{1}{2}\left[\bdAt,\bdAt\right] \right) =0 .
\end{equation}

This is because, being curvature forms, each of the two-forms in this bracket have to be basic (i.e they vanish when evaluated on any vertical tangent vector) and are thus zero if and only if they vanish on $Ker(A)$ or $Ker(\At)$. Evaluating \eqref{Hitchin6D reduction: Hitchin field equations} on any two vectors in $Ker(A)$ or $Ker(\At)$ then allows to conclude that $\gO=0$.

By construction, the curvature of a Cartan connection is the obstruction to M being locally isomorphic to one of the homogeneous model space \begin{equation}
S^3 = \left( SU(2)\times SU(2)\right) / SU(2) \qquad  or \qquad H^3 = SL(2,\C)/ SU(2)
\end{equation}
These are thus equations for 3d gravity.

On top of this flatness conditions we have field equations for $\ga$ and $\gat$:
\begin{equation}\label{Hitchin6D reduction: ga field equation}
d \ga \W\left(\frac{\eps^{abc}}{6}\; A^a\W A^b \W A^c\right)= 0, \qquad \qquad  d \gat \W \left( \frac{\eps^{abc}}{6}\; \At^a\W \At^b \W \At^c\right)= 0.
\end{equation}
Because $C$ is $SU(2)$ invariant $\ga$ and $\gat$ must be constant along the fibre, i.e $X_i \id d\ga =0$, $X_i \id d\gat=0$ for all $i \in 1,2,3$. Equation \eqref{Hitchin6D reduction: ga field equation} then implies that $\ga$ and $\gat$ are constants on $P^6$.

\subsubsection{SU(2) reduction of Hitchin Theory}

We just saw how the $\SU(2)$ reduction of a three-form $C$ together with the field equations $d C=0$, $d\Ch=0$ is just 3d-gravity. We now want to consider the associated variational principle. We again consider a $\SU(2)$ invariant three-form
\begin{equation} 
C =  -2\frac{\Tr}{3}\left( \; \ga \;\bdA \W \bdA \W \bdA + \;\gat\;\bdAt \W \bdAt \W \bdAt \right).
\end{equation}
As we just explained, $SU(2)$ invariance of $C$ implies that $\ga$ and $\gat$ are constant along the fibre $X_i \id d\ga =0$, $X_i \id d\gat=0$, i.e functions on $M$. In the positive orbit case they are two real functions, in the negative case they are conjugated complex functions. On the other hand $\SU(2)$ invariance allows us to think of $\bdgo = \left(\bdA, \bdAt\right)$ as a Cartan connection modelled on $\left(\SU(2)\times \SU(2),\SU(2)\right)$ or $\left(\SL(2,\C),\SU(2)\right)$ depending on the sign of the orbit.
We denote by $\gL$ the sign of the orbit and by convention $\sqrt{-1}=i$.

We then parametrize this Cartan connection as
\begin{equation}
\bdA= \bdW + \sqrt{\gL} \bdE,\qquad \bdAt= \bdW - \sqrt{\gL}\bdE.
\end{equation}
where $\bdE$ is an equivariant $\su(2)$-valued one-form on $P^6$ and therefore can be understood as a frame field on $M^3$,
\begin{equation}
\bdE = g^{-1} \bde g \in \gO^1\left(P^6 , \su(2)\right)
\end{equation}
and $\bdW$ is a connection one-form for the $\SU(2)$ principal bundle $P^6$. We denote by $\bdw$ the associated potential on $M^3$,
\begin{equation}
\bdW = g^{-1}dg + g^{-1}\bdw g.
\end{equation}
This is the natural parametrisation for 3D gravity, see section \ref{section: 3D Gravity in Terms of 3-Forms}.

Hitchin (`background invariant') theory consists of the following steps: first we impose closeness of $C$ and then we compute the Hitchin functional on the resulting three-form.

\paragraph{Closing $C$}\mbox{}

The above variables are the most natural. With this parametrisation, this is however not straightforward to interpret the closeness of $C$. Rather, it turns out to be useful to repackage $\ga$ and $\gat$ and parametrise them in terms of two scalar fields $k$ and $\gr$:
\begin{equation}
\ga = \frac{k}{2} \left(1 + \sqrt{\gL}\gr \right), \qquad \gat = \frac{k}{2} \left(1 - \sqrt{\gL}\gr \right)
\end{equation}
and shift $\bdW$
\begin{equation}\label{Hitchin6D reduction: W' def}
\bdW' = \bdW + \gr \sqrt{\gL}\bdE.
\end{equation}
So that the Cartan connection now reads
\begin{equation}
\bdA= \bdW' + \sqrt{\gL}\left(1 -\gr \right) \bdE,\qquad \bdAt= \bdW' - \sqrt{\gL}\left(1 +\gr\right)\bdE.
\end{equation}
This is convenient as $C$ now takes the following form
\begin{equation}\label{Hitchin6D reduction: C EW form}
C = -2\;k \; \Tr\;\left( \frac{1}{3}\bdW'\W \bdW' \W \bdW' +\gL \bdW' \W \bdE \W \bdE + \gr \sqrt{\gL}\left(1- 3\gL\right) \frac{1}{3}\bdE \W \bdE \W \bdE \right).
\end{equation}
In order to see why this is a useful parametrisation one should compare equation \eqref{Hitchin6D reduction: C EW form} with the situation in pure 3D gravity, see \eqref{3D Gravity in Terms of 3-Forms: gO matrix form EW}. The only difference is in the basic, $\Tr\bdE^3$, term and the overall scaling $k$.

Accordingly, by going from the the pure 3D gravity case discussed in section \ref{section: 3D Gravity in Terms of 3-Forms} to a general $SU(2)$ reduction of a three-form, two scalar degrees of freedom appear: the first, $k$, as an overall scalar factor and second $\gr$ as the coordinates of a basic three-form $\Tr\bdE^3$.

This point of view is useful to interpret the closure condition $dC=0$. A bit of though or a direct calculation indeed shows that the last term is always closed. This is because $k$, $\gr$ are supposed to be constant along the $\SU(2)$ fibres and that there are no basic four-forms.

Thus asking for $dC=0$ amounts to imposing the closeness of the first part of the three-form. We however already did this computation in the pure 3D gravity case, see \eqref{3D Gravity in Terms of 3-Forms: dgO}:
\begin{IEEEeqnarray}{l}
dC =  \\ \nonumber \\ \frac{dk}{k} C + k\; \Tr\left( -2\; \bdW' \W \bdW' \W \left(\bdF'_{\bdW'} +\gL \; \bdE \W \bdE \right) - \gL\; \bdW' \W \left(d_{\bdW'} \left[\bdE \W \bdE \right]\right) \right)\nonumber
\end{IEEEeqnarray}
From the above, one easily reads off the constraints as
\begin{equation}\label{Hitchin6D reduction: dC constraint}
dC=0 \qquad \Leftrightarrow \qquad k=cst,\quad \bdF'_{\bdW'} +\frac{\gL}{2}\; \left[\bdE \W \bdE\right]=0.
\end{equation}
Note that, here again, the third term in the expression above vanishes identically when the second term does.
 
The meaning of the closeness of $C$ is now straightforward: as explained in section \ref{ssection: The pure connection formulation (3D)} we can solve $\bdE$ in terms of $\bdW'$ and just as in section \ref{ssection: Hitchin Functional and the Chern-Simons three-form} we get a theory of a $\SU(2)$ connections. As opposed to the situation in the pure gravity case there is however a remaining real scalar field $\gr$. 

Making use of the constraints \eqref{Hitchin6D reduction: dC constraint} we can rewrite \eqref{Hitchin6D reduction: C EW form} as
\begin{equation}\label{Hitchin6D reduction: Chern Simon form}
C = k\left( - CS\left(\bdW'\right) -2 \gr \sqrt{\gL}\left(1- 3\gL\right) \frac{\Tr}{3} \bdE \W \bdE \W \bdE \right)
\end{equation}

It is instructive to compare again with section \ref{ssection: Hitchin Functional and the Chern-Simons three-form} where only 3D gravity was involved.  See eq \eqref{3D Gravity in Terms of 3-Forms: chern simon three-form} as compared to eq \eqref{Hitchin6D reduction: Chern Simon form}. As we already stressed, the only difference is a term proportional to $\bdE^3$ both in\eqref{Hitchin6D reduction: C EW form} and \eqref{Hitchin6D reduction: Chern Simon form}. In effect, it implements a coupling of the $\SU(2)$ connection with a scalar field.

\paragraph{Hitchin Functional and The Pure Connection Action, again.}\mbox{}

We now turn to the Hitchin functional. 
\begin{equation}
\Phi\left[C\right]=  4\;\ga\; \gat \; \left(\sqrt{\gL}\right)^3\frac{1}{3}\Tr\left( \bdA\W \bdA \W \bdA \right)\W \frac{1}{3}\Tr\left( \bdAt\W \bdAt \W \bdAt \right)
\end{equation}
We essentially already did this computation in \eqref{3D Gravity in Terms of 3-Forms: v-omega}. The end result is
\begin{align} \label{Hitchin6D reduction: hitchin functionnal}
\Phi[C] &=-8\gL \;\ga \gat\;\left(-\frac{2}{3}\Tr (\bdW^3) \right) \left(-\frac{2}{3}\Tr(\bdE^3_{\bdf'}) \right) \\
 &= -8\gL\;\left(\frac{k^2}{4} \left(1- \gL\;\gr^2 \right)\right)   \; \left(-\frac{2}{3}\Tr (\bdm^3) \right) \left(-\frac{2}{3}\Tr(\bde^3_{\bdf'}) \right).
\end{align}
So that the Hitchin functional is
\begin{align}
S\left[\bdw' , \gr \right] &= \int_{P^6} \Phi\left[C \right]\\
&= 2k^2 \left(\int_{\SU(2)} -\frac{2}{3} \Tr (\bdm^3) \right) \left(- \gL\int_{M^3} \left(1- \gL\;\gr^2 \right) v_{\bdf'} \right) \nonumber
\end{align}

Varying this action with respect to $\gr$ gives $\gr =0$. Varying \eqref{Hitchin6D reduction: hitchin functionnal} with respect to $\bdw'$ then gives the missing equations for 3D gravity.

We conclude that the $\SU(2)$ reduction of Hitchin Theory is 3D gravity in the pure connection formulation coupled with a constant scalar field.

\end{PartII}

\begin{PartIII}
\part[\\Variations on Hitchin Theory in Seven Dimensions]{Variations on Hitchin Theory \\in Seven Dimensions}\label{Part: Variations on Hitchin Theory in Seven Dimensions}

\section*{Introduction to Part 3:\\ \hspace*{1cm}Hitchin Theory and Seven Dimensions}
\counterwithout{equation}{section}		\setcounter{equation}{0}
\counterwithout{thmcnter}{chapter}		\setcounter{thmcnter}{0}
\addstarredchapter{Introduction to Part 3} \markboth{}{Introduction to Part 3}

In the first part of this thesis, quaternion geometry and its consequences played a major role. First, because of the identification $\R^4 \simeq \Hbb$ and the isomorphism
\begin{equation}
\SO(4) \simeq \U(1,\Hbb) \times \U(1,\Hbb) 
\end{equation}
that was the starting point for chiral formulations of gravity. Second because of the twistor bundle $\Hbb \inj \Hbb^2 \to \HP^1 \simeq \S^4$,
realising the isomorphism
\begin{equation}
Conf(4) \simeq \SO(5,1) \simeq \PSL(2,\Hbb).
\end{equation}
In the second part, complex (respectively para-complex) structures played a central role as a negative (resp positive) stable three-form in six dimensions reduces the structure group to 
\begin{equation}
\SL(2,\C) \qquad \left(\text{resp}\quad \SL(3,\R) \times \SLt(3,\R)\right).
\end{equation}
Equivalently, it defines two three-dimensional distributions $D \oplus \Dt = TP^6$. Then reducing the theory by a $\SU(2)$ action, in essence, amounts to choosing a diagonal embedding of $\SU(2)$ into one of these structure group.

We now turn to the seven dimensional case and mimic this discussion to see what one can expect. In seven dimensions, the geometry of octonions $\Obb$ or rather imaginary $\IObb$ will play a pivotal role. As will be discussed in chapter \ref{Chapter: Hitchin Theory in Seven Dimensions} the stabiliser of a positive stable three-form in seven dimension is the exceptional group $G_2$ which is also the automorphism group of octonions. As it stabilises the identity element of octonions it indeed naturally acts on the imaginary part $\IObb \simeq R^7$. Therefore a global choice of such a stable forms allows to identify the tangent space to a seven dimensional manifold $\P^7$ to imaginary octonions
\begin{equation}
TP^7 \simeq \IObb.
\end{equation}
Starting with a stable three-form $C$, this can be done explicitly. One first constructs a 7D metric $\go_{C}$ from the three-form (see below for details) and then makes use of this metric to lift one of the indices of the three form. The resulting operator
\begin{equation}
C^a{}_{bc} \from TP^7 \times TP^7 \to TP^7
\end{equation}
then gives a product on $TP^7$ that can be thought of as a product on imaginary octonions.
In fact, this is only one of the two possible orbits for a three-form in seven dimensions. The other orbit has stabiliser $G_2'$ which is the automorphism group of split-octonions.

The essential reason why a stable three-form $C$ in 7D defines a metric $g_{\gO}$ is because $G_2$ (reps $G_2'$) is a subgroup of $\SO(7)$ (resp $\SO(3,4)$). Then critical points in Hitchin theory have the following metric interpretation: Three-forms that are solutions of the seven-dimensional Hitchin theory then give metric with holonomy $G_2$ (see below for a general discussion on holonomy in Riemannian geometry). 

This has the following important consequences. When considering a $\SU(2)$-principal bundle $\SU(2) \inj \P^7 \to \M^4$ together with a $\SU(2)$-invariant three-form we automatically obtain a $\SU(2)$-invariant metric. It follows that $\SU(2)$-invariant three-forms give a four dimensional metric together with a $\SU(2)$-connection. 

A natural questions is to wonder whether GR can be written in these terms. This turns out not to be so simple, at least we could not achieve it. However we found that in this context, a particular "Chiral deformations of gravity" (see section \ref{section: Chiral Deformations of Gravity}) appears naturally and is associated with $G_2$ holonomy metric in eight dimensions:
\begin{Proposition}{\cite{Herfray:2016azk}}\label{Proposition: detgravity to G2}
Let $\bdA$ be a connection on a $\SU(2)$-principal bundle over a four dimensional manifold $M^4$. Then solutions to the four dimensional theory 
\begin{equation}\label{Intro partIII: det theory}
S\left[\bdA \right] = \int \left(det F\W F \right)^{1/3}
\end{equation}
can be lifted to metric with $G_2$ holonomy on the $\SU(2)$ associated vector bundle $\R^3 \inj B \to M^4$. 
\end{Proposition}

As will be made more precise in chapter \ref{Chapter: Hitchin Theory in Seven Dimensions}, the above result is essentially an embedding of a certain chiral deformation of GR, defined by \eqref{Intro partIII: det theory} into Hitchin theory in seven dimensions. Note however that the seven dimensional manifold here is \emph{not} a principal bundle but rather some associated bundle.

The full dimensional reduction of the theory turns out not to be as interesting. The essential reason is that $\SU(2)$-principal bundle over self-dual Einstein manifold do not easily yield $G_2$-holonomy manifold. In particular there is no known $G_2$ holonomy metric on the seven sphere. Instead let us consider the following `variations', one can add a kinetic term of the form $C \W dC$ the potential term $\Phi\left[C\right]$ of Hitchin theory.
\begin{equation}\label{IntroIII: kirill action}
S\left[C\right] = \int_{P^7} C \W dC - 3\Phi\left[C\right].
\end{equation}
 Note that is \emph{not} what we referred previously as `background independent Hitchin theory' but rather some theory that only is possible in this peculiar dimension. This action was first discussed in \cite{Krasnov:2016wvc,Krasnov:2017uam}. The reason why this is interesting is that field equations give `nearly parallel $G_2$ structures' and that $\SU(2)$ principal bundle over self-dual Einstein manifold have these in abundance. In particular there are nearly parallel $G_2$ structure on the seven sphere. Nearly $G_2$ structure are weaker version of manifold with constant holonomy, (they are also called weak holonomy $G_2$ in some reference). Again, see below for a general overview on holonomy in Riemannian geometry.  
 
 Surprisingly its $\SU(2)$ reduction is of $BF$ plus potential type. It however does not yield GR either but rather some tensor-scalar theory involving another `chiral deformations of gravity'. Interestingly, however, there is a regime around self-dual Einstein solutions where GR appears as an approximative case see \cite{Krasnov:2016wvc,Krasnov:2017uam}.  While the author of this thesis was not directly involved in those works, some of the aspects will be reviewed here, in order to present as complete a picture as possible. Most details, however, will be left aside and the interested reader should consult \cite{Krasnov:2016wvc,Krasnov:2017uam}.

This last part of the thesis is organised as follows. In the first chapter, we review some elementary geometry of (split-)octonions. With this in hand, we describe Hitchin theory of three-forms in seven dimensions. Finally, we prove Proposition \ref{Proposition: detgravity to G2}.

In the second chapter, we briefly describe the action for nearly $G_2$ manifold \eqref{IntroIII: kirill action}. We then take some time to give an overview of holonomy in Riemannian geometry which realtes different object that we encounter up to now: nearly Kahler, nearly parallel $G_2$ structure etc.
Finally we consider the dimensional reduction of the action \eqref{IntroIII: kirill action}.

\counterwithin{equation}{section}		\setcounter{equation}{0}
\counterwithin{thmcnter}{chapter}		\setcounter{thmcnter}{0}

\parttoc

\chapter{Hitchin Theory in Seven Dimensions}\label{Chapter: Hitchin Theory in Seven Dimensions}

In this chapter we first review the geometry of stable three-form in seven dimensions: A globally defined stable positive three-form in seven dimensions gives a $G_2$ structure. Just like almost complex structure identified the tangent space of a real, $2n$-dimensional, manifold with $\C^n$, a $G_2$ structure identifies the tangent space of seven dimensional manifold with the imaginary octonions $\IObb$.
The parallel of integrable complex structure are then $G_2$ holonomy metric. Before discussing stable forms in seven dimensions we therefore briefly review the geometry of Octonions. 

Finally we describe the seven dimensional interpretation of a certain `chiral deformation of GR' as $G_2$ holonomy metric which is the main result of this chapter.

\section{Octonions Geometry}\label{section: Octonions Geometry}

\subsection{Octonions}

We take octonions $\Obb$ to be the algebra defined by the basis $\left\{1,e_1,e_2,e_3,e_4,e_5,e_6,e_7 \right\}$ and the multiplication table \ref{Octonions Geometry: Table: Octonion table}. 

\begin{figure}

	\centering
	
	\arrayrulecolor{white}
	\arrayrulewidth=1pt
	\begin{tabular}{| >{\columncolor{blue!40!white}}c| >{\columncolor{blue!20!white}}c| >{\columncolor{blue!20!white}}c| >{\columncolor{blue!20!white}}c| >{\columncolor{blue!20!white}}c| >{\columncolor{blue!20!white}}c| >{\columncolor{blue!20!white}}c|
	>{\columncolor{blue!20!white}}c| >{\columncolor{blue!20!white}}c|		}
		\rowcolor{blue!40!white}
	$\times$ &   1 	 & $e_1$ & $e_2$& $e_3 $ & $e_4$ & $e_5$ & $e_6$ & $e_7$\\
		\hline
		  1  &   1	 & $e_1$ & $e_2$ & $e_3$ & $e_4$ & $e_5$ & $e_6$ & $e_7$ \\
		$e_1$& $e_1$ &   -1  & $e_3$ & -$e_2$& $e_5$ & -$e_4$& -$e_7$& $e_6$ \\
		$e_2$& $e_2$ & -$e_3$&   -1  & $e_1$ & $e_6$ & $e_7$ & -$e_4$& -$e_5$\\
		$e_3$& $e_3$ & $e_2$ & -$e_1$&   -1  & $e_7$ & -$e_6$& $e_5$ & -$e_4$\\
		$e_4$& $e_4$ & -$e_5$& -$e_6$& -$e_7$&   -1  & $e_1$ & $e_2$ & $e_3$ \\
		$e_5$& $e_5$ & $e_4$ & -$e_7$& $e_6$ & -$e_1$&   -1  & -$e_3$& $e_2$ \\
		$e_6$& $e_6$ & $e_7$ & $e_4$ & -$e_5$& -$e_2$& $e_3$ &   -1  & -$e_1$\\
		$e_7$& -$e_7$& -$e_6$& $e_5$ & $e_4$ & -$e_3$& -$e_2$& $e_1$ &   -1  \\
			\end{tabular}

	\captionof{table}[]{Multiplication Rules for Octonions. (
		 This table reads from left to right. e.g: $e_1 e_2 = e_3$.)}
	\label{Octonions Geometry: Table: Octonion table}

\vspace{1cm}

	\includegraphics[width=0.5\textwidth]{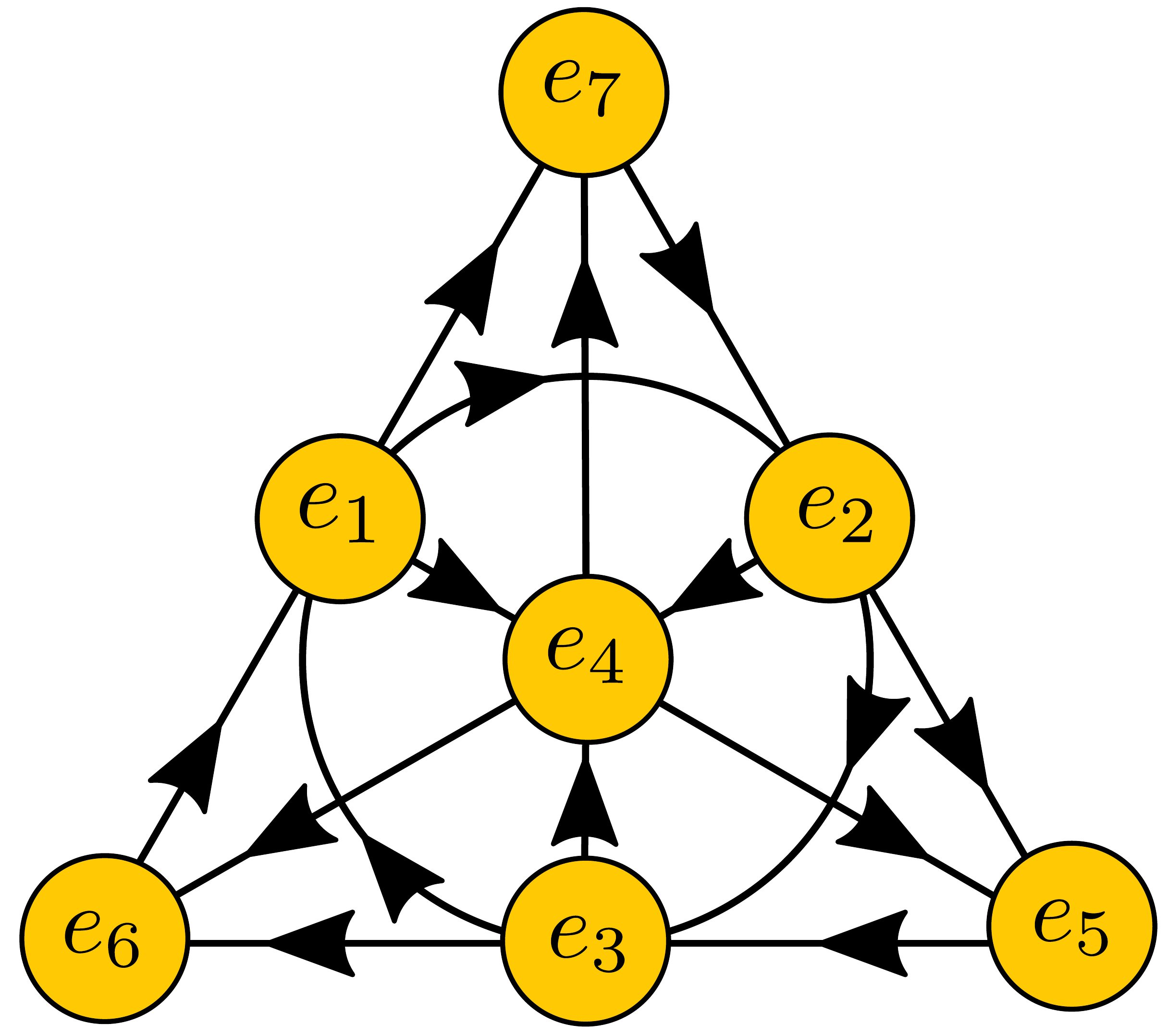}
	
	\captionof{figure}{A mnemonic for the multiplication of unit octonions: The Fano plane. (Each of the lines or circle of the plane form an associative sub-algebra e.g $e_1 (e_4 e_5)= (e_1 e_4) e_5  -1$.)}
	\label{Octonions Geometry: Fig: FanoPlane}

\end{figure}

For $X \in \Obb$ we will write,
\begin{equation}
X= \sum_{0}^{7} \;X^i e_i, \qquad \text{with}\; e_0 \coloneqq 1.
\end{equation}
The identity element of octonions plays somewhat of a special role and it will be useful to consider \emph{imaginary octonions} that is to say octonions $X\in \IObb$ such that $X^0 =0$.

In particular, the multiplication table \ref{Octonions Geometry: Table: Octonion table} restricted to $\IObb$ is skew-symmetric but for the diagonal part : elements of the basis $\{e_1 ... e_7\}$ anti-commute with each others.

Octonions are not associative which makes them somewhat unintuitive e.g
\begin{equation}
\left(e_4 e_1\right) e_2 = -e_5 e_ 2 = e_7 \quad\text{but}\qquad  e_4\left(e_1 e_2\right)=  e_ 4 e_3 =  -e_7.
\end{equation}
 However they are \emph{alternative}, i.e any sub-algebra generated by two elements is associative. Using this fact together with the `Fano plane'  (see figure \ref{Octonions Geometry: Fig: FanoPlane}) makes it easier to deal with the multiplication rule. Each line or circle on the Fano plane defines a sub-algebra generated by two element.
 
  E.g
 \begin{equation}
 e_2 e_4 = e_6, 
 \end{equation}
 By alternativity, each of these sub-algebras is associative. Then, making use of the anti-commutativity of unit quaternions and the fact that they each square to $-1$ one easily gets all the other multiplication rules.
 
  E.g
 \begin{equation}
 e_4 e_6 = e_2,\qquad e_6 e_2 = e_4, \qquad \text{etc}
 \end{equation}

 Elements of the basis $\{e_1 ... e_7\}$ all square to minus one, however if $i\neq j$ then $e_i e_j$ is never proportional to $1$. This allows to define a cross product on imaginary octonions:
 \begin{equation}\label{Octonions Geometry: Cross Product def}
 X\times Y \coloneqq\frac{1}{2}\left( X Y - Y X \right), \quad X,Y \in \IObb.
 \end{equation}
 
 Octonions also have a metric structure. Let $X \in \Obb$,
 \begin{equation}
 X = X^0 \;1 + X^i e_i\qquad i \in \{1,...,7\}.
 \end{equation}
 Define the conjugation operation
 \begin{equation}
 \overline{X} = X^0\; 1 - X^i e_i \qquad i \in \{1,...,7\}
 \end{equation}
 then
 \begin{equation}
 \bra X,Y\ket= \frac{1}{2}\left(\Xb Y + \Yb X \right)= \sum_{i = 0}^7 X^i Y^i
 \end{equation}
 
The metric and the cross product are all that is needed to recover the product on imaginary octonions (and thus on octonions):
 \begin{equation}
 XY = \bra X, Y \ket + X\times Y,\qquad X,Y \in \IObb.
 \end{equation}

Now let $i$ and $j$ take different values. This is convenient to introduce the tensor notation:
 \begin{equation}
e_i e_ j \coloneqq \times_{kij} \;e_ k .
\end{equation}

Multiplying on both side by $e_k$ we obtain,
\begin{equation}
\left(e_i e_j\right) e_k = -\times_{kij}
\end{equation}
By alternativity we can get rid of the parenthesis and making use of anti-commutativity of unit quaternions, one sees that the tensor $\times$ is completely anti-symmetric $\times_{[ijk]} = \times_{ijk}$.

Consequently we can define the following three-form on $\IObb$:
\begin{equation}
\gO\left(X,Y,Z\right) \coloneqq \bra X \times Y , Z \ket = \times_{ijk}\;  X^i Y^j Z^k, \qquad X,Y,Z \in \IObb.
\end{equation}
Let $\{e^1, ... , e^7\}$ be a dual basis of $\{e_ 1, ... , e_7\}$, then
\begin{equation}
\gO = \times_{ijk}  \; \frac{e^i \W e^j \W e^k}{6}.
\end{equation}
In what follows we will need its explicit expression:
\begin{equation}\label{Octonions Geometry: gO def}
\gO = e^1 \W e^2 \W e^3 +  e^{1}\W \left( e^4 \W e^5 - e^6 \W e^7  \right) + e^{2}\W \left( e^4 \W e^6 - e^7 \W e^5  \right) + e^{3}\W \left( e^4 \W e^7 - e^5 \W e^6  \right)
\end{equation}

\subsection{Split-Octonion}

Similarly to Octonions, Split Octonions $\Obb'$ are defined by a basis $\left\{1,e_1,e_2,e_3,e_4,e_5,e_6,e_7 \right\}$ but with a different multiplication rule see table \ref{Octonions Geometry: Table: Split Octonion table}, see also figure \ref{Octonions Geometry: Fig: FanoPlane_Split} for the associated Fano plane.
\begin{figure}
	\centering
	
	\arrayrulecolor{white}
	\arrayrulewidth=1pt
	\begin{tabular}{| >{\columncolor{blue!40!white}}c| >{\columncolor{blue!20!white}}c| >{\columncolor{blue!20!white}}c| >{\columncolor{blue!20!white}}c| >{\columncolor{blue!20!white}}c| >{\columncolor{blue!20!white}}c| >{\columncolor{blue!20!white}}c|
			>{\columncolor{blue!20!white}}c| >{\columncolor{blue!20!white}}c|		}
		\rowcolor{blue!40!white}
		$\times$ &  $1$	 & $e_1$ & $e_2$& $e_3 $ & $e_4$ & $e_5$ & $e_6$ & $e_7$\\
		\hline
		$1$  & $1$   & $e_1$ & $e_2$ & $e_3$ & $e_4$ & $e_5$ & $e_6$ & $e_7$ \\
		$e_1$& $e_1$ &  -$1$ & $e_3$ & -$e_2$& $e_5$ & -$e_4$& -$e_7$& $e_6$ \\
		$e_2$& $e_2$ & -$e_3$&  -$1$ & $e_1$ & $e_6$ & $e_7$ & -$e_4$&-$e_5$ \\
		$e_3$& $e_3$ & $e_2$ & -$e_1$&  -$1$ & $e_7$ & -$e_6$& $e_5$ &-$e_4$ \\
		$e_4$& $e_4$ & -$e_5$& -$e_6$& -$e_7$&  $\RC 1$ & \RC-$e_1$ &\RC -$e_2$ &\RC -$e_3$ \\
		$e_5$& $e_5$ & $e_4$ & -$e_7$& $e_6$ & $\RC e_1$ &  $\RC 1$ & $\RC e_3$ & \RC -$e_2$ \\
		$e_6$& $e_6$ & $e_7$ & $e_4$ & -$e_5$& $\RC e_2$ &\RC-$e_3$ &  $ \RC 1$ & \RC $e_1$ \\
		$e_7$& -$e_7$& -$e_6$& $e_5$ & $e_4$ & $\RC e_3$ & \RC $e_2$ &\RC -$e_1$ &  $\RC 1$  \\
	\end{tabular}
	\captionof{table}{Split Octonion Multiplication Rules. The elements of the table that differ from the usual octonion multiplication rule from table \ref{Octonions Geometry: Table: Octonion table} have been highlighted in red.}
	\label{Octonions Geometry: Table: Split Octonion table}
	
	\vspace{1cm}
	
	\includegraphics[width=0.5\textwidth]{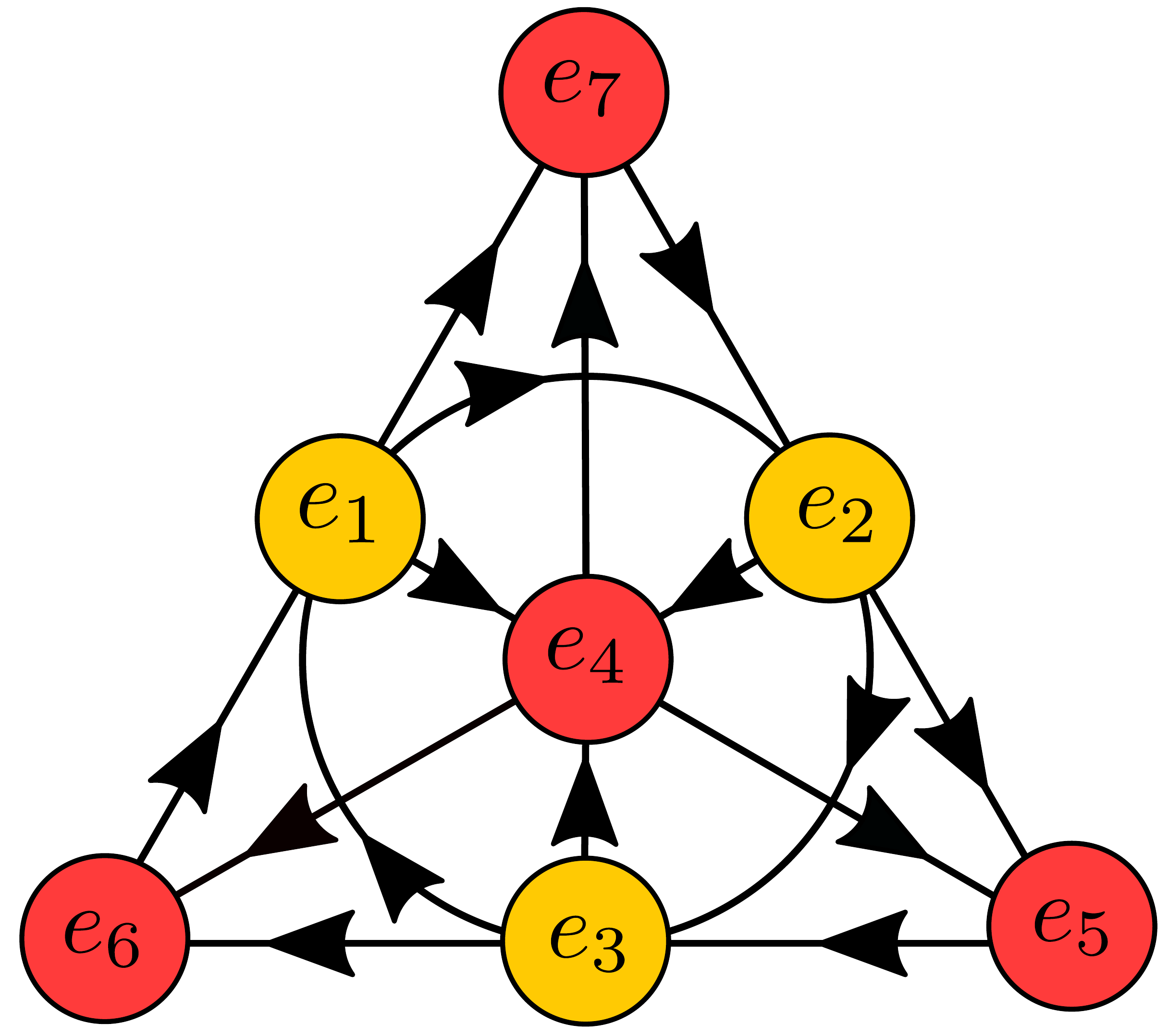}
	\captionof{figure}{A modified version of the Fano plane that encodes the multiplication of unit split-octonions: as compare to the standard octonions a minus sign should be added when multiplying two red quaternions.}
	\label{Octonions Geometry: Fig: FanoPlane_Split}
	
\end{figure}

An essential difference is the diagonal part of the table: half of the unit split-octonions square to minus one.  Consequently the metric on split-octonion has split signature $(4,4)$:
\begin{equation}
\bra X, Y \ket = \frac{1}{2}\left(\Xb Y+ \Yb X \right) = \sum_{i = 0}^3 X^i Y^i - \sum_{i = 4}^7 X^i Y^i.
\end{equation}
Just as for octonions, the different element of the basis $\{e_1 \ldots e_ 7\}$ anti-commute with each others and we can define a cross product on imaginary split octonions. What is more alternativity still holds which means that we can define a three-form on split-octonions just as we did for octonions. It however takes a different form
\begin{equation}\label{Octonions Geometry: gO' def}
\gO' = e^1 \W e^2 \W e^3 -  e^{1}\W \left( e^4 \W e^5 - e^6 \W e^7  \right) - e^{2}\W \left( e^4 \W e^6 - e^7 \W e^5  \right) - e^{3}\W \left( e^4 \W e^7 - e^5 \W e^6  \right)
\end{equation}
This reflects the fact that the multiplication rules are different as we now briefly discuss. 

 Just as for octonions, each line or circle on the Fano plane (figure \ref{Octonions Geometry: Fig: FanoPlane_Split}) gives a sub-algebra generated by two elements.

E.g
\begin{equation}
e_2 e_4 = e_6.
\end{equation}
Once again, by alternativity, each of these sub-algebras is associative.
The structure of the Fano plane for split-octonions is the same as for usual octonions which relates to the fact that split-octonions multiplication algebra only differ by signs from the multiplication rule for octonions.

The precise multiplication rules can be obtained for each sub-algebras by making use of anti-commutativity of unit octonions and the fact that they all square to $-1$ \emph{or} $1$. The presence of unit octonions that square to $1$ will lead to signs discrepancy:

E.g
\begin{equation}
e_4 e_6 = -e_2,\qquad e_6 e_2 = e_4, \qquad\text{etc}.
\end{equation}
This is encapsulated in the Fano plane of figure \ref{Octonions Geometry: Fig: FanoPlane_Split} by the presence of \emph{coloured} (red) node: any time one multiply two coloured (red) node one should add a minus sign to the usual multiplication rule. Note that the coloured octonions are just the one that square to one.

\subsection{The exceptional group $G_2$}

The exceptional group $G_2$ is a Lie group with dimension $14$. It is best thought as the automorphism group of octonions, i.e $\phi \in G_ 2$ if and only if $\phi \in End(\Obb)$ and is such that for all $X,Y  \in \Obb$
\begin{equation}
\phi\left(XY\right) = \phi\left(X\right) \phi \left(Y\right).
\end{equation}
In particular $\phi$ must stabilise the identity. Now because
\begin{equation}
XY = -\bra X,Y \ket  \;1 + X \times Y,
\end{equation}
$\phi$ has to preserve both the metric and the cross product. An immediate consequence is that $\phi$ leaves $\gO$ invariant:
\begin{equation}
\gO\left(\phi\left(X\right),\phi\left(Y\right),\phi\left(Z\right)\right) = \gO\left(X,Y,Z\right).
\end{equation}

Now it turns out that the converse is also true: Let $\gO$ be a three-form on a seven dimensional vector space such that it can be written in the form \eqref{Octonions Geometry: gO def}. Then $G_2$ can de defined as the subgroup of $GL(7)$ stabilising $\gO$. Cf \cite{Bryant:1987} for a proof. In fact, this has been known for more than a century. In particular Engel mentioned this property in 1900 as an elegant definition of $G_2$, see \cite{Agricola08G2History} for more on the history of the group $G_2$. Note that $G_2 \subset SO(7)$ as it preserves the octonions metric.

In fact the complex Lie algebra $\mathfrak{g}^{\C}_2$ has two real forms, the compact one $\mathfrak{g}_2$ generates the automorphism group of octonions, the other one called `split' $\mathfrak{g}_2'$ generates the automorphism group of split octonions $G_2'$, consequently it leaves invariant three forms of the type \eqref{Octonions Geometry: gO' def}. Note that $G_2' \subset \SO(3,4)$ as it preserves the split-metric on octonions.

\section{Geometry of Stable 3-Forms in Seven Dimensions}\label{section: Geometry of Stable 3-Forms in 7D}

\subsection{Stable 3-Forms in Seven Dimensions}

Following \cite{Hitchin:2000sk} we here apply the methods from \ref{section: Geometry of Stable Forms} to seven dimensions. Let $E$ be a seven dimensional vector space. A three-form $\gO\in \gL^3 E^*$ is called \emph{stable} if it lies in a open orbit under the action of $\GL(7)$.

For real three-forms, there are exactly two distinct open orbits of stable forms, each of which is related to one of the real form of $G_2^\C$. Let $\gs = \pm 1$ be the sign of the orbit. According to a theorem from \cite{Hitchin:2000sk}, for every such $\gO$ there exists a set $e^1,\ldots, e^7$ of one-forms such that $\gO$ can expanded in the following canonical form:
\begin{equation}\label{Geometry of Stable 3-Forms in 7D: three form canonical}
\gO = e^1\W e^2 \W e^3 +\gs e^1 \W \gSt^1
+ \gs e^2\W \gSt^2 + \gs  e^3 \W \gSt^3,
\end{equation}
where
\begin{equation}
\gSt^1 = e^{4}\W e^{5} - e^{6}\W e^{7}, \quad \gSt^2 = e^{4}\W e^{6} - e^{7}\W e^{5}, \quad \gSt^3 = e^{4}\W e^{7} - e^{5}\W e^{6}.
\end{equation}
Note that this is the same form as in \eqref{Octonions Geometry: gO def},\eqref{Octonions Geometry: gO' def}. Here the particular combinations $\gSt$'s are the same anti-self-dual two-forms that appeared in four dimension, compare with the formula \eqref{Appdx: Sigma def (tetrad)} given in Appendix.  They are related to the embedding of $\SO(3)$ into $\SO(4)\subset G_2$. The relation to anti-self-dual two-forms in 4 dimensions will be central in the construction below. Note however that there is nothing deep about the appearance of anti-self-dual two-forms rather than their self-dual counter part for $e_4 \to -e_4$ sends $\gSt \to \gS$ (see eq \eqref{Appdx: Sigma def (tetrad)}) and that this corresponds to a $GL(7)$ action: consequently the three-form
\begin{equation}
\gO = e^1\W e^2 \W e^3 +\gs e^1 \W \gS^1
+ \gs e^2\W \gS^2 + \gs  e^3 \W \gS^3,
\end{equation}
is in the same orbit as \eqref{Geometry of Stable 3-Forms in 7D: three form canonical}. It just happens that the three-form \eqref{Geometry of Stable 3-Forms in 7D: three form canonical} appears more naturally from the `usual' algebra for Octonions, see table \ref{Octonions Geometry: Table: Octonion table}.

It also follows from this discussion that the space of \emph{positive} stable three-forms is the homogeneous group manifold $\GL(7)/G_2$.

One then generalise the notion of stable forms to three-forms on a 7-dimensional differentiable manifold $M$. Stable differential forms then are differential forms that are stable at every points.

Therefore a positive (resp negative) stable three-forms in seven dimensions effectively identify, at each points $x\in M$, the tangent space $T_x M$ with the imaginary (resp split) octonions $\IObb$. This idea can be made even more precise by explicitly constructing a cross product on $T_x M$. This is done in two times: first we construct a metric $g_{\gO}$ from the three-form $\gO$, second we `raise' an indices of $\gO$.

\subsubsection{The metric and Hitchin volume form}

A basic fact about stable three-forms on a 7-dimensional manifold $M$ is that they naturally define a metric on $M$. For positive three-forms this metric is definite and of signature $\left(3,4\right)$ for negative ones. Of course none of this comes as a surprise if one has in mind the preceding discussion on octonions.  

This metric can in fact be explicitly constructed as follows. First one construct a volume-form-valued metric $\gti_{\gO} \in T^{0,2}M \times \gO^7(M)$ as
\begin{equation}\label{Geometry of Stable 3-Forms in 7D: 3form-metric}
\gti_\gO(\xi,\eta)= -\frac{1}{6}\gi_\xi \gO \W \gi_\eta \gO \W \gO \qquad \in \gO^7(M).
\end{equation}
Here $\gi_\xi$ denotes the operation of insertion of a vector into a form.

It is then a simple computation that, for a three-form taking the canonical form \eqref{Octonions Geometry: gO def} (i.e in the positive orbit) the arising metric is
\begin{equation}
\gti_\gO =  \left(\sum_{i=1}^{7} e^i \otimes e^i  \right) \otimes e^1\W \ldots \W e^7 \, ,
\end{equation}
and for the other orbit \eqref{Octonions Geometry: gO' def}
\begin{equation}
\gti_\gO =  \left(\sum_{i=1}^{3} e^i \otimes e^i - \sum_{i=4}^{7} e^i \otimes e^i \right) \otimes e^1\W \ldots \W e^7.
\end{equation}

Now, one can define a volume form, the Hitchin functionnal, out of this metric: Taking its determinant we obtain $det\left(\gti_{\gO}\right) \in \left(\gO^7(M)\right)^{9}$. All is left is then to take the ninth square root to obtain the Hitchin volume form:
\begin{equation}
\Psi_{\gO} \coloneqq \left( det\left(\gti_{\gO}\right)\right)^{1/9}.
\end{equation}
Note that, contrary to what happened in six dimensions, there is no sign ambiguity here. Accordingly, a stable three-form in seven dimension defines an orientation.

With this volume form in hand, one can define a proper metric $g_{\gO}$ as
\begin{equation}\label{Geometry of Stable 3-Forms in 7D: 3form-metric2}
 g_{\gO} \otimes \Psi_{\gO} \coloneqq \gti_{\gO}.
\end{equation}
By construction, this metric is such that its volume form coincide with the Hitchin volume. The two possible orbits for $\gO$ are then distinguished by the sign of $det\left(g_{\gO}\right)$.

The counting of components shows that three-forms contain more information than just that of a metric. Indeed, to specify a metric in 7 dimensions, we need $7\times 8/2=28$ numbers, while the dimension of the space of three-forms is $35$. Thus, there are $7$ more components in a three-form.  These correspond to components of a unit spinor. This is not really going to be a useful perspective for us here but see \cite{Witt:2009zz,Agricola:2014yma} for more details.

\subsubsection{The Hat Operator}

Having a metric in hand one can consider $*\gO$, the hodge dual of $\gO$. This is a stable four-form. Taking, $\gO$ of the form \eqref{Octonions Geometry: gO def}, it has the following form
\begin{equation}
*\gO = e^4 \W e^5 \W e^6 \W e^7 - \frac{1}{2}\eps^{ijk} e^i \W e^j \W \gSt^k \qquad i,j,k \in 1,2,3
\end{equation}
then
\begin{equation}
\Psi_{\gO} = \frac{1}{7} \gO \W *\gO = e^1 \W...\W e^7.
\end{equation}
Comparing with the relation \eqref{Geometry of Stable Forms: Psi = k/n gOh W gO} we get
\begin{equation}\label{Geometry of Stable 3-Forms in 7D: hat operator}
\gOh \coloneqq \frac{1}{3} *\gO
\end{equation}
i.e
\begin{equation}
\gd \Psi_{\gO} =\frac{1}{3} *\gO \W \gd \gO.
\end{equation}

\subsubsection{A more direct construction of the functional}

Given a stable three-form, we construct the metric and the corresponding Hitchin volume form as described above. Integrating this volume form over the manifold we get the functional
\begin{equation}
S[\gO] = \int_M \Psi_{\gO} \, .
\end{equation}
This functional can also be computed explicitly, without computing the metric, via the following construction. Let $\epst^{\ga_1\ldots \ga_7}$ be the canonical anti-symmetric tensor density that exists independently of any metric. Then construct
\begin{equation}\label{Geometry of Stable 3-Forms in 7D: object}
\gO_{\ga_1\gb_1\gc_1} \ldots \gO_{\ga_4\gb_4\gc_4}
\tilde{\gO}^{\ga_1\ldots\ga_4}\tilde{\gO}^{\gb_1\ldots\gb_4}\tilde{\gO}^{\gc_1\ldots\gc_4}
\, ,
\end{equation}
where
\begin{equation}
\tilde{\gO}^{\ga_1\ldots\ga_4} := \frac{1}{6}\epst^{\ga_1\ldots\ga_7}
\gO_{\ga_5\ga_6\ga_7}\, .
\end{equation}
Then the object \eqref{Geometry of Stable 3-Forms in 7D: object}) is of homogeneity degree $7$ in $\gO$ and has density weight $3$. Its cubic root then is the coordinate of a volume form. A direct computation shows that this volume form is a multiple of Hitchin's.

Perhaps surprisingly, the invariant \eqref{Geometry of Stable 3-Forms in 7D: object} has been known already to Engel in 1900, see \cite{Agricola08G2History}. This invariant gives a useful stability criterion: a form $\gO$ is stable iff \eqref{Geometry of Stable 3-Forms in 7D: object} is non-zero. Its sign gives another way to distinguish between the two $\GL(7)$-orbits for three-forms described above.

\subsection{Hitchin theory for three-forms in seven dimension}\label{ssection: Hitchin theory for three-forms in seven dimension}

\subsubsection{Hitchin Functional}

We already gave different ways of constructing the Hitchin functional for three-form in seven dimension:
\begin{equation}\label{Geometry of Stable 3-Forms in 7D: Hitchin Functionnal}
\Psi \left| \begin{array}{ccc}
\gO^3(M^7) & \to & \R \\
\gO & \mapsto & \int_{M} \left(det\left(\gti_{\gO}\right)\right)^{1/9}.
\end{array}\right.
\end{equation}
As explained above there is no choice involved in this construction, in particular stable three-forms define an orientation. 
 
\subsubsection{Critical Points}

As stated in theorem \ref{Theorem: Hithin Theory}, varying in a cohomology class, the critical points are three-forms that satisfy:
\begin{equation}
d\gO =0 \qquad d\gOh=0.
\end{equation}
Now because here $\gOh \propto *\gO$ this is equivalent to
\begin{equation}\label{Geometry of Stable 3-Forms in 7D: hitchin field eqs}
d\gO= 0 \qquad d*\gO=0.
\end{equation}
Therefore critical points of Hitchin theory in seven dimensions are three-forms that are closed and co-closed (for the metric they define).

\subsubsection{Holonomy reduction}

The fundamental result due to M.Fernandez and A.Gray, \cite{GrayFernandez1982}, states: Let $\gO\in \gO^3\left(M\right)$
be a three-form on a 7-manifold. Then $\gO$ is parallel with respect to the Levi-Civita
connection of $g_\gO$ iff $d\gO=0$ and $d{}^*\gO=0$. In other words, the
condition of $\gO$ being parallel with respect to the metric it defines is equivalent
to the conditions of $\gO$ being closed and co-closed, where co-closeness is again
with respect to the metric it defines.

As a result of the holonomy principle (see below and theorem \ref{Theorem: Holonomy principle} for more on holonomy on Riemannian manifolds) if a metric has a parallel positive (reps negative) stable three-form $\gO$ this implies that the holonomy group is included in the stabiliser of $\gO$. Consequently, the holonomy group has to be included in $G_2$ (reps $G_2'$). 

 Here we will not be concerned whether the holonomy group is all of $G_2$ or is just contained in it, and will simply refer to 7-manifolds $M$ with three-forms satisfying $d\gO=0$ and $d{}^*\gO=0$ as $G_2$-holonomy manifolds. Techniques for proving that the holonomy equals $G_2$ can be found in \cite{BryantSalamon:1989}.

Combining the above characterisation of critical points of Hitchin functionnal, Gray's result and this last remark with one get the following theorem:

\begin{Theorem}{Hitchin (\cite{Hitchin:2000sk})}\label{Theorem: 7D hitchin Theory}

Let $M$ be a closed 7-manifold with a metric with holonomy G2, with
defining three-form $\gO$. Then $\gO$ is a critical point of the functional \eqref{Geometry of Stable 3-Forms in 7D: Hitchin Functionnal} restricted to the
cohomology class $[\gO
] \in H^3(M,\R)$.
Conversely, if $\gO$ is a critical point on a cohomology class of a closed oriented 7-manifold M such that $\gO$ is everywhere positive, then $\gO$
defines on $M$ a metric with holonomy $G_2$.
\end{Theorem}

\section{$G_2$ holonomy manifold from `gravity' in 4D} \label{section: G2 holonomy from 4D}

The history of $G_2$-geometry is almost as old as that of the exceptional Lie group $G_2$ itself, see \cite{Agricola08G2History} for a nice exposition. For a long time the existence of metrics of $G_2$ holonomy was an open problem. Their existence was first proven in \cite{Bryant:1987}. This paper also gave a construction of the first explicit example. Several more examples, among them complete, were constructed in \cite{BryantSalamon:1989}. The first compact examples where obtained in \cite{Joyce96}. 
More local examples can be obtained by evolving 6-dimensional $\SU(3)$
structures, see \cite{Hitchin:2000sk}. These examples, as well as many other things, are reviewed in \cite{Salamon:2002}. Metrics of $G_2$ holonomy are of importance in physics as providing the internal geometries for compactification of M-theory down to 4 space-time dimensions, while preserving super-symmetry. A nice mathematical exposition of this aspect of $G_2$ geometry is given in \cite{Witt:2009zz}.

We here show a result of a different type. We demonstrate that solutions of certain 4D gravity theory can be lifted to $G_2$-holonomy metrics.
The gravity theory in question is \emph{not} General Relativity, but rather a certain other theory of the `Chiral deformation of GR' type ( see section \ref{ssection : Chiral Deformations of Gravity} for more on these deformations). The $G_2$-holonomy lift that we describe in the following indeed singles out one of them, and it is distinct from GR. We describe this theory in details below.

A suggestion as to the existence of a link between some theory in 7 dimensions (referred to as topological M-theory) and theories of gravity in lower dimensions was made in \cite{Vafa:2004te}. That paper reinterpreted the constructions \cite{BryantSalamon:1989} of 7D metrics of $G_2$ holonomy from constant curvature metrics in 3D and self-dual Einstein metrics in 4D as giving evidence (among other things) for the existence of such a link.
The construction given here is similar in spirit, but we present a much stronger evidence linking 4D and 7D structures. Thus, our construction lifts any solution of a certain 4D gravity theory with local degrees of freedom to a $G_2$ metric. The main difference with the previous examples is that the theory that one is able to lift to 7D
is no longer topological. We find this result to be interesting as it interprets the full-fledged 4D gravity as a subspace of solutions of a theory of differential forms in 7 dimensions so that very different type of theory are related.

\subsection{Bryant--Salamon construction}
\label{ssection: Bryant--Salamon construction}

We now review the construction of \cite{BryantSalamon:1989} using a notation compatible with ours.

\subsubsection{Ansatz}

Let $(M,g)$ be a self-dual Einstein 4-manifold, and let $\gSt^i$, $i=1,2,3$, be a basis of anti-self-dual two-forms of the form \eqref{Appdx: Sigma def (tetrad)}. They satisfy
\begin{equation}\label{G2 holonomy from 4D: sigma property}
\gSt^i \W \gSt^j = -2 \gd^{ij} e^0\W e^1 \W e^2 \W e^3
\end{equation}
 Let $\At^i$ be the anti-self-dual part of the Levi-Civita connection. This is the $\SO(3)$-connection
that satisfies 
\begin{equation}
	d_{\At} \gSt^i =0\, .
\end{equation}
The self-dual Einstein equations read
\begin{equation}\label{G2 holonomy from 4D: Instatons eqs}
\Ft^i = \frac{\gL}{3} \gSt^i
\end{equation}
Here the metric has scalar curvature $4\gL$, see Appendix \ref{Chapter : Appdx 4d Space-Time Conventions} for more on our conventions.

An arbitrary anti-self-dual two-form can be written as $\gSt(y) = \gSt^i y^i$, and so the $y^i$ are the fibre coordinates in the bundle of anti-self-dual two-forms over $M$. We make the following ansatz for the calibrating three-form:
\begin{equation}\label{G2 holonomy from 4D: BS ansatz}
	\gO = \frac{1}{6} \ga^3\eps^{ijk} d_{\At} y^i \W d_{\At} y^j \W d_{\At} y^k  +
	2\ga\gb^2  d_{\At} y^i \W \gSt^i\, ,
\end{equation}
where $d_{\At} y^i = d y^i + \eps^{ijk} \At^j y^k$ is the covariant derivative with respect to $\At$, and $\ga$ and $\gb$ are functions of $y^2 = y^i y^i$ only. In particular this implies
\begin{equation}
d \ga = 2\frac{\pa \ga}{\pa y^2} y^i d y^i =2\frac{\pa \ga}{\pa y^2} y^i d_{\At} y^i
\end{equation}

\subsubsection{Closing $\gO$}

We now require the form $\gO$ to be closed. When differentiating the first term, we only need to differentiate the quantities $d_A y^i$, as differentiating $\ga$ would lead to exterior products of four one-forms from the triple $\{ d_A y^i \}_{i \in 1,2,3}$, which are zero. In the second term, we do not need to apply the derivative to $\gS^i$ because it is covariantly closed. We also do not need to differentiate $d_A y^i$ since this produces a multiple of $\eps^{ijk} F^j y^k \W \gS^i$, which is equal to zero due to \eqref{G2 holonomy from 4D: Instatons eqs}. We thus get
\begin{equation}
	d\gO = \frac{1}{2} \ga^3 \eps^{ijk} \eps^{ilm} \Ft^l y^m \W d_{\At} y^j
	\W d_{\At} y^k + 2 \left( \ga\gb^2 \right)' \left( 2y^i d_{\At} y^i \right) \W
	\left( d_{\At} y^j \W \gSt^j \right)\, .
\end{equation}
We now use \eqref{G2 holonomy from 4D: Instatons eqs} and decompose the product of two epsilon tensors into products of
Kronecker deltas. We obtain
\begin{equation}
	d\gO = \left[ - \frac{\gL}{3} \ga^3 + 4 \left( \ga\gb^2 \right)' \right] \left( y^i
	d_{\At} y^i \right) \W \left( d_{\At} y^i \W \gS^j \right) \, .
\end{equation}
Thus, we must have
\begin{equation}\label{G2 holonomy from 4D: BS-eqn-1}
	4 \left( \ga\gb^2 \right)' = \frac{\gL}{3} \ga^3
\end{equation}
in order for the form to be closed.

\subsubsection{Canonical form}

We now compute the metric defined by $\gO$, as well as its Hodge dual. The easiest way to do this is to write the three-form in the canonical form, so that the metric and the dual form are immediately written. Thus, let $\tht^1, \ldots, \tht^7$ be a set of one-forms such that the three-form $\gO$ is
\begin{equation}\label{G2 holonomy from 4D: canonical form}
\gO = \tht^1\W \tht^2 \W \tht^3 + \tht^1 \W \gSt^1
+ \tht^2\W \gSt^2 + \tht^3 \W \gSt^3,
\end{equation}
where
\begin{equation}
\gSt^1 = \tht^{4}\W \tht^{5} - \tht^{6}\W \tht^{7}, \quad \gSt^2 = \tht^{4}\W \tht^{6} - \tht^{7}\W \tht^{5}, \quad \gSt^3 = \tht^{4}\W \tht^{7} - \tht^{5}\W \tht^{6}.
\end{equation}
Then the one-forms $e$ are an orthonormal frame for the metric determined by $\gO$
\begin{equation}
	g_\gO = \left( \tht^1 \right)^2 + \ldots + \left( \tht^7 \right)^2\, ,
\end{equation}
and the Hodge dual ${}^*\gO$ of $\gO$ is given by
\begin{equation}
*\gO = \tht^4 \W \tht^5 \W \tht^6 \W \tht^7 - \frac{1}{2}\eps^{ijk} \tht^i \W \tht^j \W \gSt^k \qquad i,j,k \in 1,2,3.
\end{equation}

\subsubsection{Calculation of the metric and the dual form}

We now put ansatz \eqref{G2 holonomy from 4D: BS ansatz} into the canonical form \eqref{G2 holonomy from 4D: canonical form}, and compute the associated metric and the dual form. The canonical frame is easily seen to be
\begin{equation}
	\tht^{i} = \ga d_A y^i, \qquad \tht^{4+I} = \gb \sqrt{2} e^I, \qquad I=1,2,3,4\, ,
\end{equation}
where $\{e^I\}_{ I\in 0,1,2,3}$ is an orthonormal frame on the base manifold. The metric is then
\begin{equation}
	g_\gO = \ga^2 \sum_{i=1}^3 \left( d_{\At} y^i \right)^2 + 2\gb^2 \sum_{I=0}^3 \left( e^I
	\right)^2\, ,
\end{equation}
and the dual form is
\begin{equation}\label{G2 holonomy from 4D: BS*}
	*\gO = -\frac{2}{3} \gb^4 \gSt^i \W \gSt^i - \gb^2 \ga^2
	\eps^{ijk}\; \gSt^i \W d_{\At} y^j \W d_{\At} y^k \, .
\end{equation}

\subsubsection{Co-closing $\gO$}

We now demand the 4-form \eqref{G2 holonomy from 4D: BS*} to be closed as well. The first point to note is that when we apply the covariant derivative to the factor $\gb^2 \ga^2$ in the second term, we generate a 5-form proportional to the volume form of the fibre. There is no such
term arising anywhere else, and we must demand
\begin{equation}\label{G2 holonomy from 4D: BS-eqn-2}
	\ga\gb = {\rm const }
\end{equation}
in order for \eqref{G2 holonomy from 4D: BS*} to be closed. Differentiation of the rest of the terms gives
\begin{equation}
	d*\gO = -\frac{2}{3} \left( \gb^4 \right)' \left( 2y^i d_{\At} y^i \right) \W
	\gSt^j\W \gSt^j - 2 \gb^2\ga^2 \eps^{ijk} \gSt^i \W \eps^{jlm}
	\Ft^l y^m \W d_{\At} y^k \, .
\end{equation}
We now use \eqref{G2 holonomy from 4D: sigma property} and \eqref{G2 holonomy from 4D: Instatons eqs} to get
\begin{equation}
	d*\gO = -\frac{2}{3} \left[ \left( \gb^4 \right)' - \frac{\gL}{3}\gb^2\ga^2 \right]
	\left( 2y^i d_{\At} y^i \right) \W \gSt^j \W \gSt^j \, ,
\end{equation}
and so we must have
\begin{equation}\label{G2 holonomy from 4D: BS-eqn-3}
	\left( \gb^4 \right)' = \frac{\gL}{3}\gb^2\ga^2\, .
\end{equation}

\subsubsection{Determining $\ga$ and $\gb$}

The overdetermined system of equations \eqref{G2 holonomy from 4D: BS-eqn-1}, \eqref{G2 holonomy from 4D: BS-eqn-2} and
\eqref{G2 holonomy from 4D: BS-eqn-3} is nevertheless compatible. Without loss of generality, we can simplify
things and rescale $y^i$ (and therefore $\ga$) so that
\begin{equation}
	\ga\gb =1 \, .
\end{equation}
With this choice, we have only one remaining equation to solve, which gives
\begin{equation}
	\gb^4 = k + \frac{\gL}{3} y^2,
\end{equation}
where $k$ is an integration constant. We can rescale the base metric to get $\gL/3 = \gs = \pm 1$. We can then further rescale $y$ and $\gb$,
keeping $\ga\gb=1$, to set $k = \pm 1$ at the expense of multiplying the three-form $\gO$ by a constant. After all these rescalings, we get the following incomplete solutions:
\begin{equation} \begin{array}{ll}
		\gs = 1\, , \quad &\gb = (y^2 - 1)^{1/4}\, , \quad y^2>1 \, , \\
		\gs = -1\, , \quad &\gb = (1-y^2)^{1/4}\, , \quad y^2<1\, ,
	\end{array}
\end{equation}
as well as a complete solution for the positive scalar curvature:
\begin{equation}
	\gs = 1, \quad \gb = (1+y^2)^{1/4} \, .
\end{equation}
The two most interesting solutions, the incomplete solution for $\gs=-1$ and the
complete solution for $\gs=+1$, can be combined together as
\begin{equation}
	\gb = (1+\gs y^2)^{1/4} \, .
\end{equation}

\subsection{A `natural' alternative to Einstein gravity}
\label{ssection: A `natural' alternative to Einstein gravity}

At the end of the first chapter of this thesis, see section \ref{section: Chiral Deformations of Gravity}, we described a family of modified theories of gravity that we referred to as `chiral deformations of GR'. This infinite family of theories was parametrised by a function $f$, see \ref{ssection : Chiral Deformations of Gravity}. Even though there exists freedom in choosing this function, there exists a mathematically natural choice (which is \emph{not} GR) as we now describe. When lifted to seven dimension, solutions to this particular theory will have the interpretation of G2 holonomy manifold.

As we already described in \ref{ssection: Urbantke metric}, the Urbantke metric can be defined as
\begin{equation}\label{G2 holonomy from 4D: Chiral deformation: Urbantke metric}
\gti_{\bdF} \left(X,Y\right)=  \Tr\left( \bdF \W \left[\bdF_X \W \bdF_Y \right] \right) \in \gO^4\left(M^4\right)
\end{equation}
from this definition the Urbantke metric is a \emph{volume-form-valued} metric. Starting  with this object, we can easily construct a volume form: After taking its determinant we indeed obtain 
\begin{equation}
det\left(\gti_{\bdF}\right) \in \left(\gO^4(M)\right)^6,
\end{equation}
 and the sixth square root of this determinant then gives a proper volume form
\begin{equation}\label{G2 holonomy from 4D: Chiral Deformation: natural volume form}
v_{\bdF} \coloneqq \left(det\left(\gti_{\bdF}\right)\right)^{\frac{1}{6}} \in \gO^4(M).
\end{equation}
It gives a most natural functional:
\begin{equation}\label{G2 holonomy from 4D: Chiral Deformation: natural functionnal}
S_{Natural}\left(\bdA \right) = \int_M v_{\bdF} \in \R \;.
\end{equation}

Now, there is in fact a sign ambiguity for $v_{\bdF}$ due to the sixth square root which makes this functional ill-defined. However, as we already described in section \ref{ssection: Definite Connections}, a definite connection provides an orientation (in which $\Xt = F \W F$ is positive definite). In turn this orientation allows us to make a choice of square root (I.e we take $v_{\bdF}$ to be in the orientation class defined by the connection). With the resulting choice of volume form, the functional \eqref{G2 holonomy from 4D: Chiral Deformation: natural functionnal} is well defined.

We now have the following proposition,

\begin{Proposition}\label{Proposition: natural functionnal explicit form}
	Let $\bdA$ be a definite $\SU(2)$-connection and let $\Xt$ be defined by the relation $F^i \W F^j = \Xt^{ij} d^4x$. Then
	\begin{equation}\label{G2 holonomy from 4D: detX action}
	v_{\bdF} = \frac{1}{2}\left(det\left(\Xt \right)\right)^{1/3} d^4x
	\end{equation}
\end{Proposition}

\begin{proof}
	In order to prove the proposition above we will	need the following 
	\begin{Lemma}
		If $\left\{\gS^i\right\}_{i\in 1,2,3}$ is an orthonormal basis of self-dual two-forms i.e $* \gS^i = \gS^i\; \forall i$ and $\gS^i \W \gS^j = \gd^{ij} \frac{\gS^k\W \gS^k}{3}$ then 
		\begin{equation}\label{G2 holonomy from 4D: v_gS = gS w gS}
		v_{\gS} = \frac{\gS^k\W \gS^k}{6}.
		\end{equation} 
		In fact, if $g$ is the Urbantke metric with volume $\frac{\gS^k\W \gS^k}{6}$ then
		\begin{equation}
		\frac{1}{6}\Tr\left( \bdgS \W \left[\bdgS_X \W \bdgS_Y \right] \right) = \frac{\gS^k \W \gS^k}{6} \; g\left(X,Y\right)
		\end{equation}
		from which \eqref{G2 holonomy from 4D: v_gS = gS w gS} follows.
	\end{Lemma}
	This lemma can be directly obtained in coordinates by making use of the algebra \eqref{Appdx: Sigma algebra} and \eqref{Appdx: Sigma self-duality}.
	
	We now come to the proof of proposition \ref{Proposition: natural functionnal explicit form}:\\ 
	The Urbantke metric allows to define the bundle of self-dual two-forms. In section \ref{ssection: Definite Connections} we explained how one can identify this bundle with an associated bundle (with structure group given by our original $\SU(2)$-bundle).  It was also explained that, once this identification is properly made and once we chose the `positive' square root for $\Xt$, we have the relation
	\begin{equation}\label{G2 holonomy from 4D: F= XgS}
	F^i = \gs \sqrt{\Xt}^{ij} \gS^j
	\end{equation}
	where $\gs$ is the sign of the connection and $\left\{\gS^i \right\}_{i \in 1,2,3}$ is an orthonormal basis of self-dual two-forms. This implies in particular
	\begin{equation}
	F^i \W F^j = \Xt^{ij} \;\frac{\gS^k \W \gS^k}{3}.
	\end{equation}
	
	With this in hand \eqref{G2 holonomy from 4D: Chiral deformation: Urbantke metric} can be rewritten
	\begin{equation}
	\gti_{\bdF}\left(X,Y\right) = \gs \; det\left(\sqrt{\Xt}\right) \Tr\left( \bdgS \W \left[ \bdgS_{X}\W \bdgS_{Y}\right]\right)
	\end{equation}
	Then, because of the lemma above:
	\begin{equation}\
	v_{\bdF} = det\left(\sqrt{\Xt}\right)^{\frac{4}{6}} \frac{\gS^k \W \gS^k}{6}
	\end{equation}
	which proves the proposition.
\end{proof}

The field equations associated with the functional \eqref{G2 holonomy from 4D: detX action} are
\begin{equation}\label{G2 holonomy from 4D: detX field eqs}
d_A\left[\left( \det \Xt \right)^{1/3} \Xt^{-1} F \right] =0.
\end{equation}
Note that the expression under the covariant derivative is homogeneity degree zero in $\Xt$ and does not depend on a particular representative.

\subsection{New local example of $G_2$ manifolds from a `gravity theory'}
\label{ssection: New $G_2$ manifold from a `gravity theory'}

We now give details of our generalisation of the Bryant--Salamon construction. The construction presented here is a local one. Global aspects will not be discussed here.

\subsubsection{Ansatz and closure}

We parametrise the three-form by an $\SO(3)$ connection in an $\R^3$ bundle over $M^4$:
\begin{equation}\label{G2 holonomy from 4D: our-3}
	\gO = \frac{1}{6} \ga^3\eps^{ijk} d_A y^i \W d_A y^j \W d_A y^k  +
	2\gs \ga\gb^2  d_A y^i \W F^i\, ,
\end{equation}
where the factor $\gs=\pm 1$ is the sign of the definite connection (see \ref{ssection: Definite Connections}). It is introduced in the ansatz so that \eqref{G2 holonomy from 4D: our-3}  reduces to \eqref{G2 holonomy from 4D: BS ansatz} for instantons \eqref{G2 holonomy from 4D: Instatons eqs}.
It is then easy to see, using the Bianchi identity $d_A F^i=0$, that the condition of closure of \eqref{G2 holonomy from 4D: our-3} is unmodified and is still given by \eqref{G2 holonomy from 4D: BS-eqn-1}.

\subsubsection{The canonical form and the metric}

We now put \eqref{G2 holonomy from 4D: our-3} into the canonical form \eqref{G2 holonomy from 4D: canonical form}. To this end, we use
the parametrisation \eqref{G2 holonomy from 4D: F= XgS} of the curvature,
\begin{equation}\label{G2 holonomy from 4D: F= XgS2}
F^i = \gs X^{ij} \gS^j
\end{equation}
Up to this point we do not have to choose any scale i.e
\begin{equation}
F^i \W F^j = X^{ij} \frac{\gS^k \W \gS^k}{3} = X^{ij} d^4x
\end{equation}
but $d^4x$ is some volume form which is left unspecified. 

 It is then clear that the one-forms $\tht^{4+i}$ are some multiples of $\ga \sqrt{X}^{ij} d_A y^j$. The correct factors are easily found. We have
\begin{equation}
\begin{array}{clc}
	\tht^{i} & = \left( \det X \right)^{-1/6} \ga \left( \sqrt{X} \right)^{ij} d_A y^j, &\; i \in 1,2,3 \\ \\
	 \tht^{4+I} &= \gb \sqrt{2} \left( \det X \right)^{1/12} e^I, &\; I \in 0,1,2,3
\end{array}
\end{equation}
Here, $\{e^I\}_{ I \in 0,...,3}$ is a co-frame for the Urbantke metric \eqref{G2 holonomy from 4D: Chiral deformation: Urbantke metric} with volume form $d^4x$. 
Note that because the particular homogeneity (with respect to $\Xt$) of the different terms, the precise choice of $d^4x$ does not matter.

The metric determined by \eqref{G2 holonomy from 4D: our-3}  is then
\begin{equation}\label{G2 holonomy from 4D: metric-om}
	g_\gO = \ga^2  \left( \det X \right)^{-1/3} d_A y^i X^{ij} d_A y^j + 2\gb^2
	\left( \det X \right)^{1/6} \sum_{I=0}^3 \left( e^I \right)^2\, .
\end{equation}

\subsubsection{The dual form and the co-closure}

The dual form reads
\begin{equation}
	*\gO = -\frac{2}{3} \gb^4 \left( \det X \right)^{1/3} \left( X^{-1} F \right)^i
	\W F^i - \gs \gb^2 \ga^2  \left( \det  X \right)^{1/3} \left( X^{-1} F
	\right)^i \eps^{ijk} \W d_A y^j \W d_A y^k\, ,
\end{equation}
where again we expressed all anti-self-dual two-forms on the base in terms of the curvature two-forms using \eqref{G2 holonomy from 4D: F= XgS2}. Note that, in both terms, the curvature appears either as itself, or in
the combination $\left( \det X \right)^{1/3} \left( X^{-1} F \right)^i$. It is now easy to see that the same steps we followed in the Bryant-Salamon case can be repeated provided
\begin{equation}
	d_A \left[ \left( \det X \right)^{1/3} \left(X^{-1} F\right)^i \right] = 0\, .
\end{equation}

\subsubsection{G2 holonomy and `gravity'}

It follows from the above considerations that, starting from 
\begin{equation}\label{G2 holonomy from 4D: omega}
\gO = \frac{1}{6}(1+\gs y^2)^{-3/4} \eps^{ijk} d_A y^i \W d_A y^j \W
d_A y^k + 2\gs(1+\gs y^2)^{1/4} d_A y^i \W F^i.
\end{equation}
we have the following
\begin{Theorem} \label{Theorem : g2 from gravity}
	If $A$ is a definite connection of sign $\gs$ which is a critical point of 
	\begin{equation}
		v_{\bdF} = \frac{1}{2}\left(det\left( F \W F \right)\right)^{1/3} 
	\end{equation}	
	i.e satisfying the second order PDE:
	\begin{equation}\label{G2 holonomy from 4D: det-eqs}
	d_A \left[ (\det X)^{1/3} \left(X^{-1} F\right)^i \right] =0\, ,
	\end{equation}
	then the three-form \eqref{G2 holonomy from 4D: omega} is stable, in the positive orbit, closed ($d\Omega=0$) and co-closed ($d
	{}^*\Omega=0$), and hence defines a metric of $G_2$ holonomy. The metric is complete (in the fibre direction) for $\gs=+1$.
\end{Theorem}

\subsubsection{Complete indefinite $G_2$ metrics for $\gs=-1$}

We can modify our construction by not putting the sign $\gs$ in front of the second term in \eqref{G2 holonomy from 4D: our-3}. Then all of the construction goes unchanged except that $\gs$ does not appear either in $\gO$ or in $*\gO$. The differential equations for $\ga$ and $\gb$ then give $\gb=(1+y^2)^{1/4}$, and the metric is then complete in the fibre direction for either sign. But the price one pays in this case is that sign of the orbit of $\gO$ is not necessarily positive any-more but rather coincide with the sign of the connection $\gs$. The main consequence is that for $\gs = -1$, the second term in \eqref{G2 holonomy from 4D: metric-om} will appear with a minus sign. This will give a complete (in the fibre direction) metric of $G_2$ holonomy, but of signature $(3,4)$ rather than a Riemannian metric.

\subsubsection{Metric induced on the base}

The three-form \eqref{G2 holonomy from 4D: our-3} defines the metric \eqref{G2 holonomy from 4D: metric-om} on the total space of the bundle. The metric induced on the base is of Urbantke type i.e it that makes the curvature two-forms $F^i$ anti-self-dual. Its exact form can be read off from \eqref{G2 holonomy from 4D: metric-om}. The corresponding volume form is
\begin{equation}
	v_\gO = 4 \left( 1 + \gs y^2 \right) \left( \det X \right)^{1/3} d^4x.
\end{equation}
Thus the induced metric is a multiple of the metric that we encountered in the context of diffeomorphism-invariant $\SU(2)$ gauge theory, see section \ref{ssection: A `natural' alternative to Einstein gravity}. 

An interesting remark is that, in the context of the above $\SU(2)$ gauge theory (and more generally for any chiral deformations of GR), the metric interpretation is possible, but nothing forces us to introduce this metric, as the theory itself is about gauge fields. The metric is a secondary object and there is no clear interpretation of the field equations \eqref{G2 holonomy from 4D: detX field eqs} in metric terms. However, after embedding the theory \eqref{G2 holonomy from 4D: detX action} into 7D, we see that the connection is a field that parametrises the closed three-form \eqref{G2 holonomy from 4D: omega} and that it naturally defines a metric in the total space of the bundle. In facts it defines a $G_2$ structure which is more. The field equation have a 7D metric interpretation as forcing the metric to have G2 holonomy. This is as opposed to the four dimensional case where there are typically no metric interpretation of the field equations for the chiral deformation of GR. Consequently, in the context of 7D theory, the metric arises more naturally and unavoidably. Since this 7D metric induces a metric on the base, the 7D construction can be seen as an explanation for why the metric should also be considered in the context of 4D chiral deformations.

\subsubsection{Relation between the 7D and 4D action functionals}

As we already discussed in \ref{ssection: Hitchin theory for three-forms in seven dimension}, the co-closure condition $d{}^*\gO$ is naturally obtained as the equations for Hitchin theory: critical point of the Hitchin functional are closed and co-closed.

For practical purpose, this functional is just the volume of the 7D manifold computed using the metric defined by $\gO$. For our ansatz \eqref{G2 holonomy from 4D: our-3}), the metric is given by \eqref{G2 holonomy from 4D: metric-om}. The fibre part gives the volume element $\ga^3 (dy)^3$, while the base part gives $4\gb^4 \left( \det  X \right)^{1/3} d^4x$.

Thus the Hitchin functional evaluated on our ansatz \eqref{G2 holonomy from 4D: our-3} is
\begin{equation}
	S[\gO] = 4 \int d^3y  \left( 1 + \gs y^2 \right)^{1/4} \int_M \left( \det X
	\right)^{1/3} d^4x\, .
\end{equation}
This is proportional to the action \eqref{G2 holonomy from 4D: detX action} of the $\SO(3)$ gauge theory on the base. In the incomplete case $\gs=-1$, the integral over the fibre (from $y=0$ to $y=1$) is finite. We get
\begin{equation}\label{G2 holonomy from 4D: relation}
	S_{\gs=-1}[\gO] = \frac{16 \sqrt{\pi}\, \Gamma^2(1/4)}{21\sqrt{2}} \int_M \left(
	\det X \right)^{1/3} d^4x \, .
\end{equation}
In either case, the volume functional for the three-form \eqref{G2 holonomy from 4D: our-3}) in 7 dimensions is a
multiple of the volume functional for the $\SO(3)$ connection in 4D. Thus, there
is a relation not only between solutions of the two theories, but also between the action
functionals.

Let us note that we can also get relation \eqref{G2 holonomy from 4D: relation} to work in the case $\gs=+1$ at the expense of making the 7D metric indefinite of signature $(3,4)$. This is achieved just by putting the minus sign in front of the second term in \eqref{G2 holonomy from 4D: our-3})
also for the $\gs=+1$ case. The 7D metric is then indefinite, but induces a Riemannian signature metric on the base. In this case, the function $\gb = \left( 1 - y^2 \right)^{1/4}$, and so we get an incomplete metric in the fibre direction, and a finite multiple relation \eqref{G2 holonomy from 4D: relation} between the volumes.

\chapter{Some more Variations on Hitchin Theory in 7d}

In this final chapter we describe a new functional for three-forms in seven dimensions describing nearly parallel $G_2$ structures and consider its reduction to four dimensions. The resulting theory is some scalar-tensor gravity in $BF$ type formulation. These results were originally presented in \cite{Krasnov:2016wvc,Krasnov:2017uam}. Note that the author of this thesis was not directly involved in this work. It is reviewed here to give a more complete picture. We also review general facts on special holonomy on Riemannian manifold that will serve as a conceptual unifying tool between this part and the previous one. 

\section{New functionals in seven dimensions}\label{section: New functionals in 7D}

\subsection{$G_2$ and weak holonomy $G_2$ functionals}

In section \ref{section: New Actions in 6d} we considered functionals of the type \eqref{New Actions in 6d: background invariant Hitchin theory} that we referred to as `background independent Hitchin theory'. In seven dimensions and four three-forms it reads
\begin{equation}\label{Geometry of Stable 3-Forms in 7D: BI hitchin action}
S\left[B,C\right] = \int_{M^7} B\W dC - 3\Phi\left[C\right]
\end{equation}
Where $B \in \gO^3\left(M^7\right)$ and $C \in \gO^3\left(M^7\right)$ are three forms in $M^7$. Here the $B$ field just implements the constraint that $C$ is closed, on the other hand the field equations for $C$ say that $\Ch$ is exact:
\begin{equation}
dB=\Ch.
\end{equation}
In the previous chapter we saw (see \eqref{Geometry of Stable 3-Forms in 7D: hat operator}) that for three-forms in seven dimension, $\Ch = \frac{1}{3} *C$. Here the hodge dual is given by the metric $g_{C}$ constructed from the three-form as in \eqref{Geometry of Stable 3-Forms in 7D: 3form-metric}, \eqref{Geometry of Stable 3-Forms in 7D: 3form-metric2}. Altogether the field equations for \eqref{Geometry of Stable 3-Forms in 7D: BI hitchin action} are
\begin{equation}
dC = 0, \qquad *C= dB.
\end{equation}
In particular this implies that the solutions describe $G_2$ holonomy metric see Theorem \ref{Theorem: 7D hitchin Theory}.

Now in seven dimension the Lagrange multiplier $B$ is a three-form as well. It is thus just as natural to consider the functional
\begin{equation}\label{Geometry of Stable 3-Forms in 7D: NG2 hitchin action}
S\left[C\right] = \int_{M^7} \frac{1}{2}C\W dC -3k\; \Phi\left[C\right].
\end{equation}
Here $k$ is a parameter that can always be put to one, up to adding an overall factor in front of the action, by a field redefinition. There is no constraint any-more and the resulting field equations are
\begin{equation}
dC = k *C.
\end{equation}
These are known as the field equations for \emph{nearly parallel $G_2$ structures}, see below for a discussion on these structures in the natural context of holonomy on Riemannian manifold. The resulting seven dimensional metric is known to be Einstein. What is more it is known that one can construct a cone over solutions to the above equations such that the resulting eight dimensional metric has $\Spin(7)$ holonomy. Once again, see below for more on holonomy on a Riemannian manifold. 

\subsection{Example: The weak holonomy $G_2$ structure of the spin bundle over Instantons}

We now give a nice example of solutions to the above equations related to construction discussed in the second chapter. This can be thought as a variant of the twistor construction. Let us consider the $\SU(2)$(spin) bundle over a four dimensional manifold $M^4$
\begin{equation}
\SU(2) \inj P^7 \to M^4.
\end{equation}
We take $\bde \in\gO^1\left(P^7,\su(2)\right)$ to be the connection-form of the (self-dual part of) the Levi Cevita connection. As we already discussed in the 6D context (see section \ref{section: 3D Gravity in Terms of 3-Forms}), this is an $\SU(2)$-equivariant one-form which, when restricted to each $\SU(2)$ fibre is just the Maurer-Cartan frame. This can also be related to the spinor notation of section \ref{section: Euclidean metric Twistor Space} as\footnote{Implicit in this notation is that $\SU(2)\simeq \S^3$ is embedded into $C^2$. Taking $R^2= \pp$ one can indeed check that the flat metric on $\C^2$ can be rewritten
	\begin{equation}
\frac{1}{2}D\p_{A'} \odot D \ph^{A'} = \left(dR\right)^2 + R^2 \; e^{A'B'} \odot e_{A'B'}.
	\end{equation}
so that $e^{A'B'}$ is a frame field for the three-sphere.}
	
\begin{equation}
e^{A'B'} =\frac{1}{\pp} \left(\ph^{(A'} D \p^{B')} - \p^{(A'}D\ph^{B')}\right).
\end{equation}
The curvature of the connection is
\begin{equation}
\bdF = d \bde + \frac{1}{2}\left[\bde, \bde \right].
\end{equation}
We can now consider
\begin{equation}\label{New functionals in 7D: C eg def}
C= \kt^{-3}\left(R^3\; \frac{1}{6}\;\eps^{ijk} e^i \W e^j \W e^k + R \; e^i \W \gS^i \right)
\end{equation}
 as three-form on $P^7$. Here $R$ is the radius of the three-sphere bundle that we leave as a parameter. With this notation the three-form is already in the canonical form \eqref{Geometry of Stable 3-Forms in 7D: three form canonical} so that
 \begin{equation}
 *C= \kt^{-4}\;\left(\frac{1}{6}\gS^i \W \gS^i + R^2\; \frac{\eps^{ijk}}{2} e^i\W e^j \W \gS^k. \right)
 \end{equation}
 
On the other hand,
\begin{equation}
dC =k^3\left( R\; F^i \W \gS^i + \left(\frac{\eps^{ijk}}{2} e^j \W e^k \right) \left( R^3 F^i + R \gS^i \right)\right)
\end{equation}
It follows that  $dC = k *C$ if and only if the base manifold is an instanton $\bdF = \frac{\gL}{3} \bdgS$. What is more one needs,
\begin{equation}
2R \gL = k/ \kt, \qquad R^3 \gL/3 + R = k/\kt\; R^2
\end{equation}
eliminating $k/\kt$ in the second equations, this system is easily seen to be equivalent to
\begin{equation}
R\;\kt = k \frac{6}{5},\qquad R^2 \gL = 3/5.
\end{equation}
One can geometrically interpret these equations as follows. Looking at \eqref{New functionals in 7D: C eg def} one sees that changing $\kt$ amounts to a rescaling of the seven dimensional metric. Making use of the first equation to solve $\kt$ as a function of $R$ means that we parametrize the global scaling of the metric in terms of the radius $R$ of the three-sphere fibres. The second equation is the true equation as it relates uniquely this radius to the curvature of the base manifold $R^2 \gL = 3/5$.

Note that for the round metric on $\S^7$ the relation between the fibre radius and the base curvature is $R^2 \gL = 3 $. The resulting metric here is thus some sort of `squashed metric' on $\S^7$ i.e it is obtained from the round metric on $\S^7$ by squashing the fibre. Because it gives the manifold a nearly $G_2$ parallel structure, in particular this squashed metric is Einstein. It is known that it can be obtained as the natural metric on the homogeneous space $\Spin(7)/G_2$, see e.g \cite{Salamon:1989}.

The above construction can be thought of as some sort of variant of the Bryant Salamon construction but for nearly parallel $G_2$ manifold instead of $G_2$ manifold.

\section{Looking back: holonomy as a unifying theme}\label{section: Holonomy} 

In the preceding chapters, we respectively encountered \emph{nearly Kahler} structures, \emph{$G_2$ holonomy} manifolds and \emph{nearly parallel $G_2$} structures.  All these structures are natural in the context of holonomy on a Riemannian manifold that we now review.

The material described in this subsection is standard see for example \cite{Hitchin:2002ea}, \cite{Besse} or \cite{Salamon:1989}. 

\subsection{Holonomy on a Riemannian manifold}
Associated with any Riemannian manifold $\left(M,g\right)$ we have a unique torsionless connection, the Livi-Civita connection $\N$. By means of this connection one can define parallel transport as follows:
Let $\gc \from [0,1] \to M $ be a path connecting two points $x_0 = \gc(0)$ and $x_1 = \gc(1)$. The pull-back connection $\gc^*D$ is necessarily flat over the image of the path and can therefore be trivialised by flat sections. In other terms, for any $v_0\in T_{x_0}M$ there is a unique vector field $v(t) \in T_{\gc(t)} M$ such that $\gc^*D v =0$ and $v(0) = v_0$. This defines an isomorphism, the \emph{parallel transport along the path $\gc$}:
\[ 
\begin{array}{lcccl}
P_{\gc} \from &T_{x_0} M &\to& T_{x_1} M \\
&v_0 & \mapsto & v(1)&.
\end{array} \]

For any $x\in M$ the \emph{holonomy group} $Hol_x(M,g)$ is the group of all parallel transports $P_{\gc}$ along closed paths $\gc \from [0,1] \to M$ with $\gc(0) =\gc(1) = x$. If $x$ and $y$ can be connected then by a path $\gc$ the holonomy group are conjugate, $Hol_y(M,g) = P_{\gc} \circ Hol_x(M,g) \circ P^{-1}_{\gc}$, and thus isomorphic. A further subtlety appears if $M$ is not simply connected. One then make a difference between the holonomy group at $x$ and the \emph{restricted holonomy group} at x, $Hol^0_x(M,g) \subset Hol_x(M,g)$, of parallel transport $P_{\gc}$ along contractible path $\gc$. In the following we will assume that $M$ is simply connected.

Now, because the Levi-Civita connection is compatible with the metric $\N g=0$, the holonomy group at $x$ of a Riemannian manifold is a subgroup of $O(T_x M)$ (Cf Proposition \ref{Theorem: Holonomy principle}).  Finally one can define the holonomy group of a (connected, simply connected) Riemannian manifold $Hol(M,g)$ as a subgroup of $O(n)$ up to conjugation.

A natural question in this context is: What groups $Hol(M,g) \subset O(m)$ can arise? For non-symmetric spaces (symmetric spaces are a special case of homogeneous spaces) there is a classification theorem due to Berger. Because the holonomy a of product $\left(M,g\right) = \left(M_1,g_1\right)\times \left(M_2,g_2\right) $ is the product $Hol(M_1,g_1) \times Hol(M_2,g_2)$, this classification is only possible when $(M,g)$ is \emph{irreducible} , i.e cannot locally be written as a product. 

\begin{Theorem}{\emph{`Berger's List' c.f \cite{BergerList}}} \mbox{}\\
	Let $\left(M,g\right)$ be a simply connected, irreducible Riemannian manifold of dimension $n$, further assume that $(M,g)$ is not locally symmetric. Then the holonomy group $Hol(M,g)$ is isomorphic to one of the following list : \\
	\begin{center}
		\arrayrulecolor{black}
		\begin{tabular}[c]{ccl} \label{Berger's List}
			Dimension & $Hol \subset \SO(n)$ & Geometry\\
			\hline
			n & $\SO(n)$ & generic \\
			2m & $\U(m)$& Kähler manifold \\
			2m, $m\geq3$ & $\SU(m)$ & Calabi-Yau manifold \\
			4m & $\Sp(m)$ & hyperKähler manifolds \\
			4m, $m\geq2$ &$\Sp(m) \times \Sp(1)$& quaternionic Kähler \\
			7 & $G_2$ & exceptional \\
			8 & $\Spin(7)$ & exceptional
		\end{tabular}
	\end{center}
\end{Theorem}

\subsection{Fundamental Principle of holonomy}

Consider $E \to M$ a tensor or spinor bundle.  We have a natural action of the Livi-Cevita connection $\N$ on this bundle. We define \emph{invariant section} (i.e under parallel transport) as sections $\ga$ such that for every loop $\gc:[0,1] \to M$, $\gc(0)=\gc(1)=x$,  $P_{\gc}\ga = \ga(x)$. By definition it means that the group of holonomy stabilises $\ga_0 \in E_x$. Taking the problem the other way round let us now suppose that $hol_x(m,g)$ stabilises $\ga_0 \in E_x$. One can then define an invariant section as $\ga(y) = P_{\gc(1)}\ga_0$, where $\gc$ is any path starting at $x$ and finishing at $y$. The precise choice of path does not matter precisely because $\ga_0$ was taken to be invariant under the holonomy group. 

In principle \emph{invariant sections} could differ from \emph{covariantly constant sections} which are the sections $\ga$ such that $\N \ga =0$. In fact, the two notions are equivalent. This is the essential content of the holonomy principle:
\begin{Theorem}{`Holonomy principle' cf Besse p282 }\label{Theorem: Holonomy principle} \mbox{}\\
	The following three properties are equivalent
	\begin{enumerate}
		\item there exists a tensor $\ga_x \in E_x$ which is invariant under $hol_x(M,g)$
		\item there exists $\ga$ an invariant section by parallel transport $P_{\gc} \ga = \ga$ for all path $ \gc$
		\item there exists $\ga$ a covariantly constant section $\N \ga =0$
	\end{enumerate}
\end{Theorem}

Looking back at Berger's list \eqref{Berger's List} one can characterise each of the entry by the tensors that it stabilises:

\paragraph{$SO(n)$ holonomy and orientability} This is the simplest application of the holonomy principle. If a manifold $(M,g)$ is orientable then there exists a globally defined nowhere-zero volume form. By properly rescaling this volume-form one obtains a covariantly constant section. From the holonomy principle it follows that $Hol(g)$ stabilises a volume form i.e $Hol(g) \subset SO(n)$. Starting with a metric whose holonomy group is a subgroup of $SO(n)$ one can construct a global nowhere-vanishing volume-form by parallel transport and the manifold is thus orientable.

\paragraph{$U(m)$ and $SU(m)$ holonomy: Kähler and Calabi-Yau manifold } The case where $Hol(g) \subset \U(m) \subset O(2n)$ corresponds to having a covariantly constant almost complex structure $\J$, compatible with the metric, i.e such that $g = g(\J,\J)$. It follows that the Kähler form $\go = g(\J.,.)$ is covariantly constant and thus closed. Finally one can show (see \cite{Huygbrechts} p215) that this is enough to imply integrability. Thus metric with holonomy $U(m)$ are just Kähler metric. On the other hand metrics with holonomy $SU(m)$ should also admit a covariantly constant top-form. Altogether this is the definition of Calabi-Yau manifold.

\paragraph{$SP(m)$ holonomy: hyperkähler manifold} We have $\Sp(m) \subset \SU(2m) \subset \SO(4m)$ so a metric with holonomy $\Sp(m)$ is in particular Calabi-Yau. However one can show that a metric has holonomy $Sp(m)$ if and only if it admits two covariantly constant almost complex structure $I,J$ such that $IJ =-JI$. It follows that $K = IJ$ is also a parallel almost complex structure. In fact for any $(x,y,z)$ such that $x^2+y^2+ z^2=1$, $xI + yJ + zK$ is a parallel almost complex structure so that we have a whole $S^2$ of them. As the notation suggest, Hyperkähler manifolds are related to the geometry of quaternions (see our discussion on quaternionic structure in section \ref{ssection : The Flat Case from Quaternions}) , i.e there is as sense in which they are `quaternion manifolds'. This turns out however not to be as useful a concept as the one of complex manifolds.\\

The remaining holonomies of Berger's list are less traditional. The $\Sp(m)\times \Sp(1)$ case is usually refer to as `quaternion Kähler'. This can be thought as an analogue of the Hyperkähler manifold but with non-zero scalar curvature.  We won't consider it here. Finally,

\paragraph{$G_2$ and $Spin(7)$ holonomy} sometimes dubbed `exceptional holonomy' and are specific to seven and eight dimensions. They respectively stabilise particular three-forms (see \eqref{Octonions Geometry: gO def}) and 4-forms. We already encountered such $G_2$ holonomy metrics in the context of Hitchin theory and we saw that $G_2$ structure essentially identifies the tangent space to seven dimensional manifold of imaginary octonions.

\subsection{Parallel and Killing Spinors}
It is also natural to look for Manifolds admitting covariantly constant spinors $\N \Psi =0$. The question then arises: When does a Riemannian manifold admits such spinors ? By the holonomy principle such manifolds must have a non-generic holonomy group and should thus appear as a sub-case of Berger's List. We indeed have the following result:
\begin{Theorem}{\emph{Wang \cite{Wang:1989}}} \mbox{}\\
	Irreducible, simply connected manifolds admitting parallel spinors $\N \Psi =0$ are those with one the following holonomy group:
	\begin{enumerate}
		\item $\SU(m)$, n=2m, $m\geq3$ : Calabi-Yau 
		\item $\Sp(m)$, n=4m : hyperkahler
		\item $G_2$, n=7 : exceptional
		\item $\Spin(7)$, n=8 : exceptional.
	\end{enumerate}
\end{Theorem}
Related to the existence of parallel spinors is the fact that all metrics listed above have vanishing Ricci tensor.

An even more interesting concept for us is that of \emph{Killing spinors}, spinors $\Psi$ which verify $\N_X \Psi = \gl X.\Psi $ for $\gl \in \C$. There is also a classification of metric admitting Killing spinors as displayed in Table \ref{Tab: Killin Spinor metric}. One sees that in this list appear the already familiar \emph{nearly Kähler} structures and \emph{nearly parallel $G_2$} structures.

 Metrics admitting Killing spinors are Einstein with positive scalar curvature.  Such metric are closely tied up to special (i.s non-generic) holonomy by the following: In \cite{Bar:1993} it was shown that Killing spinors on a manifold $M$ extend to become parallel spinors on the n+1 dimensional cone over $M$. Thus Killing spinors in dimension $n$ are related to special holonomies in $n+1$ dimensions. \\
\begin{table}[h] \caption{Metric admitting Killing spinors} \label{Tab: Killin Spinor metric}
	\begin{center}
		\begin{tabular}{cll}
			Dimension& Geometry & Cone Geometry \\ \hline
			n& round sphere & flat \\
			4m-1 & 3-Sasaki & hyperKähler \\
			4m $\pm$ 1& Sasaki-Einstein & Calabi-Yau \\
			6 & nearly Kähler & $G_2$ holonomy \\
			7 & nearly parallel $G_2$ & $Spin(7)$ holonomy
		\end{tabular}
	\end{center}
\end{table}

\section{Reduction form 7D to 4D}

\subsection{Reduction of the three-form}

We now briefly consider the $\SU(2)$-reduction of \eqref{Geometry of Stable 3-Forms in 7D: NG2 hitchin action}, see \cite{Krasnov:2016wvc,Krasnov:2017uam} for details.

 We take $P^7$ to be a seven dimensional manifold and, just as in section \ref{section: 3D Gravity in Terms of 3-Forms}, we suppose that $\SU(2)$ acts freely on $P^7$ such that $P^7$ has the structure of a principal bundle with base $M^4 = P^7 / \SU(2)$:
\begin{equation}
\SU(2) \inj P^7 \to M^4.
\end{equation}
What is more, we restrict ourselves to considering three-forms that are invariant under this action, $R^*C=C$. This necessarily implies that $C$ can be interpreted in terms of fields on $M^4$. 

In order to have an idea of this field content, let us consider the following reasoning. As we already discussed several times, a stable three-form $C$ in seven dimensions gives a $G_2$ structure, in particular it defines a seven dimensional metric $g_{C}$. See \eqref{Geometry of Stable 3-Forms in 7D: 3form-metric}, \eqref{Geometry of Stable 3-Forms in 7D: 3form-metric2} for the explicit construction. Requiring the $\SU(2)$ invariance of $C$ necessarily means the $\SU(2)$ invariance of the metric. It implies that the field content of the $\SU(2)$-invariant three-form should contain a four-dimensional metric $g_4$, a $\SU(2)$-connection $\bdW$ and a $\SU(2)$-metric $g_3$:

\begin{equation}\label{Geometry of Stable 3-Forms in 7D: metric reduction}
g_{C} = \Mtx{ g_3 & & \bdW \\ \\ \bdW^t & & g_4}.
\end{equation}
In fact a $G_2$ structure is more than a seven dimensional metric (rather it is equivalent to a seven dimensional metric plus a 7D unit spinor see \cite{Witt:2009zz,Agricola:2014yma}) so that we will in fact use a different parametrisation for $C$. The above reasoning will however serve as a motivation for introducing a $\SU(2)$-connection to help parametrizing $C$.

Let us thus introduce a $\SU(2)$-connection on the principal bundle $P^7 \to M^4$ and let 
\begin{equation}
\bdW \in \gO^1\left(P^7 \right) \times \su(2)
\end{equation}
 be the associated connection-form. It can also be written in terms of its potential $\bdw$, the two being related by
\begin{equation}
\bdW = g^{-1}dg + g^{-1} \bdw g.
\end{equation}
see section \ref{section: 3D Gravity in Terms of 3-Forms} for our principal bundle notations.

Introducing this connection-form is useful for one can now decompose $C$ as (we do not need to assume invariance of C for now)
\begin{equation}
C = \Tr\left(\;\phi \;\bdW \W \bdW \W \bdW + \bdW \W \bdW\W \bdA + \bdW \W \bdB\right) + c
\end{equation} 
Where $\phi \in \gO^0\left(P^7\right)$ is a scalar function and  
\begin{equation}
\bdA \in \gO^1\left(P^7\right) \times \su(2), \quad \bdB \in \gO^2\left(P^7\right) \times \su(2), \quad c  \in \gO^3(P^7)
\end{equation}
 are \emph{basic} forms, i.e they vanish on vertical vectors. There is no loss of generality with this parametrisation: we are just choosing to use a connection as a parameter. Had we start from another connection, this would just have shifted the connection-form $\bdW \to \bdW + \bda$ by a $\SU(2)$-equivariant, $\su(2)$-valued one-form $\bda \in \gO^1\left(P^7\right) \times \su(2)$ (then all other forms should be shifted accordingly).

Let us now suppose that $C$ is $\SU(2)$-equivariant. This implies that $\bdA$, $\bdB$ are respectively equivariant basic $\su(2)$-valued one- and two-forms. In particular, they are uniquely associated with one- and two- forms on the base $M^4$, $\bda\in \gO^1\left(M^4, \su(2)\right)$ and $\bdb \in  \gO^2\left(M^4, \su(2)\right)$ taking values in section the associated bundle $\left(P^7 \times \su(2)\right) /\SU(2)$.  For concreteness
\begin{equation}
\bdA = g^{-1}\bda g,\qquad \bdB = g^{-1}\bdb g.
\end{equation}
(again see section \ref{section: 3D Gravity in Terms of 3-Forms} for our principal bundle notations). What is more equivariance of $C$ implies that $\phi$ and $c$ respectively are the pull back by the projection operator of a scalar field and a three-form on $M^4$.

One then easily sees that, by making an appropriate shift of the connection-form, one can always suppose that the second term $\bdW' \W \bdW'\W \bdA$ is absent. Let now $\bdW$ be the connection-form achieving this: 
\begin{equation}
C = \Tr\left(\;\phi\; \bdW \W \bdW \W \bdW + \bdW \W \bdB\right) + c.
\end{equation}
The above parametrisation shows that the field content of the $\SU(2)$-reduced theory is: a scalar field $\phi \in \gO^0\left(M^4\right)$, a $\SU(2)$ connection (with potential $\bdw$) on the principal bunlde, a $\su(2)$-valued two-form $ \bdb \in \gO^1\left(M^4,\su(2)\right)$ and a three form $c \in \gO^3\left(M^4\right)$. 

In the following it will actually be more convenient to parametrize $C$ as
\begin{equation}\label{Geometry of Stable 3-Forms in 7D: C parametrisation}
C = -2\Tr\left(\;\frac{\phi^3}{3}\; \bdW \W \bdW \W \bdW + \phi \;\bdW \W \bdB\right) + c.
\end{equation}

\subsection{Reduction of the Action}

Evaluating the kinematic term  of \eqref{Geometry of Stable 3-Forms in 7D: NG2 hitchin action} on \eqref{Geometry of Stable 3-Forms in 7D: C parametrisation}, we obtain\footnote{Here, $\bdm = g^{-1}dg$ is the Maurer-Cartan frame on $\SU(2)$.}
\begin{equation}\label{Geometry of Stable 3-Forms in 7D: Kin term reduced}
\int_{P^7} \frac{1}{2}\;C dC = \left(\int_{\SU(2)} \frac{-2}{3} \Tr(\bdm^3)\right) \times \int_{M^4} -2\Tr \left(\phi^4\; \bdB \bdF + \frac{\phi^2}{2}\; \bdB \W \bdB \right) + \phi^3\; dc
\end{equation}
Which is just a BF theory! Field equations reads,
\begin{equation}
d \phi = 0,\quad  dc= 0 \quad \text{and}\quad \phi^2 \; \bdF = -\bdB.
\end{equation}
We now come to the potential term, it is proportional Hitchin's volume \begin{equation}\label{Geometry of Stable 3-Forms in 7D: 7D hitchin volume}
\Phi\left[C\right] = \left(det\left( \gti_{C}\right)\right)^{1/9}.
\end{equation}
In order to evaluate \eqref{Geometry of Stable 3-Forms in 7D: 7D hitchin volume} on \eqref{Geometry of Stable 3-Forms in 7D: C parametrisation} one first needs to compute the seven dimensional metric in terms of \eqref{Geometry of Stable 3-Forms in 7D: C parametrisation}.

In order to do so, it will be convenient to introduce a conformal metric on the base manifold $M^4$. The natural motivation for this is that, as we already discussed, $\SU(2)$-invariant three-forms define a four dimensional metric \eqref{Geometry of Stable 3-Forms in 7D: metric reduction}. 

Following the strategy exposed at the beginning of this thesis, see section \ref{section: Chiral Formulations of GR - Fundations}, a (conformal) metric $\gti_{\bdB}$ can be directly constructed from a $\SU(2)$-valued two-form $\bdB$ through Urbantke formula (see eq \eqref{Chiral Formulations of GR - Fundations: Urbantke metric}). Let $\left(e^I_{\bdB}\right)_{I \in 0...3}$ be a conformal tetrad for the Urbantke metric $\gti_{\bdB}$. Let $\bdgSt = \gS^i \gs_i$ be the $\su(2)$-valued basis of orthogonal two-form constructed as in \eqref{Appdx: Sigma def (tetrad)}. Here $\gti_{\bdB}$ has density weight $-1/2$, $e_{\bdB}$ has density weight $-1/4$ and $\bdgSt$ has density weight $-1/2$.

 Then 
\begin{equation}
\bdB = \left( \sqrt{X} ^{ij}\; \gSt^j \right) \gs_j
\end{equation}
where $X$ has density weight $1$. Here
\begin{equation}
\bdB \W \bdB = -2\;Vol_{\bdgSt} \; \left(X^{ij}\right) \gs_i \otimes \gs_j
\end{equation}
and
\begin{equation}
Vol_{\bdgSt} \coloneqq -6\;\gSt^i \W \gSt^i
\end{equation}
is a four-form with density weight $-1$.

It is also convenient to parametrise the three-form $c \in \gO^3\left(M^4\right)$ in terms of $\gti_{\bdB}$ and a vector field $v$ as
\begin{equation}
c= -2\; \left(det(X)\right)^{1/4} \; \gi_{v} Vol_{\bdgSt}
\end{equation}
The pre-factor here is for future convenience. With these definitions, $v$ has density weight $-1/4$. It follows that $|V|^2 =\gti_{\bdB}\left(v,v\right)$ has density weight $0$ and is therefore a proper scalar.

A direct calculation of the seven dimensional conformal metric using \eqref{Geometry of Stable 3-Forms in 7D: 3form-metric} then gives the following matrix form for $\gti_C$,
\begin{equation}\
\Mtx{\phi^(5/2) \sqrt{X} & 0 \\
	0 & \phi^{3/2} \left(det(X)\right)^{1/4} e_{\bdB}} \; \Mtx{\Id & \gi_{v} \gS \\ \gi_{v} \gS & \Id
	} \; \Mtx{\phi^(5/2) \sqrt{X} & 0 \\
	0 & \phi^{3/2} \left(det(X)\right)^{1/4} e_{\bdB}}.
\end{equation}
Under this form it is easy to compute the determinant,
\begin{equation}
det \Mtx{ \Id & \gi_{v} \gS \\
	\gi_v \gS & \Id} = 1 - |v|^2
\end{equation}
 and 
 \begin{equation}
 det \Mtx{\phi^(5/2) \sqrt{X} & 0 \\
 	0 & \phi^{3/2} \left(det(X)\right)^{1/4} e_{\bdB}} = \phi^{27} \; \left(det(X)\right)^3 det\left(g_{\bdB}\right)
 \end{equation}
 so that (here and in what follows we take $k=1$ in \eqref{Geometry of Stable 3-Forms in 7D: NG2 hitchin action})
 \begin{equation}\label{Geometry of Stable 3-Forms in 7D: Pot term reduced}
 	\int_{P^7} -3\;\Phi\left[C\right] = -3\; \left( \int_{\SU(2)}-\frac{2}{3} \Tr\left(\bdW^3\right)  \right) \int_{M^4} \phi^3 \; \left(1-|v|^2\right)^{1/3}\;\left(\left(det(X)\right)^{1/3}\; Vol_{\bdgS}\right).
 	\end{equation}
 	Few remarks are in order, first all density weight compensate each others here so that the integrand is a proper volume form. Second the last term  $\left(det(X)\right)^{1/3}\; Vol_{\bdgS}$ looks just like \eqref{G2 holonomy from 4D: detX action}.
 	
 	Combining the kinematic term \eqref{Geometry of Stable 3-Forms in 7D: Kin term reduced} and the potential term \eqref{Geometry of Stable 3-Forms in 7D: Pot term reduced} and factoring out by the volume of $\SU(2)$ the reduction of \eqref{Geometry of Stable 3-Forms in 7D: NG2 hitchin action} is
 	\begin{equation}\label{Geometry of Stable 3-Forms in 7D: Kirill's reduced action}
\begin{array}{ll}
 	S\left[\bdW, \bdB, \phi, v\right] = \\ \\
\begin{array}{ll}
 	\int_{M^4} &-2\Tr\left(\phi^4 \bdB \bdF + \left(\phi^2/2\right) \bdB \bdB \right) + 6 \phi^2 \left(det(X)\right)^{1/4} \left(v^{\mu} \pa_{\mu}\phi \right) Vol_{\bdgS} \\& -3\; \phi^3 \; \left(1-|v|^2\right)^{1/3}\;\left(\left(det(X)\right)^{1/3}\; Vol_{\bdgS}\right).
\end{array}
\end{array}
 	\end{equation}

\subsection{Interpretation}

A first point is that if one takes $\phi=cst$, the above action is a particular chiral deformation of GR of the general form $BF$ plus potential $V(BB)$ (see \eqref{Chiral Deformations of Gravity: S[A,B]} and \cite{Krasnov:2009iy}). This therefore describes two propagating degrees of freedom of gravity type (spin two).

Another interesting feature of the above equation is that for $\phi=const$ the value of the effective 4D cosmological constant is determined by $\phi$. Moreover, one can show that for values $\phi\approx 1$ the 4D cosmological constant is arbitrarily small {\it and} the deviations from General Relativity for curvatures smaller than Planckian are negligible. See below for more on this.

In \cite{Krasnov:2009ik}, it was explained how these BF-type gravity theories can be explicitly recast into metric form. See also section \ref{ssection : Variational Principles}. As already discussed, the result is a gravitational Lagrangians starting with the Einstein-Hilbert term, but corrected with an infinite number of higher powers of the curvature terms, see \cite{Krasnov:2009ik} for details. 

Now in the full action \eqref{Geometry of Stable 3-Forms in 7D: Kirill's reduced action} (i.e when one does not froze $\phi=cst$ and $v=0$) there is also a scalar field on top of $\bdB$ and $\bdW$. Following the same steps as in \cite{Krasnov:2009ik} one can envisage eliminating from the Lagrangian all fields apart from the metric and the scalar field $\phi$, and obtaining a scalar-tensor theory of a specific type. 

Prior to eliminating any fields, the action \eqref{Geometry of Stable 3-Forms in 7D: Kirill's reduced action} is first-order in derivatives. Now, the Euler-Lagrange equation for $v^\mu$ that follows from \eqref{Geometry of Stable 3-Forms in 7D: Kirill's reduced action} are an algebraic equations for $v$ in terms of the derivatives $\partial \phi$. Solving this equation, while difficult explicitly, is possible in principle. In that sense, the vector field $v$ is an auxiliary field needed to put the second-order scalar field Lagrangian into a first-order form. Eliminating $v$ in this fashion, one obtains the Lagrangian for $\phi$ of the type 
\begin{equation}
{\cal L} = K( \phi^3, |\partial_\mu \phi^3|^2).
\end{equation}

This type of scalar theories has been studied under the name of "K-essence" in \cite{ArmendarizPicon:1999rj}.

Let us now briefly show that there is a regime, where this theory is close to GR. Essentially, we consider the Lagrangian \eqref{Geometry of Stable 3-Forms in 7D: Kirill's reduced action} around an anti-self-dual Einstein background i.e around 
\begin{equation}
v =0, \qquad \phi = cst, \qquad B^i \W B^j = \gd^{ij}\frac{ B^i \W B^i}{3}. \qquad 
\end{equation}

Remember that with our definition
\begin{equation}
B^i \W B^j = -2\; X^{ij} \; Vol_{\bdB}
\end{equation}
where the volume $Vol_{\bdB}$ was left undetermined. We now take this volume to be such that $\Tr \sqrt{X} = 3$. Practically this amounts to taking
\begin{equation}
Vol_{\bdB} = \frac{1}{18}\left(\Tr \left(\sqrt{ B \W B}\right)\right)^2.
\end{equation}
We can then parametrise $X$ as
\begin{equation}
\sqrt{X}^{ij} = \gd^{ij}+ \Psi^{ij},
\end{equation}
where $\Psi^{ij}$ is traceless. We now take $\Psi$ to be small in Planck units, $\Psi \ll 1$. 

Then
\begin{equation}
\left(det(X) \right)^{1/3} = \left( det\left( \gd^{ij}+ \Psi^{ij}\right) \right)^{2/3} \simeq 1 + \Oc(\Psi^2)
\end{equation}
so that to first order in $\Psi$ and taking $\phi=cst$, $v=0$ the action \eqref{Geometry of Stable 3-Forms in 7D: Kirill's reduced action} can be rewritten
\begin{equation}
S\left[\bdW, \bdB\right] \propto \int_{M^4} -2\Tr\left(\bdB \W \bdF + \frac{\phi^{-2}}{2} \bdB \W \bdB \right) - 3\;\;\phi^{-1} \; Vol_{\bdB}.
\end{equation}
or
\begin{equation}
S\left[\bdW, \bdB \right] \propto \int_{M^4} B^i \W F^i -\;\frac{1}{3}\phi^{-1}\;\frac{\left(\Tr \left(\sqrt{ B \W  B}\right)\right)^2}{2}  +   \phi^{-2}\; \frac{B^i \W B^i}{2}
\end{equation}
which is just of the form \eqref{Chiral Deformations of Gravity: S_GR[A,B]} with 
\begin{equation}
\eps= - \phi^{-2}, \qquad \frac{\eps- \gL}{3} = -\frac{\phi^{-1}}{3}
\end{equation}

 and therefore describes usual GR with cosmological constant
\begin{equation}
\gL = \frac{\phi -1}{\phi^{-2}}
\end{equation}
 (see \cite{Herfray:2015rja} for a proof).
 
In particular a very nice feature of this is that for $\SU(2)$-fibre of Planck size $\phi \simeq 1$ the cosmological constant is small in Planck units. This is both interesting and surprising as it is exactly what one would like to have for the relation between the size of compact dimensions and the value of the cosmological constant.

\end{PartIII}

\begin{Conclusion}
\bookmarksetup{startatroot}
\addtocontents{toc}{\protect\partbegin}

\chapter*{Conclusion}
\addstarredchapter{Conclusion} \markboth{Conclusion}{}

All the approaches developed in this thesis were aspects of a search for a new perspective on gravity. On the one hand, we reconsidered the twistor formulation of (Euclidean) four dimensional GR and clarified that it fits in the broader scheme of chiral formulations of gravity. On the other hand, we proposed a new point of view on GR. We showed that three and four dimensional gravity theories can be obtained as $\SU(2)$-reduction of Hitchin's theories of differential forms in six and seven dimensions.

We demonstrated that there is a nice interplay between chiral formulations of gravity and twistors. This is particularly true of the self-dual sector where `perfect connections', satisfying $F^i \W F^j\propto \gd^{ij}$, directly correspond to one-form $\gt$ on the twistor space satisfying $\gt \W d\gt \W d\gt = 0$. Both equations were known to describe self-dual Einstein gravity, see \cite{Capovilla:1990qi, Mason&Wolf09}, but we clarified the explicit relation between the two pictures. This culminated in an (Euclidean) version of the non-linear graviton theorem where perfect connections on space-time are explicitly obtained from holomorphic data on twistor space.

 What is more these results suggest a change of perspective on twistor theory. The traditional approach to twistor theory emphasises the duality between (conformal) metric on space-time and complex structure on twistor space. From our presentation it is however clear that it is just as legitimate to take the duality between self-dual connections on space-time and one-form on twistor space as fundamental. In the `metric' twistor perspective see \cite{Penrose:1976js, Ward:1980am} this one-form was understood as `additional' to the complex structure, in the dual (space-time) picture this corresponded to fixing conformal freedom by choosing a self-dual connection. In our `connection' twistor perspective, that was first discussed in \cite{Herfray:2016qvg}, one-forms are the starting point and complex structures are derived objects. This parallels the pure connection formulation of gravity where Einstein metrics are constructed from self-dual connections.

 Our original hope was that this change of perspective might lead to new insight on the googly problem: Perfect connections have twistor interpretation and can be related to the non-linear graviton theorem. Einstein connections exist, do they have a twistor interpretation? After all, the pure connection formulation of gravity gives a strikingly compact description of full gravity in chiral terms. There however does not seems to be a simple answer to this question. 
 
 This difficulty with obtaining a twistor description of full GR motivated us to look for a relation between Hitchin's theories of forms in six and seven dimensions and twistors, in the hope of finding a realisation of the full GR along these lines. Our first result is a demonstration that the $\SU(2)$ reduction of the six dimensional theory is just 3D gravity coupled with a (constant) scalar field. The sign of the orbit in the space of three-forms then corresponds to the sign of the cosmological constant. This result has first been described in \cite{Herfray:2016std}. The 6D Hitchin theory is topological in the sense that its one-loop partition function, computed in \cite{Pestun:2005rp}, is a ratio of holomorphic Ray-Singer torsions. One therefore expects that the dimensional reduction of this theory will also be topological and this is confirmed by our result. The one-loop partition function of 3D gravity is also known and is also given by an appropriate Ray-Singer torsion. It would be interesting to understand the relation between these two results, but this has not been dealt with in this thesis.

Our work on Hitchin's theory in six dimensions also led us to introduce new 6D theories of differential forms. Unlike the original case of Hitchin where the Lagrangian only depends on a three-form field, these are theories of two- and three-form fields. We named these theories `background independent Hitchin theories'. Here `background independent' refers to the fact that one does not pick by hand a particular cohomology class inside which to vary our forms. Instead, the closedness of the three-form field is now imposed as a dynamical equation obtained by varying with respect the two-form field. These theories were explicitly proved to be topological in \cite{Herfray:2017imd}, by carrying out the Hamiltonian analysis and exhibiting the constraints. We also constructed a theory of two- and three-forms whose critical points are nearly Kähler manifolds.

 An interesting direction for future work is that, in \cite{Hitchin:2004ut}, Hitchin described a generalisation of the volume functional $\Phi\left[C\right]$ to all odd or even polyforms
in 6D. There is thus a generalisation of all the theories described in the second part of this thesis to polyforms, necessarily involving forms of all degree. It would be interesting to study these theories, and characterise them in terms of the degrees of freedom they propagate as well as their dimensional reduction. It would also be very interesting if this theory,  whose dimensional reduction describes three dimensional gravity in six dimensions could be related to double field theory (see \cite{Hohm:2013bwa} for a review). In this respect looking for a relationship between some version of eight-dimensional Hitchin theory and four dimensional gravity would also seem very natural. Such a relationship does not sound completely impossible but, at this point, is however highly hypothetical. 

Finally we remark that, at least for a positive cosmological constant, three dimensional quantum gravity is reasonably well understood and can be constructed using the Turaev-Viro state sum model \cite{Reshetikhin:1991tc}. The gravity partition function is then the square of the Chern-Simons one, as expected from the action. Our interpretation of 3D gravity as sitting inside the 6D Hitchin theory suggests that it should also be possible to construct the 6D quantum theory. It is likely that the case of three-forms of positive type, which is related to 3D gravity with positive cosmological constant, should be the simplest starting point. It would be very interesting to attempt to define the quantum theory by some state sum model in 6D, so that this reduces to the Turaev-Viro model when the 6D manifold is of the product form $P = \SU(2) \times M$. In turn, the 3D understanding may help to construct the 6D quantum theory.

Turning to the seven-dimensional version of the Hitchin's story, we found that a certain chiral deformation of gravity in four dimensions can naturally be lifted to solutions to Hitchin's equations and thus give $G_2$ holonomy manifold. This result had first been described in \cite{Herfray:2016azk}. However, since the critical points of the 7D Hitchin theory are G2 holonomy manifolds and these are Ricci flat, there is no natural setup in which this theory would be dimensionally reduced to 4D on a compact manifold and give rise to a version of 4D gravity. Thus, this theory is not very promising from the perspective of obtaining a 4D gravity by the dimensional reduction on a compact 3D manifold. \todo{ref to joel's work}

What turned out to be more promising is the addition of a kinematical term $CdC$ to the Hitchin action $\Phi\left[C\right]$. The resulting action then described nearly parallel $G_2$ structures. This is more promising because the dimensional reduction of the nearly G2 structure on the round seven-sphere gives the round metric on the four-sphere. This is just the 7D version of the Hopf fibration that views S7 as the S3 fibre bundle over S4. What is more the construction naturally extends to $\SU(2)$-principal bundle over self-dual Einstein solutions. 

Considering the $\SU(2)$-reduction of this theory one obtains what can be interpreted as another chiral deformation of gravity together with a scalar field, see \cite{Krasnov:2016wvc,Krasnov:2017uam}.

The physical interpretation -if any- of this theory however is unclear:
There are clearly no matter degrees of freedom described by this action, definitely no fermionic degrees of freedom. Thus, if this set of ideas is ever to be developed into a physical theory, one must define how other known bosonic fields (e.g. gauge fields) and fermions couple to this type of gravity. The fact that it seems to be possible to describe gravity with differential forms suggests that one should try to use the same formalism for describing all other building blocks of Nature. How or if this can be done remain completely open questions.

The other open question is whether the theory under consideration reduces to General Relativity in some regime. As we already discussed, it is possible to get a 4D theory that is arbitrarily close to General Relativity by
dimensionally reducing the theory on a three sphere of a fixed size $\phi = cst$ and tuning this constant appropriately. `Physical' implication of this type of deformation have already been discussed for spherically symmetric and anisotropic cosmological solutions, see \cite{Krasnov:2007ky} and  \cite{Herfray:2015fpa} respectively. In particular it was found that this type of simple modifications easily shows `singularity resolution' mechanism. Typically the metric is ill defined or degenerate around the GR singularity but the $\SU(2)$-connection, for which we solve the field equations, is completely smooth.

 By considering this theory we however artificially froze the scalar degree of freedom $\phi$. This is most likely inconsistent. At least this is the case with the more familiar Kaluza-Klein theories, see e.g. \cite{Duff:1986hr}. The right approach should be to allow this field to be dynamical, and let it settle dynamically to some value. However, as  was shown in \cite{Krasnov:2017uam}, the natural values are $\phi = 5/6$ for the squashed seven sphere and $\phi = 2$ for the round sphere solutions so that none of this is the $\phi=1$ case that would give an approximately flat 4D base. So, overall, there appears to be no solution of the reduced theory that approximates General Relativity. This is disappointing but maybe not very surprising as this sort of difficulty is shared with the more traditional Kaluza-Klein theories where explaining what tunes the size of the extra dimensions to a phenomenologically acceptable value is already problematic.
 
 Finally, one should point that a generic difficulty of the type of reformulations we discussed is to deal with the Lorentzian signature. As we already already pointed out it indeed does not seem easy, but may be not impossible, to extend our twistor `connection approach' to complexified space-time - and indeed such space already appeared naturally in our version of the non-linear graviton theorem. In fact it might be that this is the way forward and that the difficulty that we encounter when trying to describe the full GR in this setting is related to our emphasis on the fibre bundle setup which pertain the the Euclidean signature. On the other hand the `chiral deformations' that typically appeared in our $\SU(2)$-reduction reductions of Hitchin theory make perfect sense on complexified space-time but it is unclear what reality conditions one should pick up to recover the Lorentzian signature. This is as opposed to the Euclidean case where, as we emphasised all along this thesis, the notion of `definite connections' is the good reality condition. 
 
All the results described, while not giving a fully satisfactory higher dimensional perspective on 4D General Relativity, do show convincingly that the four-dimensional metric structure of GR can be encoded by a differential form in a higher-dimensional space, be that the contact one-form in our twistor story, or the three-form in our version of the Hitchin's story. This does suggest a new perspective on 4D GR, as promised in the title of this thesis. Whether this new perspective will turn out to be useful remains to be seen. But whatever the future developments may bring, we believe our results show that there is still a lot of hidden structure in our usual, four-dimensional, non-supersymmetric, General Relativity.

\end{Conclusion}

\begin{Appendix}
\appendix

\cleardoublepage
\addtocontents{toc}{\begingroup\def\protect\cftpartpresnum{}}
\addstarredpart{Appendix}
\addtocontents{toc}{\endgroup}
\part*{Appendix}

\chapter{General Conventions}

\section{Density and All That}\label{Section : Appdx Density }

Let $Id^n \in \W^n TM \otimes \gO^n(M)$ be the invariantly defined tensor given by the identity of n-vector
\begin{equation}
Id_{\W^n TM} \from \W^n TM \to \W^n TM.
\end{equation}
Alternatively this is the only tensor $Id^n \in \W^n TM \otimes \gO^n(M)$ that gives one when contracted with itself.

In tensorial notation it reads\footnote{$\epst$ and $\epsut$ are completely antisymmetric tensor of weight 1 and -1 respectively.}
\begin{equation}
Id^n =\left(\frac{1}{n!}\right)^2\; dx^n\; \epst^{\mu_1 ... \mu_n } \pa_{\mu_1}\W...\W\pa_{\mu_n} = \left(\frac{1}{n!}\right)^2\; dx^{\mu_1}\W...\W dx^{\mu_n} \epsut_{\mu_1 ... \mu_n } \; \pa^n
\end{equation}
where $dx^n \coloneqq dx^1\W ... \W dx^n$ and $\pa^n \coloneqq \pa_1 \W...\W \pa_n \in \W^n TM$.\\

\begin{ExtraComputation}
	\begin{framed}
\begin{align*}
Id^n &= \frac{1}{n!}\; dx^n \; \pa^n  \\
 &= \frac{1}{n!}\; \frac{dx^{\mu_1}\W...\W dx^{\mu_n}}{n!} \epsut_{\mu_1 ... \mu_n }\; \epst^{\nu_1 ... \nu_n } \frac{\pa_{\nu_1}\W...\W\pa_{\nu_n}}{n!} \\
 &= \frac{1}{n!}\;  \epsut_{\mu_1 ... \mu_n }\;dx^{\mu_1}\otimes...\otimes dx^{\mu_n}\; \epst^{\nu_1 ... \nu_n } \pa_{\nu_1}\otimes...\otimes\pa_{\nu_n}
\end{align*}
	\end{framed}
\end{ExtraComputation}

This tensor is useful to give a concrete form to the isomorphism $\gO^k(M) \simeq TM^{n-k}(M) \otimes \gO^n(M)$:

\begin{tcolorbox}
\begin{equation}\label{Appendix: Weight isom}
\begin{array}{ccc}
\gO^k(M) & \to & TM^{n-k}(M) \otimes \gO^n(M) \\ \\
\gr & \mapsto &  \grt = \frac{n!}{k!}\; id^n \id \gr
\end{array}
\end{equation}
\end{tcolorbox}
Using the tensorial notation,
\begin{equation}
\gr = \frac{1}{k!}\gr_{\mu_1 ...\mu_k} dx^{\mu_1}\W ... \W dx^{\mu_k},\qquad \grt = \frac{1}{(n-k)!} \grt^{\mu_{1} ... \mu_{n-k}} \;dx^n \otimes \pa_{\mu_{1}} \W ... \W \pa_{\mu_{n-k}}.
\end{equation}
So that in coordinates, this operation reads:
\begin{tcolorbox}
\begin{equation}\label{Appendix: Weight def}
\grt^{\mu_{1} ... \mu_{n-k}} \coloneqq \frac{1}{k!} \;\epst^{\mu_{n-k+1} ...\mu_n \;\mu_{1} ... \mu_{n-k}} \gr_{\mu_{n-k+1} ...\mu_n}
\end{equation}
\end{tcolorbox}
The numerical coefficients are chosen such that the inverse operation takes a similar form:
\begin{equation}\label{Appendix: Weight def inverse}
\gr_{\mu_{1} ... \mu_{k}} \coloneqq \frac{1}{(n-k)!} \;\epsut_{\mu_1 ... \mu_n} \grt^{\mu_{k+1} ...\mu_n}.
\end{equation}

\section{The Lie algebra of $\SU(2)$ and its representations}\label{Appdx: The Lie Algebra of SU(2)}

\subsection*{Intrinsic definition}
$\su$2 is the three dimensional Lie algebra defined by
\begin{tcolorbox}
\begin{tabular}{ccc}
	 $\su2=Span(\gs_i), i\in(1,2,3)$& and the algebra &
$[\gs^i,\gs^j]=\epsilon^{ijk}\gs^k$.
\end{tabular}
\end{tcolorbox}

For convenience, everywhere in the text we take the overall factor of the Killing metric on $\su(2)$ to be defined by
\begin{tcolorbox}
\begin{equation*}
K\left(\gs_i, \gs_j\right) = \gd_{ij}.
\end{equation*}
\end{tcolorbox}

\begin{ExtraComputation}
	\begin{framed}
\noindent We have the alternative basis:
\begin{equation*}
\gs_+=-i\gs_1-\gs_2, \qquad 
\gs_-=i\gs_1-\gs_2 \qquad 
\gs_0=-i\gs_3.
\end{equation*}
The associated algebra is
\begin{equation}
\begin{array}{lll}
[\gs_-,\gs_+]=2\gs_0 \qquad \qquad & [\gs_0,\gs_+]=\gs_+ \qquad \qquad & [\gs_0,\gs_-]=-\gs_- \qquad
\end{array}
\nonumber
\end{equation}
Using a spinor notation
\footnote{$\gs_i{}^{A}{}_{B}$ is defined by the relation
	\[ V^i=\{x,y,z\}, \qquad  v^i\gs_i{}^{A}{}_{B} = \frac{1}{2i}\Mtx{z & x-iy \\
		x+iy & -z}\], }, we can also write these relations as 
\[ 
\gs_{AB} = \gs_i \gs^i{}_{AB}   = \frac{1}{2} \Mtx{ i\gs_1 - \gs_2& -i\gs_3\\
	-i\gs_3	& -i\gs_1 - \gs_2} = \frac{1}{2} \Mtx{\gs_- & \gs_0 \\ \gs_0 & \gs_+},
\]
and for any $M\in \su2$,
\[ 
M = M^i \gs_i = M^{AB} \left( 2\gs_{AB}\right).
\]
	\end{framed}
\end{ExtraComputation}

\subsection*{Spin $1/2$ representation of $\su(2)$}

In the fundamental representation, elements of $\su2$ corresponds to hermitian tracefree matrices. The explicit isomorphism is:
\begin{tcolorbox}
\begin{equation}\label{Appdx: Spin 1/2 representation of su2}
\gt\text{:}\left \{
\begin{array}{llc}
su2 & \longmapsto & \text{Hermitian tracefree}\\
v^i \gs_i &\qquad & v^i \gs_i{}^{A'}{}_{B'} 
\end{array} \right.
\end{equation}
where \begin{equation*}
v^i=\{x,y,z\}, \qquad  v^i\gs_i{}^{A}{}_{B} = \frac{1}{2i}\Mtx{z & x-iy \\
	x+iy & -z}
\end{equation*}
\end{tcolorbox}
One can indeed directly check the algebra:\footnote{spinor indices A,B... are raised and lowered with $\eps_{AB}$ according to the usual spinor covention.}
\begin{equation*}
\gs^i_{AC} \gs^j{}^C{}_B = \frac{\gd^{ij}}{4} \eps_{AB}  + \frac{\eps^{ijk}}{2} \gs^k_{AB}.
\end{equation*}
So that, in particular (here matrix multiplication is implied)
\begin{tcolorbox}
\begin{equation*}
\left[ \gs_i , \gs_j \right]= \eps_{ij}{}_{k} \gs_k, \qquad -2\;\Tr\left( \gs_i \gs_j\right) = \gd_{ij}.
\end{equation*}
\end{tcolorbox}
To be clear, the second equation reads,
\begin{equation*}
-2 \gs_i{}_{AB} \gs_j{}^{BA} = \gd_{ij}.
\end{equation*}
So that we have for any $M\in \su(2)$:
\begin{equation*}
\begin{array}{lll}
M_{AB}=M^i \gs_i{}_{AB}
&\qquad \Leftrightarrow \qquad&
M^i = M_{AB} \gs^{AB}{}^i.
\end{array}
\end{equation*}

\begin{ExtraComputation}
	\begin{framed}
\paragraph{NB:} With this notation the spin $\frac{1}{2}$ representation, $M^{A'}{}_{B'}=M^i \gs_i{}^{A'}{}_{B'}$, of a Lie algebra element $M= M^i \gs_i $ coincide with its coordinates in the creation-annihilation basis, $M= M^{A'B'} \left( 2\gt_{A'B'}\right)$.
\\
We indeed have,
\[ 
M= M^{i} \gs_{i}= 2M^{A'B'} \gs^i{}_{A'B'} \gs_{i}= M^{A'B'} \left( 2\gs_{A'B'}\right).
\]
	\end{framed}
\end{ExtraComputation}

\subsection*{Spin $1$ representation of $\su(2)$}
The adjoint representation of $\mathfrak{su}2$ is also the defining representation of the Lie algebra of $SO(3)$. It corresponds to the $3\times3$ antisymmetric matrices such that:
\begin{equation*}
exp(T^k{}_i) \; \delta_{kl} \; exp( T^l{}_j) =  \delta_{ij} \Rightarrow T_{ij}+T_{ji}=0 .
\end{equation*}
The explicit isomorphism now is:
\begin{tcolorbox}
\begin{equation}\label{Appdx: Spin 1 representation of su2}
\gs\text{:}\left \{
\begin{array}{llc}
su2 & \longmapsto & \text{3$\times$3 antisymmetric}\\
v^i \gs_i &\qquad & v^i \gs_i{}^{j}{}_{k} 
\end{array} \right.
\end{equation}
where
\begin{equation*}
\left( \gs_i \right)_{jk}=-\epsilon_{i}{}_{jk}.
\end{equation*}
\end{tcolorbox}
Again, one can check that the algebra (matrix multiplication implied) is:
\begin{tcolorbox}
\begin{equation*}
\left[ \gs_i , \gs_j \right]= \epsilon_{ij}{}_{k} \gs_k, \qquad -\frac{1}{2}\Tr\left(\gs_i , \gs_j \right) = \gd_{ij}
\end{equation*}
\end{tcolorbox}
For clarity the second equation more explicitly reads
\begin{equation*}
-\frac{1}{2}\gs_{iab} \gs_j{}^{ba} =\gd_{ij}.
\end{equation*}
So that we have for any $M\in \su(2)$:
\begin{equation*}
\begin{array}{lll}
M_{ab}=M^i \gs_i{}_{ab} = -M^i\epsilon_{i}{}_{ab}
&\qquad \Leftrightarrow \qquad&
M^i = -\frac{1}{2}\epsilon^{i}{}^{ab}M_{ab}.
\end{array}
\end{equation*}

\subsection*{Self-dual two-forms in 4D as $\su2$ representation}

One can also represent $\su(2)$ elements by self-dual two-forms in four dimension. The explicit isomorphism now is:
\begin{tcolorbox}
	\begin{equation}\label{Appdx: self-dual representation of su2}
	\gS\text{:}\left \{
	\begin{array}{llc}
	su2 & \longmapsto & \text{self-dual two-forms}\\
	v^i \tau_i &\qquad & v^i \frac{\gS^i }{2}
	\end{array} \right.
	\end{equation}
	where
	\begin{equation*}
	\gS^i = -e^0 \W e^i - \frac{\eps^{ijk}}{2} e^j\W e^k
	\end{equation*}
\end{tcolorbox}

One can indeed check the algebra,
\begin{equation*}
\gS^i{}_{IK} \gS^j{}^K{}_J = -\gd^{ij} g_{IJ} +\eps^{ijk}\gS^k_{IJ}.
\end{equation*}
So that in particular
\begin{tcolorbox}
\begin{equation*}
\left[\gS^i ,\gS^j \right] = 2\eps^{ijk} \gS^k, \qquad \frac{1}{4}\Tr\left(\gS^i, \gS^j\right) = \gd^{ij}.
\end{equation*}	
\end{tcolorbox}

\subsubsection*{Self-dual two-forms in 4D and spinor notation.}

Let $\gL = \gL^i \gs^i \in \su(2)$, let $\gL^i \frac{\gS^i}{2}$ be the associated self-dual two-form.

\begin{tcolorbox}
 Depending on the notation (tetras or spinor notation) it can be written 
\begin{equation*}
\gL = \gL^i \frac{\gS^i}{2} = \gL_{A'B'} \;\gS^{A'B'} = -\gL_{A'B'}\eps_{AB}\;\frac{e^{A'A}\W e^{B'B}}{2}
\end{equation*}
What's more if $M$ and $N$ are two self-dual two-forms our conventions are such that the algebra works out properly:\\

In the standard (tetrad) notation is explicitly the $1/2\otimes 0$ representation of $\SO(4)$:
	\[ \left\{
	\begin{array}{ll}
	\gL &\longmapsto \gL^i \frac{\gS^i}{2}  = \gL^i \gS^i{}_{IJ}\frac{ e^I \W e^J}{4}  
	\\ \\ \relax
	[M,N] &\longmapsto \left( \eps^{ijk} M^iN^j\right) \frac{\gS^k}{2} = \left( M_{IK}N^{K}{}_{J} - N_{IK}M^{K}{}_{J}\right)\frac{ e^I \W e^J}{4}
	\end{array}\right.
	\]
	
While in a spinor notation, the coefficient of the two-form are again the spin $1/2$ representation of $\su2$:
	\begin{equation*}
	\left\{
	\begin{array}{ll}
	\gL &\longmapsto \gL_{A'B'} \gS^{A'B'} = \gL^i\left(-\gt^i{}_{A'B'}\eps_{AB} \right) \frac{ e^{AA'} \W e^{BB'}}{2}
	\\ \\ \relax
	[M,N] &\longmapsto \left( M_{A'C'}N^{C'}{}_{B'} - N_{A'C'}M^{C'}{}_{B'} \right) \gS^{A'B'}
	\end{array}\right.
	\end{equation*}
\end{tcolorbox}

This can be checked by a direct computation:
\begin{equation*}
\begin{array}{ll}
[M,N] &= \left(M_{A'C'}N^{C'}{}_{B'} - N_{A'C'}M^{C'}{}_{B'}  \right) \gS^{A'B'} \\ \\ 
&= \left(\gt^i{}_{A'C'}\gt^j{}^{C'}{}_{B'} - \gt^j{}_{A'C'}\gt^i{}^{C'}{}_{B'} \right)M^i N^j\; \gS^{A'B'} \\ \\
&= \eps^{ijk}\gt^k{}_{A'B'} M^i N^j\; \gS^{A'B'}\\ \\
&= \left( \eps^{ijk}\;M^i N^j \; \right)\frac{\gS^k}{2} =\frac{1}{2} \left( \eps^{ijk}\gS^k{}_{IJ} \right)\;M^i N^j \;\frac{e^I \W e^J}{2} \\ \\
&= \frac{1}{4}\left( \gS^i_{IK}\gS^j{}^{K}{}_{J} -\gS^i{}_{IK}\gS^j{}^{K}{}_{J} \right)M^i N^j \frac{ e^I \W e^J}{2}\\ \\
&=\left( M_{IK}N^{K}{}_{J} - N_{IK}M^{K}{}_{J}\right)\frac{ e^I \W e^J}{2}.
\end{array}
\end{equation*}

It follows from this discussion that the self-dual basis $\left\{ \frac{1}{2}\gS^i = \gS^{A'B'} \gs^i_{A'B'} \right\}$ can be thought of as the isomorphism between  $\mathfrak{su}2 $ and self-dual two-forms (which is the $1/2 \otimes 0$ representation of $\su2$):
\begin{tcolorbox}
\begin{equation*}
\frac{1}{2}\gS\in \gO^2 \otimes \su(2)^\ast\text{:}\left \{
\begin{array}{llc}
su2 & \longmapsto &\gO^2 \\
v^i \gs_i &\qquad & v^i \frac{\gS_{i}}{2} = v^{A'B'} \gS_{A'B'} .
\end{array} \right. \nonumber
\end{equation*}
\end{tcolorbox}

Alternatively $\left\{ \frac{1}{2}\gS^i = \gS^{A'B'} \gs^i_{A'B'} \right\}$ can be thought of as an $\mathfrak{su}2 $-valued two form :
\begin{equation*}
\gS \in \gO^2 \otimes \mathfrak{su}2 = \gS^i \gs_i =  \gS^{A'B'} 2\gs_{A'B'}. 
\end{equation*}

\chapter{4d Space-Time Conventions}\label{Chapter : Appdx 4d Space-Time Conventions}

\section{Decomposition of the Riemann Curvature Tensor in Coordinates}\label{section: Appdx Decomposition of the Curvature}

In this appendix we prove, using coordinates, the different claims made in the first part of chapter \ref{Chapter: Chiral formulation of gravity}.

In this appendix we use freely the isomorphism $\so(4)\simeq \gO^2$ to represent elements of $\so(4, \R)$ as two-forms. I.e, we pick up a basis of one-forms, $\left\{e^I\right\}_{I\in \{0...3\}}$ compatible with the metric, $ds^2= e^I \otimes e^I$, and write  for $\boldsymbol{b} \in \su(2)$ as $ \boldsymbol{b} = b_{IJ} \frac{e^I \W e^J}{2} $, with abuse of notation . The metric allows to raise and lower $I,J,K...$ indices. With this notations, the Lie bracket reads,
\begin{equation}
\boldsymbol{a},\boldsymbol{b}\in \so(4), \qquad \left[\boldsymbol{a},\boldsymbol{b}\right] = \left( a_{I}{}^{K} b_{KJ} - b_{I}{}^{K} a_{KJ} \right)\frac{e^I \W e^J}{2}.
\end{equation}
Then for any $\boldsymbol{b} \in \su(2)$ the decomposition $\so(4) = \su(2)\oplus \su(2)$ reads:
\begin{equation}\label{Appdx: two-form decomposition}
b_{IJ} \frac{e^I\W e^J}{2}= B^i \frac{\gS^i}{2} + \widetilde{B}^i \frac{\gSt^i}{2}. 
\end{equation} 
Where the sigma tensors coincide with the one described at the end of the first chapter, see section \ref{ssection: Two useful tensors}. In terms of the tetrad they take the explicit form:
\begin{tcolorbox}
\begin{align}\label{Appdx: Sigma def (tetrad)}
&\left\{\gS^i = -e^0 \W e^i - \frac{\eps^{ijk}}{2} e^j \W e^k\right\}_{i\in 1,2,3}, 
&\left\{\gSt^i = e^0 \W e^i - \frac{\eps^{ijk}}{2} e^j \W e^k\right\}_{i\in 1,2,3}.
\end{align}

They form a basis of self-dual and anti-self-dual two-forms respectively. This basis is orthogonal for the wedge product:
\begin{equation}\label{Appdx: Sigma orthogonality}
\gS^i \W \gS^j = -\gSt^i \W \gSt^j = 2 \gd^{ij} e^0\W e^1 \W e^2 \W e^3, \qquad \gS^i \W \gSt^j = 0.
\end{equation}
\end{tcolorbox}
As was already stated in the main part of this thesis, the decomposition of Lie algebra $\so(4)=\su(2)\oplus \su(2)$  corresponds to the decomposition $\gO^2 = \gO^2_+ \oplus \gO^2_-$ of two-forms:
\begin{equation}
 \left[\gS^i ,\gS^j \right] = 2\eps^{ijk} \gS^k, \quad \left[\gSt^i ,\gSt^j \right] = 2\eps^{ijk} \gSt^k, \quad \left[\gS^i ,\gSt^j \right] = 0.
\end{equation}

In what follows we will make an important use of the tensors, $\gS^i_{IJ}$, $\gSt^i_{IJ}$ defined by $\gS^i = \gS^i_{IJ} \frac{e^I\W e^J}{2}$, $\gSt^i = \gSt^i_{IJ} \frac{e^I\W e^J}{2}$. They verify the algebra,
\begin{tcolorbox}
\begin{equation}\label{Appdx: Sigma algebra}
\begin{array}{ccc}
\gS^i{}_{IK} \gS^j{}^K{}_J = -\gd^{ij} g_{IJ} +\eps^{ijk}\gS^k_{IJ}, \qquad \gSt^i{}_{IK} \gSt^j{}^K{}_J =-\gd^{ij} g_{IJ} +\eps^{ijk}\gSt^k_{IJ} ,\\ \\ \gS^{i}{}_{IK} \gSt^{j}{}^K{}_J = s^{ij}_{IJ}.
\end{array}
\end{equation}
\end{tcolorbox}
Where $s^{ij}_{IJ}$ is a tensor with the following symmetries:
\begin{equation}
s^{ij}_{[IJ]}=0, \quad s^{[ij]}_{IJ}=0.
\end{equation}
Note that (anti)-self-duality explicitly reads
\begin{tcolorbox}
\begin{equation}\label{Appdx: Sigma self-duality}
\gS^i{}^{IJ} = \frac{\eps^{IJKL}}{2} \gS^i_{KL},\qquad \qquad \gSt^i{}^{IJ} = -\frac{\eps^{IJKL}}{2} \gSt^i_{KL}.
\end{equation}
\end{tcolorbox}

\paragraph{Decomposition of the Curvature tensor in coordinates} \mbox{} \\
Consider a 4d Riemannian manifold $\{g, M\}$, $\{e^I\}_{I\in 0..4}$ an orthonormal co-frame. The Levi-Civita connection, $\nabla $, then naturally is a $SO(4)$-connection. We will write its potential one-form $\boldsymbol{a}$ and curvature two-form $\boldsymbol{f}$ as
\begin{equation}
a^I{}_J = a^{I}{}_{J\;K} e^K, \qquad f^I{}_J = d a^I{}_J + a^I{}_K \W a^K{}_J = f^I{}_J{}_{KL} \frac{e^K \W e^L}{2}.
\end{equation}
Note that the Riemann curvature $\bdf$ here is a $\so(4)$-valued two-form.

Now we can use the decomposition $\so(4)= \su(2) \oplus \su(2)$, concretely realised as \eqref{Appdx: two-form decomposition}, to define the chiral connections $\left(D,\Dt \right)$ with potential $\left(A,\At \right)$ as
\begin{equation}\label{Appdx: LeviCivita split}
a^I{}_J = A^i \frac{\gS^i{}^I{}_J}{2} + \At^i \frac{\gSt^i{}^I{}_J}{2}.
\end{equation}
These connections naturally are $SU(2)$-connections.  In chapter \ref{Chapter: Chiral formulation of gravity} we stated that these connections are compatible with $\gS^i$, $\gSt^i$ in the following sense:
\begin{equation}\label{Appdx: Sigma/A compatibility}
d_A \gS^i = 0, \qquad d_{\At} \gSt^i=0.
\end{equation}
We can prove this by a direct computation:
\begin{proof}
	
\begin{align*}
d_A \gS^i &= \frac{1}{2} d_A \left( \gS^i_{IJ} e^I \W e^J \right)\\
&=\frac{1}{2} d_A \left( \gS^i_{IJ}\right) e^I \W e^J\\ 
& = \frac{e^I \W e^J}{2} \W  \left( \eps^{ijk} A^j \gS^k_{IJ} - 2\gS^i_{IK} a^K{}_J  \right) \\
&= \frac{e^I \W e^J}{2} \W  \left( \eps^{ijk} A^j \gS^k_{IJ} - 2A^j \gS^i_{IK}  \frac{\gS^j{}^K{}_J}{2} -2 \At^j \gS^i_{IK}\frac{\gSt^j{}^K{}_J}{2}  \right)\\
&=0
\end{align*}
where in step 1 we used the torsion freeness of $a$ (i.e $d_a e^I =0$), step 3 is just the decomposition of the Levi Civita connection into its chiral parts, (i.e, eq\eqref{Appdx: LeviCivita split}) and at step 4 we made use the algebra \eqref{Appdx: Sigma algebra}.
\end{proof}

As already stated in the main body of this thesis, the relations \eqref{Appdx: Sigma/A compatibility} can be used as an alternative way of defining $A$(resp $\At$) as the unique $SU(2)$-connection compatible with $\gS^i$(resp $\gSt^i$). 

\begin{proof}
	
Let us suppose that $A$ and $A' = A + M$ are are both $\SU(2)$-connections compatible with $\gS^i$. It follows that
\begin{equation}
d_{A'} \gS^i - d_A \gS^i = \eps^{ijk} M^j \gS^k =0,
\end{equation}
or equivalently, by making use of the self-duality of $\gS$,
\begin{equation}\label{Appdx: proof A unicity (1)}
M_{\nu}^{[i} \gS^{j]}{}^{\mu\nu}=0.
\end{equation}
By multiplying  this expression by another sigma symbol we have
\begin{align*}
0&= \eps^{ijk} M^j_{\nu} \gS^k{}^{\mu\nu} \gS^l_{\mu\gr}\\
 &= \eps^{ijk} \eps^{klm} M^j_{\nu} \gS^m{}^{\nu}{}_{\gr} + \eps^{ijl} M^j_{\gr} \\
&=\gd^{il} M^k_{\nu} \gS^k{}^{nu}{}_{\gr} -M^l_{\nu}\gS^i{}^{\nu}{}_{\gr} +  \eps^{ijl} M^j_{\gr} 
\end{align*}
where we made use of the algebra \eqref{Appdx: Sigma algebra} and the identity $\eps^{abm} \eps^{ijm} = \gd^{ai}\gd^{bj} - \gd^{aj}\gd^{bi}$. Anti-symmetrising this last expression in the $i,l$ indices  and making use of \eqref{Appdx: proof A unicity (1)} we obtain
\begin{equation}
0=-M^{[l}_{\nu}\gS^{i]}{}^{\nu}{}_{\gr}+ \eps^{ilj} M^j_{\gr} = \eps^{ilj} M^j_{\gr}.
\end{equation}
Which conclude the proof that there is a unique connection satisfying \eqref{Appdx: Sigma/A compatibility}.
\end{proof}

In complete parallel with \eqref{Appdx: LeviCivita split} we define the `self-dual part of the Curvature' $F$ and the `anti-self-dual part of the Curvature' $\Ft$ as
\begin{equation}\label{Curvature two-form split}
f^I{}_J = F^i \frac{\gS^i{}^I{}_J}{2} + \Ft^i \frac{\gSt^i{}^I{}_J}{2},
\end{equation}
and these are naturally $\su(2)$-valued two-forms.
In fact we have,
\begin{equation}
F^i = dA^i +\frac{\eps^{ijk}}{2} A^j \W A^k, \qquad \Ft^i = d\At^i +\frac{\eps^{ijk}}{2} \At^j \W \At^k,
\end{equation}
as can be seen using the algebra \eqref{Appdx: Sigma algebra}. I.e, the (anti-)self-dual part of the curvature is the curvature of the (anti-)self-dual connection.

Now $F^i$, $\Ft^i$ being ($\su(2)$-valued) two-forms, we can decompose them into self-dual and anti-self-dual pieces: 
\begin{equation}
F^i = F^{ij} \gS^j + G^{ij} \gSt^j, \qquad \Ft^i = \Gt^{ij} \gS^j + \Ft^{ij} \gSt^j.
\end{equation}
This is just another way of writing the bloc decomposition \eqref{Chiral Formulations of GR - Fundations: F, Ft decomposition}.
The Riemann curvature now reads
 \begin{equation}
f^I{}_J = \frac{1}{2}\left(  F^{ij}\; \gS^j \; \gS^i{}^I{}_J + G^{ij} \; \gS^j \;\gSt^i{}^I{}_J +  \Gt^{ij} \;\gSt^j \;\gS^i{}^I{}_J + \Ft^{ij} \;\gSt^j \; \gSt^i{}^I{}_J \right).
\end{equation}
Again, this is just another version of the bloc decomposition \eqref{Chiral Formulations of GR - Fundations: Riemann decomposition 1}.
To get the final form of this decomposition we write 
\begin{equation}
F^{ij} = \frac{\gL}{3} \gd^{ij} +\Psi^{ij}, \qquad \Ft^{ij} = \frac{\gLt}{3} \gd^{ij} +\Psit^{ij},
\end{equation}
with $\Psi$, $\Psit$ some traceless tensors and $\gL = tr F$, $\gLt =tr\Ft$. Finally, we can write:

\begin{tcolorbox}
\begin{align}\label{Appdx: Decomposition of the Riemann tensor}
f^I{}_J &= \underbrace{\frac{\gLt}{3} \;\gSt^i \; \frac{\gSt^i{}^I{}_J}{2} + \frac{\gL}{3} \;\gS^i \; \frac{\gS^i{}^I{}_J}{2}}_{\text{Scalar Part}}    \nonumber\\
&+ \underbrace{ \frac{1}{2}\left(G^{ij} \; \gS^j \;\gSt^i{}^I{}_J +  \Gt^{ij} \;\gSt^j \;\gS^i{}^I{}_J+\Psi^{[ij]} \;\gS^j \; \gS^i{}^I{}_J + \Psit^{[ij]} \;\gSt^j \; \gSt^i{}^I{}_J\right)}_{\text{Ricci Part}} \nonumber\\& + \underbrace{\Psi^{(ij)} \;\gS^j \; \frac{\gS^i{}^I{}_J}{2} + \Psit^{(ij)} \;\gSt^j \; \frac{\gSt^i{}^I{}_J}{2}}_{\text{Weyl Part}} 
\end{align} 
This is just a hands-on way of rewriting the bloc decomposition \eqref{Chiral Formulations of GR - Fundations: Riemann decomposition 1}. One can identify the following elementary brick of the Riemann tensor:
\begin{equation}\label{Appdx: Decomposition of the Riemann tensor 2}
\begin{array}{ll}
W_{IJKL}= \frac{1}{2}\Psi^{(ij)} \;\gS^i_{IJ} \; \gS^j{}_{KL} \quad  &\text{is the self-dual part}\\ & \text{of the Weyl~tensor}, \\ \\
\Wt_{IJKL} = \frac{1}{2}\Psit^{(ij)} \;\gSt^i_{IJ} \; \gSt^j{}_{KL} \quad  &\text{is the anti-self-dual part } \\ &  \text{of the Weyl~tensor,}\\ \\
R = 2\gL +2\gLt \quad &\text{is the Scalar curvature,} \\ \\
\text{and the traceless Ricci tensor is},& \\ \\ R^{I}{}_{J} = \underbrace{\frac{1}{2} \left(G^{ji}+\Gt^{ij} \right) \gS^i{}^{KI}\; \gSt^j{}_{KJ}}_{\text{symetric traceless Ricci}} & \underbrace{-\frac{1}{2}\left(\Psi^{[ij]} \;e^{ijk}\; \gS^k{}^I{}_{J} \; +\Psit^{[ij]} \;e^{ijk}\; \gSt^k{}^I{}_{J}\right)}_{\text{anti-symetric Ricci}}   \\ \\

\end{array}
\end{equation}
\end{tcolorbox}
These can be related to the usual definitions
\begin{align*}
R = f^{KL}{}_{KL}, \qquad R^I{}_J = f^{KI}{}_{KJ}-\frac{1}{4}R\;\gd^I{}_{J}, \;\text{with}\; F^I{}_J = F^I{}_{JKL} \frac{e^K \W e^L}{2}
\end{align*}
 by contracting indices in \eqref{Appdx: Decomposition of the Riemann tensor} and using the algebra \eqref{Appdx: Sigma algebra}.

\begin{ExtraComputation}
	\begin{framed}
More explicitly:
\begin{align*}
f^{KL}{}_{KL} &= \frac{1}{2}\Psi^{(ij)} \;\gS^i_{KL} \; \gS^j{}^{KL}+\frac{1}{2}\Psit^{(ij)} \;\gSt^i_{KL} \; \gSt^j{}^{KL} \\
&=  2 \;\Tr \Psi + 2\; \Tr \Psit
\end{align*}

\begin{align*}
R^I{}_{J} &\coloneqq f^{KI}{}_{KJ}-\frac{1}{4}R\;\gd^I{}_{J} \\
&=  \frac{1}{2}\left(G^{ij} \;\gSt^i{}^{KI}\; \gS^j_{KJ} \; +  \Gt^{ij} \;\gS^i{}^{KI}\; \gSt^j_{KJ}+\Psi^{[ij]} \; \gS^i{}^{KI} \;\gS^j_{KJ} + \Psit^{[ij]} \; \gSt^i{}^{KI} \;\gSt^j_{KJ} \right) \\
&=  \frac{1}{2}\left(\left(G^{ji} +  \Gt^{ij}\right) \;\gS^i{}^{KI}\; \gSt^j_{KJ}+\Psi^{[ij]} \; \gS^i{}^{KI} \;\gS^j_{KJ} + \Psit^{[ij]} \; \gSt^i{}^{KI} \;\gSt^j_{KJ} \right) \\
&=  \frac{1}{2}\left(\left(G^{ji} +  \Gt^{ij}\right) \;\gS^i{}^{KI}\; \gSt^j_{KJ} \right)-\frac{1}{2}\left(\Psi^{[ij]} \;e^{ijk}\; \gS^k{}^I{}_{J} \; +\Psit^{[ij]} \;e^{ijk}\; \gSt^k{}^I{}_{J}\right) 
\end{align*}
\danger Note that by using the $\SL(2,\C) \times \SL(2,\C)$ action $G^{ji} +  \Gt^{ij}$ can always be chosen to be symmetrical.

	\end{framed}
\end{ExtraComputation}

Consequently Einstein equations are equivalent to
\begin{equation}
\Psi^{[ij]}=0,\quad \Psit^{[ij]}=0, \quad G^{ij} + \Gt^{ji}=0.
\end{equation}

As stated in the main body of this paper the torsion freeness of the Levi-Civita connection implies that the Riemann tensor has some internal symmetries (usually called first Bianchi identity):
\begin{equation}
0= d_A d_A e^{I} = F^{I}{}_J \W e^J \; \Leftrightarrow \; F^{I}{}_{[JKL]}=0.
\end{equation}
Together with the skew symmetries of the Riemann tensor it implies $f_{IJKL} = f_{KLIJ}$. It leads to further simplifications:
\begin{equation}
f_{IJKL} = f_{KLIJ} \quad \Rightarrow \quad \Psi^{ij}=\Psi^{(ij)},\quad \Psit^{ij}=\Psit^{(ij)}, \quad \Gt^{ij}=G^{ji}.
\end{equation}
 \begin{equation}
f_{I[JKL]}=0 \quad \Leftrightarrow\quad  f_{NIKL} \eps^{NJKL}=0 \quad \Rightarrow \quad \gL = \gLt.
\end{equation}
The second relation follows from using the (anti)-self duality (see eq \ref{Appdx: Sigma self-duality}) of the sigma tensors.
\begin{ExtraComputation}
	\begin{framed}
\begin{align*}
f_{NIKL} \eps^{NJKL} =  \left(G^{ji} -  \Gt^{ij}\right) \;\gS^i{}_{NI}\; \gSt^j{}^{NJ}-\Psi^{[ij]} \;e^{ijk}\; \gS^k{}^I{}_{J} \; +\Psit^{[ij]} \;e^{ijk}\; \gSt^k{}_I{}^{J} + \left(\gL-\gLt\right)\gd_I{}^J
\end{align*}
	\end{framed}
\end{ExtraComputation}

With those symmetries, Einstein equations
\begin{equation}
R_{IJ} = \gL g_{IJ}, 
\end{equation}
 are equivalent to
\begin{equation}\label{Appdx: Chiral Einstein equations}
F^i = \left( \Psi^{ij} + \frac{\gL}{3} \gd^{ij} \right) \gS^{j}
\end{equation} 
(i.e  $G=0$ ) and we therefore need only one half of the Riemann tensor to state them.

\section{Spinor conventions} \label{Section : Appdx Spinor conventions} 

\subsection{Spinors and $\su(2)$}

\indent We convert $\su(2)$ lie algebra indices into spinor notations according to the rule:\footnote{
	\danger\; As a convention, we use $\gs_i^{AB}$ to convert "spatial indices" $i\in \{1,2,3\}$ into "unprimed spinor indices" and $\gsb_i^{A'B'}$, its complex conjugate, to convert "spatial indices" into "primed spinor indices".}
\begin{tcolorbox}
\begin{equation}\label{Appdx: su(2)/Spinor indices conversion}
\begin{array}{l}
V = V^i \gs_i \in \su(2), \quad V^i=\{x,y,z\} \\ \Leftrightarrow  \\  V^i \gs_i^{A}{}_{B} = V^{A}{}_{B} = \frac{1}{2i}\Mtx{z & x-iy \\
	x+iy & -z} \in \su(2).
\end{array}
\end{equation}
\end{tcolorbox}
Where the $\gs$'s are such that $\left[\gs^i, \gs^j \right] = \eps^{ijk} \gs^k$. Latin indices are raised and lowered with the flat metric $\gd^{ij}$, spinor indices are raised and lowered as usual using the antisymmetric $\eps^{A'B'}$. This is done according to the conventions from \cite{Penrose_vol1,Penrose_vol2}. Primed and unprimed spinor are treated in a completely symmetric way:\footnote{See however the preceding footnote.}

\begin{align}
\eps_{AB}=\eps^{AB}= \eps_{A'B'}=\eps^{A'B'} = \Mtx{0 & 1 \\ -1& 0},
\end{align} 

\begin{align}
\ga^A = \eps^{AB}\ga_B, &\quad  & \ga_A = \ga^B \eps_{BA}, & \quad  &
\ga^{A'} = \eps^{A'B'}\ga_{B'}, &\quad & \ga_{A'} = \ga^{B'} \eps_{B'A'} .
\end{align}

\begin{align}
\eps^{CB}\eps_{CA}= \eps_A{}^B = \gd_A{}^B, & & \eps^{AB}\eps_{AB} = 2.
\end{align}

We will also use the following shorthand for contraction of spinors,
\begin{equation}
\ga.\gb \coloneqq \ga_{A'}\gb^{A'},\qquad \ga.\gb \coloneqq \ga_{A}\gb^{A}.
\end{equation}
In the Euclidean setting spinors are equipped with a quaternionic structure, i.e an anti-linear map $\circonf : S \to S$ that squares to minus one. This is equivalent to equipped spinors with a Hermitian metric:
\begin{align}
\bra\ga, \gb\ket  \coloneqq \gah_{A'}\gb^{A'}  = \gah.\gb \geq 0, \qquad 
\bra\ga, \gb\ket  \coloneqq \gah_{A}\gb^{A}   = \gah.\gb \geq 0 .
\end{align}

We go from one type of indices to the other as follows:
\begin{tcolorbox}
\begin{equation}
V^{AB} = V^i \,\gs_i^{AB} \qquad \Leftrightarrow \qquad V^i = 2\gs^i_{AB} V^{AB}.
\end{equation}
\end{tcolorbox}

Finally the $\gs$ matrices satisfy the following algebra
\begin{tcolorbox}
\begin{equation}
\gs^i_{AC} \gs^j{}^C{}_B = \frac{\gd^{ij}}{4} \eps_{AB}  + \frac{\eps^{ijk}}{2} \gs^k_{AB}
\end{equation}
\end{tcolorbox}
In particular
\begin{equation}
V^i U^i =  2 V^{AB}  U_{AB}.
\end{equation}
This is sometimes also useful to know that
\begin{equation}
\gs_i{}^{AB} \gs_i{}_{CD} = \frac{1}{2} \eps^{(A}{}_{C} \;\eps^{B)}{}_{D}.
\end{equation}

\subsection{Spinors, space-time indices and two-forms}
\subsubsection{Null tetrad and spinors}

In order to convert space-time indices into spinor ones we use the convention:
\begin{tcolorbox}
\begin{equation} 
V^I e_I^{AA'} = \frac{1}{i\sqrt{2}}\Mtx{-it+z & x-iy \\  x+iy & -it-z}, \qquad  V^I = \{t,x,y,z\}.
\end{equation}
\end{tcolorbox}
With this convention,
\begin{equation}
V^{AA'}= V^I e_I^{AA'} \qquad \Leftrightarrow \qquad V^I = e^I_{AA'}V^{AA'} 
\end{equation}
The $e^{AA'}_I$ symbol satisfies the Clifford-like algebra
\begin{tcolorbox} \vspace{-0.5cm}
	\begin{align}\label{(e,e)}
	\{e_I, e_J\}^{AB} = 2 e_{(I}{}^{A}{}_{C'} e_{J)}{}^{B}{}^{C'} = g_{IJ} \eps^{AB}
	&, & 	\{e_I, e_J\}^{A'B'} = 2 e_{(I}{}^{A'}{}_{C} e_{J)}{}^{B'}{}^{C} = g_{IJ} \eps^{A'B'}.
	\end{align}
\end{tcolorbox}
\noindent In particular:
\begin{equation}\
g_{IJ} = e_I{}^{AA'} e_J{}^{BB'} \eps_{AB}\eps_{A'B'} =  e_I{}^{AA'} e_J{}_{AA'}.
\end{equation}
We also define,
\begin{tcolorbox} 
	\begin{equation}\label{[e,e]}
	[e_I, e_J]^{AB}= 2 e_{[I}{}^A{}_C' e_{J]}{}^{BC'} = - 2\gS_{IJ}{}^{AB}.
	\end{equation}
\end{tcolorbox}

\subsubsection{Two-forms and spinors}

Let $\gL$ be a general two-form, in spinor notation:
\begin{equation}
\begin{array}{llll}
\gL &= \gL_{AA'BB'} \frac{e^{AA'}\W e^{BB'}}{2} \\ \\
&=\left( \gL{}_{E}{}^{E}{}_{A'B'} \;\eps_{AB}+ \gL{}_{E'}{}^{E'}{}_{AB} \;\eps_{A'B'} \right)\frac{e^{AA'}\W e^{BB'}}{4} \\ \\
&= -\left( \gL{}_{A'B'}\right) \frac{e_{C}{}^{A'}\W e^{CB'}}{2} - \left( \gL{}_{AB}\right) \frac{e_{C'}{}^{A}\W e^{C'B}}{2}.
\end{array}
\end{equation}
ie,\footnote{Note that this is coherent with the algebra \eqref{[e,e]}:
	\begin{equation*}
	[e_I, e_J]^{AB}= 2 e_{[I}{}^A{}_{C'} e_{J]}{}^{BC'} = - 2\gS_{IJ}{}^{AB}.
	\end{equation*}}
\begin{tcolorbox}
\begin{equation}
	\begin{array}{ll}
	\gL= \gL_{A'B'} \gS^{A'B'} +\gL_{AB} \gSt^{AB}, \qquad \qquad &	
	\gS^{A'B'}=\frac{e^{A'C}\W e^{B'}{}_{C}}{2}, \qquad  \gSt^{AB}=\frac{e{}^{AC'}\W e^{B}{}_{C'}}{2}.
	\end{array}
\end{equation}
\end{tcolorbox}
\noindent where 
\[ 
\begin{array}{ll}
\gL{}_{A'B'}=-\frac{\gL{}_{E}{}^{E}{}_{A'B'}}{2}, & \gL{}_{AB}=-\frac{\gL{}_{E'}{}^{E'}{}_{AB}}{2}.
\end{array} \]
Converting the $AB$ indices into spatial indices:,
\begin{align*} 
\gSt^i&=2\gSt^{AB}\gs^i{}_{AB}& &  \gSt^{AB}=\gSt^i \gs_i{}^{AB}=-\frac{e^{A}{}_{C'}\W e^{BC'}}{2},
\end{align*}
one can rewrite this decomposition in a usual tetrad:
\begin{tcolorbox}
	\[ 
	\gL=\gL_{IJ} \frac{e^{I}\W e^{J}}{2} = \gL^i \frac{\gS_i}{2} + \gLt^i \frac{\gSt_i}{2}.
	\]
\end{tcolorbox}
This can be taken as a definition for the $\gS^i$:
\begin{tcolorbox}
	\begin{equation}
\begin{array}{ccc}
 \gS^i = - e^0 \W e^i -\frac{\eps^{ijk}}{2} e^j \W e^k & \qquad &
\gSt^i =  e^0 \W e^i -\frac{\eps^{ijk}}{2} e^j \W e^k
\end{array}
	\end{equation}
\end{tcolorbox}
The $\gS^i$ span the space of self-dual two forms while the $\gSt^i$ span the space of self-dual two forms:
\begin{equation}
\begin{array}{ccc}
\ast\gS^i =  \gS &\qquad &\ast\gSt^i =  -\gSt
\end{array}
\end{equation}
with $ \star \gL_{IJ}= \frac{1}{2} \eps_{IJKL} \gL^{KL}$.

\chapter{Some more on Twistors and the Riemann sphere}

\section{Geometry of the Riemann sphere}\label{section: Appdx Geometry of the Riemann sphere}
\subsection{The Riemann sphere and its Fubini-Study metric}
The Riemann Sphere is the one dimensional complex projective space $\CP^1$, i.e the projective version of $S' \simeq \C^2$ (here $S'$ stands for the space of primed spinors),
\[ 
\CP^1 \coloneqq \left\{\text{One dimensional subspaces of } S' \right\}.
\]
as such it is at the same time the simplest non trivial Riemann surface and the simplest complex projective space (in particular it is a Kähler manifold) and is a nice non trivial starting point into the realm of complex geometry.

As for every projective space it will be very convenient to use the so called "homogeneous coordinates" to represent elements of $\CP^1$. That is we will use non zero vectors of $S'$ to represent the vector space that they generates: $\left[\p^{A'}\right]\in \CP^1$.
As we are working in coordinates here, we implicitly assumed that we chose an orthonormal basis of $S'$. Importantly the notation with a free indices reminds us that there is a natural $SU(2)$ action on $CP^1$ induced by the unitary transformation of $S'$. As usual , depending on the point of view this action can either be though as a "active action" on the Riemann sphere or a "passive" change of homogeneous coordinates.

Alternatively, we can use the "inhomogeneous coordinates" i.e coordinates of the Riemann sphere as a one dimensional complex manifolds. Consider the charts and mappings\footnote{The reason for this precise choice of charts, which we take from \cite{Mason:1991rf}, is that then $\left[ \p_{A'}\right]=\sMtx{ 1 \\ \gz }=\sMtx{ \gz' \\ 1 } $ on $U\cap U'$.} 
:
\[ 
\begin{array}{ll}
\phi:& \left \{
\begin{array}{llc}
\C &\mapsto & U\coloneqq \left\{\left[\p^{A'}\right] \text{such that } \p^{1'}\neq0 \right\}\\
\gz &\mapsto& \sMtx{\gz \\ -1 }
\end{array} \right.
\\ \\
\phi':& \left \{
\begin{array}{llc}
\C &\mapsto & U'\coloneqq \left\{\left[\p^{A'}\right] \text{such that } \p^{0'}\neq0 \right\}\\
\gz' &\mapsto& \sMtx{1\\ -\gz'}.
\end{array} \right.
\end{array}
\]
With transition map:
\[\begin{array}{ll}
\phi'^{-1} \circ \phi&  \left \{
\begin{array}{llc}
\C &\mapsto & \C\\
\gz &\mapsto& \gz'=\frac{1}{\gz}.
\end{array} \right.
\end{array} \]

In explicit computations we will mostly use homogeneous coordinates $\left[\p^{A'}\right]$ to emphasize the global nature of the different constructions while we will restrict the open subspace $U \subset \CP^1$ and use the inhomogeneous coordinate $\gz$ in the rare case when we want to clarify a how things look like locally from the point of view of the manifold. One can then freely go from to two the other through the relation $\left[\p^{A'}\right] = \sMtx{\gz \\ -1}$.

\paragraph{Euclidean Conjugation}
The "Euclidean Conjugation" acts on $\CP^1$ in an obvious way, $\circonf : \left[\p^{A'}\right] \mapsto \left[\ph^{A'}\right] $. This is now an involution but importantly it has no fix points. Its representation in inhomogeneous coordinates makes it clear that this is just antipodal map: $\hat{\gz}=\frac{-1}{\gzb}$. 

\paragraph{Kahler structure}
As one dimensional complex projective space, the Riemann sphere has a natural $SU(2)$-invariant Kähler structure. 
\begin{align}\label{Kahler structure on CP1}
\hspace*{-1.25cm}
J&= id\p^{A'}\pa_{A'} -i d\ph^{A'} \pah_{A'} = i d\gz \pa_{\gz} - i d\gzb \pa_{\gzb},\\ \nonumber \\
\go &= \frac{4iR^2}{2}\frac{d\p_{A'} \W d\ph^{A'}}{ \pp}=\frac{4iR^2}{2} \frac{d\gz \W d\gzb}{\left( 1 + \gz \gzb\right)^2},\\ \nonumber \\
g&= \frac{4R^2}{2} \frac{d\p_{A'} \odot d\ph^{A'}}{ \pp} =  \frac{4R^2}{2} \frac{d\gz \odot d\gzb}{\left( 1 + \gz \gzb\right)^2}.
\end{align}
Where $\frac{1}{R}$ is the scalar curvature. The notation is coherent with the fact that the Riemannian metric $g$ is the round metric of a sphere with radius $R$ \footnote{Note that its volume is $Vol_{\S^2}= \int \go  = \int \sqrt{g} d^2x  =  4R^2\;\int_0^{\infty} \frac{2\pi r dr}{\left( 1 + r^2\right)^2} = 4\pi R^2$, as desired}.
A direct computation shows that $\rho= 4R^2 log \left( 1+ \gz\gzb \right)$ acts as a local Kahler potential, $\go = \frac{i}{2} \pa \pab \rho$. \footnote{Formally we could also write $\rho = 4R^2 log \left( \pp \right) $. It is then a matter of taste: this is a slightly misleading as $\pp$ really is a section of $\Oc\left(1,1\right)$ and taking its logarithm would not make much sense, on the other hand one could say that the homogeneous notation nicely suggests by itself that the Kahler potential can only be defined locally.}.

\subsection{Holomorphic line bundles over $\CP^1$}
A holomorphic vector bundle bundle is a complex vector bundle over a complex manifold such that the total space is a complex manifold and the projection operator is holomorphic. Practically, this is equivalent to the fact that transition functions between different trivialisations have to be holomorphic with respect to the complex structure of the base.
Holomorphic line bundles are the most simple vector bundles and arguably the most simple geometrical structure that one can construct that is compatible with the complex structure. Interestingly the space of of holomorphic line bundle (up to isomorphism) form a group, the Picard group:

\begin{Prop*}{\emph{The Picard Group}}\footnote{See for ex \cite{Huygbrechts} p69} \mbox{} \\
	The tensor product and the dual endow the set of all isomorphism class of Holomorphic bundles on a complex manifold X with the structure of an abelian group (with product given by tensor product and inverse given by the dual). This group is the Picard group Pic(X) of X. 
\end{Prop*}
The tensor product $L_1\otimes L_2$ of line bundles and the dual $L^*$ of a line bundle being line bundles all one needs to do to prove this proposition is to see that $L^* \otimes L$ is isomorphic to the trivial bundle. By considering the transition map of this bundle however one easily see that they are trivial.

In the special case of the Riemann sphere, holomorphic line bundles up to isomorphisms happened to be classified by only one topological invariant $n\in\Z$, the "Chern number". The Picard group therefore is isomorphic to group of integers, $Pic\left( \CP^1\right) \simeq\Z$.  The equivalence class of holomorphic line bundle over $\CP^1$ with Chern number $n$ is referred to as $\Oc(n)$. We have $\Oc\left( n_1\right)\otimes \Oc\left( n_2\right)\simeq \Oc\left(n_1+n_2 \right)$, $\Oc\left(n \right)^* \simeq \Oc\left(-n \right)$. $\Oc\left(0 \right)$ beeing the trivial bundle.

In some way, $\Oc(n)$ bundles are the elementary building blocs of the geometry of the vector bundles on Riemann sphere. This statement is made precise by 
\begin{Thrm*}{\emph{"Grothendieck's lemma"}\footnote{see \cite{Huygbrechts} p244}}\\
	Every holomorphic vector bundle E on $\CP^1$ is isomorphic to a holomorphic vector bundle of the form $\bigoplus\Oc\left(a_i \right)$. What's more the ordered sequence $a_1\geq a_2\geq...\geq a_r$ is uniquely determined.
\end{Thrm*} 

We now describe explicit realisations of the $\Oc(n)$ bundles. The most condensed definition uses of homogeneous coordinates:
\[ 
\Oc(n)\coloneqq S'\times \C \diagup \left\{\left( \p^{A'}, \chi\right) \sim \left( \gl\p^{A'}, \gl^n\chi\right)  \right\}.
\]
Then the projection clearly is 
\[ \begin{array}{ll}
\Pi:& \left\{
\begin{array}{lll}
\Oc(k) &\rightarrow &\CP^1 \\
\left[ \p^{A'}, \chi \right] &\rightarrow & \left[ \p^{A'}\right]
\end{array}\right.
\end{array}
\]
However the definition "in inhomogneous coordinates" i.e in terms of trivialisation and transition map is also enlightening and is good to keep in mind:
\[ 
\begin{array}{ll}
\psi:& \left \{
\begin{array}{llc}
\C\times\C &\mapsto & \Pi^{-1}\left( U\right)\\
\left( \gz, \chi\right) &\mapsto& \left[ \Mtx{\gz\\ -1}, \chi \right]
\end{array} \right.
\end{array}
\qquad, \qquad 
\begin{array}{ll}
\psi':& \left \{
\begin{array}{llc}
\C\times\C &\to & \Pi^{-1}\left( U'\right)\\
\left( \gz', \chi'\right) &\mapsto& \left[  \Mtx{1\\ -\gz'}, \chi' \right]
\end{array} \right.
\end{array}
\]
\[\begin{array}{ll}
\psi'^{-1} \circ \psi&  \left \{
\begin{array}{llc}
\C\times\C &\to & \C\times\C\\
\left( \gz, \chi\right) &\mapsto& \left( \gz', \chi'\right)=\left( \frac{1}{\gz}, \gz^{-n}\chi \right).
\end{array} \right.
\end{array} \]

Practically, a section of $\Oc(k)$ is represented by a function $f$ on $S'$ with holomorphic homogeneity degree $n$, i.e such that:  $f\left( \gl\p^{A'}\right) = \gl^n f\left( \p^{A'}\right) $. Then the associated section is,
\[ 
s_f: \left\{\begin{array}{lll}
\CP^1&\to&\Oc(n) \\ \\
\left[\p^{A'}\right] &\mapsto& 
\left[\p^{A'},f\left( \p^{A'}\right) \right]=
\left[\gl\p^{A'},\gl^n f\left( p^{A'}\right) \right]=
\left[\gl\p^{A'}, f\left( \gl\;\p^{A'}\right) \right]
\end{array}\right.
\]
or using the definition in terms of chart
\[\begin{array}{ll}
\psi'^{-1} \circ \psi&  \left \{
\begin{array}{clc}
\C\times\C &\mapsto & \C\times\C\\
\left( \gz, f\left( \gz o - \oh \right) \right) &\mapsto& \left( \gz', f\left( o-\gz' \oh \right) \right)=\left( \frac{1}{\gz}, \gz^{-n}f\left( \gz o - \oh \right)\right).
\end{array} \right.
\end{array} \]
Holomorphic global sections of $\Oc(n)$ are represented by holomorphic functions on $S'$. One can easily show that for $n\geq0$ such holomorphic sections are in fact represented by symmetric spinors\footnote{Essentially it suffices to use differentiation $n$ times to get a section of $\Oc(0)$, then holomorphicity implies that this section has to be a constant map.}: $f\left(\p^{A'}\right)=\Psi_{A'_1 \ldots A'_n } \p^{A'_1}\ldots\p^{A'_n}$. On the other hand there are no holomorphic global section of $\Oc(n)$ for $n<0$.

Some of these holomorphic line bundles are of particular importance and deserve a name:\\
The \emph{"tautological bundle"}, $\Oc(-1)$,  is the natural line bundle over $\CP^1$ such that its total space identifies with $S'$:
\[ 
\begin{array}{lll}
\Oc(-1)&\mapsto& S'\\
\left[ \p^{A'}, \chi\right]= \left[ \gl\p^{A'}, \gl^{-1}\chi\right]
&\mapsto& \p^{A'}\chi
\end{array}
\]
Alternatively, in terms of trivialisation:
\begin{IEEEeqnarray*}{llllllll}
	\psi:& \left \{
	\begin{array}{llc}
		\C\times\C &\mapsto & \Pi^{-1}\left( U\right)\subset S'\\
		\left( \gz, \chi\right) &\mapsto& \chi\left(\gz o-\oh \right)
	\end{array} \right.
	&,\qquad &
	\psi':& \left \{
	\begin{array}{llc}
		\C\times\C &\mapsto & \Pi^{-1}\left( U'\right)\subset S'\\
		\left( \gz', \chi'\right) &\mapsto& \chi'\left(o-\gz'\oh \right)
	\end{array} \right.
	&
\end{IEEEeqnarray*}
\begin{IEEEeqnarray*}{llllllll}	
	\psi'^{-1} \circ \psi&  \left \{
	\begin{array}{llc}
		\C\times\C &\mapsto & \C\times\C\\
		\left( \gz, \chi\right) &\mapsto& \left( \gz', \chi'\right)=\left( \frac{1}{\gz}, \gz\chi \right)
	\end{array} \right. .
\end{IEEEeqnarray*}

The \emph{"hyperplane bundle"}, $\Oc(1)$, simply is the dual of the tautological bundle.\\
Two other important holomorphic bundles are the \emph{"holomorphic tangent bundle"} $T^{\left(1,0 \right)}\CP^1\simeq\Oc(2)$ and the \emph{"holomorphic cotangent bundle"} $T^*_{\left(1,0 \right)}\CP^1\simeq\Oc(-2)$ (Note that because the Riemann sphere is a one dimensional complex manifold, the \emph{holomorphic cotangent bundle} coincide with the \emph{canonical bundle} which is the bundle of holomorphic top form). This is better seen in charts:

\begin{IEEEeqnarray*}{llllllllllll}
	\psi:& \left \{
	\begin{array}{llc}
		\C\times\C &\mapsto & \Pi^{-1}\left( U\right)\\
		\left( \gz, \chi\right) &\mapsto& \left( \gz, \chi \pa_{\gz} \right)
	\end{array} \right.
	&, \qquad &
	\psi':& \left \{
	\begin{array}{llc}
		\C\times\C &\mapsto & \Pi^{-1}\left( U'\right)\subset S'\\
		\left( \gz', \chi'\right) &\mapsto& \left( \gz', -\chi' \pa_{\gz'} \right)
	\end{array} \right.
	&,
\end{IEEEeqnarray*}
\begin{IEEEeqnarray*}{llllllllllll}	
	\psi'^{-1} \circ \psi&  \left \{
	\begin{array}{llc}
		\C\times\C &\mapsto & \C\times\C\\
		\left( \gz, \chi\right) &\mapsto& \left( \gz', \chi'\right)=\left( \frac{1}{\gz}, \gz^{-2}\chi \right)
	\end{array} \right. .
\end{IEEEeqnarray*}

From this, it is clear how to make a similar construction for $T_{1,0}^*\CP^1=\Oc(-2)$.\\

\paragraph{$\Oc(n,m)$-bundles}
Up to now we discussed holomorphic line bundles over $\CP^1$ but one can also consider \emph{non-holomorphic} line bundles, in particular we define
\[ 
\Oc(n,m)\coloneqq S'\times \C \diagup \left\{\left( \p^{A'}, \chi\right) \sim \left( \gl\p^{A'}, \gl^n \bar{\gl}^m\chi\right)  \right\}.
\]
Note that the transition functions are indeed non-holomorphic,
\[\begin{array}{ll}
\psi'^{-1} \circ \psi&  \left \{
\begin{array}{llc}
\C\times\C &\to & \C\times\C\\
\left( \gz, \chi\right) &\mapsto& \left( \gz', \chi'\right)=\left( \frac{1}{\gz}, \gz^{-n}\bar{\gz}^{-m}\chi \right).
\end{array} \right.
\end{array} \]
The above discussion straightforwardly generalises to these bundles. Note in particular that $T^{\left(0,1 \right)}\CP^1\simeq\Oc(0,2)$ and  $T^*_{\left(0,1 \right)}\CP^1\simeq\Oc(0,-2)$

\begin{ExtraComputation}
	\begin{framed}
Finally it is also interesting to note that the \emph{Normal bundle of $\CP^1$ in $\C^2$}, $N_{\CP^1 | C^2}$, is the trivial bundle $\Oc(0)$.
\todo{Add details on the construction}
	\end{framed}
\end{ExtraComputation}

\paragraph{Hermitian metric on the $\Oc(n)$ bundles}
The total space of the $\Oc(-1)$line bundle $\Pi:S' \mapsto \CP^1$ being equipped with an hermitian metric, $g=\frac{1}{2} d\p_{A'}\odot d\ph^{A'}$ it induces a metric on the fibers. This is done by restricting $g$ to the vertical tangent subspace $\Vc$:
\[ \Vc = Ker\left(\Pi \right) = Span\left(E,\Eb\right), \qquad h_{(-1)}= g\left(E,\Eb\right) = \pp,\]
where we introduced the "Euler vectors" , $E=\p^{A'} \pa_{A'}$,$ \Eb=\ph^{A'} \pah_{A'}$.
By tensor products and dual we obtain a hermitian metric $h_{(n)}$ on each $\Oc(n)$ bundle:
\begin{equation}\label{Hermitian metric on O(n) over CP1}
h_{(n)}= \left(1+\gz\gzb \right)^{-n}.
\end{equation}

Clearly this metric is a section of the $\Oc(n,n)$ bundle.

\paragraph{Covariant derivative on $\Oc(n)$ bundle}
We now introduce the Chern connection associated with the Hermitian metric \eqref{Hermitian metric on O(n) over CP1}:
\begin{equation}\label{Chern connection}
a_{(n)} = -n \;\frac{\gzb}{1+\gz\gzb}d\gz.
\end{equation}
If $\ga'(\gz)$ is any $\Oc(n,m)$-valued k-form on $\CP^1$, then from \eqref{Chern connection}, we can define its covariant derivative as
\begin{equation}\label{covariant derivative on CP1 in inhomogeneous coordinates}
d_a \ga' = d\ga' + a_{(n)}\W \ga' + \bar{a}_{(m)}\W \ga'
\end{equation}

As we already discussed, $T^{(1,0)}\CP^1=\Oc(2)$, and for $n=2$ this covariant derivative coincide with the Levi-Cevita connection of the Fubini-Study (kahler)metric:  
\[ 
\N \left( f\pa_{\gz}\right) = \left( df + a_{(-2)}f\right)\pa_{\gz}.
\]
It curvature is,
\begin{equation}\label{Chern curvature}
f_{n} = da_{(n)} = n \;\frac{d\gz \W d\gzb}{\left( 1+\gz\gzb\right)^2} =\frac{n}{2iR^2} \go
\end{equation}
where $\go$ is the Kahler form \eqref{Kahler structure on CP1}.\\

\paragraph{Chern connection}
The connection \eqref{Chern connection} has the following property:  it is compatible with the complex structure, in the sense that $a_{(n)}\Big|_{T^{(0,1)}}=0$.
It is also compatible with the Hermitian metric \eqref{Hermitian metric on O(n) over CP1} in the sense that 
\[ 
d_{a} h_{n}= d\left(\left(1+\gz\gzb \right)^{-n} \right) + \left( a_{(n)}+\ab_{(n)} \right) \left(1+\gz\gzb \right)^{-n} =0
\]
This uniquely defines the Chern connection as follows from a standard result of complex geometry\footnote{ see for example \cite{Huygbrechts} p177}:
\begin{Prop*}{\emph{Chern connection}} \mbox{}\\
	On any hermitian holomorphic bundle over a complex manifold there is a unique connection compatible with both the metric and complex structure. This connection is called the \emph{Chern connection}. In local coordinates, if $h$ is the hermitian metric then $a=\hb^{-1}\pa \hb$.
\end{Prop*}
A direct computation shows that \eqref{Chern connection} is indeed of the form $\left(h_{n}\right)^{-1} \pa h_{n}$. Note that the mere fact that the Levi-Cevita connection and the Chern connection for the holomorphic tangent bundle coincide is equivalent to the statement that $\CP^1$ is Kahler.\\
Once we have the Chern connection on a vector bundle, the first Chern class is proportional to the cohomology class of its curvature.
\[ 
c1\left[\Oc(n)\right]\coloneqq \left[\frac{i}{2\p} f_{n}\right] = \left[\frac{n}{4\p R^2} \go\right]
\]
And the Chern number of the bundle is the integral of the first chern class
\[ 
n = \int_{\CP^1} \frac{n}{4\p R^2} \go.
\]
It turns out that the first Chern Class in fact is a topological invariant of the manifold, here $S^2$.

\section{The Twistor Space of Complexified Anti-Self-Dual Space-Times}\label{Section : Appdx Twistor Space of Complexified Space-Times}

We here review how to construct the twistor space $\T(M)$ of an anti-self-dual complexified space-time (cf \cite{Penrose:1976js}, \cite{Ward:1980am} for the original references, \cite{Ward:1990vs},\cite{Mason:1991rf} for pedagogical presentations). We especially emphasise how the self-dual connection $D= d + A$ on space-time gives a $\Oc(2)$-valued one-form $\gt$ on the associated twistor space.

We first define the so-called ``correspondence space" $\F(M)$ as the primed 2-spinor bundle over $M$: $\C^2 \inj \F(M) \xto{\pi} M$. We will also make use  of its projective version $\PF(M)$: $\CP^1 \inj \PF(M) \xto{\pi} M$.  As we take $M$ to be a 4d complex manifold, $\F(M)$ naturally is a 6d complex manifold, $\PF(M)$ a 5d complex manifold. We will use coordinates adapted to the fibre bundle structure, $(x^{\mu},\p_{A'})$. For $\PF(M)$, $\p^{A'}$ should be understood as homogeneous coordinates.

We first consider the following distribution on $\PF(M)$, ``the distribution of $\ga$-planes":
\begin{equation}\label{alpha plane distribution}
D_{\ga-plane} = \text{Span}\left\{ \p^{A'} D_{AA'} = \p^{A'}\left( \frac{\pa}{\pa x^{AA'}} -\p_{C'} A_{AA'}{}^{C'}{}_{D'} \frac{\pa}{\pa \p_{D'}} \right) \right\}_{A \in 0,1}
\end{equation}
where $D_{\mu}$ is the horizontal lift of $\frac{\pa}{\pa x^{\mu}}$ with respect to the Levi-Cevita connection. A direct calculation shows that this distribution is integrable if and only if the space-time that we started with is anti-self-dual. If we assume this to be the case then, at least locally, we can consider the integral surfaces of this distribution, referred to as $\ga$-surfaces. The projective twistor space associated with $M$ is then defined as the space of $\ga$-surfaces. Importantly there is only one $\ga$-surface that passes through each point in $\PF(M)$. Therefore $\PF(M)$ has the structure of a fibre bundle over $\PT(M)$. We note $\Pi \from \PF(M) \to \PT(M)$ the projection operator.

We then have the classical double fibration picture:
\begin{center}
	\begin{tikzpicture}
	
	\node(x1) {$\PF(M)$};
	\node(x2) [left = 2cm of x1] {};
	\node(x3) [right = 2cm of x1]{};
	
	\node(y2) [below = 1.5cm of x2] {$\PT(M)$};
	\node(y3) [below = 1.5cm of x3] {$M$};
	
	\draw [<-] (y2) -- node[left,pos=0.6]{$\Pi$ \;} (x1) ;
	\draw [<-] (y3) -- node[right, pos= 0.6] {\;$\pi$} (x1) ;
	
	\end{tikzpicture}
\end{center}

To define the (non projective) twistor space $\T(M)$, we now consider the distribution \eqref{alpha plane distribution} as a distribution on $\F(M)$. The integrability condition is unchanged and each of the resulting integral surfaces now contains the geometrical data of an $\ga$-surface in $M$ together with a particular scaling for $\p^{A'}$. This space of integral surfaces in $\F(M)$ thus naturally is a the total space of a complex line bundle over $\PT(M)$ referred to as the twistor space. We finally get the following commutative diagram :

\begin{center}
	\begin{tikzpicture}
	\node(x01){$\F(M)$};
	\node(x11) [below = 1cm of x01] {$\PF(M)$};
	\node(x02) [left = 2cm of x01] {};
	\node(x03) [right = 2cm of x01]{};
	\node(y02) [below = 0.5cm of x02] {$\T(M)$};
	\node(y12) [below = 1cm of y02] {$\PT(M)$};   
	\node(y03) [right = 4.5cm of y12] {$M$};
	
	\draw [<-] (x11) -- node[left,pos=0.6]{} (x01) ;
	\draw [<-] (y12) -- node[left,pos=0.6]{} (y02) ;
	
	\draw [<-] (y02) -- node[left,pos=0.95]{} (x01) ;
	\draw [<-] (y12) -- node[left,pos=0.6]{} (x11) ;
	
	\draw [<-] (y03) -- node[right, pos= 0.6] {} (x01) ;
	\draw [<-] (y03) -- node[right, pos= 0.6] {} (x11) ;
	\end{tikzpicture}
\end{center}
We also refer to the complex line bundle over $\C \inj \T(M) \to \PT(M)$ as $\Oc(-1)$. The dual of this line bundle will be referred to as $\Oc(1)$. We also define $\Oc(n) = \Oc(1)^n$ and $\Oc(-n) = \Oc(-1)^n$ by taking tensor products.

Let $p\in \T(M)$ be a point in the twistor space, it corresponds to a certain integral surface in $\F(M)$ that we note $\hat{p}$ (again this is equivalent to an $\ga$-surface in $M$ together with a primed spinor $\p^{A'}$). A tangent vector to $p$ then corresponds to a certain vector field on $\hat{p}$ that ``connects" $\hat{p}$ to an other infinitesimally close integral surface :

\begin{equation}\label{local twistor: connecting vector field}
X(x) = V(x){}^{AA'} D_{AA'} + \gb(x){}^{B'} \frac{\pa}{\pa \p^{B'}} 
\end{equation}
Being a ``connecting vector field", it is defined up to elements of $D_{\ga-plane}$, i.e $V^{AA'}$ is defined up to $\gl^{A} \p^{A'}$. On the other hand the ``local twistor field" $\left(\ga^{A},\gb_{A'} \right) (x)=\left( V^{AA'}(x)\p_{A'},\gb_{A'}(x)\right)$ is well defined. In fact, because \eqref{local twistor: connecting vector field} is a connecting vector field, it has to satisfy \[ \left[X^{AA'}, \gl^{B} \p^{B'} \right]=0, \qquad \forall \gl^{B}, \] from which it follows that $\ga$ and $\gb$ are not independent. One can gather these relations in the so-called ``local twistor transport equations":
\begin{align*}
& \p^{A'}\nabla_{AA'} \ga^B + \gd^{B}_A \p^{A'}\gb_{A'} =0 \\
&\p^{A'}\nabla_{AA'} \gb_{B'} + P_{ABA'B'} \p^{A'} \ga^{B} =0 
\end{align*}
with $P_{ab} = \Phi_{ab} -\gL g_{ab}$ and $\Phi_{ab}$ the trace-free part of the Ricci tensor.

Now we can define the following one-form on $\F(M)$:
\begin{equation}\label{section3 : tau def}
\gt = \p_{A'} \left(d\p^{A'} + A^{A'}{}_{B'} \p_{B'} \right)
\end{equation}
by construction it annihilates horizontal vectors, $\gt\left(D_{a}\right)=0$. Contracting this form with the connecting vector field \eqref{local twistor: connecting vector field} we get a scalar field on $\hat{p}$,
\[ 
\gt(X) = \gb_{A'}\p^{A'}.
\]
Now $\gt$ defines a one-form on $\T(M)$ if and only if this scalar field is constant along $\hat{p}$ i.e
\[ 
\p^{A'}\nabla_{AA'} \left( \gt(X) \right) =0.
\]
Making use of the ``local twistor transport equations", we can show that \[ \p^{A'}\nabla_{AA'} \left( \gt(X) \right) = \Phi_{ABA'B'} \;\ga^B \p^{A'}\p^{B'}. \]
We thus see that $\gt$ defines a one-form on $\T(M)$ if and only if Einstein equations are verified.

Therefore we see that for complexified anti-self-dual Einstein space-time the ``self-dual part" of the Levi-Cevita connection is directly related to the one-form \eqref{section3 : tau def} the associated twistor space. As \eqref{section3 : tau def} is homogeneous degree 2 in $\p^{A'}$ it descends to a $\Oc(2)$-valued one form on $\PT(M)$.

 \chapter{3d Gravity Conventions}\label{Section : Appdx 3d gravity Conventions}
 
 In this section, we review some basic facts about 3D gravity. Our notations are standard for the gravity literature.
 
 \section{Einstein-Cartan frame formalism in 3D} \mbox{}
 
 Let $\left(e^{i}\right)_{i\in\{1,2,3\}}$ be orthogonal frame field so that the 3D metric is 
 \begin{equation}
 ds^2 =e^i \otimes e^j \eta_{ij},
 \end{equation}
 where $\eta_{ij}={\rm diag}(1,1,1)$. We raise and lower indices with the metric $\gd_{ij}$, and the spin-connection is the set of one-forms $w^{ij}=w^{[ij]}$. The anti-symmetry is the statement that the connection is $\gd_{ij}$ metric compatible. Let $f^{ij}$ be the curvature
 \begin{equation}
 f^{ij} = dw^{ij} + w^{ik} \W w_k{}^j.
 \end{equation}
The action for 3D gravity with cosmological constant $\gL$ is
 \begin{equation}\label{s-wt}
 S[e,w]= - \frac{1}{4} \int_M \left( e^i \W f^{jk}  - \frac{\gL}{3} \, e^i\W e^j\W e^k \right) \eps_{ijk} .
 \end{equation}
 The orientation implied here is that of the three-form $e^i\W e^j\W e^k \eps_{ijk}$. The minus sign in front of the action is the usual choice for the all plus signature. We work in units in which the 3D Newton's constant satisfies $4\pi G=1$. Varying this action with respect to $w$ we get the torsion-free condition
 \begin{equation}\label{torsion-free}
 \N_w e^i \equiv de^i + w^i{}_j \W e^j=0.
 \end{equation}
 It says that the connection $w$ is the unique $e$-compatible connection. Substituting this connection into (\ref{s-wt}) we find
 \begin{equation}
 S[e,w(e)] = - \frac{1}{4}\int_M (R-2\gL) v_g,
 \end{equation}
 where $R$ is the Ricci scalar of the metric, and the integration is carried out with respect to the metric volume element $v_g$. 
 
 The connection matrix $w^{ij}$ being anti-symmetric, we can write 
 \begin{equation}
 w^{ij} = \eps^{ikj} w^k,
 \end{equation}
 which defines the new connection one-forms $w^i$. We then have for the curvature 
 \begin{equation}
 f^{ij} = \eps^{ikj} f^k, \qquad f^i = dw^i + \frac{1}{2} \eps^{ijk} w^j \W w^k.
 \end{equation}
 
 \section{Matrix notations}\mbox{}
 
 It is very convenient to get rid of the internal $i,j,\ldots$ indices at the expense of making all objects $\su(2)$-valued. The Lie algebra generators $\left(\gs^{i}\right)_{i \in \{1,2,3\}}$ are taken such that they satisfy
 \begin{equation}
 \Tr(\gs_i \gs_j) = - \frac{1}{2} \gd_{ij}, \qquad
 [\gs_i, \gs_j]  = \eps_{ij}{}^k \gs_k.
 \end{equation}
 
 We then form a $\su(2)$-connection
 \begin{equation}
 \bdw := w^i \gs_i.
 \end{equation}
 In what follows we will always denote a matrix-valued object by a bold-face letter. The matrix valued curvature $\bdf \coloneqq f^i \gs_i$ is computed as
 \begin{equation}
 \bdf = d\bdw + \bdw\W \bdw.
 \end{equation}
 We also form anti-hermitian frame field one-forms
 \begin{equation}
 \bde \coloneqq e^i \gs_i,
 \end{equation}
 in terms of which the metric is
 \begin{equation}
 ds^2 = - 2\, \Tr (\bde\otimes \bde).
 \end{equation}
 In terms of the matrix-valued fields the torsion-free condition \eqref{torsion-free} takes the form
 \begin{equation}\label{nabla}
 \N\bde \equiv d\bde + \bdw\W \bde + \bde\W \bdw = 0.
 \end{equation}
 The field equations obtained by varying the action (\ref{s-wt}) with respect to $\bde$ takes the form
 \begin{equation}
 \bdf=-\gL\; \bde\W \bde.
 \end{equation}
 The action itself takes the form
 \begin{equation}\label{s-matrix}
 S[\bde,\bdw] = - \int_M \Tr\left( \bde\W \bdf + \frac{\gL}{3} \bde\W\bde\W \bde\right).
 \end{equation}
 
 \section{Chern-Simons formulation}\mbox{}
 
 The two sets of equations $\N \bde=0, \bdf=\bde\W\bde$ can be combined as the real and imaginary parts of a single complex-valued equation by introducing $\sl(2,\C)$-valued fields
 \begin{equation}\label{CS-connection}
 \bda_{\pm} \coloneqq \bdw \pm \sqrt{\gL} \bde.
 \end{equation}
 Here and in what follows $\sqrt{\gL}$ will stand for $i\sqrt{|\gL|}$ when $\gL<0$ so that in this particular case $\bda_{+}$ and $\bda_{-}$ are complex conjugated.
 The field equations of 3D gravity then combine into the statement that the curvature of the two $\SL(2,\C)$-connections $\bda_{\pm}$ are zero
 \begin{equation}
 0=\bdf_{+} \equiv d\bda_{+} + \bda_{+}\W \bda_{+},\qquad  0=\bdf_{-} \equiv d\bda_{-} + \bda_{-}\W \bda_{-}.
 \end{equation}
 These are the field equations following from the Chern-Simons Lagrangian. Alternatively, we can write the Einstein-Cartan Lagrangian (\ref{s-matrix}) (with $\gL=-1$), modulo a surface term, as
 \begin{equation}\label{grav-cs}
 S[\bde,\bdw]=  -\frac{\sqrt{\gL}}{4 \gL} \int_M  CS[\bda_{+}] - CS[\bda_{-}],
 \end{equation} 
 where
 \begin{equation}
 CS[\bda]:= \Tr\left( \bda\W d\bda + \frac{2}{3} \bda\W \bda\W \bda\right)
 \end{equation}
 is the Chern-Simons three-form for $\bda$. 

\end{Appendix}

\begin{Bibliography}

\bibliographystyle{apalike}
\bibliography{./Biblio/Biblio_Approches_to_Quantum_Gravity,./Biblio/Biblio_HitchinFunctionnal,./Biblio/Biblio_LQG,./Biblio/Biblio_Math,./Biblio/Biblio_pure_connection,./Biblio/Biblio_Twistor,./Biblio/Biblio_MyPapers}

\begin{thebibliography}{}

\bibitem[Aad et~al., 2012]{Aad:2012tfa}
Aad, G. et~al. (2012).
\newblock {Observation of a new particle in the search for the Standard Model
  Higgs boson with the ATLAS detector at the LHC}.
\newblock {\em Phys. Lett.}, B716:1--29.

\bibitem[Abbott et~al., 2016]{LIGGO2016}
Abbott, B.~P. et~al. (2016).
\newblock {Observation of Gravitational Waves from a Binary Black Hole Merger}.
\newblock {\em Phys. Rev. Lett.}, 116(6):061102.

\bibitem[Adamo, 2013]{Adamo:2013cra}
Adamo, T. (2013).
\newblock {Twistor actions for gauge theory and gravity}.

\bibitem[Adamo et~al., 2011]{Adamo:2011pv}
Adamo, T., Bullimore, M., Mason, L., and Skinner, D. (2011).
\newblock {Scattering Amplitudes and Wilson Loops in Twistor Space}.
\newblock {\em J.Phys.}, A44:454008.

\bibitem[Adamo and Mason, 2014]{Adamo:2013tja}
Adamo, T. and Mason, L. (2014).
\newblock {Conformal and Einstein gravity from twistor actions}.
\newblock {\em Class.Quant.Grav.}, 31(4):045014.

\bibitem[Ade et~al., 2016]{Ade:2015lrj}
Ade, P. A.~R. et~al. (2016).
\newblock {Planck 2015 results. XX. Constraints on inflation}.
\newblock {\em Astron. Astrophys.}, 594:A20.

\bibitem[Agricola, 2008]{Agricola08G2History}
Agricola, I. (2008).
\newblock {Old and new on the exceptional group $G_2$}.

\bibitem[Agricola et~al., 2015]{Agricola:2014yma}
Agricola, I., Chiossi, S.~G., Friedrich, T., and Höll, J. (2015).
\newblock {Spinorial description of $SU(3)$-and G$_2$-manifolds}.
\newblock {\em J. Geom. Phys.}, 98:535--555.

\bibitem[Alexandrov et~al., 2005]{Alexandrov:2004cp}
Alexandrov, B., Friedrich, T., and Schoemann, N. (2005).
\newblock {Almost Hermitian 6-manifolds revisited}.
\newblock {\em J. Geom. Phys.}, 53:1--30.

\bibitem[Armendariz-Picon et~al., 1999]{ArmendarizPicon:1999rj}
Armendariz-Picon, C., Damour, T., and Mukhanov, V.~F. (1999).
\newblock {k - inflation}.
\newblock {\em Phys. Lett.}, B458:209--218.

\bibitem[Ashtekar, 1986]{Ashtekar:1986yd}
Ashtekar, A. (1986).
\newblock {New Variables for Classical and Quantum Gravity}.
\newblock {\em Phys. Rev. Lett.}, 57:2244--2247.

\bibitem[Atiyah et~al., 2017]{Atiyah:2017erd}
Atiyah, M., Dunajski, M., and Mason, L. (2017).
\newblock {Twistor theory at fifty: from contour integrals to twistor strings}.

\bibitem[Atiyah et~al., 1978]{Atiyah:1978wi}
Atiyah, M., Hitchin, N.~J., and Singer, I. (1978).
\newblock {Selfduality in Four-Dimensional Riemannian Geometry}.
\newblock {\em Proc.Roy.Soc.Lond.}, A362:425--461.

\bibitem[Baez, 2002]{Baez:2001dm}
Baez, J.~C. (2002).
\newblock {The Octonions}.
\newblock {\em Bull. Am. Math. Soc.}, 39:145--205.

\bibitem[Bar, 1993]{Bar:1993}
Bar, C. (1993).
\newblock {Real Killing spinors and holonomy }.
\newblock {\em Comm. Math. Phys.}, 154:509--521.

\bibitem[Bengtsson, 1991]{Bengtsson:1991bq}
Bengtsson, I. (1991).
\newblock {Selfduality and the metric in a family of neighbors of Einstein's
  equations}.
\newblock {\em J. Math. Phys.}, 32:3158--3161.

\bibitem[Berger, 1955]{BergerList}
Berger, M. (1955).
\newblock {Sur les groupes d'holonomies des variétés à connection affine et
  des variétés Riemanniennes. }.
\newblock {\em Bull.Soc.Math.France}, 83:279--330.

\bibitem[Bern et~al., 2015]{Bern:2015xsa}
Bern, Z., Cheung, C., Chi, H.-H., Davies, S., Dixon, L., and Nohle, J. (2015).
\newblock {Evanescent Effects Can Alter Ultraviolet Divergences in Quantum
  Gravity without Physical Consequences}.
\newblock {\em Phys. Rev. Lett.}, 115(21):211301.

\bibitem[Besse, 1987]{Besse}
Besse, A. (1987).
\newblock {\em {Einstein Manifolds}}.

\bibitem[Bryant, 1987]{Bryant:1987}
Bryant, R. (1987).
\newblock {Metrics with exceptionnal holonomy}.
\newblock {\em Ann. of Math}, 126:525--576.

\bibitem[Bryant and Salamon, 1989]{BryantSalamon:1989}
Bryant, R. and Salamon, S. (1989).
\newblock {On the construction of some complete metrics with exceptional
  holonomy}.
\newblock {\em Duke Math. J.}, 58:829--850.

\bibitem[Cachazo et~al., 2004]{Cachazo:2004kj}
Cachazo, F., Svrcek, P., and Witten, E. (2004).
\newblock {MHV vertices and tree amplitudes in gauge theory}.
\newblock {\em JHEP}, 09:006.

\bibitem[Capovilla et~al., 1990]{Capovilla:1990qi}
Capovilla, R., Jacobson, T., and Dell, J. (1990).
\newblock {GRAVITATIONAL INSTANTONS AS SU(2) GAUGE FIELDS}.
\newblock {\em Class. Quant. Grav.}, 7:L1--L3.

\bibitem[Capovilla et~al., 1991a]{Capovilla:1991kx}
Capovilla, R., Jacobson, T., and Dell, J. (1991a).
\newblock {A Pure spin connection formulation of gravity}.
\newblock {\em Class. Quant. Grav.}, 8:59--73.

\bibitem[Capovilla et~al., 1991b]{Capovilla:1991qb}
Capovilla, R., Jacobson, T., Dell, J., and Mason, L.~J. (1991b).
\newblock {Selfdual two forms and gravity}.
\newblock {\em Class. Quant. Grav.}, 8:41--57.

\bibitem[Carlip et~al., 2015]{Carlip:2015asa}
Carlip, S., Chiou, D.-W., Ni, W.-T., and Woodard, R. (2015).
\newblock {Quantum Gravity: A Brief History of Ideas and Some Prospects}.
\newblock {\em Int. J. Mod. Phys.}, D24(11):1530028.

\bibitem[Celada et~al., 2016]{Celada:2016iah}
Celada, M., González, D., and Montesinos, M. (2016).
\newblock {Plebanski-like action for general relativity and anti-self-dual
  gravity}.
\newblock {\em Phys. Rev.}, D93(10):104058.

\bibitem[Chalmers and Siegel, 1996]{Chalmers:1996rq}
Chalmers, G. and Siegel, W. (1996).
\newblock {The Selfdual sector of QCD amplitudes}.
\newblock {\em Phys.Rev.}, D54:7628--7633.

\bibitem[Chatrchyan et~al., 2012]{Chatrchyan:2012xdj}
Chatrchyan, S. et~al. (2012).
\newblock {Observation of a new boson at a mass of 125 GeV with the CMS
  experiment at the LHC}.
\newblock {\em Phys. Lett.}, B716:30--61.

\bibitem[Dijkgraaf et~al., 2005]{Vafa:2004te}
Dijkgraaf, R., Gukov, S., Neitzke, A., and Vafa, C. (2005).
\newblock {Topological M-theory as unification of form theories of gravity}.
\newblock {\em Adv. Theor. Math. Phys.}, 9(4):603--665.

\bibitem[Duff et~al., 1986]{Duff:1986hr}
Duff, M.~J., Nilsson, B. E.~W., and Pope, C.~N. (1986).
\newblock {Kaluza-Klein Supergravity}.
\newblock {\em Phys. Rept.}, 130:1--142.

\bibitem[Einstein, 1915]{Einstein:1915ca}
Einstein, A. (1915).
\newblock {The Field Equations of Gravitation}.
\newblock {\em Sitzungsber. Preuss. Akad. Wiss. Berlin (Math. Phys.)},
  1915:844--847.

\bibitem[Einstein, 1918]{Einstein:1918btx}
Einstein, A. (1918).
\newblock {Über Gravitationswellen}.
\newblock {\em Sitzungsber. Preuss. Akad. Wiss. Berlin (Math. Phys.)},
  1918:154--167.

\bibitem[Fefferman and Graham, 1985]{FeffermanG}
Fefferman, C. and Graham, C.~R. (1985).
\newblock {Conformal invariants}.
\newblock In {\em The Mathematical Heritage of Elie Cartan, Asterisque, Numero
  Hors Serie}, volume~57, pages 95--116.

\bibitem[Fernandez and Gray, 1982]{GrayFernandez1982}
Fernandez, M. and Gray, A. (1982).
\newblock Riemannian manifolds with structure group $g_2$.
\newblock {\em Annali di Math. Pura Appl}, 32.

\bibitem[Fine, 2011]{Fine:2011}
Fine, J. (2011).
\newblock {A Gauge Theoretic Approach to anti-self-dual Einstein Equations}.

\bibitem[Fine et~al., 2016]{Fine:2015hef}
Fine, J., Herfray, Y., Krasnov, K., and Scarinci, C. (2016).
\newblock {Asymptotically hyperbolic connections}.
\newblock {\em Class. Quant. Grav.}, 33(18):185011.

\bibitem[Fine et~al., 2014]{Fine:2013qta}
Fine, J., Krasnov, K., and Panov, D. (2014).
\newblock {A gauge theoretic approach to Einstein 4-manifolds}.
\newblock {\em New York J.Math.}, 20:293--323.

\bibitem[Fine and Panov, 2008]{Fine:2008}
Fine, J. and Panov, D. (2008).
\newblock {Symplectic Calabi–Yau manifolds, minimal surfaces and the
  hyperbolic geometry of the conifold}.

\bibitem[Goroff and Sagnotti, 1985]{Goroff:1985sz}
Goroff, M.~H. and Sagnotti, A. (1985).
\newblock {QUANTUM GRAVITY AT TWO LOOPS}.
\newblock {\em Phys. Lett.}, 160B:81--86.

\bibitem[Green et~al., 1988]{Green:1987sp}
Green, M.~B., Schwarz, J.~H., and Witten, E. (1988).
\newblock {\em {SUPERSTRING THEORY. VOL. 1: INTRODUCTION}}.
\newblock Cambridge Monographs on Mathematical Physics.

\bibitem[Hawking and Penrose, 1970]{Hawking:1969sw}
Hawking, S.~W. and Penrose, R. (1970).
\newblock {The Singularities of gravitational collapse and cosmology}.
\newblock {\em Proc. Roy. Soc. Lond.}, A314:529--548.

\bibitem[Herfray, 2017]{Herfray:2016qvg}
Herfray, Y. (2017).
\newblock {Pure Connection Formulation, Twistors and the Chase for a Twistor
  Action for General Relativity}.
\newblock {\em J. Math. Phys.}, 58(11):112505.

\bibitem[Herfray and Krasnov, 2015]{Herfray:2015rja}
Herfray, Y. and Krasnov, K. (2015).
\newblock {New first order Lagrangian for General Relativity}.

\bibitem[Herfray and Krasnov, 2017]{Herfray:2017imd}
Herfray, Y. and Krasnov, K. (2017).
\newblock {Topological field theories of 2- and 3-forms in six dimensions}.
\newblock {\em J. Math. Phys.}, 58(8):082304.

\bibitem[Herfray et~al., 2017]{Herfray:2016std}
Herfray, Y., Krasnov, K., and Scarinci, C. (2017).
\newblock {6D Interpretation of 3D Gravity}.
\newblock {\em Class. Quant. Grav.}, 34(4):045007.

\bibitem[Herfray et~al., 2016a]{Herfray:2016azk}
Herfray, Y., Krasnov, K., Scarinci, C., and Shtanov, Y. (2016a).
\newblock {A 4D gravity theory and G2-holonomy manifolds}.

\bibitem[Herfray et~al., 2016b]{Herfray:2015fpa}
Herfray, Y., Krasnov, K., and Shtanov, Y. (2016b).
\newblock {Anisotropic singularities in chiral modified gravity}.
\newblock {\em Class. Quant. Grav.}, 33:235001.

\bibitem[Hitchin, 2002]{Hitchin:2002ea}
Hitchin, N. (2002).
\newblock {Special holonomy and beyond}.
\newblock In {\em {Strings and geometry. Proceedings, Summer School, Cambridge,
  UK, March 24-April 20, 2002}}, pages 159--175.

\bibitem[Hitchin, 2003]{Hitchin:2004ut}
Hitchin, N. (2003).
\newblock {Generalized Calabi-Yau manifolds}.
\newblock {\em Quart. J. Math.}, 54:281--308.

\bibitem[Hitchin, 2000]{Hitchin:2000sk}
Hitchin, N.~J. (2000).
\newblock {The Geometry of three forms in six-dimensions}.
\newblock {\em J.Diff.Geom.}, 55:547--576.

\bibitem[Hitchin, 2001]{Hitchin:2001rw}
Hitchin, N.~J. (2001).
\newblock {Stable forms and special metrics}.

\bibitem[Hohm et~al., 2013]{Hohm:2013bwa}
Hohm, O., Lüst, D., and Zwiebach, B. (2013).
\newblock {The Spacetime of Double Field Theory: Review, Remarks, and Outlook}.
\newblock {\em Fortsch. Phys.}, 61:926--966.

\bibitem[Huggett and Tod, 1986]{Huggett:1986fs}
Huggett, S. and Tod, K. (1986).
\newblock {An introduction to Twistor Theory}.

\bibitem[Huygbrechts, 2005]{Huygbrechts}
Huygbrechts, D. (2005).
\newblock {\em {Complex Geometry - an introduction}}.

\bibitem[Jacobson and Smolin, 1988]{Jacobson:1988yy}
Jacobson, T. and Smolin, L. (1988).
\newblock {Covariant Action for Ashtekar's Form of Canonical Gravity}.
\newblock {\em Class. Quant. Grav.}, 5:583.

\bibitem[Jiang, 2008]{Jiang08}
Jiang, W. (2008).
\newblock {Aspects of Yang-Mills Theory in Twistor Space}.

\bibitem[Joyce, 1996]{Joyce96}
Joyce, D. (1996).
\newblock {Compact Riemannian manifolds with holonomy $G_2$ I \& II}.
\newblock {\em J. Diff. Geom.}, 43:291--328 and 329--375.

\bibitem[Krasnov, 2008]{Krasnov:2008zz}
Krasnov, K. (2008).
\newblock {On deformations of Ashtekar's constraint algebra}.
\newblock {\em Phys.Rev.Lett.}, 100:081102.

\bibitem[Krasnov, 2009a]{Krasnov:2009iy}
Krasnov, K. (2009a).
\newblock {Gravity as BF theory plus potential}.
\newblock {\em Int. J. Mod. Phys.}, A24:2776--2782.

\bibitem[Krasnov, 2009b]{Krasnov:2008fm}
Krasnov, K. (2009b).
\newblock {Plebanski gravity without the simplicity constraints}.
\newblock {\em Class.Quant.Grav.}, 26:055002.

\bibitem[Krasnov, 2010]{Krasnov:2009ik}
Krasnov, K. (2010).
\newblock {Effective metric Lagrangians from an underlying theory with two
  propagating degrees of freedom}.
\newblock {\em Phys. Rev.}, D81:084026.

\bibitem[Krasnov, 2011a]{Krasnov:2011up}
Krasnov, K. (2011a).
\newblock {Gravity as a diffeomorphism invariant gauge theory}.
\newblock {\em Phys. Rev.}, D84:024034.

\bibitem[Krasnov, 2011b]{Krasnov:2009pu}
Krasnov, K. (2011b).
\newblock {Plebanski Formulation of General Relativity: A Practical
  Introduction}.
\newblock {\em Gen.Rel.Grav.}, 43:1--15.

\bibitem[Krasnov, 2011c]{Krasnov:2011pp}
Krasnov, K. (2011c).
\newblock {Pure Connection Action Principle for General Relativity}.
\newblock {\em Phys.Rev.Lett.}, 106:251103.

\bibitem[Krasnov, 2015]{Krasnov:2014eza}
Krasnov, K. (2015).
\newblock {GR uniqueness and deformations}.
\newblock {\em JHEP}, 10:037.

\bibitem[Krasnov, 2016]{Krasnov:2016wvc}
Krasnov, K. (2016).
\newblock {General Relativity from Three-Forms in Seven Dimensions}.

\bibitem[Krasnov, 2017]{Krasnov:2017uam}
Krasnov, K. (2017).
\newblock {Dynamics of 3-Forms in Seven Dimensions}.

\bibitem[Krasnov and Shtanov, 2008]{Krasnov:2007ky}
Krasnov, K. and Shtanov, Y. (2008).
\newblock {Non-Metric Gravity. II. Spherically Symmetric Solution, Missing Mass
  and Redshifts of Quasars}.
\newblock {\em Class.Quant.Grav.}, 25:025002.

\bibitem[Lebrun, 2004]{LeBrun04}
Lebrun, C. (2004).
\newblock {Geometry of Twistor Spaces}.
\newblock {\em Simons Workshop Lecture, 7/30/04}.

\bibitem[Livine et~al., 2012]{Livine:2011vk}
Livine, E.~R., Speziale, S., and Tambornino, J. (2012).
\newblock {Twistor Networks and Covariant Twisted Geometries}.
\newblock {\em Phys. Rev.}, D85:064002.

\bibitem[Mason, 2005]{Mason05}
Mason, L. (2005).
\newblock {Twistor actions for non-self-dual fields: A Derivation of
  twistor-string theory}.
\newblock {\em JHEP}, 0510:009.

\bibitem[Mason and Skinner, 2010]{Mason:2008jy}
Mason, L. and Skinner, D. (2010).
\newblock {Gravity, Twistors and the MHV Formalism}.
\newblock {\em Commun.Math.Phys.}, 294:827--862.

\bibitem[Mason and Wolf, 2009]{Mason&Wolf09}
Mason, L. and Wolf, M. (2009).
\newblock {Twistor Actions for Self-Dual Supergravities}.
\newblock {\em Commun.Math.Phys.}, 288:97--123.

\bibitem[Mason and Woodhouse, 1991]{Mason:1991rf}
Mason, L. and Woodhouse, N. (1991).
\newblock {\em {Integrability, Self-Duality, and Twistor Theory}}.

\bibitem[McDuff, 2004]{McDuff&Salamon2004}
McDuff, D.~Salamon, D. (2004).
\newblock {\em {J-holomorphic curves and sympleptic topology}}, volume~52 of
  {\em "Colloquium publications"}.
\newblock AMS.

\bibitem[Nicolai, 2014]{Nicolai:2013sz}
Nicolai, H. (2014).
\newblock {Quantum Gravity: the view from particle physics}.
\newblock {\em Fundam. Theor. Phys.}, 177:369--387.

\bibitem[Niedermaier, 2007]{Niedermaier:2006ns}
Niedermaier, M. (2007).
\newblock {The Asymptotic safety scenario in quantum gravity: An Introduction}.
\newblock {\em Class. Quant. Grav.}, 24:R171--230.

\bibitem[Peldan, 1992]{Peldan:1991mh}
Peldan, P. (1992).
\newblock {Connection formulation of (2+1)-dimensional Einstein gravity and
  topologically massive gravity}.
\newblock {\em Class. Quant. Grav.}, 9:2079--2092.

\bibitem[Peldan, 1994]{Peldan:1993hi}
Peldan, P. (1994).
\newblock {Actions for gravity, with generalizations: A Review}.
\newblock {\em Class. Quant. Grav.}, 11:1087--1132.

\bibitem[Penrose, 1965]{Penrose:1964wq}
Penrose, R. (1965).
\newblock {Gravitational collapse and space-time singularities}.
\newblock {\em Phys. Rev. Lett.}, 14:57--59.

\bibitem[Penrose, 1969]{Penrose:1969ae}
Penrose, R. (1969).
\newblock {Solutions of the zero-rest-mass equations}.
\newblock {\em J. Math. Phys.}, 10:38--39.

\bibitem[Penrose, 1976]{Penrose:1976js}
Penrose, R. (1976).
\newblock {Nonlinear Gravitons and Curved Twistor Theory}.
\newblock {\em Gen.Rel.Grav.}, 7:31--52.

\bibitem[Penrose, 1999]{Penrose:1999cw}
Penrose, R. (1999).
\newblock {The Central programme of twistor theory}.
\newblock {\em Chaos Solitons Fractals}, 10:581--611.

\bibitem[Penrose, 2014]{Penrose:2014nha}
Penrose, R. (2014).
\newblock {On the Gravitization of Quantum Mechanics 1: Quantum State
  Reduction}.
\newblock {\em Found. Phys.}, 44:557--575.

\bibitem[Penrose and Rindler, 1985]{Penrose_vol1}
Penrose, R. and Rindler, W. (1985).
\newblock {\em {Spinors and space-time. Vol 1. Two spinor calculus and
  relativistic fields}}.

\bibitem[Penrose and Rindler, 1986]{Penrose_vol2}
Penrose, R. and Rindler, W. (1986).
\newblock {\em {Spinors and space-time. Vol. 2: Spinor and Twistor methods in
  space-time geometry}}.

\bibitem[Pestun and Witten, 2005]{Pestun:2005rp}
Pestun, V. and Witten, E. (2005).
\newblock {The Hitchin functionals and the topological B-model at one loop}.
\newblock {\em Lett. Math. Phys.}, 74:21--51.

\bibitem[Plebanski, 1977]{Plebanski:1977zz}
Plebanski, J.~F. (1977).
\newblock {On the separation of Einsteinian substructures}.
\newblock {\em J. Math. Phys.}, 18:2511--2520.

\bibitem[Polchinski, 2007]{Polchinski:1998rq}
Polchinski, J. (2007).
\newblock {\em {String theory. Vol. 1}}.
\newblock Cambridge University Press.

\bibitem[Reshetikhin and Turaev, 1991]{Reshetikhin:1991tc}
Reshetikhin, N. and Turaev, V.~G. (1991).
\newblock {Invariants of three manifolds via link polynomials and quantum
  groups}.
\newblock {\em Invent. Math.}, 103:547--597.

\bibitem[Rovelli, 2000]{Rovelli:2000aw}
Rovelli, C. (2000).
\newblock {Notes for a brief history of quantum gravity}.
\newblock In {\em {Recent developments in theoretical and experimental general
  relativity, gravitation and relativistic field theories. Proceedings, 9th
  Marcel Grossmann Meeting, MG'9, Rome, Italy, July 2-8, 2000. Pts. A-C}},
  pages 742--768.

\bibitem[Rovelli, 2004]{Rovelli:2004tv}
Rovelli, C. (2004).
\newblock {\em {Quantum gravity}}.
\newblock Cambridge Monographs on Mathematical Physics. Univ. Pr., Cambridge,
  UK.

\bibitem[Salamon, 1989]{Salamon:1989}
Salamon, S. (1989).
\newblock {\em {Riemannian geometry and holonomy groups}}, volume 201 of {\em
  Pitman Research Notes in Mathematics Series}.

\bibitem[Salamon, 2002]{Salamon:2002}
Salamon, S. (2002).
\newblock {A tour of exceptional geometry}.
\newblock {\em Milan J. Math.}

\bibitem[Schwarz, 1978]{Schwarz:1978cn}
Schwarz, A.~S. (1978).
\newblock {The Partition Function of Degenerate Quadratic Functional and
  Ray-Singer Invariants}.
\newblock {\em Lett. Math. Phys.}, 2:247--252.

\bibitem[Schwarz, 1979]{Schwarz:1979ae}
Schwarz, A.~S. (1979).
\newblock {The Partition Function of a Degenerate Functional}.
\newblock {\em Commun. Math. Phys.}, 67:1--16.

\bibitem[Sharpe, 1997]{Sharp}
Sharpe, R. (1997).
\newblock {\em {Differential Geometry: Cartan's Generalisation of Klein's
  Erlangen Program}}.

\bibitem[Speziale and Wieland, 2012]{Speziale:2012nu}
Speziale, S. and Wieland, W.~M. (2012).
\newblock {The twistorial structure of loop-gravity transition amplitudes}.
\newblock {\em Phys. Rev.}, D86:124023.

\bibitem[Starobinsky, 1980]{Starobinsky:1980te}
Starobinsky, A.~A. (1980).
\newblock {A New Type of Isotropic Cosmological Models Without Singularity}.
\newblock {\em Phys. Lett.}, 91B:99--102.

\bibitem[Thiemann, 2008]{Thiemann:2007zz}
Thiemann, T. (2008).
\newblock {\em {Modern canonical quantum general relativity}}.
\newblock Cambridge University Press.

\bibitem[Urbantke, 1984]{Urbantke:1984eb}
Urbantke, H. (1984).
\newblock {On integrability properties of su(2) Yang-Mills Fields}.

\bibitem[van~de Ven, 1992]{vandeVen:1991gw}
van~de Ven, A. E.~M. (1992).
\newblock {Two loop quantum gravity}.
\newblock {\em Nucl. Phys.}, B378:309--366.

\bibitem[Wang, 1989]{Wang:1989}
Wang, M.~Y. (1989).
\newblock {Parallel spinors and parallel forms. }.
\newblock {\em Ann. Global Anal. Geom.}, 7-1:59--68.

\bibitem[Ward, 1977]{Ward:1977ta}
Ward, R. (1977).
\newblock {On Selfdual gauge fields}.
\newblock {\em Phys.Lett.}, A61:81--82.

\bibitem[Ward, 1980]{Ward:1980am}
Ward, R. (1980).
\newblock {Self-dual space-times with cosmological constant}.
\newblock {\em Commun.Math.Phys.}, 78:1--17.

\bibitem[Ward and Wells, 1990]{Ward:1990vs}
Ward, R. and Wells, R. (1990).
\newblock {Twistor Geometry and Field Theory}.

\bibitem[Weinberg, 1980]{Weinberg:1980gg}
Weinberg, S. (1980).
\newblock {ULTRAVIOLET DIVERGENCES IN QUANTUM THEORIES OF GRAVITATION}.
\newblock In {\em General Relativity: An Einstein Centenary Survey}, pages
  790--831.

\bibitem[Wells, 2008]{Wells}
Wells, R.~O. (2008).
\newblock {\em {Differential Analysis on Complex Manifolds}}.

\bibitem[Witt, 2009]{Witt:2009zz}
Witt, F. (2009).
\newblock {Gauge theory in dimension 7}.
\newblock {\em AIP Conf. Proc.}, 1093:180--195.

\bibitem[Wolf, 2007]{Wolf:2007tx}
Wolf, M. (2007).
\newblock {Self-Dual Supergravity and Twistor Theory}.
\newblock {\em Class.Quant.Grav.}, 24:6287--6328.

\bibitem[Woodard, 2009]{Woodard:2009ns}
Woodard, R.~P. (2009).
\newblock {How Far Are We from the Quantum Theory of Gravity?}
\newblock {\em Rept. Prog. Phys.}, 72:126002.

\bibitem[Woodhouse, 1985]{Woodhouse85}
Woodhouse, N. (1985).
\newblock {Real methods in twistor theory}.
\newblock {\em Class.Quant.Grav.}, 2:257--291.

\end{thebibliography}
\end{Bibliography}
\end{document}